\documentclass[a4paper, twoside]{report}

\usepackage[british]{babel}
\usepackage[utf8]{inputenc}
\usepackage[T1]{fontenc}

\usepackage[a4paper, top=3cm, bottom=2cm, left=3cm, right=3cm, marginparwidth=1.75cm]{geometry}

\IfFileExists{./main.aux}{
\usepackage[backend=biber, style=imperialvancouver, sorting=none, backref=true, defernumbers=true, natbib]{biblatex}
}{
\usepackage[backend=biber, style=imperialvancouver, sorting=none, backref=true, defernumbers=false, natbib]{biblatex}
}

\usepackage{afterpage}
\usepackage{amsmath}
\usepackage{amsthm}
\usepackage{amssymb}
\usepackage{enumitem}
\usepackage{graphicx}
\usepackage{lipsum}
\usepackage{booktabs}
\usepackage[frozencache=true,cachedir=_minted-main]{minted}
\usepackage{xcolor}
\usepackage{wrapfig}
\usepackage[hypcap=false]{caption}

\definecolor{green}{rgb}{0.0, 0.5, 0.0}
\definecolor{red}{rgb}{0.69, 0.0, 0.25}
\definecolor{yellow}{rgb}{0.75, 0.55, 0.13}

\usepackage{listings}
\usepackage[labelfont=bf,justification=centering]{caption}

\usepackage[ruled,vlined]{algorithm2e}
\makeatletter
\renewcommand{\SetKwInOut}[2]{%
  \sbox\algocf@inoutbox{\KwSty{#2}\algocf@typo:}%
  \expandafter\ifx\csname InOutSizeDefined\endcsname\relax
    \newcommand\InOutSizeDefined{}\setlength{\inoutsize}{\wd\algocf@inoutbox}%
    \sbox\algocf@inoutbox{\parbox[t]{\inoutsize}{\KwSty{#2}\algocf@typo:\hfill}~}\setlength{\inoutindent}{\wd\algocf@inoutbox}%
  \else
    \ifdim\wd\algocf@inoutbox>\inoutsize%
    \setlength{\inoutsize}{\wd\algocf@inoutbox}%
    \sbox\algocf@inoutbox{\parbox[t]{\inoutsize}{\KwSty{#2}\algocf@typo:\hfill}~}\setlength{\inoutindent}{\wd\algocf@inoutbox}%
    \fi%
  \fi
  \algocf@newcommand{#1}[1]{%
    \ifthenelse{\boolean{algocf@inoutnumbered}}{\relax}{\everypar={\relax}}%
    {\let\\\algocf@newinout\hangindent=\inoutindent\hangafter=1\parbox[t]{\inoutsize}{\KwSty{#2}\algocf@typo:\hfill}~##1\par}%
    \algocf@linesnumbered
  }}%
\makeatother

\usepackage{bm}

\usepackage{fancyhdr}
\usepackage{tipa}
\usepackage{tcolorbox}
\definecolor{mdgrey}{rgb}{0.8, 0.8, 0.8}
\usepackage[framemethod=tikz]{mdframed}
\usepackage{lipsum}
\newtheoremstyle{defi}
  {\topsep}%
  {\topsep}%
  {\normalfont}%
  {}%
  {\bfseries}%
  {:}%
  {.5em}%
  {\thmname{#1}\thmnote{~(#3)}}%
\theoremstyle{defi}
\newmdtheoremenv{definitioni}{Definition}
\newmdtheoremenv[
hidealllines=true,
leftline=true,
innertopmargin=0pt,
innerbottommargin=0pt,
linewidth=4pt,
linecolor=gray!40,
innerrightmargin=0pt,
]{definitionii}{Definition}
\newmdtheoremenv[
roundcorner=5pt,
innertopmargin=0pt,
innerbottommargin=5pt,
linewidth=4pt,
linecolor=gray!40,
]{definitioniii}{Definition}

\usepackage[colorlinks=true,allcolors=blue]{hyperref}
\usepackage[nameinlink]{cleveref}

\renewcommand{\digamma}{\psi_0}

\usepackage{csquotes}

\title{Communication-free and Parallel Simulation of Neutral Biodiversity Models}

\author{Momo Langenstein [they/them]}

\bibliography{bibs/bibliography.bib}

\DeclareBibliographyCategory{cited}
\AtEveryCitekey{\addtocategory{cited}{\thefield{entrykey}}}

\makeatletter
\DeclareCiteCommand{\citepalias}
  {\usebibmacro{prenote}}
  {\usebibmacro{citeindex}%
  \printtext[bibhyperref]{\@citealias{\thefield{entrykey}}}}
  {\multicitedelim}
  {\usebibmacro{postnote}}

\makeatletter
\let\blx@rerun@biber\relax
\makeatother

\begin{document}
\begin{titlepage}

\newcommand{\HRule}{\rule{\linewidth}{0.5mm}} 


\includegraphics[width=8cm]{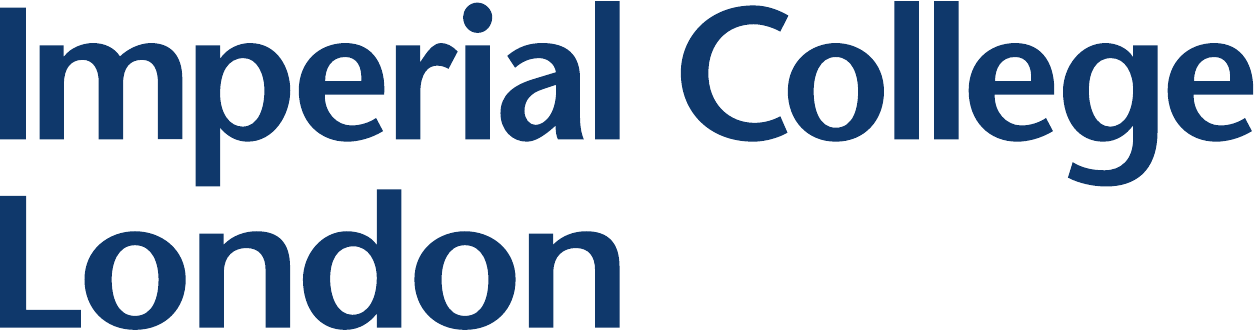}\\[1cm] 
 

\center 


\textsc{\LARGE MEng Individual Project}\\[1.5cm] 
\textsc{\Large Imperial College London}\\[0.5cm] 
\textsc{\large Department of Computing}\\[0.5cm] 

\makeatletter
\HRule \\[0.6cm]
{ \huge \bfseries \@title}\\[0.6cm] 
\HRule \\[1.5cm]
 

\begin{minipage}{0.4\textwidth}
\begin{flushleft} \large
\emph{Author:}\\
\@author 
\end{flushleft}
\end{minipage}
~
\begin{minipage}{0.4\textwidth}
\begin{flushright} \large
\emph{Supervisor:} \\
Dr. Anthony Field  \\[1.2em] 
\emph{Second Marker:} \\
Prof. Paul Kelly \\[2.4em] 
\emph{External Advisors:} \\
Dr. James Rosindell \\
Prof. Ryan A. Chisholm
\end{flushright}
\end{minipage}\\[2cm]
\makeatother



{\large 14th June 2021}\\[2cm] 

\vfill 

\end{titlepage}

\begin{abstract} \setcounter{page}{2}

\noindent Biodiversity simulations are used in ecology and conservation to predict the effect of habitat destruction on biodiversity. We present a novel \textbf{communication-free} algorithm for individual-based probabilistic neutral biodiversity simulations. The algorithm transforms a neutral Moran ecosystem model into an embarrassingly parallel problem by trading off inter-process communication at the cost of some redundant computation. \\

\noindent Specifically, by careful design of the random number generator that drives the simulation, we arrange for evolutionary parent-child interactions to be modelled \textbf{without requiring knowledge of the interaction, its participants, or which processor} is performing the computation. Critically, this means that \textbf{every individual can be simulated entirely independently}. The simulation is thus \textbf{fully reproducible} irrespective of the number of processors it is distributed over. With our novel algorithm, a simulation can be (1) split up into independent batch jobs and (2) simulated across any number of heterogeneous machines -- all without affecting the simulation result. \\

\noindent We use the Rust programming language to build the extensible and statically checked simulation package \texttt{necsim-rust}. We evaluate our parallelisation approach by comparing three traditional simulation algorithms against a CPU and GPU implementation of our Independent algorithm. These experiments show that as long as some local state is maintained to cull redundant individuals, our Independent algorithm is as efficient as existing sequential solutions. The GPU implementation further outperforms all algorithms on the CPU by a factor ranging from $\sim 2$ to $\sim 80$, depending on the model parameterisation and the analysis that is performed. Amongst the parallel algorithms we have investigated, our Independent algorithm provides the only non-approximate parallelisation strategy that can scale to large simulation domains. For example, while Thompson's 2019 \texttt{necsim} simulation takes more than 48 hours to simulate $10^{8}$ individuals, we have been able to simulate $7.1\cdot 10^{10}$ individuals on an HPC batch system within a few hours.

\end{abstract}

\renewcommand{\abstractname}{Acknowledgements}
\begin{abstract} \setcounter{page}{3}

\noindent This project started on July 27th 2020, when I reached out to James, asking if we could collaborate on my individual project. I have known James since my amazing 2019 UROP with him at Silwood Campus. Last year, James welcomed me back with open arms and five suggestions for an MEng project, out of the first two of which this project was born. Two days later, James introduced me to Ryan, and we all had our first meeting together on August 3rd. The same day, I also first contacted Tony, who later agreed to supervise this project. On August 6th, I talked with my friend and Maths student Philipp about a very early attempt at solving what would later become a core part of this project: jumping around inside homogeneous Poisson point processes. \\

\noindent Therefore, my first round of thanks goes to my three supervisors. From the beginning, their ideas and experience have inspired and streamlined this project whilst also allowing me to slowly direct it exactly where I wanted it to go. I could not be more grateful for \textbf{Tony}, \textbf{James} and \textbf{Ryan}, who have met with me (almost) every week throughout this project and have brought their diverse science backgrounds to push this project from all sides. Thank you for always encouraging me to think further, plan ahead, not go for the most outlandish approach first, and keep track of my mental health throughout this process.

Besides my supervisors, I also want to express my gratitude to my many mentors at ICL. Thank you to \textbf{Francesco}, \textbf{Philippa}, \textbf{Holger} and \textbf{Paul}, who have all given me invaluable advice and feedback on this project and life in general. Paul's Advanced Computer Architecture course still remains my favourite class to this day. My special thanks go to \textbf{Chris} and \textbf{Elizabeth}, who have helped and formed me in so many more ways than they could imagine. Finally, this project would have crashed and burned without the help of \textbf{CSG}, in particular \textbf{Lloyd}, who have always supported me.

Next, I would like to thank all Biodiversity Lab group members who have supported me both personally and academically throughout this challenging year. In particular, I would like to thank \textbf{Francis} for his critical thinking, \textbf{Pokman} and \textbf{Olivia} for their feedback, and \textbf{Rach} for just being an inspiration. Most importantly, though, I am enormously grateful for \textbf{Sam}, who wrote the original \texttt{necsim} simulation that this project is based on, and \textbf{Lucas}, who has always been a saviour and source of encouragement (and Maths help). It was such an honour to work with you during my UROP and now this master's project.

Last but not least, I want to thank my family and friends. After living through the first lockdown in London, I am so grateful that I could spend this past year back home. At the start of the second year, ICL gifted us rubber ducks to listen to our computing problems. Without any doubt, though, \textbf{my parents} have been the best ducks I could have asked for. After one year of swamping them with my thoughts and fears on countless walks, I can only apologise for talking so much about this project. I want to further thank my second family back in the UK, \textbf{Freddie}, \textbf{George}, \textbf{Jack}, \textbf{Jessica} and \textbf{Steve}, who opened their arms and hearts to me many years ago and who have since been my rock abroad. I would also like to express my gratitude to \textbf{Alexander}, \textbf{Declan}, \textbf{Isabella}, \textbf{Iurii}, \textbf{Philipp} and \textbf{Tiger}, who have been great pillars of emotional support. Finally, my eternal gratitude goes to my friend \textbf{Jamie}, without whom I would not have gotten through this year's dark times. \\

\noindent This project is a labour of much love, sweat and tears. I would not have been able to do it without my amazing support system around me, and I thank all of them from the bottom of my heart.

\end{abstract}

\setcounter{page}{4}
\renewcommand{\contentsname}{Table of Contents}
\tableofcontents


\chapter{Introduction}


Probabilistic individual-based models (IBMs) are an integral part of scientific research, with applications in many areas such as transportation, particle physics, population genetics, and ecology. This project focuses on the modelling of ecosystem biodiversity, where the objective is to predict the species richness of a landscape. Ecologists use these predictions, for example, to evaluate the effects of habitat destruction (see, e.g. \cite{Thompson2019}) or different area-based habitat protection schemes.

The usual modelling approach is to simulate the ecosystem forwards in time, starting from some initial state. However, this approach can be highly inefficient when we are only interested in studying a small part of the system. For example, suppose the objective is to study the evolution of species in a nature reserve, then, depending on the nature of the model. In that case, we may only need to simulate the ancestors of the individuals who inhabit the reserve today. In order to determine this set of ancestors, we first require some mechanism to simulate backwards in time.

An important class of ecosystem models that facilitate time-reversed simulations are the so-called neutral models \cite{Hubbell2001}. They make the simplifying assumptions that there are no species-specific traits and that there is no feedback from individuals to the environment. Consequently, an individual's behaviour is neither influenced by its species identity nor any individuals that are not its ancestors. In 2008, Rosindell, Wong and Etienne used neutral models to trace the ancestry of a set of individuals backwards in time to discover their species identities \cite{Rosindell2008}. More recently, Thompson developed a simulation framework, \texttt{necsim}, to manage and run these reverse-time models \cite{ThompsonPhd}. \\


\noindent The primary goal of this project is the parallelisation of reverse-time neutral simulations. When these simulations grow too large to fit onto a single computer, they need to be split up and run in parallel over multiple machines. The traditional approach to parallelising IBMs is to maintain a globally consistent model state \cite{Kunz2012, Komarov2012, Ridwan2004, Arjunan2020, Bauer2015}. However, this requires processors to communicate and synchronise, thereby limiting the scaling of parallel simulations as communication costs grow.

The key idea of this project is to model each individual independently instead. We can then perform all interactions between individuals without any inter-partition / inter-processor communication, though at the expense of some redundant computation. In this thesis, our key research question is whether the saved communication costs outweigh the additional costs of redundant computation. We explore this question in several hybrid algorithm variants. \\


\noindent For our proposed algorithm to work, the simulated trajectory of every individual must be reproducible regardless of which processor performs the simulation. We achieve this by developing both a novel random number generation scheme and next-event-time sampling method. In combination, they make the random trajectories dependent only on an individual's time and location. Ensuring that any such generator is statistically robust is a key challenge.

Our algorithm builds on Salmon et al.'s counter-based pseudo-random number generators (CBRNGs) \cite{Salmon2011}, and subsequent work by Hill and Jun et al. on making probabilistic simulations reproducible \cite{Hill2015, Jun2020}. It also incorporates Phillips, Anderson and Glotzer's approach to limit communication between individuals who are known to interact \cite{Phillips2011}. However, we go one step further and ensure that our algorithm requires no knowledge about an interaction at all.


\noindent The main contributions of this Master's Thesis are:
\begin{enumerate}
    \item \texttt{necsim-rust}, a type-safe modular neutral simulation software framework that is written in Rust. Using Rust's trait system \cite{Schaerli2003}, we have designed a simulation component system that is statically checked for component compatibility (\ref{architecture-component-system}). We have also extended Rust's borrowing rules to improve the safety of sharing data between the CPU and GPU (\ref{implementation-cuda}). Both the component system and safer GPU interaction have applications beyond this project.
    \item We provide various sequential implementations of existing algorithms for neutral simulations. Most notably, we design a non-approximate event-skipping Gillespie algorithm that excels in sparsely sampled models (\ref{independence-coalescence-gillespie}). We also implement different existing parallelisation strategies for these algorithms using the \textbf{M}essage \textbf{P}assing \textbf{I}nterface (\ref{implementation-monolithic-parallelisation}).
    \item An implementation of the new Independent algorithm and its associated random number generator. In particular, we implement the algorithm on the CPU (\ref{implementation-independent}), the GPU (\ref{implementation-cuda}), and parallelise the CPU version using MPI (\ref{implementation-independent-parallelisation}). Furthermore, we demonstrate that the neutral simulation can now be partitioned into independent batch system jobs.
    \item A detailed evaluation of the correctness (\ref{analysis-functional-correctness}) and performance of all algorithms (\ref{analysis-algorithm-scalability}) and parallelisation strategies (\ref{analysis-parallelism}). The analysis shows that the new event-skipping Gillespie algorithm outperforms existing solutions on the CPU on sparse models with a small area. More importantly, the new Independent algorithm is faster than existing solutions and even outperforms the event-skipping Gillespie variant on dense models. When partitioned over multiple CPUs communicating via MPI, the Independent algorithm is the only viable non-approximate parallelisation strategy. In simulations with many individuals and species creation rates above $10^{-7}$, the GPU implementation of the Independent algorithm performs even better. Depending on the model parameterisation and analysis performed, it further outperforms all CPU variants by a factor ranging from $\sim 2$ to $\sim 80$ (\ref{analysis-event-reporting}, \ref{analysis-domain-speciation-scalability}).
\end{enumerate}

\noindent We have also significantly increased the scope of ecosystem models that can be simulated. The starting point for this project was the single-threaded sequential \texttt{necsim} library. While \texttt{necsim} takes 16 hours to simulate $10^7$ individuals with a a per-generation speciation probability of $\nu = 10^{-6}$, our \textbf{Independent} algorithm only takes a few hours to simulate $7.1 \cdot 10^{10}$ individuals on an HPC batch system (\ref{analysis-limit}). In the worst case, simulation times scale inversely linear with $\nu$. Even still, we have been able to simulate $10^{8}$ individuals with $\nu = 10^{-12}$ within 27 hours using our event-skipping Gillespie algorithm (\ref{analysis-limit}). \\


\noindent This thesis is divided into three main parts, including an ethical discussion in \Cref{ethical-discussion}:

First, there is an extensive background section. \Cref{background-scientific} covers the \underline{Neutral Theory of Bio-} \underline{diversity} (\ref{background-neutral-biodiversity}) and the necessary foundations for the novel algorithm including Poisson processes (\ref{background-poisson-point-processes}), \underline{random number generation} (\ref{background-rng}) and the Gillespie Algorithm (\ref{background-gillespie}). \Cref{background-technical} explores the technical frameworks this project uses to implement and parallelise the simulation, in particular MPI (\ref{background-mpi}) and CUDA (\ref{background-cuda}). Readers who are already familiar with the Rust programming language and its type system can skip \cref{background-rust}. Finally, \cref{background-related} introduces prior work related to this project, including the \texttt{necsim} library (\ref{background-necsim}) and CBRNGs (\ref{background-cbrngs}) which this thesis builds on.

The core contribution of this thesis, the \underline{novel Independent algorithm}, is presented in \cref{independence}. Next, \cref{architecture} shows how we have used the Rust programming language to design an extensible and safe simulation architecture. Readers who are less interested in Rust should still read \cref{architecture-component-system}, which introduces the core \underline{components of the biodiversity simulation}. Finally, \cref{implementation} shows in detail how the Independent algorithm is implemented on the CPU (\ref{implementation-independent}) and GPU (\ref{implementation-cuda}). It also describes how we have \underline{parallelised} all algorithms (\ref{implementation-parallelisation}) using MPI (\ref{implementation-mpi}).

In \cref{analysis-evaluation}, we evaluate all algorithms and parallelisation strategies that we have implemented. In particular, \cref{analysis-parallelism} explores the \underline{scaling} of all algorithms for increasing parallelism. Finally, \cref{conclusion} concludes this thesis and highlights some opportunities for \underline{future work}. \\


\noindent Overall, we hope that our simulation's improved performance and ability to simulate much larger models will support ecologists to predict global biodiversity loss more efficiently and help guide area-based conservation measures to protect the biodiversity on planet Earth.

\chapter{Review of Scientific Background} \label{background-scientific}

\section{Biodiversity Loss and Conservation}

Planet Earth is currently in a biodiversity crisis. The rate at which species are disappearing has risen dramatically in the 200 years since the Industrial Revolution. Over the past century, in particular, this rate of disappearance has been more than 100 times higher than the historical average \cite{Ceballose2015}. Habitat loss and fragmentation are the leading causes for this drastic decline in biodiversity \cite{Woodley2019}. To fight this decline, area-based conservation initiatives have increasingly been used as a ``key policy and practical solution to biodiversity loss'' \cite[p.~32]{Woodley2019}. In 2010, 20 targets to conserve biodiversity were passed by the 196 Parties to the Convention on Biological Diversity. Target 11 of these Aichi Biodiversity Targets was explicitly focused on protecting ``at least 17 per cent of terrestrial and inland water, and 10 per cent of coastal and marine areas'' \cite{aichitargets}. However, several countries, including the United Kingdom, have failed to reach this target by the 2020 deadline \cite{Mrema2020, RSPB2020}. Instead of being created in the ``places important for halting biodiversity loss'' \cite[p.~37]{Woodley2019}, protected areas have often been cheaply established in areas with low economic appeal \cite{Smith2013}. While simulations of biodiversity models will not alone solve the dire threat of species extinction and climate crisis \cite{OttoPortner2021}, they are valuable tools to predict how changes to the landscape will affect biodiversity. These predictions can then inform conservationists' and policymakers' decision-making.

\section{The Neutral Theory of Biodiversity} \label{background-neutral-biodiversity}

The Neutral Theory of Biodiversity was popularised in 2001 by Stephen P. Hubbell \cite{Hubbell2001}. It makes several simplifying assumptions to present a unified model of biodiversity:
\begin{enumerate}
    \item \textbf{Neutral assumption:} There are no species-specific traits, which means that individuals cannot exhibit any species-dependent behaviour that would affect their chance of survival\footnote{The survival of species can still depend on environmental factors like the quality of their habitat, though.} \cite[p.~14]{Hubbell2001}. In nature, species specific-traits can clearly influence survival. However, the Neutral Theory has already successfully been used to predict static biodiversity patterns such as species-area curves (\ref{background-neutral-scenarios}) and endemic species on islands \cite{Rosindell2012}. Thompson, Chisholm and Rosindell have further employed neutral models to make analytical predictions about the long term extinction debt caused by habitat loss \cite{Thompson2019}. Overall, neutral models are advantageous when decisions about conservation policy have to be made without knowing about species-specific traits \cite{Rosindell2012}.
    \item \textbf{Zero-Sum:} When an individual dies, it is immediately replaced by another individual's offspring. This direct coupling of birth-death events means that the number of individuals remains constant as long as the simulation parameters do not change \cite[p.~14]{Hubbell2001}.
    \item \textbf{Asexual Reproduction:} The creation of new offspring requires only one parent individual.
\end{enumerate}

\noindent This neutral model can best be described as a simple algorithm:

\begin{minted}[linenos]{python}
def simulate_biodiversity(landscape, nu):
    # Initial population of individuals and their given species identities
    population = landscape.generate_initial_population()

    while not population.has_reached_steady_state():
        individual = population.pop_random()

        if sample_random_event(nu):
            # Speciate with probability nu
            population.push(new Individual(new Species())
        else:
            # Individual dies with probability 1-nu and is replaced by a new
            #  offspring of another individual (the model is zero-sum)
            population.push(new Individual(
                population.peek_random().species
            ))

    species_richness = len(set(i.species for i in population))

    return species_richness
\end{minted}
The algorithm starts from some initial population. At every step, an individual dies and is replaced by another individual's offspring. With probability $\nu$, this offspring mutates and creates a new unique species. After the simulation has reached equilibrium, the landscape's biodiversity can be measured as its species richness, i.e. the number of unique species.

\label{background-moran-fisher} In the above algorithm, individuals can die at any point in time, which is indicative of a Moran Model\footnote{Moran Models also assume that generation lengths are distributed according to an exponential distribution \cite[p.~62]{Moran1958}.} in population genetics \cite{Moran1958}. However, one could also assume that different generations do not overlap. In such a Wright-Fisher Model, the offspring replace the entire population at fixed intervals \cite{Wright1942}. For instance, while humans reproduce throughout the year, annual plants such as peas have exactly one generation within each growing season. Building on Moran's work \cite[p.~63]{Moran1958}, Yahav, Danino and Shnerb have shown that the Wright-Fisher model results in approximately twice the species richness as the Moran model \cite[p.~1]{Yahav2017}.

\section{A Neutral Coalescence-based Simulation} \label{background-neutral-simulation}

As we have seen above, the fundamental algorithm of the neutral model of biodiversity can be trivially implemented in a probabilistic IBM. However, this algorithm has two inefficiencies:

Firstly, it often has to simulate extraneous individuals. Say we want to determine the biodiversity of a small patch of forest, i.e. we are interested in the species identities of the individuals that live in this sample patch at equilibrium. However, we do not know where the ancestors of those individuals lived at the start of the simulation. If we only simulate individuals living in this small patch, we cannot account for immigration from outside. Instead, we have to simulate a much larger piece of the landscape to ensure that we capture all ancestors. Overall, we have to waste computation on individuals whose lineages either die out or end up outside the forest patch.

The second problem is that the above algorithm assumes a steady-state can be reached\footnote{As ``ecological interactions and random dispersal'' \cite{Denk2020} can cause chaotic behaviour, however, this termination condition might not be reached in all models.} and only terminates once it has done so. However, there is no metric to detect equilibrium immediately. Instead, a conservative heuristic has to monitor the simulation state and guess when it has become stable. Therefore, the simulation still has to run throughout this conservative monitoring period even after a steady-state has been reached, wasting computation.

\begin{figure}[ht]
    \centering
    \includegraphics[width=0.8\textwidth]{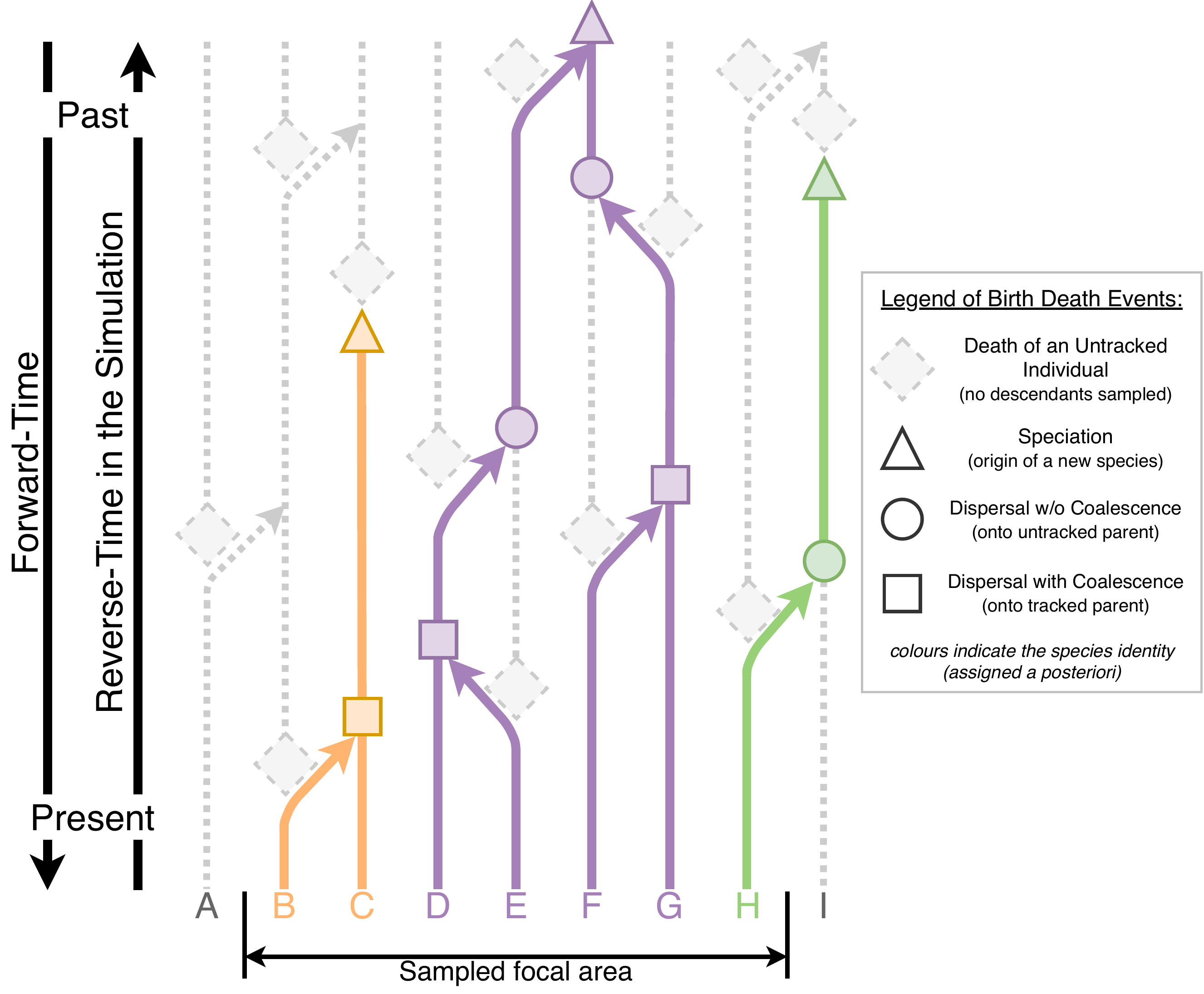}
    \caption{An example of a neutral coalescence simulation with nearest-neighbour dispersal (adapted from \cite[p.~31]{ThompsonPhd}). It shows the event traces of the individuals in a one-dimensional landscape. While a forwards simulation has to track all lineages and produce all events, the reverse-time coalescence simulation only needs to model the lineages of the sampled individuals. Specifically, the coalescence simulation does not compute the events and traces with dotted lines.}
    \label{fig:coalescence-lineages}
\end{figure}

\noindent Rosindell, Wong and Etienne presented a solution for both of these problems in 2008 \cite{Rosindell2008}. Instead of running the simulation forwards in time, we now perform it in reverse from the present back to the past. How does this approach work and produce statistically equivalent results?

\noindent The Neutral Theory of Biodiversity assumes that there are no species-specific traits, i.e. individuals cannot display any species-dependent behaviour. Therefore, no knowledge about the species identity of an individual is required to simulate its behaviour. In fact, the behaviour of individuals only consists of dispersing or speciating at every birth-death event. Both of these actions must be exchangeable, i.e. reversing them and the order of all events must not change the joint probability distribution over the simulation results.

In the forwards simulation, an individual speciates with probability $\nu$ to create a new species identity which then passes on to its offspring. The speciation process can also be looked at in reverse. When an individual speciates, it must create a unique species that all of its offspring (which have not mutated since) are a part of. Therefore, speciation represents the base case of the reverse-time algorithm. Since the ancestor individual is assigned a new and unique species identity, this individual does not need to be simulated any further.

With probability $1 - \nu$, another individual gives birth to their offspring that disperses (jumps away) and replaces our individual. To reverse this action, the dispersal kernel must be inverted, such that it now describes the probability of nearby individuals being the parent of our child individual. After reverse dispersal, there are two different cases in the reverse-time simulation:

\label{background-coalescence-pruning} \begin{enumerate}
    \item An individual can disperse to an already occupied location, causing coalescence. In \Cref{fig:coalescence-lineages}, this case is represented by rhombi. By the semantics of our reverse-time simulation, this coalescence means that the child just collided with its parent and must share the same species. Therefore, we only need to continue simulating the parent individual.
    \item In the second case, the location is unoccupied, meaning that the simulation is not currently tracking the parent individual. In \Cref{fig:coalescence-lineages}, this case is represented by circles. Why does this happen? The simulation only tracks individuals who it already knows to have descendants in the sample area. For instance, this child might be the youngest of its siblings who are also related to the sampled individuals. However, now that a direct lineage to the sample area has been found, the simulation needs to start tracking the parent individual. This is achieved by simulating the child individual in place of its parent, i.e. renaming the child to be its parent. Therefore, the child-now-parent individual continues to be simulated in this second case.
\end{enumerate}

\noindent This reverse-time process is repeated until all individuals have either speciated or coalesced. The speciation events that were observed represent the past ancestors who originated the entire present species diversity. The computed coalescence tree can be traversed to determine the species identity of all original individuals. The species richness of the simulated landscape, and by proxy its biodiversity, is equal to the number of observed speciation events. \\

\noindent This coalescence based approach has several advantages:
\begin{enumerate}
    \item It continuously prunes the simulation space by only simulating the individuals who are related to the present time population and have, therefore, contributed to its biodiversity.
    \item The coalescence of lineages in common ancestors ensures that no extraneous simulation work is performed while exploring the search space.
    \item The reverse-time coalescence algorithm has a clearly defined stopping criterion. Unlike in the forwards simulation, the simulation does not have to simulate until a steady-state has been found. Instead, it goes backwards in time until the species identities of all individuals have been determined, i.e. all individuals have either coalesced or speciated.
\end{enumerate}

\section{Three Neutral scenarios} \label{background-neutral-scenarios}

The Neutral Theory of Biodiversity can be used to describe many different model scenarios \cite{Rosindell2021}. This section briefly introduces three such scenarios. These three scenarios and their analytical solutions, which are summarised in \Cref{appendix:neutral-scenarios}, are crucial to verify the implementation correctness of the \texttt{necsim-rust} simulation library in \cref{analysis-functional-correctness}.

\subsection{Non-Spatial} \label{background-non-spatial-point-speciation}

The non-spatial scenario describes a closed community of $J$ individuals, such as an island. Dispersal inside this community is homogeneous, i.e. dispersal from any location is equally likely to all possible locations. In this and most other neutral models, speciation is modelled as an instantaneous point process, meaning every speciation event creates a new and unique species. Consequently, the present state might contain some very young species, which only include a small number of individuals. \Cref{appendix:protracted-speciation} explains how this limitation has been addressed by Rosindell et al. \cite{Rosindell2010}.

\subsection{Spatially Implicit}

Hubbell's spatially implicit scenario expands the model by adding migration \cite{Hubbell2001}. We now have a small local community that is closed by a larger external metacommunity. The primary source of biodiversity in the local community is migration from the metacommunity. As this migration is assumed to dominate speciation, this model ignores speciation in the local community \cite{Rosindell2021}.

Instead, the migration probability function $m(A)$ is introduced, where $A$ is the `area' of the local community. It describes the per capita probability of migration from the metacommunity to the local community. In the reverse-time simulation, $m(A)$ can be interpreted as a per individual, per generation probability that the individual's parent lived in the metacommunity instead of the local community. There are two biologically easily justifiable ways to define $m(A)$ \cite{Chisholm2016}. It can either be set to some constant $m(A) = m_k$ such that $M \propto A$, i.e. the total number of immigrants per generation scales linearly with $A$. The other option is to set $m(A) = m_k / \sqrt{A}$ such that $M \propto \sqrt{A}$, i.e. the number of immigrants is proportional to the perimeter of $A$. It is important to note that dispersal in the local community remains homogeneous.

The larger metacommunity follows non-spatial dynamics and can be specified using $J_{meta}$ and $\nu_{meta}$. We assume that in comparison to the local community, the metacommunity is static and effectively of infinite size\footnote{In personal discussions, Ryan Chisholm has suggested that an infinite static metacommunity might be equivalent to a finite dynamic metacommunity. We briefly test this hypothesis in \cref{analysis-scenario-spatially-implicit}.}. As the local and metacommunity are separate but connected through migration, the model has a spatial aspect. Therefore, this scenario is called spatially implicit.

\subsection{Spatially Explicit}

The neutral model can also describe landscapes in which an individual's location does matter, i.e. which are spatially explicit. This scenario requires an explicit description of the habitat distribution and the dispersal kernel across the landscape. In the first two scenarios, the analytical formulas calculate the species richness on a finite island, though the spatially implicit model has also been extensively applied to contiguous landscapes \cite{Hubbell2001}. In the spatially explicit case, a large, potentially infinite landscape is modelled instead. Therefore, we introduce a smaller survey subarea $A$. Only the present-time species identities of individuals in this sample area count towards the biodiversity.

\section{Homogeneous Poisson Point Processes} \label{background-poisson-point-processes}

Point processes are random elements that spread points across a space. In this section, the properties of homogeneous Poisson point processes are summarised, focusing only on $\mathbb{R}^{+}_{0}$, i.e. $[0; \infty)$.

\subsection{Properties of Homogeneous Poisson Point Processes} \label{background-poisson-point-properties}

We start with some point process $\eta$ on $\mathbb{R}^{+}_{0}$,  $\eta(\mathbb{R}^{+}_{0})$, which produces points $T_{1}, ..., T_{n}$ where $n \in \mathbb{N}$. Without loss of generality, we enumerate $T_{i}$ in sorted order such that $T_{i} \leq T_{i + 1}$. We also set $T_{0} = 0$. $\eta(\mathbb{R}^{+}_{0})$ is a homogeneous Poisson point process iff the distances between adjacent points, i.e. $X = T_{i + 1} - T_{i}$, are independent and exponentially distributed with a constant rate parameter $\lambda$, i.e. $X \sim \textrm{Exp}(\lambda)$ \cite[p.~59]{Last2017}. From this definition, one can derive three important further properties of $\eta(\mathbb{R}^{+}_{0})$:
\begin{enumerate}
    \item \textbf{Poisson distributed point counts:} Let us rename $\mathbb{R}^{+}_{0} = B$. For every subset $B_{i} \subseteq B$, $\eta(B_{i})$ is still a homogeneous Poisson point process. Furthermore, the distribution of the number of points in $B_{i}$, $N_{i}$, is Poisson, i.e. $N_{i} \sim \textrm{Poi}(\lambda \cdot \mid B_{i} \mid)$, where $\mid B_{i} \mid$ describes the length of the subinterval $B_{i}$ \cite[p.~19]{Last2017}\cite[p.~41]{Chiu2013}.
    \item \textbf{Independent Scattering:} If $B$ is partitioned into disjoint subintervals $B_{1}, ..., B_{m}$ (with $m \in \mathbb{N}$) such that $B = \biguplus_{i = 1}^{m} B_{i}$, then all $\eta(B_{i})$ are independent, i.e. all $N_{i}$ are independent \cite[p.~19]{Last2017}\cite[p.~41]{Chiu2013}. As $\eta(B)$ satisfies this property, it is called completely independent \cite[p.~19]{Last2017}, or purely/completely random \cite[p.~41]{Chiu2013}. Note that because $\eta(B_{i})$ are independent Poisson point processes, this property can be applied recursively on the subintervals of $B_{i}$.
    \item \textbf{Uniform Point Distribution:} The points of the homogeneous Poisson point process $\eta(B)$ are uniformly distributed across $B$. Specifically, if we condition $\eta(B)$ on an exact number $N$ of points, the conditioned $\eta(B)$ is equivalent to a binomial point process on $B$ \cite[p.~43]{Chiu2013}. A binomial point process $\phi(B, N)$ is defined as a point process of $N$ independent points which are uniformly distributed across $B$ \cite[pp.~36-37]{Chiu2013}. This property can be used to sample the spatial distribution of point from $\eta(B)$. First, $\textrm{Poi}(\lambda \cdot \mid B \mid)$ can be sampled to obtain $N$, the number of points in $B$. Then, $N$ points can be uniformly distributed across $B$ \cite[pp.~38-39, p.~53]{Chiu2013}.
\end{enumerate}

\subsection{Properties of the Exponential and Poisson distribution} \label{background-exp-poi-properties}

The prior section has shown that the (negative) exponential distribution $\textrm{Exp}(\lambda)$ is a continuous distribution that, by definition, describes the inter-event times between events coming from a homogeneous Poisson point process over an interval $[a, b)$ \cite[p.~25]{Last2017}. Furthermore, the number of events is distributed discretely according to the Poisson distribution $\textrm{Poi}(\lambda \cdot \mid b - a \mid)$ \cite[p.~19]{Last2017}\cite[p.~41]{Chiu2013}. This section goes into more detail about the properties of these probability distributions.

For $\lambda > 0$ and $x \geq 0$, the exponential distribution's pdf (probability density function) and cdf (cumulative distribution function), i.e. $P(X \leq x)$, are \cite[pp.~21-22]{Thomopoulos2017}:
\begin{equation}
    f(x) = \lambda \cdot {e}^{- \lambda \cdot x} \quad \quad \quad \quad F(x) = 1 - {e}^{- \lambda \cdot x}
\end{equation}
For $k \in \mathbb{N}_{0}$, the Poisson distribution's pmf (probability mass function) and cdf are \cite[p.~143]{Thomopoulos2017}:
\begin{equation}
    f(k) = \frac{{\lambda}^{k} \cdot {e}^{-\lambda}}{k!} \quad \quad \quad \quad F(k) = {e}^{-\lambda} \cdot \sum_{i = 0}^{k}\frac{{\lambda}^{i}}{i!}
\end{equation}
It can also be shown that $X \sim \textrm{Exp}(\lambda)$ has the mean $E[X] = \frac{1}{\lambda}$ \cite[p.~21]{Thomopoulos2017}, while $Y \sim \textrm{Poi}(\lambda)$ has the mean $E[Y] = \lambda$ \cite[p.~144]{Thomopoulos2017}. The following useful properties can be derived for the distributions:
\begin{enumerate}
    \item \textbf{Exponential Memorylessness:} \label{background-exp-memoryless} The time left to wait for the next event is unaffected by how long we have already waited for the event: $Pr(X > t + s \mid X > t) = Pr(X > s)$ for $X \sim \textrm{Exp}(\lambda)$ \cite[p.~24]{Thomopoulos2017}.
    \item \textbf{Exponential Minimum:} \label{background-exp-minimum} The time until the first of many independent event streams is exponentially distributed with the sum of all event rates: $\textrm{min} \{X_{1}, ..., X_{n}\} \sim \textrm{Exp}\left( \sum_{i = 1}^{n}{\lambda_{i}} \right)$ where $X_{i} \sim \textrm{Exp}(\lambda_{i})$ and all $X_{i}$ are independent. The probability that the first event came from stream $i$ is $\frac{\lambda_{i}}{\sum_{i = 1}^{n}{\lambda_{i}}}$ \cite[pp.~181-182]{Takahara2017}.
    \item \textbf{Poisson superposition:} Similarly, the total number of events coming from multiple independent Poisson processes is Poisson distributed with the sum of all event rates: $\sum_{i=1}^{n} \textrm{Poi}(\lambda_{i})$ $= \textrm{Poi}(\sum_{i = 1}^{n}{\lambda_{i}})$ \cite[pp.~57-58]{Gupta2010}.
    \item \textbf{Geometric distribution:} \label{background-exp-geo} The floor of an exponential distribution is geometrically distributed: $\lfloor \textrm{Exp}(\lambda) \rfloor = \textrm{Geo}(1 - e^{-\lambda})$ \cite[p.~37]{Brereton2015}\cite{PiR8} where $\textrm{Geo}(p)$ has the pmf $f(k) = (1 - p)^{k} p$ and cdf $F(k) = 1 - (1 - p)^{k + 1}$ for $k \in \mathbb{N}_{0}$, as well as the mean $E[Z] = \frac{1 - p}{p}$ for $Z \sim \textrm{Geo}(p)$ \cite[p.~128]{Thomopoulos2017}.
\end{enumerate}

\section{Random Number Generation} \label{background-rng}

Random numbers are required to simulate probabilistic models. However, the sampling of true physical randomness is costly \cite[p.~303]{vanderLeest2012}. This section goes over several different methods to artificially generate numbers that appear random. Please refer to \Cref{appendix:distribution-sampling} on how to use these random numbers to sample different probability distributions.

\subsection{Hash functions and the Avalanche Effect} \label{background-avalanche}

Hash functions deterministically map values from a potentially infinite domain to a finite image \cite{Estébanez2014, Ticki2016}. They are often used in Computer Science to calculate a fixed-size fingerprint of some value. These fingerprints can then be compared for non-equivalence: if the hashes of two values $a, b$ differ, the values cannot be the same. If the hashes match, then either $a = b$ or the hash function has a collision. Ideally, hash functions minimise the probability of collisions, $h(a) = h(b)$ for $a \neq b$, to occur. The collision probability should be minimal independent of the distribution of the input values \cite{Estébanez2014, Ticki2016}.

The goal to minimise collisions can best be achieved when the hash function maps input values uniformly to its output domain, i.e. when the output values appear random. Informally, any small change in the input value $a$ should have a large and seemingly random effect on the hash $h(a)$. This property is called the avalanche effect and can be formally described using two criteria:
\begin{enumerate}
    \item \textbf{Strict avalanche criterion:} If an input bit is flipped, each output bit should change with a probability of $50\%$ \cite[p.~524]{Webster1986}.
    \item \textbf{Bit independence criterion:} The changes in the output bits should be pairwise independent \cite[p.~526]{Webster1986}.
\end{enumerate}
Therefore, the uniformity of a hash function can be evaluated by testing that any change in an input bit changes $\frac{H}{2}$ outputs bits on average where the hash value is $H$ bits long \cite{Sateesan2020}. The correlation of any two random variables describing the change of output bits $A$ and $B$ should be $0$ \cite[p.~527]{Webster1986}.

\subsection{Pseudo-Random Number Generation} \label{background-pseudo-random}

Pseudo-random number generators, PRNGs, are deterministic functions that produce a sequence of seemingly random numbers. In general, PRNGs consist of two parts: a \textbf{state transition} function $f: s_{k} \rightarrow s_{k+1}$ and an \textbf{output} function $g: s_{k} \rightarrow X_{k}$. Good PRNGs have at least the following four important qualities \cite{Bhattacharjee2018}:

\begin{enumerate}
    \item \textbf{Uniformity:} The samples that a PRNG generates are uniformly spread out over its output space. While non-uniform PRNGs exist as well, \Cref{appendix:distribution-sampling} shows how uniform PRNGs can easily be used to sample other distributions.
    \item \textbf{Independence:} RNGs are often used in cryptographic applications, in which the next random numbers must not be easily predictable from prior ones. If independence is not required, quasi-random number generators based on low-discrepancy sequences \cite{Niederreiter1978} can be used instead, which often offer superior uniformity in comparison to PRNGs \cite{Roberts2018}.
    \item \textbf{Large Period:} Every pseudo-random sequence is cyclic and will at some point repeat itself. The length of this cycle is called the period of the PRNG. Ideally, a PRNG with an internal state of $b$ bits should have a period of $2^b$.
    \item \textbf{Reproducibility:} PRNGs are seeded with their initial starting value $X_0$. If this seed is known, the following sequence of pseudo-random numbers can be regenerated deterministically.
\end{enumerate}

\noindent Several randomness tests suites including TestU01 \cite{LEcuyer2007}, PractRand \cite{PractRand}, Dieharder \cite{dieharder} and Ent \cite{ENT} have been developed. They try to disprove the null hypothesis that a generated sequence is statistically indistinguishable from a truly random sequence. For a more comprehensive introduction to PRNGs and their history, the reader is referred to \cite{Bhattacharjee2018} and \cite{LEcuyer2017History}, respectively.

\section{The Gillespie Algorithm} \label{background-gillespie}

The Gillespie algorithm was first introduced in 1976 to accurately simulate stochastic systems of reacting chemical agents \cite{Gillespie1976}. During each execution, one realisation of the probabilistic system is computed. The mean of multiple independent executions converges to the exact result of the modelled problem, making the Gillespie algorithm a variant of the Monte Carlo Method \cite{Gillespie1977, Weinzierl2000}.

The algorithm works with a set of molecules that react with each other probabilistically. Unlike in Wright-Fisher models, the Gillespie algorithm samples the next reaction that will occur, modifies the state $\textbf{X}$ accordingly, and then repeats until the system has reached a steady-state. Formally, the algorithm is based on sampling the distribution of $P(\tau, j \mid \textbf{x}, t) d \tau$ \cite[p.~39]{Gillespie2007}. This formula describes the probability that ``given $\textbf{X}(t) = \textbf{x}$, that the next reaction in the system will occur in the infinitesimal time interval $[t + \tau, t + \tau + d \tau)$, and [that it] will be an $R_j$ reaction'' \cite[p.~39]{Gillespie2007}.

The reactions are modelled as Poisson processes, i.e. the inter-reaction times are exponentially distributed. Each of the $M$ reaction processes is characterised by its per-capita event rate $\lambda$. The joint probability function for drawing the next event and its time is \cite[p.~39]{Gillespie2007}:
\begin{equation} \label{eq:gillespie}
    P(\tau, j \mid \textbf{x}, t) = a_{j}(\textbf{x}) \cdot e^{-a_{0}(\textbf{x}) \tau} \quad \text{where } a_{0}(\textbf{x}) = \sum_{i = 1}^{M} a_{i}(\textbf{x})
\end{equation}
As all events come from Poisson processes, \Cref{eq:gillespie} implies that $\tau \sim \textrm{Exp}(a_{0}(\textbf{x}))$, and that $j$ is distributed according to $P(j) = \frac{a_{j}(\textbf{x})}{a_{0}(\textbf{x})}$ (see \ref{background-exp-poi-properties}). Since the introduction of the Gillespie Algorithm, several methods have been proposed to efficiently perform the algorithm.

\subsection{The ``Direct'' Method}

In the ``Direct'' Method, the distributions of $\tau$ and $j$ are sampled directly. The samples determine the time until the next event and which type of event it is \cite[pp.~417-419]{Gillespie1976}. This method uses two uniform random numbers and has $O(M)$ complexity per step.

The primary performance limitation of the ``Direct'' Method is the linear complexity to sample $j$ on every step. To sample $j$, we need to iterate over the list of $M$ reactions to find the smallest $j$ such that $\sum_{i = 1}^{j} a_{i}(\textbf{x}) \geq \textrm{U}(a_{0}(\textbf{x}))$. One optimisation that has been proposed is to sort this list in decreasing order of their $a_{i}(\textbf{x})$, thereby reducing the average depth of the linear search required to find $j$ \cite{Yang2004, McCollum2006}.

To avoid sorting the list after every step, an incremental bubble sort can be used instead. Specifically, when a reaction fires, it is bubbled up to the next lower spot in the list. Therefore, the list becomes sorted eventually and can adapt to changing event rates at an $O(1)$ cost \cite{McCollum2006}.

However, the Sorting ``Direct'' Method also has a fundamental accuracy problem. By accumulating $a_{i}(\textbf{x})$ in decreasing order, rounding errors are more likely to exclude the events with the lowest $a_{i}(\textbf{x})$ from being sampled at all \cite[p.~44]{Gillespie2007}. Instead, the processes should be sorted in increasing order to minimise this error. Gillespie later proposed to split the reaction processes into a lower and an upper family, which are sampled separately, as an alternative solution \cite[p.~44]{Gillespie2007}.

\subsection{The ``First-reaction'' and ``Next-reaction'' Methods}

The ``First-reaction'' Method \cite[pp.~419-421]{Gillespie1976} uses the fact that all events from the reaction Poisson processes are independent. Therefore, the next time for each of them is $\tau_{i} \sim \textrm{Exp}(a_{i}(\textbf{x}))$ and can be sampled independently as $\tau_{i} = -\frac{\textrm{log}(\textrm{U}(0,1))}{a_{i}(\textbf{x})}$. Then, $\tau_{j} = \textrm{min}(\tau_{i})$, i.e. $j$ is the $i$ of the minimum $\tau_{i}$. With $\tau_{j}$ and $j$, the event is now applied, all $\tau_{i}$ are discarded, and the method is repeated for the next step. This method uses $M$ random numbers per step to sample all $\tau_{i}$ for all of the $M$ reaction processes.

The discarding of all unused $\tau_{i}$ at every step is very inefficient, of course. The ``Next-reaction'' Method \cite{Gibson2000} instead only discards and recalculates those inter-event times $\tau_{i}$ whose reaction $R_{i}$ was affected by the application of the previous event. In practice, all next-event times $t_{i}$ are kept in a priority queue sorted in increasing order of $t_{i}$. At the beginning of each step, the smallest $t_{i}$ is popped off the queue as $t_{j}$, and its corresponding event is applied. Then $a_{j}(\textbf{x})$ is recalculated and $t_{i} \sim t_{j} + \textrm{Exp}(a_{j}(\textbf{x}))$ is reinserted into the queue. If any other reactions $R_{k}$ were affected by the event, their $t_{k} \sim t_{j} + \textrm{Exp}(a_{k}(\textbf{x}))$ are also recalculated and reordered in the priority queue. In the best case, this method only uses one random number per step. If a binary heap is used to implement the priority queue, each step's complexity is $O(\textrm{log}(M))$. Slepoy, Thompson and Plimpton have presented an improved algorithm that uses an adaptive version of Walker's alias method \cite{Slepoy2008}. Their method reduces the per-step complexity to $O(\textrm{log}_2(\frac{\textrm{max}(a_{i})}{\textrm{min}(a_{i})}))$, which is independent of the number of reactions $M$ and, therefore, usually constant.

\subsection{Tau Leaping}

The performance of the Gillespie algorithm is fundamentally bounded by having to simulate every event. $\tau$-leaping \cite{Gillespie2001} instead leaps forward in predetermined jumps of length $\tau$, during which multiple events can occur. The number of events which occur for each reaction type during $\tau$ are Poisson distributed according to $\textrm{Poi}(a_{i}(\textbf{x}) \cdot \tau)$. This approach can be significantly faster if the results of many events can be applied to the system at once. For instance, the change of $M$ reactant quantities from $N$ reactions can be updated in $O(M) < O(N)$.

However, $\tau$-leaping can only approximate the exact results from the Gillespie algorithm. More-over, it assumes that all events occurring within $\tau$ are independent, which is not necessarily the case. Therefore, $\tau$ must be chosen carefully to minimise the effect of missing interference whilst also generating enough events that sampling the Poisson distribution for every reaction process is beneficial \cite[p.~46]{Gillespie2007}. Furthermore, extra care must be taken to ensure that no invalid system states are generated, for instance, using a reactant more often in $\tau$ than it was available in $\textbf{x}$.

\chapter{Review of Technical Background} \label{background-technical}

\section{The Rust Programming Language} \label{background-rust}

The Rust Programming Language was created in 2006 by Graydon Hoare \cite{Perkel2020}. He designed Rust to become a high-level system language, which should provide safety and performance. Hoare first announced Rust at the 2010 Mozilla Annual Summit, after which development of the language increased \cite{Hoare2010}. The first full stable version, Rust 1.0, was released in 2015 and became known as the 2015 edition \cite{Rust2015}. In 2018, the second edition of Rust, Rust 2018, was released alongside version 1.31 \cite{Rust2018}. As of May 2021, Rust 1.52 has been released \cite{Rust1.52}, and there have been discussions to create a third, Rust 2021, edition \cite{Rust2021}. Even though the Rust Programming Language is still very young, it has been rated as the most loved programming language in each annual StackOverflow Developer Survey since 2016 \cite{StackOverflow2020}. To learn more about Rust, the reader is referred to \cite{Cameron2021} for an introduction to Rust from a C++ programmer's perspective.

\subsection{The Rust Type System}

Rust is a statically typed language in which the type system is enforced at compile time. The type system is split up into primitive (e.g. \mintinline{rust}{bool}, \mintinline{rust}{i32}) and custom (e.g. \mintinline{rust}{struct} and \mintinline{rust}{enum}) types. Please refer to \Cref{appendix:rust-syntax} and \Cref{appendix:rust-safety} for a short introduction into the syntax of Rust's basic type system, and memory safety in Rust, respectively.

\subsubsection{Composition and the Trait System} \label{background-rust-traits}

Languages such as C++ and Java use inheritance to enable reusing behaviour between classes. Rust has neither classes nor inheritance and only allows the composition of types in \mintinline{rust}{struct}s. Instead, Rust uses a Trait system first proposed Sch{\"a}rli et al. to define the reuse of functionality \cite{Schaerli2003}. Traits are stateless interface specifications that define which methods a type must provide. Traits can also be composed together to specify a dependency graph:
\begin{minted}[linenos,escapeinside=@@]{rust}
// A simple trait specifying the `waddle` method.
trait @\textcolor{yellow}{\textbf{Waddle}}@ {
    fn waddle(&mut self, destination: Location);
}

// A simple trait specifying the `quack` method.
trait @\textcolor{yellow}{\textbf{Quack}}@ {
    fn quack(&mut self) -> String;
}

// A composed `Duck` trait which requires both `Waddle` and `Quack`.
trait @\textcolor{yellow}{\textbf{Duck}}@: Waddle + @\textcolor{blue}{\textbf{Quack}}@ {}
\end{minted}

\subsubsection{Generic Rust Types} \label{background-rust-generic}

Rust also allows compound types, traits and functions to be parameterised by one or more type parameters. For instance, instead of having \texttt{null} values, Rust provides the generic  \mintinline{rust}{Option<T>} type:
\begin{minted}[linenos]{rust}
enum Option<T> {
    Some(T),
    None
}
\end{minted}
\mintinline{rust}{Option<T>} can then be specialised for any Rust type. The combination of generic types and traits result in the expressiveness of the Rust type system. Traits can bound the type parameters to require that the substituted type provides the requested functionality. It is also important to note that, like C++ templates, Rust traits are specialised at compile time into unique, substitution-specific, monomorphised implementations.

\subsection{Verification}

Rust was designed to be a safe language and provide an expressive type system that can statically encode many guarantees to be checked at compile time. There have been several approaches to expand both the verification of and using the Rust language:
\begin{itemize}
    \item Evans et al. performed an analysis of popular Rust crates (libraries) and surveyed developers to study the use of unsafe code. They found that while only less than 30\% of libraries use unsafe code directly, many rely on it through dependencies \cite{Evans2020}.
    \item \textbf{RustBelt} is a formal safety proof of Rust's borrowing and mutability system \cite{Jung2017, JungPhD}.
    \item \textbf{Stacked Borrows} is an operational semantics for safe memory aliasing. It has been implemented as an extension to Rust's mid-level \textbf{I}ntermediate \textbf{R}epresentation interpreter MIRI to dynamically check mutability and aliasing guarantees in unsafe code \cite{Jung2018, JungPhD}.
    \item \textbf{Prusti} is a static validation tool that uses the Rust type system and user-provided Hoare triples to prove properties like overflow and panic freedom, and the correctness of functional specifications. It has been implemented as a plugin for the Rust compiler \cite{Astrauskas2019}.
    \item \textbf{contracts} is a Rust crate that allows programmers to write pre- and postconditions which can be checked at runtime \cite{RustContracts}. The library can also be instructed to output annotations for:
    \item \textbf{MIRAI} is an abstract interpreter for Rust's MIR. It can statically verify the implementation of protocols and check functional specifications, e.g. those provided by the contracts crate \cite{MIRAI}.
\end{itemize}

\section{Different Types of Parallelisation} \label{background-parallelisation-types}

\subsection{Task vs Data Parallelism}

In 1995, Ian Foster proposed a four-stage design methodology to parallelise programs \cite{FosterIan1995}. The first step is to partition the problem into parts that can be computed in parallel. This decomposition can be applied at two different levels. First, the problem can be split into different sub-tasks which perform different functions, called functional decomposition. For instance, a monolithic application could be split into small distinct microservices. In this case, the degree of parallelism is restricted to the number of independent sub-tasks we can extract and run in parallel.

Second, the domain of a problem can be decomposed. In this data parallelism approach, the same computation is applied to different subsets of the input data. When little to no interaction between the sub-computations is required, the data-parallel problem is called embarrassingly parallel\footnote{Cleve Moler cites \cite{Cleve1986} in \cite{Cleve2013} and \cite[28:01-28:28]{Cleve2007} as the first publication in which they coined the term.}, and the degree of parallelism is equivalent to the cardinality of the input data.

For a probabilistic simulation, data parallelism can be exploited both externally and internally \cite{Kunz2012}. The former refers to running \textbf{multiple} differently seeded instances of the same model in parallel, for instance using multiple independent processes. Internal parallelism, on the other hand, requires the computation itself to be decomposed. The input data of just \textbf{one} model is then distributed. Multithreading, for instance, can be used to perform the sub-simulations \cite{Christophe2014}.

\subsection{Data Sharing and Communication}

If we have decomposed a problem and distributed its computations amongst multiple threads, processes, or even machines, we often need some communication primitives to share and exchange data. As a simplification, the degree of sharing can be described on a scale \cite{Porobic2012, Ravela2010}:
\begin{enumerate}
    \item \textbf{Shared-Everything:} In this model, all computing units can access the same shared storage, for example, shared memory. However, this approach also requires careful protection of memory accesses to avoid data races. In particular, inconsistent reads or writes to data, which another process has just changed, must not occur.
    \item \textbf{Hardware Islands:} This model exploits that computing units are usually arranged in some spatial layout, which favours data sharing between physically close cores. For instance, CPU cores on the same socket usually share the same L2 cache, while all cores on one machine can access local RAM faster than a remote CPU. Therefore, data is shared on small hardware islands, while messages are used to communicate between islands. Most computations over a large set of input data can exploit some spatial locality in that data. Therefore, hardware islands can be a suitable compromise solution.
    \item \textbf{Shared-Nothing:} In this approach, all computing units are effectively independent and do not share any data. This method requires no protection to access any data, as it is always owned by just one process. However, if data needs to be exchanged, costly messages have to be sent between the participants. Furthermore, care must be taken to avoid any deadlocks.
\end{enumerate}

\subsection{Shared Memory Parallelism: CUDA} \label{background-cuda}

The Shared Memory Model gives programmers an intuitive mental model to organise distributed collaboration. Every participating process is given access to some shared global address space that can be manipulated just like regular memory. This abstraction is very implicit and transparent, as the programmer does not need to handle the details of how the program state is shared. However, the programmer needs to take care to maintain the consistency of the program state. Since all participants can modify any part of the state in any order, it is up to programmers to add protections to uphold consistency. Therefore, shared memory parallelism is error-prone when data access and update patterns are complicated.

Graphics Processing Units were initially created as fixed-function accelerators for rendering. However, with the addition of programmable shaders around 2001, they were increasingly extended to support general-purpose programming \cite{Tamasi2008}. The Compute Unified Device Architecture (CUDA) is a programming model and \textbf{A}pplication \textbf{P}rogramming \textbf{I}nterface for NVIDIA GPUs. CUDA was introduced in 2006. It has since allowed programmers to utilise the GPU as a multiprocessor with a vast number of cores \cite{Sanglard2020}.

\subsubsection{Computing Model} \label{background-cuda-model}

The CUDA computing model fully embraces data parallelism. Instead of large general-purpose cores and caches, modern GPUs have several thousands of tiny, more limited computing units \cite[p.~2]{CUDACpp}. The programmer provides CUDA with a kernel which is then invoked on every core in parallel. This paradigm shift can be thought of in the following way: let us consider an application that applies a \texttt{map()} function over a collection of elements. This functionality would be implemented on the CPU by iterating over the collection and applying \texttt{map()} to every element. In CUDA, we instead launch one kernel invocation for every element in the collection. The map function is then executed independently in parallel, and every thread only applies the \texttt{map()} function to one particular element.

Cores and the threads that run on them are grouped in groups of 32 (on NVIDIA GPUs), which are called warps \cite[pp.~104-106]{CUDACpp}. Each warp is one \textbf{S}ingle-\textbf{I}nstruction \textbf{M}ultiple-\textbf{D}ata unit of execution, i.e. all cores in the warp have to either execute in lockstep or sleep.\footnote{Since the release of the Volta GPU architecture in 2007, NVIDIA GPUs have been able to schedule threads independently inside each warp \cite[p.~2-3]{CUDAVolta}. This improvement has made it easier to translate CPU to GPU programs, as less performance is lost when threads diverge.} The user groups threads into thread blocks. The threads inside a block can be addressed using a virtual 1D, 2D or 3D block index. The final spatial abstraction is a 1D/2D/3D grid of thread blocks, the dimension of which the user specifies when launching a kernel. The combination of the per-block thread index, the block size, the block index, and the grid size are used to calculate a unique index for every thread that is executed during one launch of the kernel.

For an introduction into the resources a CUDA kernel can access, including the memory hierarchy on NVIDIA GPUs, please refer to \Cref{appendix:cuda-recources}.

\subsubsection{CUDA Kernel compilation} \label{background-cuda-kernel}

CUDA kernels are usually written in an extended C / C++ like language. The kernels can be either written and compiled separately or interleaved with the CPU host code and compiled together \cite[pp.~1-2]{nvcc}. The NVIDIA CUDA compiler \texttt{nvcc} can compile kernels into Parallel Thread Execution (PTX) code, which is an assembly like intermediate virtual instruction set for parallel thread execution architectures \cite[p.~8]{nvcc}\cite[pp.~1-2]{ptxisa}. At runtime, the PTX code is then loaded by the CUDA driver and compiled for the specific GPU that the user has. Since 2011/12, the open-source LLVM compiler toolchain has been used as the middle-end of \texttt{nvcc} \cite{Grover2009, Holewinski2011}. This enables programmers to write CUDA kernels in other languages, which can compile to the LLVM IR (intermediate representation) \cite{Harris2012}.

In particular, a PTX backend has been available in Rust since the \texttt{nightly-2017-01-XX} version \cite{RustNVPTX}. This version includes the unstable \texttt{abi\_ptx} feature to export non-generic Rust functions as CUDA kernels \cite{RustPtxAbi}. Since then, there have been several projects to create a full build pipeline for CUDA kernels written in Rust and provide an interface to the CUDA API:
\begin{enumerate}
    \item \textbf{rust-ptx-linker} is a custom linker for the \texttt{nvptx64\-nvidia\-cuda} target of the Rust compiler \cite{RustPtxLinker}. It is built on LLVM and does the final cleanup work of the LLVM IR generated by the Rust compiler. Afterwards, it uses LLVM's PTX backend to generate valid PTX code.
    \item \textbf{rust-ptx-builder} is a helper library which can be used in Rust \texttt{build.rs} scripts \cite{RustPtxBuilder}. It instructs \texttt{rustc} to compile a Rust crate (library) as a CUDA PTX kernel. The PTX instructions can then be embedded into the host code at compile-time, such that the CUDA kernel is part of the CPU executable.
    \item \textbf{Accel} is a high-level GPGPU framework that is written for a CUDA backend \cite{Accel}. It contains the CUDA driver API to launch kernels, as well as bindings to additional runtime features such as just-in-time compilation and profiling. It is worth noting that \texttt{accel} allows the programmer to interleave host and kernel code.
    \item \textbf{RustaCUDA} is a high-level interface to the CUDA driver API, which can launch kernels \cite{RustaCUDA}. Unlike \texttt{accel}, \texttt{rustacuda} does not take care of compiling a kernel but expects to read the PTX from a string or file. Therefore, its kernel launch interface is lower-level than \texttt{accel} and does not provide the same safety guarantees.
\end{enumerate}

\subsection{Message Passing Parallelism: MPI} \label{background-mpi}

Shared memory parallelism shares data by sharing access to a global address space, which exposes the user to potential consistency issues. Message Passing instead aims to provide programmers with a safe programming model for parallelism. Different processes can only communicate with each other by exchanging messages, which move data from the source process' address space to the destination's address space.

The Message Passing Interface (MPI) was designed to standardise message passing between processes running on heterogeneous MIMD (multiple-instruction multiple-data) machines. While the MPI standard has been under development since 1992, the initial draft version was presented at the 1993 ACM/IEEE conference on Supercomputing \cite{MPI1993}. Since the release of MPI v1.0 in 1994, the standard has been updated to v3.1 in 2015, with v4.0 being under development \cite{MPIv4}.

\begin{figure}[h]
    \centering
    \includegraphics[width=0.8\textwidth]{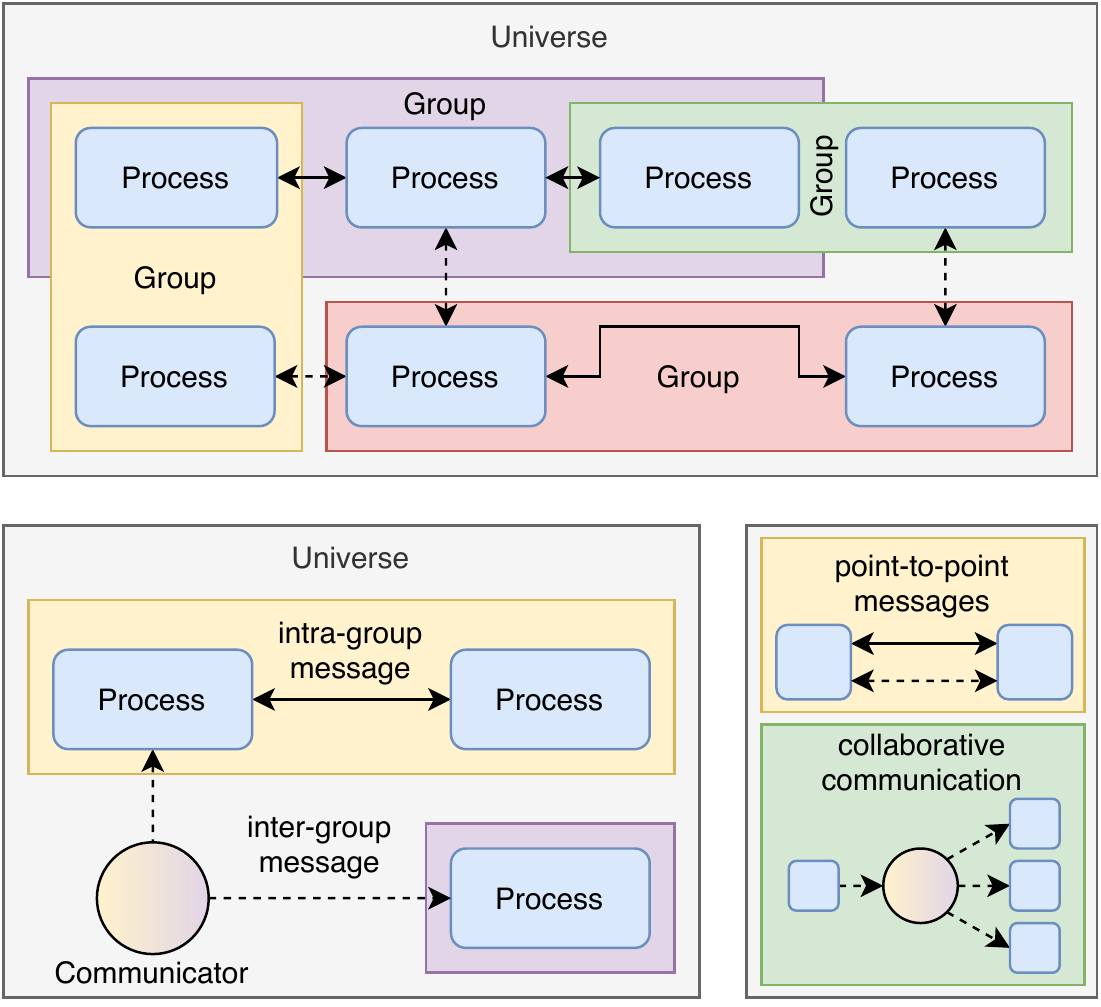}
    \caption{Simplified overview of the MPI programming model. MPI allows processes within the same universe to communicate both point-to-point and to collaborate collectively.}
    \label{fig:mpi-overview}
\end{figure}

\noindent MPI provides a standardised interface for processes to communicate with each other. \Cref{fig:mpi-overview} shows a simplified overview of MPI's process-based programming model. Processes are part of one or more group, which are ordered sets. Groups can be split and combined using the standard set semantics such as union and intersection \cite[p.~226,230-234]{MPIv3.1}. Processes can send messages inside one group, or across different groups using communicators. Collective communication, on the other hand, always uses a communicator that spans all participating groups \cite[pp.~224-228]{MPIv3.1}. Both sending and receiving messages can be blocking or non-blocking. Point-to-point messaging also offers one synchronous and multiple asynchronous variants. Last but not least, MPI also supports explicit barrier synchronisation, as sending and receiving messages does not provide any synchronisation guarantees \cite[p.~142,147]{MPIv3.1}. \\

\noindent To avoid the mixup and misinterpretation of messages, MPI provides several additional features:
\begin{enumerate}
    \item Messages from the same source maintain their sending order upon receipt \cite[p.~41,56]{MPIv3.1}.
    \item Point-to-point messages can also be tagged with an arbitrary integer \cite[p.~27]{MPIv3.1}.
    \item The user can create entirely distinct universes, which act as a global tag on all messages that are passed around inside the universe \cite[p.~27]{MPIv3.1}.
    \item MPI also allows programmers to define custom structured data types, which can then be checked and interpreted correctly on the receiving end \cite[p.~83,93]{MPIv3.1}.
\end{enumerate}

\chapter{Review of Related Work} \label{background-related}

\section{The \texttt{necsim} library} \label{background-necsim}

Before \texttt{necsim} was developed, spatially explicit neutral simulations were lacking \cite[p.~27]{ThompsonPhd}. \texttt{necsim} was created to allow ecologists to quickly tinker with different neutral models without having to write their own implementation. In addition to the C++ library, Thompson, Chisholm and Rosindell also developed the \texttt{pycoalescence} and \texttt{rcoalescence} interfaces for Python and R \cite{Thompson2020}.

The \texttt{necsim} library implements a spatially explicit individual-based simulation that is performed on a structured grid. Each grid cell provides habitat for a well-mixed group of individuals, which is called a deme. To simulate the individuals living in such a landscape, \texttt{necsim} is usually used in two steps. First, the neutral coalescence simulation described in \cref{background-neutral-simulation} builds a coalescence tree of all sampled individuals. Second, \texttt{pycoalescence} and \texttt{rcoalescence} provide analysis functionality and calculate metrics such as the species richness or the species abundance distribution over the coalescence tree \cite[p.~28]{ThompsonPhd}. This architecture allows scientists to perform any posterior analysis on the tree. However, \texttt{necsim} lacks live monitoring to debug the simulation or directly perform analysis without building the costly intermediary coalescence tree.

The \texttt{necsim} library was developed to support a plethora of model scenarios. These include the non-spatial, spatially-implicit\footnote{The spatially-implicit model is implemented as a combination of the non-spatial model and an optional metacommunity. In \texttt{necsim}, a metacommunity can be specified for any of the simulation scenarios \cite[p.~33]{ThompsonPhd}.} and spatially-explicit scenarios, which are described in \cref{background-neutral-scenarios}. \texttt{necsim} also implements a novel partly spatial archipelago scenario consisting of $N$ non-spatial islands connected through migration \cite[p.~30]{ThompsonPhd}. In addition to these model scenarios, \texttt{necsim} is also able to simulate contiguous heterogeneous landscapes, which can be finite or infinite. The habitat, dispersal, and local turnover (birth-death) rates on each landscape can vary both spatially and temporally. The user can also specify which individuals should be sampled using a sampling map. Finally, \texttt{necsim} offers both point (\ref{background-non-spatial-point-speciation}) and protracted (\ref{appendix:protracted-speciation}) speciation \cite[p.~33]{ThompsonPhd}.

While \texttt{necsim} enables the simulation of a broad range of scenarios, the C++ library was not built to be trivially extensible. For instance, the coalescence algorithm, the potential wrapping of coordinates, and the output to an SQLite database are all combined in the \texttt{SpatialTree}\footnote{\href{https://bitbucket.org/thompsonsed/necsim/src/c824201/SpatialTree.cpp}{https://bitbucket.org/thompsonsed/necsim/src/c824201/SpatialTree.cpp}} class. This tight coupling between the simulation's parts complicates code reuse when adding new algorithms or implementations. Furthermore, \texttt{necsim} was only designed to run on one single-threaded machine. Thus, it does not offer the required code modularity to be easily parallelised.

\section{The \texttt{msprime} library} \label{background-msprime}

\texttt{msprime} \cite{Kelleher2016} is one of many population genetics coalescent simulators that can reconstruct the evolution of genomes. Specifically, it uses a coalescent algorithm that produces a genetic genealogy, i.e. a set of correlated coalescence trees along a genome. The trees are stored in a bespoke sparse data format, which was developed for more efficient analysis and reduced data storage requirements \cite{Kelleher2018}. While many population genetics simulations are non-spatial, Kelleher, Barton and Etheridge have also investigated spatially continuous extensions to the algorithm \cite{Kelleher2013}. The \texttt{msprime} library also comes with a Python frontend, enabling the simple configuration of simulations and integration with other tools \cite{Kelleher2020}. While ecology poses different research questions than population genetics, they share similar coalescence algorithms, and insights are often transferable.

\section{Parallel Random Number Generation}

This section discusses several strategies to give every partition in a parallel probabilistic simulation its own \textbf{independent} stream of \textbf{uniformly distributed} random numbers \cite[p.~2]{Schoo2005}.

\subsection{Splitting Random Number Generators}

The first approach is to split one RNG over $P$ partitions. The two primary techniques to generate random numbers for the partition with index $i$ are cycle splitting and parameterisation \cite[p.~3]{Schoo2005}.

\subsubsection{Cycle Splitting}

The full random sequence is partitioned on the sequence index. In \textbf{Block Splitting}, the $i^{\text{th}}$ partition samples random numbers from the $i^{\text{th}}$ disjoint continuous subsequence. At initialisation, this method requires an efficient, sublinear method to jump ahead in the sequence \cite[p.~11]{Dammertz2010}. In \textbf{Leap-Frogging}, the $i^{\text{th}}$ partition samples random numbers from all sequence indices $j$ where $j \textrm{ mod } M = i$ \footnote{Eddy has suggested that random leaps might also be worth exploring \cite[p.~68]{Eddy1990}.}${}^{,}$\footnote{Tan also lists a shuffling leap frog variant, but further details on this method are lacking in the literature \cite[p.~695]{Tan2002}.}. This variant requires an $O(1)$ method to jump $M$ steps ahead in the sequence \cite[p.~12]{Dammertz2010}. All cycle splitting methods require that the RNG has a sufficiently large period. L'Ecuyer and Simard recommend that a period of $N^{3}$ should be used to generate $N$ samples \cite{LEcuyer2001}.

\subsubsection{Parameterisation}

Independent random number streams can also be generated by parameterising each partition's RNG differently. In \textbf{Seed Parameterisation}, a partition-dependent seed is constructed from both $i$ and the global simulation seed. This is the least invasive option to split an RNG and can be used dynamically at runtime \cite[p.~10]{LEcuyer2017}. However, the RNG can also be instead specifically designed to include a sequence number -- this variant is called \textbf{Iteration Function Parameterisation}. For instance, the PCG algorithm uses an increment of $2i + 1$ in its Linear congruential generator stage \cite[p.~1,24]{ONeill2014}. Claessen and Pa\l{}ka instead designed an RNG that injectively encodes splits and iteration into a sequence of bits, which is then encrypted to get a pseudo-random output \cite{Claessen2013}.

\subsection{Counter-based Pseudo-Random Number Generators} \label{background-cbrngs}

In 2011, Salmon et al. introduced counter-based PRNGs in their paper ``Parallel random numbers: as easy as 1, 2, 3'' \cite{Salmon2011}. They propose the following construction:
\begin{equation*}
    \text{state transition: } s_{k + 1} = s_{k} + 1 \quad \quad \quad \quad \text{output: } g(s_{k}) = hash(s_{k})
\end{equation*}
As the RNG's state is just an integer counter, the hashing output function is the only source of randomness. Therefore, the authors initially test cryptographical hash functions such as AES and Threefish, which both fulfil the strict avalanche criterion (\ref{background-avalanche}) for security purposes. However, using a cryptographically secure hash function is not necessary for generating random numbers. Therefore, the authors also introduce three more performant non-cryptographic variations: ARS (Advanced Randomisation System), Threefry, and Philox, which is optimised explicitly for GPU applications. Finally, the authors also show that all of their RNGs pass the TestU01 test suite \cite{LEcuyer2007}.

CBRNGs have several key advantages. First, they can easily be used non-sequentially as their state is just an integer counter. Therefore, cycle splitting or arbitrary jumping in the random sequence are supported at \textbf{no} extra cost. Second, CBRNGs can trivially produce independent streams by changing the key of a keyed hash function. Third, the authors noted that \cite[p.~10]{Salmon2011}:
\begin{displayquote}
    ``If [a] simulation also requires random forces, counter-based PRNGs can easily provide the same per-atom random forces on all the processors with copies of that atom by using the atom identiﬁer and timestep value [as part of the counter]''.
\end{displayquote}

\subsection{Reproducible Random Number Generation} \label{background-reproducible-rng}

Hill notes that many parallel probabilistic simulations are not reproducible \cite{Hill2015}, i.e. that their results depend on how many threads or processes are used. As a counter-measure, Hill suggests that each individual should get its own independent and consistently parameterised random stream, irrespective of whether the simulation is run sequentially or in parallel.

CBRNGs are well suited to generate many independent streams \cite[p.~10]{Salmon2011}. For example, Lang and Prehl use one CBRNG per particle performing a random walk on a fractal \cite{Lang2017}. Each particle simply uses its index as part of the hashing key. Martineau and McIntosh-Smith employ the same approach in their neutral Monte Carlo particle transport application, in which they also observe that parallelising over events instead of particles is faster \cite{Martineau2017}. In population genetics, Lawrie uses a similar method for independent mutations in a forwards-time Wright-Fisher (\ref{background-moran-fisher}) simulation \cite{Lawrie2017}. They also remark on the tradeoff between simulating mutations with embarrassing parallelism on the GPU, and synchronising to clean up finished mutations. Notably, Jun et al. go one step further and also allow interactions between the particles in their transport simulation \cite{Jun2020}. When two particles interact, they derive a new CBRNG from the interaction's attributes.

Reproducibility can also be used to reduce communication. Phillips, Anderson and Glotzer use a hashing PRNG to apply the same pseudo-random force to two interacting particles without communication \cite{Phillips2011}. In between synchronisations, every particle is simulated independently using a PRNG that has been initialised with the hash of the particle's index, the simulation time step and the global seed. When two particles are known to interact, they both hash the combination of their indices to sample the same pseudo-random interaction force. Note, however, that regular synchronisation is still required to give all particles the knowledge about their interaction partners.

\section{Random Number Scrambling} \label{background-rng-scrambling}

Scrambling is a technique that is used to further randomise and improve the statistical quality of an RNG. Scrambling can be applied either to RNG output as an extra output function, or to the sequence index in a low-discrepancy sequence. Several methods have been proposed for scrambling:
\begin{enumerate}
    \item \textbf{Random Digit Scrambling} flips each bit of the input with an independent uniform probability \cite[p.~538]{Matousek1998}, e.g. by XORing the sample with a partition-dependent seed value \cite{Kollig2002}.
    \item \textbf{Owen Scrambling} defines a nested uniform permutation where each output digit only depends on more or equally significant input digits \cite{Owen1995}. This method can be implemented as a keyed hash function that only avalanches to more significant digits, as was proposed by Owen \cite{Owen2003}, implemented by Laine and Karras \cite{Laine2011} and refined by Burley \cite[pp.~9-11]{Burley2020}.
    \item \textbf{Hashing} the random sample can also significantly improve the quality of an RNG. For instance, O'Neill applies simple avalanching hash functions in their PCG RNG family \cite{ONeill2014}.
\end{enumerate}

\section{Parallelising the Gillespie Algorithm}

The Gillespie Algorithm is a universally used Monte-Carlo algorithm, which is often applied to simulate simulations with a large number of particles / individuals. This section provides an overview of existing methods that parallelise the algorithm and several of its variants.

\subsection{Parallelisation on a GPU}

Kunz parallelises the event simulation on an internal and external level \cite{Kunz2012}. They use the GPU as a coprocessor and pipeline the execution of multiple events. The CPU performs the scheduling of events from different instances of the same simulation. Komarov and D'Souza also parallelise the simulation both internally and externally \cite{Komarov2012}. In contrast to Kunz, they move both layers into the GPU to exploit the device's multi-level thread architecture. Specifically, all threads in a warp coordinate one execution, while different warps run different parameterisations of the simulation.

Dittamo and Cangelosi parallelise the $\tau$-leaping method by generating every reaction's $\tau_{i}$ in parallel using the GPU \cite{Dittamo2009}. The minimum $\tau_{j}$ is then sent back to the CPU, which continues the simulation. On the GPU, they use a parallel Mersenne Twister RNG. Nobile et al. also parallelise the Gillespie algorithm on a GPU using $\tau$-leaping \cite{Nobile2014}. However, unlike Dittamo and Cangelosi, Nobile et al. do not just use the GPU as a parallel random number generator. Instead, they split the algorithm into four disjoint stages: (1) calculate the $\tau$s, (2) perform $\tau$-leaping for some individuals, (3) compute Gillespie's ``Direct'' Method for other individuals, and (4) perform a combined termination check on all individuals. Note that only the fourth step requires synchronisation. This decomposition decreases the GPU kernel's complexity and improves thread occupancy.

\subsection{Parallelisation in a \textbf{H}igh-\textbf{P}erformance \textbf{C}omputing environment} \label{background-gillespie-hpc}

Ridwan, Krishnan and Dhar parallelise the Gillespie algorithm's ``Direct'' method by decomposing the simulation domain \cite{Ridwan2004}. To maintain consistency, they use message passing to exchange individuals dynamically between partitions. The authors also propose an approximate method that averages out the well-mixed reactant population at special synchronisation steps. It is important to note that this method is designed explicitly for non-spatial simulations. \Cref{implementation-gillespie-averaging} shows how we have adapted this method for this project to handle spatially-explicit migrations.

Arjunan et al. implement a highly parallel version of the ``Direct'' method, which is decomposed into hexagonal voxels \cite{Arjunan2020}. Each subdomain is responsible for a subset of these voxels. In this method, the authors limit dispersal between voxels to nearest-neighbour only. Therefore, every subdomain only needs to exchange messages with its direct neighbours. Every subdomain also caches a read-only version of adjacent edge voxels, called ghost voxels, for better performance.

Bauer describes how parallel discrete event simulations can either progress in discrete or stochastic time steps \cite{Bauer2015}. In the former method, they assume that events are independent during each time step. In the latter strategy, which we adapt in \cref{implementation-monolithic-optimistic}, the simulation has to be run optimistically and potentially be paused or rolled back to maintain global consistency. Jeschke et al. explore a similarly optimistic parallelisation of the spatial ``Next-Subvolume'' method \cite{Jeschke2008}.

\section{Parallelising Spatial Simulations}

Exploiting spatial locality is crucial in spatial simulations. Harlacher et al. propose to dynamically partition a static simulation domain using a space-filling curve \cite{Harlacher2012}. Notably, their method only requires knowledge of the global amount of work and can be implemented using MPI reductions.

Partitioned algorithms often require communication between the simulation subdomains, which can limit performance. Thus, further localising computation is beneficial. Field, Kelly and Hansen propose to delay the evaluation of MPI reduction operations so that several can be dynamically fused at runtime \cite{Field2002}. On the GPU, we can instead combine entire threads. Stawinoga and Field show how a GPU compiler can statically predict the optimal level of such thread coarsening \cite{Stawinoga2018}.

There has also been increasing work on developing domain-specific compilers to translate mathematical systems into parallelised code directly. For instance, \cite{Mudalige2013} parallelises unstructured meshes for highly heterogeneous systems, i.e. CPUs and GPUs \cite{Bertolli2003}. The authors' proposed framework exploits both data independences and architectural features separately for every loop.

Finally, note that the \texttt{necsim} library simulates all individuals on a regular cartesian grid. However, the simulation could also be extended to and parallelised using unstructured meshes.

\chapter{The Declaration of Independence} \label{independence}

In \cref{background-scientific}, we have reviewed the Neutral Model of Biodiversity (\ref{background-neutral-biodiversity}) and how we can use a reverse-time coalescence algorithm (\ref{background-neutral-simulation}) to simulate it. This algorithm, which is implemented in the \texttt{necsim} simulation library (\ref{background-necsim}), has one significant downside, however. At each step, the simulation has to check whether a dispersing individual collides and coalesces with a different individual. As this check requires globally consistent knowledge of the location of all individuals, it limits the scalability and parallelisation of the simulation.

This chapter describes the core idea of this thesis: trading communication for redundancy. Specifically, a novel algorithm is presented that simulates individuals independently with embarrassing parallelism whilst also maintaining consistency across the whole simulation. First, \cref{independence-coalescence-gillespie} shows how the Gillespie algorithm can be used to run the coalescence simulation. Next, \cref{independence-independent-algorithm} introduces the novel Independent algorithm. In particular, \cref{independence-hashing-rng} and \cref{independence-align-rng} show how a hashing pseudo-random number generator can direct individuals to follow the same trajectory after they have coalesced independently. \Cref{independence-exponential-generations} then demonstrates how the novel Independent algorithm can still generate exponentially distributed inter-event times.

\section{Coalescence à la Gillespie} \label{independence-coalescence-gillespie}

The Gillespie algorithm is an exact probabilistic algorithm that is summarised in \cref{background-gillespie}. The coalescence algorithm implemented in \texttt{necsim} (\ref{background-necsim}) can be seen as a special case of the ``First-reaction'' Method. \texttt{necsim} uses geometrically distributed inter-event times to approximate a single global Poisson point process which produces the next event for any individual in the simulation\footnote{\href{https://bitbucket.org/thompsonsed/necsim/src/c824201/Tree.cpp\#lines-592}{https://bitbucket.org/thompsonsed/necsim/src/c824201/Tree.cpp\#lines-592}}. At every step, the algorithm then picks the next active individual $j$ using rejection sampling\footnote{\href{https://bitbucket.org/thompsonsed/necsim/src/c824201/SpatialTree.cpp\#lines-1099}{https://bitbucket.org/thompsonsed/necsim/src/c824201/SpatialTree.cpp\#lines-1099}}${}^{,}$\footnote{\href{https://bitbucket.org/thompsonsed/necsim/src/c824201/ActivityMap.cpp\#lines-83}{https://bitbucket.org/thompsonsed/necsim/src/c824201/ActivityMap.cpp\#lines-83}}, taking the local turnover rate $\lambda$ into account. This turnover rate specifies the distribution of lifetimes $X \sim \textrm{Exp}(\lambda)$, i.e. the times between the births and deaths of the simulated individuals.

Since we are investigating a neutral model, this turnover rate only depends on the current location but not on the properties of an individual or species. Therefore, we shall use $\lambda_{x, y}$ from now on, where $(x, y)$ specifies the current location of an individual. As neutral biodiversity models are zero-sum, any death at a location $(x, y)$ is immediately followed by a birth to the same location. Therefore, at each location, there is an event stream produced by an infinite homogeneous Poisson point process $\textrm{Poi}(\lambda_{x, y})$, which describes the births / deaths occurring at the location. Note that as we are going backwards in time, we use a Poisson point process on $\mathbb{R}^{-}_{0}$.

In \texttt{necsim}'s coalescence simulation, however, events are generated for individuals, not locations. So what is the distribution of inter-event times of a single lineage? A single individual produces an event whenever it reverses its birth and disperses back (in)to its parent. The next event in this lineage would then be the birth event of the parent individual. Therefore, the time between events is the time between a child's birth and its parent's birth. It clearly is the case that $t_{birth(parent)} < t_{birth(child)} < t_{death(parent)}$, and that $X = (t_{death(parent)} - t_{birth(parent)}) \sim \textrm{Exp}(\lambda_{x, y})$. We now use the memoryless property of the exponential distribution (\ref{background-exp-memoryless}) to show that $(t_{birth(child)} - t_{birth(parent)}) \sim \textrm{Exp}(\lambda_{x, y})$ as well.

\label{independence-exponential-inter-event} Intuitively, the memoryless property states that `even when we have already waited for a while, the time until the next event is still exponentially distributed precisely the same as when we started waiting'. When individuals coalesce in the reverse-time simulation, the parent has been waiting for its birth event ever since it reversed its death. When one of its children rewinds its birth, the distribution of the remaining time until the parent's birth is still exponentially distributed at the same rate. Specifically, $(t_{birth(child)} - t_{birth(parent)}) \sim \textrm{Exp}(\lambda_{x, y})$ as well. In summary, both the turnover and inter-event times at location $(x, y)$ are exponentially distributed with rate $\lambda_{x, y}$. \\

\begin{figure}[h]
    \centering
    \includegraphics[width=0.8\textwidth]{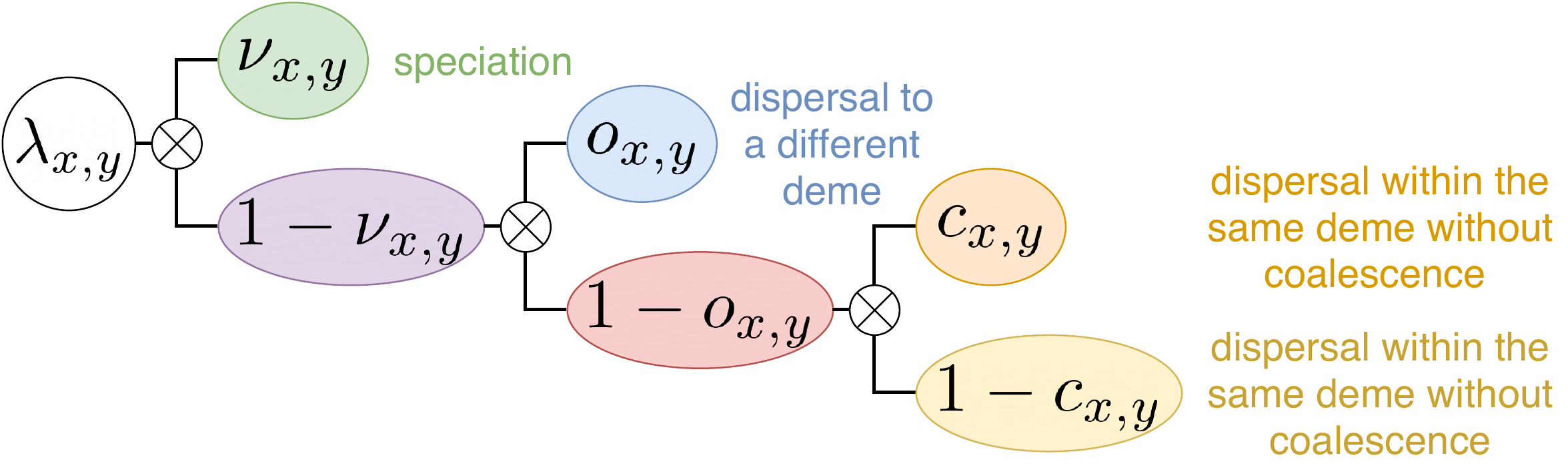}
    \caption{Breakdown of the different event types and their probabilities.}
    \label{fig:event-breakdown}
\end{figure}

\noindent What is the benefit of knowing that inter-generation, i.e. per-individual inter-event times, are exponentially distributed? At every generation, an individual can speciate with probability $\nu$ or disperse with probability $1 - \nu$. By using exponentially distributed inter-generation times, we can now also define the rate of the different types of events by using the minimum of exponential random variables (\ref{background-exp-minimum}). For instance, speciation at location $(x, y)$ has a rate of $\nu_{x, y} \cdot \lambda_{x, y}$. \Cref{fig:event-breakdown} shows the breakdown of all the events types. As in \texttt{necsim} (\ref{background-necsim}), a deme represents a well-mixed group of individuals that all live at the same location $(x, y)$.

\noindent While the coalescence algorithm in \texttt{necsim} uses rejection sampling to handle varying turnover rates, the ``Next-reaction'' Gillespie method provides a more elegant solution. Since we now assign every individual its own Poisson point process, they can also have different event rates. These Poisson point processes can be split up and combined to optimise the sampling of the next event:

First, we can group $k$ individuals with the same event rate $\lambda$ together and simulate them as one combined Poisson point process with rate $k \cdot \lambda$. When this process produces the next event, one of the individuals in the group can be chosen uniformly as the one who executes this next event.

Second, not all types of events change the state of the simulation. Let us assume that an individual currently has turnover rate $\lambda_{x, y}$, speciation probability $P(speciation) = \nu_{x, y}$, probability $P(outside) = o_{x, y}$ to jump to a different location on dispersal, and probability $P(coalescence)$ $= c_{x, y}$ to coalesce within the current deme. \Cref{fig:event-breakdown} shows a breakdown of the four different event types between which we differentiate\footnote{Many thanks to James Rosindell, Lucas Dias Fernandes and Samuel Thompson, with whom I worked out the mathematical details of this event rate split during my 2019 UROP.}.

Of these events, only the first three have an effect on the simulation state, while dispersal within the same deme (same $(x, y)$) without coalescence mostly wastes computation. With exponential event rates, we can simply ignore this last type of event by subtracting its rate from the original $\lambda_{x, y}$, producing the following event-skipping rate:
\begin{equation*}
    \lambda_{x, y}' = \lambda_{x, y} \cdot (1 - P(dispersal \land \overline{outside} \land \overline{coalescence})) = \lambda_{x, y} \cdot (1 - (1 - \nu_{x, y}) \cdot (1 - o_{x, y}) \cdot (1 - c_{x, y}))
\end{equation*}
\label{independence-event-skipping}This event skipping mechanism can reduce the runtime of the algorithm to $\frac{\lambda_{x, y}'}{\lambda_{x, y}}$. We have implemented this optimisation in the \textbf{SkippingGillespie} algorithm (\ref{implementation-monolithic-parallelisation}). It is especially potent when simulating scenarios in which the probability of dispersing back to the same location is large (\ref{analysis-domain-speciation-scalability}). For instance, this occurs when the landscape is small but has large demes, i.e. when many individuals can co-inhabit the same location.

\section{The Independent Algorithm} \label{independence-independent-algorithm}

\begin{figure}[h]
    \centering
    \includegraphics[width=0.9\textwidth]{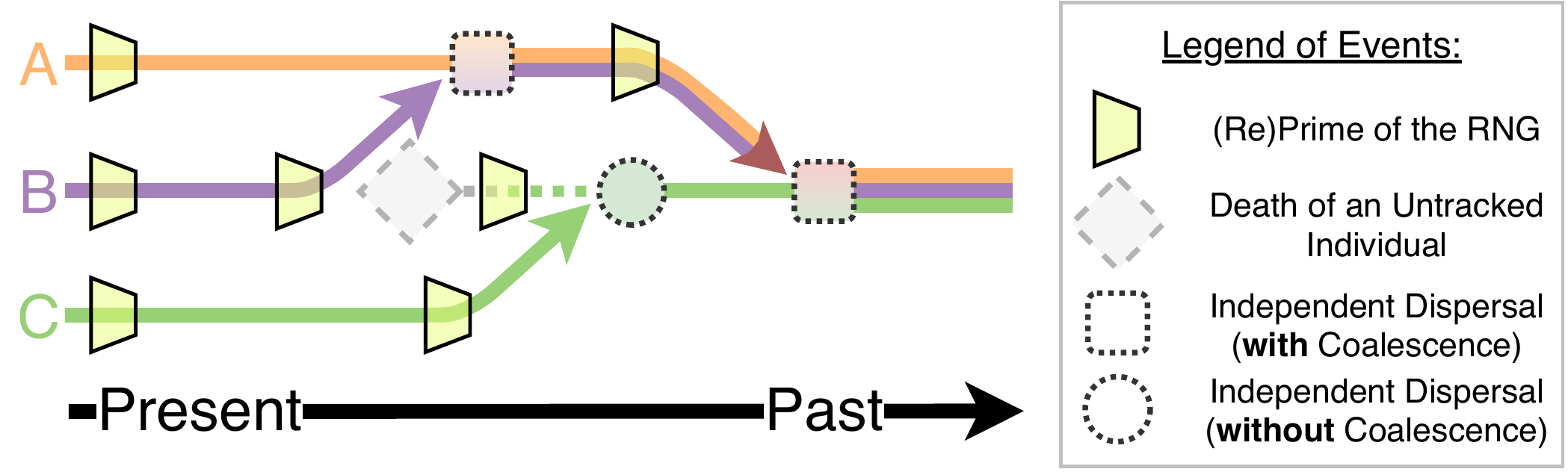}
    \caption{Example of RNG repriming during an independent coalescence simulation. In the Independent algorithm, an individual reprimes its random number generator at least once before every event. This process ensures that the next event is consistent across `coalesced' individuals.}
    \label{fig:rng-reprime}
\end{figure}

In the coalescence-based biodiversity simulation, coalescence is the only direct interaction between individuals. By definition, two individuals coalesce when one individual disperses and then collides with the other. Semantically, the child individual jumps back (in)to its parent individual when rewinding its birth. In the original \texttt{necsim} algorithm, we stop simulating the child individual after coalescence. In the Independent algorithm, we want to avoid the communication that is required to know the locations of all other individuals. Well, how would we expect the child to behave if we kept simulating them after their collision? One way to think about coalescence is to think about Matryoshka dolls: when a child coalesces with its parent, it simply jumps `back into' the parent individual. Consequently, after coalescence, both individuals should take precisely the same steps at precisely the same times, as is shown in \Cref{fig:rng-reprime} in their combined trajectories.

\noindent This is the core idea of the Independent algorithm:
\begin{enumerate}
    \item We do not maintain any globally consistent simulation state to search for coalescence.
    \item Instead, every individual is simulated entirely independently.
    \item If we later look at the trajectories of two independently simulated individuals and see that the individuals collided, i.e. inhabited the same location for some non-zero time span, we guarantee that both individuals have produced the same events from the point of collision.
    \item Therefore, we can detect coalescence a posteriori by searching for redundancy in the combined event traces of multiple individuals.
    \item In summary, we trade off communication for redundant computation.
\end{enumerate}
The following subsections describe in detail how these properties are achieved.

\subsection{Environmental RNG Priming} \label{independence-hashing-rng}

The prior section has briefly introduced the core idea of the Independent algorithm. This section focuses on how two independently simulated individuals can be directed to follow the same trajectory from the point of their collision.

Since the model we are simulating is neutral, we know that the behaviour of each individual is only determined by its environment, i.e. location and time, but not by its species identity. Therefore, we can design a deterministic function that pseudo-randomly selects the next event based solely on an individual's current location and time. If the parent and child individual are already aligned in space and time, this function ensures they behave in lockstep and sample the same events from this point forward. However, a child and parent collide at the time of the child's birth, which is an event coming from the child's event stream, which is entirely independent of the parent's. \Cref{independence-align-rng} explains how the initial alignment at collision is reached independently. The following paragraphs lay out how we can use strong hashing to define the next-event function. \\

\noindent As introduced in \cref{background-avalanche}, hash functions deterministically map values from their input domain to a finite image. In this application, we want a hash function that maps the current location and time of an individual to the time and properties of the next event. Effectively, this can be achieved by resetting the individual's random number generator such that its state is determined only by this environmental information, for instance, by reseeding the RNG. However, many RNGs only apply a weak hash function to initialise their internal state. Since this algorithm resets the RNG multiple times at every step, relying on such a weak mechanism would expose the algorithm to statistically low-quality, non-random or highly correlated streams of `random' numbers. Instead, we define a new method for the RNG, called \texttt{prime()}. Similar to Phillips, Anderson and Glotzer's work \cite{Phillips2011}, we apply a strong hash function to an individual's current time and location, and then reset the RNG's internal state accordingly (\ref{background-reproducible-rng}). Crucially, this hash function must satisfy the avalanche effect (\ref{background-avalanche}) so that its hashes are distributed uniformly even when its inputs are highly correlated.

The definition of this separate \texttt{prime()} method has another benefit compared to reseeding. While the latter forgets the initial global seed of the simulation, the \texttt{prime()} method can combine this original seed into the hash. Therefore, this RNG can produce the same stream of random numbers for different, equally-primed individuals in the same simulation but produce different pseudo-random streams across differently seeded simulations.

Just as any other RNG, the environmentally primed RNG has some internal state. A state transition function updates this state to generate a stream of pseudo-random numbers at a particular location-time tuple. However, this internal state is only one part of the effective state of the environmentally primed random number generator. To be precise, the complete state of this RNG consists of its constant original seed, its variable location, its variable time, and its variable inner state. Consequently, this design introduces a non-trivial RNG state update function, which involves jumps in space, and the addition of exponential inter-event times. This location-dependent update function could have a higher resistance to short RNG periods. However, the quality of this state update function heavily depends on the simulation setup. If, for instance, an individual is no longer allowed to disperse, then only its time and internal state remain variable. Therefore, care must be taken that the primeable RNG works even under such extreme conditions. \\

\noindent This section has described how strong, avalanching hashing can be used to prime an individual's random number generator based on its current location and time. If multiple individuals can co-inhabit $(x, y)$, we extend the \texttt{prime()} location to the triple $(x, y, i)$ to distinguish their positions in the deme. We can now ensure that two independent individuals follow the same path after coalescence, as long as they agree on the first event directly following their collision. The following two subsections describe how to generate exponential inter-event times with this method.

\subsection{Aligning the RNGs of Colliding Individuals} \label{independence-align-rng}

Each individual draws the inter-event time $\delta t$ until its next event $t_{next}$ directly after its latest event $t_{last}$. A parent individual draws the time of its birth event $t_{birth(parent)}$ after rewinding its death at $t_{death(parent)}$. A child individual coalesces with this parent at the child-chosen time $t_{birth(child)}$ when the child individual disperses back to its parent's location. At this time, the child individual must draw a matching $\delta t$ such that $t_{next} = t_{birth(parent)}$. Therefore, we need a method to independently sample two continuous inter-event time distributions $X_1, X_2$ at different continuous-time points $t_{death(parent)}, t_{birth(child)}$ such that both give the same next event time.

We approach this problem by simplifying it first. For now, let us divide time into discrete time steps $t_0, t_{-1}, ...$. For instance and without loss of generality, we can only allow times to occur at integer time points $t \in \mathbb{N}^{-}_{0}$. We now also designate an event probability $P(t_i) = p$, which decides whether an event occurs at time point $t_i$. When an individual at time $t_{last} = t_i$ wants to sample the time of its next event, it iterates through $t_{j} \in t_{i-1}, t_{i-2}, ...$, reprimes its RNG at each $t_j$, performs a Bernoulli trial with probability $P(t_j)$, and stops at the first succeeding trial time $t_{next}$. Note that the inter-event times generated by this simple approach are distributed according to the geometric distribution $X \sim 1 + \textrm{Geo}(p) = 1 + \lfloor \textrm{Exp}(-\textrm{ln}(1 - p)) \rfloor$ (\ref{background-exp-geo}).

Why does this method work? Let us assume that the parent and child individual coalesce at time $t_{birth(child)}$. There are now two cases of when the parent's next event $t_{birth(parent)}$ can occur. If $t_{birth(parent)} \geq t_{birth(child)}$, then what we call the parent individual cannot actually be the child's parent. To be specific, it would have already executed its next event, its birth $t_{birth(parent)}$, before the potential collision with the child at $t_{birth(child)}$. This would break the assumption that we are indeed talking about the child and its parent. Therefore, we know that all Bernoulli trials from $t_{death(parent) - 1}$ up to and including $t_{birth(child)}$ must have failed. As we now know that $t_{birth(parent)} < t_{birth(child)}$, both the parent and child individual have to prime and test at times $t_{birth(child) - 1}, ...$. They then both find the same $t_{next} = t_{birth(parent)}$ as the first $t_j < t_{birth(child)}$ at which the Bernoulli trial succeeds. Therefore, both individuals have now independently aligned their next event time and successfully coalesced without any communication.

So far, we have only achieved independent coalescence using geometrically distributed inter-event times. However, the Gillespie algorithm and Moran Model require exponentially distributed inter-event times. The simplistic approach not only provides the wrong distribution of $\delta t$s but can also significantly change the simulation's outcome if we choose large time steps $\Delta t = t_{i} - t_{i-1}$ \footnote{To be specific, geometrically distributed inter-event times can be seen as an intermediate model in between the Moran Model (exponential inter-event times / $\Delta t \rightarrow 0$) and the Wright-Fisher Model (fixed inter-event times / $\Delta t \gg {\lambda_{x, y}}^{-1}$), which were introduced in \cref{background-moran-fisher}}.

Therefore, we now extend the simplistic method to continuous inter-event times. We retain the discrete time steps $t_0, t_{-1}, t_{-2}, ...$ but reinterpret them as bounds for the intervals $..., (t_{-2}; t_{-1}],$ $(t_{-1}, t_{0}]$ which partition $\mathbb{R}^{-}_{0}$. Now, when an individual at time $t_{last} \in (t_{i-1}; t_{i}]$ and location $(x, y)$ samples its next event, it first primes on $(x, y, t_i)$. Next, it searches for its next event inside the same interval. If the next event is not in $(t_{i-1}; t_{i}]$, the individual reprimes on $(x, y, t_{i - 1})$ and tries again inside $(t_{i-2}; t_{i-1}]$. This process is repeated until $t_{next} < t_{last}$ is found.

Since this method also uses discrete time points $t_i$ to align the RNGs of different individuals and bring them into lockstep, its correctness can be shown just as we have done for the simplistic approach above. The only remaining problem is the sampling of exponentially distributed inter-event times based around the repriming at time steps $t_i$, for which the following section presents two mathematically equivalent yet computationally different approaches.

\subsection{Exponential inter-event Times} \label{independence-exponential-generations}

\subsubsection{Memoryless sums of exponentials}

Let us assume that we start at some continuous time point $t_{last} \in (t_{i-1}; t_{i}]$. Suppose we then draw a random variable $X \sim \textrm{Exp}(\lambda_{x, y})$. There are now two cases to analyse. In the base case where $(t_{last} - X) > t_{i-1}$, we have just found the time of the next event $t_{next}$ such that $t_{i-1} < t_{next} <$ $t_{last} \leq t_{i}$. If $(t_{last} - X) \leq t_{i-1}$, we now know that $t_{next} \leq t_{i-1}$ and we can apply the memoryless property (\ref{background-exp-memoryless}). It states that the time left to wait for the next event is unaffected by how long we have already waited. This means that in order to draw the exponentially distributed time between $t_{last}$ and $t_{next}$, we can simply draw a second random variable $X' \sim \textrm{Exp}(\lambda_{x, y})$ and add it to $t_{last} - t_{i-1}$ without changing the distribution of $X$, which now is $X = (t_{last} - t_{i-1}) + X'$. This procedure can be applied recursively to sample $X'$ inside $(t_{i-2}; t_{i}]$ with $t' = t_{i-1}$ until the base case is reached.

What if multiple events occur between $t_{i-1}$ and $t_{i}$, i.e. multiple exponential samples $X_j \sim \textrm{Exp}(\lambda_{x, y})$ have to be summed up to exceed $(t_{i} - t_{i-1})$? For instance, say we have $t_{i-1} < t_{birth(parent)}$ $< t_{last} < t_{death(parent)} \leq t_{i}$, where $t_{last} = t_{birth(child)}$ is the previous event of the coalescing child individual. When the parent searches for $t_{birth(parent)}$ after $t_{death(parent)}$, it first primes at $t_{i}$, the start of its current interval. Next, it samples $X_1 \sim \textrm{Exp}(\lambda_{x, y})$, which produces $t_{i} - X_1 = t_{death(parent)}$. Since $t_{death(parent)} \not < t_{death(parent)}$, it tries again, samples $X_2 \sim \textrm{Exp}(\lambda_{x, y})$ and gets to $t_{i} - X_1 - X_2 = t_{birth(parent)}$. If $t_{i} - X_1 - X_2 < t_{i - 1}$, the parent would have had to restart its search inside $(t_{i-2}, t_{i-1}]$. Since this is not the case, the parent has found its $t_{next} = t_{birth(parent)}$. Now the child individual jumps to the parent's location at $t_{birth(child)}$, which is entirely independent from the parent. It also primes at $t_{i}$, checks $X_1$ and finds $t_{i} - X_1 = t_{death(parent)}$ which is larger than $t_{birth(child)}$ and, therefore, not its next event. Next, it samples $X_2$ and also finds $t_{i} - X_1 - X_2 = t_{birth(parent)} = t_{next}$. Therefore, both the parent and child individual have now coalesced independently with exponentially distributed inter-event times using this process.

\subsubsection{Homogeneous Poisson Point Process}

\begin{figure}[h]
    \centering
    \includegraphics[width=0.9\textwidth]{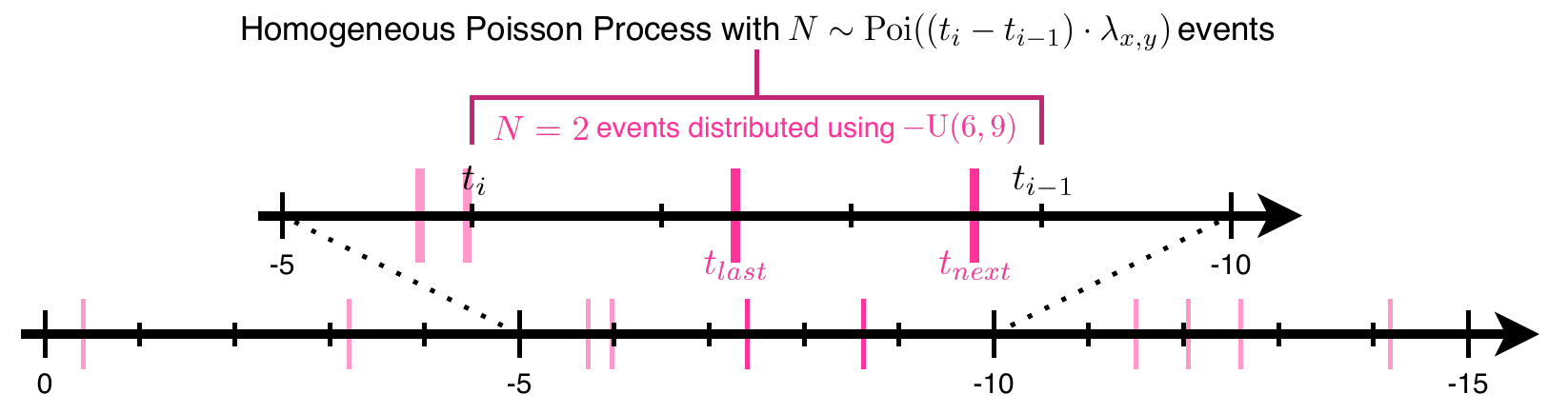}
    \caption{Homogeneous Poisson process over $\mathbb{R}^{-}_{0}$ with $X \sim \textrm{Exp}(0.5)$ inter-event times.}
    \label{fig:exp-poisson}
\end{figure}

In \cref{independence-exponential-inter-event}, we have already established that the distribution of inter-event times at a location $(x, y)$ is equal to the distribution of the local turnover times. We now look at the even stream at a single location, which comes from a homogeneous Poisson point processes (\ref{background-poisson-point-properties}) with rate $\lambda_{x, y}$ and inter-event times $X \sim \textrm{Exp}(\lambda_{x, y})$. \Cref{fig:exp-poisson} shows an example Poisson point process event stream. We now zoom in on the time interval $[6; 9)$, or $(t_{i-1}; t_{i}]$ in general. As the whole process is homogeneous, the events in $(t_{i-1}; t_{i}]$ also come from a homogeneous Poisson point process. Furthermore, there are $N \sim \textrm{Poi}(\lambda_{x, y} \cdot (t_{i} - t_{i-1}))$ events inside the interval (\ref{background-poisson-point-properties}). We also know that these $N$ events are uniformly spread across the interval (\ref{background-poisson-point-properties}). If we now want to find the next event time $t_{next}$ before $t_{last} \in (t_{i-1}; t_{i}]$, we first reprime the RNG at $t_i$ and sample $N$. Next, we distribute $N$ event times uniformly across the interval $(t_{i-1}; t_{i}]$. If there is a $\textrm{max} \{j \mid t_j < t_{last} \}$, we have found our $t_{next}$. Otherwise, we repeat the same process for the interval $(t_{i-2}; t_{i-1}]$.

\noindent This process generates exponentially distributed inter-event times by construction. Inside each interval $(t_{i-1}, t_{i}]$, this holds as the events are distributed uniformly (\ref{background-poisson-point-properties}). Furthermore, we can recombine the independent homogeneous Poisson point processes from all intervals into one global homogeneous Poisson point process (\ref{background-poisson-point-properties}). As inter-event times inside a homogeneous Poisson point process are distributed exponentially, we can conclude that all inter-event times, whether inside one interval or across multiple intervals, are distributed exponentially with rate $\lambda_{x, y}$.

\subsection{Summary of the Independent Algorithm}

This chapter has shown how the ``Next-reaction'' Gillespie method can be used to execute the coalescence algorithm. We have then introduced the novel Independent algorithm, which trades off communication for redundancy by simulating every individual independently whilst maintaining a posteriori analysis consistency. It combines RNG repriming at discrete time steps with either one of the two inter-event time sampling methods described above to generate the time and type of the next event, which is guaranteed to be the same for individuals that have collided.

\begin{minted}[linenos]{python}
def independent_algorithm(individual, rng, landscape):
    while True:
        # Advance the individual to the next event
        individual.last_event_time -= rng.primed_sample_inter_event_time(
            individual.location, landscape
        )

        rng.prime_with(individual.location, individual.last_event_time)

        if rng.sample_random_event(landscape.nu(individual.location)):
            return individual.speciate()

        # Disperse the individual if it has not yet speciated
        individual.location = rng.sample_dispersal(landscape, individual.location)
\end{minted}

\noindent Following this introduction to the Independent algorithm, the next two chapters describe how we have designed, implemented and parallelised the \texttt{necsim-rust} simulation package.

\chapter{The Simulation Architecture Design} \label{architecture}

From the beginning, this project set out to build and improve upon the \texttt{necsim} simulation library (\ref{background-necsim}). While \texttt{necsim} supports very extensive configuration, the C++ library was not designed to be extended or parallelised. Therefore, the core of the simulation has been reimplemented in the Rust simulation package \texttt{necsim-rust} to achieve the following three design goals:

\begin{enumerate}
    \item \textbf{Extensible:} The simulation system should clearly separate functionality into different components, which can easily be swapped out in code to run a different type of simulation. There should also be strong functional guarantees between the different components, which can be further extended to accommodate more restrictive component implementations. Overall, this system should minimise code duplication.
    \item \textbf{Analysable:} The simulation should report any state update to some external analysis code. This analysis must be just as flexible as the simulation itself. The analysis system should support both statically compiled and dynamically loaded analysis routines. Most importantly, any inactive analysis should not hamper the simulation's performance.
    \item \textbf{Parallelisable:} The simulation system should be easily parallelisable. Parallelisation should only require exchanging those components which communicate with the outside world. The system should not impose a particular parallelisation method and should also be extensible.
\end{enumerate}

\noindent The following three sections explain in detail how \texttt{necsim-rust} has been designed to achieve these goals using the expressive and statically typed Rust programming language (\ref{background-rust}).

\section{The Statically Checked Component System} \label{architecture-component-system}

The fundamental coalescence algorithm, which was introduced in \cref{background-neutral-simulation}, can easily be implemented in a succinct and simple procedure. In this project, however, we are considering a whole landscape of simulation algorithm. In particular, this thesis compares four different algorithms, simulates four different scenarios and explores different parallelisation strategies. Therefore, the \texttt{necsim-rust} simulation package defines \textbf{one core component system} architecture, which \textbf{supports all of the different simulation variants}. Different variants are defined by simply swapping in some specialised component implementations, whilst still allowing most code to be reused. The following functional components are defined in the \texttt{necsim-core} package as Rust traits (\ref{background-rust-traits}):

\begin{figure}
    \centering
    \includegraphics[width=1.0\textwidth]{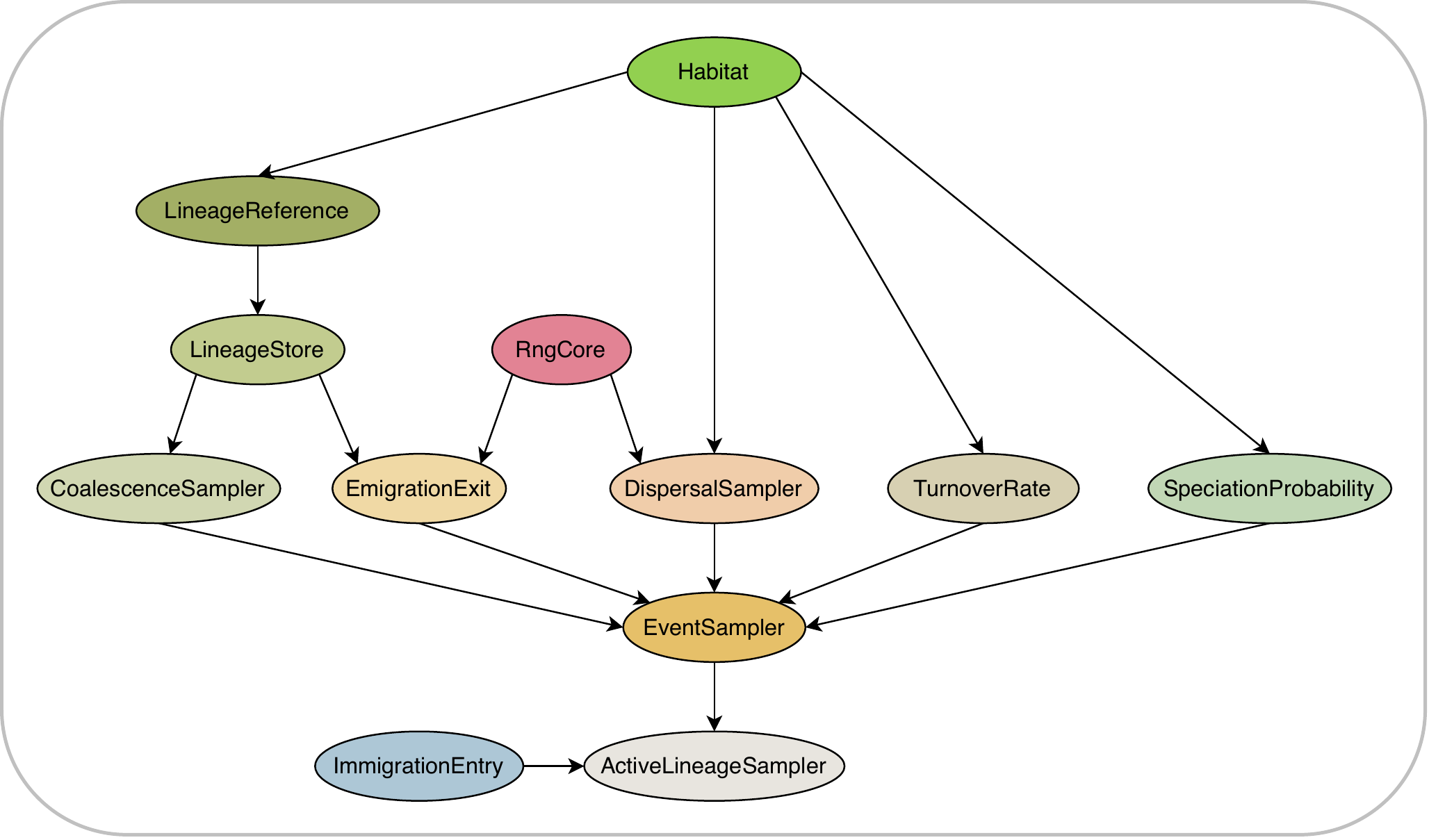}
    \caption{A Hasse-Diagram-like overview of \texttt{necsim-rust}'s component system architecture. Components with no incoming dependency arrows are fundamental, while components further down in the graph are composite and provide higher-level functionality.}
    \label{fig:simulation-components}
\end{figure}

\begin{itemize}
    \item The \texttt{\textcolor{yellow}{\textbf{RngCore}}} trait implements the core functionality to create a new seeded random number generator and draw random \texttt{\textcolor{purple}{u64}}s from it. It also provides common distribution sampling through the \texttt{\textcolor{yellow}{\textbf{RngSampler}}} trait which is automatically implemented on top of \texttt{\textcolor{yellow}{\textbf{RngCore}}}. The core RNG can be extended with the \texttt{\textcolor{yellow}{\textbf{PrimeableRng}}} and \texttt{\textcolor{yellow}{\textbf{SplittableRng}}} traits, which provide RNG priming and substream splitting, respectively.
    \item The \texttt{\textcolor{yellow}{\textbf{Habitat}}} is the foundation of every simulation as it defines which parts of a landscape are habitable and how many individuals can live in the deme at each location.
    \item The \texttt{\textcolor{yellow}{\textbf{TurnoverRate}}} trait returns the location-dependent turnover rate $\lambda_{x, y}$. It must only be queried for habitable locations.
    \item The \texttt{\textcolor{yellow}{\textbf{SpeciationProbability}}} trait returns the location-dependent speciation probability $\nu_{x, y}$. It must only be queried for habitable locations.
    \item The \texttt{\textcolor{yellow}{\textbf{DispersalSampler}}} randomly samples the dispersal from a habitable origin location $(x_o, y_o)$ to a habitable target location $(x_t, y_t)$. Every habitable location must allow dispersal, though it may just be back to the original location $(x_o, y_o)$.
    \item Every individual must have a unique ID in its current local simulation subdomain. The \texttt{\textcolor{yellow}{\textbf{LineageReference}}} trait is a marker trait to specify the data type of these individual IDs. Across subdomains, the globally consistent \texttt{\textcolor{blue}{\textbf{GlobalLineageReference}}} data type is used instead.
    \item The \texttt{\textcolor{yellow}{\textbf{LineageStore}}} stores the set of all individuals which are (currently) being simulated in the local subdomain. As the \texttt{\textcolor{yellow}{\textbf{LineageStore}}} is indexed by the \texttt{\textcolor{yellow}{\textbf{LineageReference}}} type, both types are usually tightly coupled.
    \item The \texttt{\textcolor{yellow}{\textbf{CoalescenceSampler}}} randomly samples the sublocation intra-deme index $i$ to which an individual disperses. Some implementations are also able to determine if the dispersing individual has coalesced with another individual.
    \item The \texttt{\textcolor{yellow}{\textbf{EmigrationExit}}} is a dispersal event sink that consumes individuals who have dispersed out to a different subdomain of the simulation.
    \item The \texttt{\textcolor{yellow}{\textbf{EventSampler}}} randomly samples the next event for an individual by combining the lower-level \texttt{\textcolor{yellow}{\textbf{SpeciationProbability}}}, \texttt{\textcolor{yellow}{\textbf{DispersalSampler}}}, \texttt{\textcolor{yellow}{\textbf{EmigrationExit}}} and \texttt{\textcolor{yellow}{\textbf{Coales-}}} \break \texttt{\textcolor{yellow}{\textbf{cenceSampler}}} components.
    \item The \texttt{\textcolor{yellow}{\textbf{ImmigrationEntry}}} is a dispersal event source that produces individuals who have dispersed in from a different subdomain of the simulation. It is responsible for producing these immigrating individuals in the correct order such that they can be consistently integrated with the local sub-simulation. Together with its counterpart, the \texttt{\textcolor{yellow}{\textbf{EmigrationExit}}}, these two traits can be used to link up different partitions of a landscape through migration.
    \item The \texttt{\textcolor{yellow}{\textbf{ActiveLineageSampler}}} randomly samples the time of the next event and the individual that will execute it. The \texttt{\textcolor{yellow}{\textbf{ActiveLineageSampler}}} is the highest level component, on top of which the generic \texttt{\textcolor{blue}{\textbf{Simulation}}} type and its generic simulation event loop are built.
\end{itemize}

\noindent \Cref{fig:simulation-components} highlights the direct dependencies between these components. If the design goal of this simulation component system had been runtime flexibility, dynamic dispatch could have been used to combine these components at runtime. However, this project aims to build a high-performance simulation in which functional correctness is of higher importance than runtime flexibility. Therefore, the component system uses the Rust type and trait systems (\ref{background-rust-traits}) to its full extent to enable static verification of the compatibility of different component implementations:

\begin{enumerate}
    \item The component traits are generic over the concrete types of the lower-level components. For instance, the \texttt{\textcolor{yellow}{\textbf{DispersalSampler}}} trait is defined as follows:

\begin{minted}[linenos,escapeinside=@@]{rust}
pub trait @\textcolor{yellow}{\textbf{DispersalSampler}}@<H: Habitat, G: RngCore> {
    #[requires(habitat.contains(location), "location is inside habitat")]
    #[ensures(old(habitat).contains(&ret), "target is inside habitat")]
    fn sample_dispersal_from_location(
        &self,
        location: &Location,
        habitat: &H,
        rng: &mut G,
    ) -> Location;
}
\end{minted}

    \item Both the component core package, \texttt{necsim-core}, and specific implementations can define subtraits building on top of the existing component system. By replacing the generic trait dependencies with more specific ones or even concrete types, individual implementations can reify their specific requirements in the type system without changing the core component system. For instance, the \texttt{\textcolor{blue}{\textbf{IndependentActiveLineageSampler}}}, which is used in the novel Independent algorithm, requires that the RNG type \texttt{\textcolor{blue}{\textbf{G}}} implements \texttt{\textcolor{yellow}{\textbf{PrimeableRng}}}, a subtrait of \texttt{\textcolor{yellow}{\textbf{RngCore}}}, and that the \texttt{\textcolor{yellow}{\textbf{LineageReference}}} type  \texttt{\textcolor{yellow}{\textbf{R}}} is equal to \texttt{\textcolor{blue}{\textbf{GlobalLineageReference}}}.

    \item The \texttt{\textcolor{blue}{\textbf{Simulation}}} type is generic over all component types and has the following (unfortunately slightly intimidating) type signature:

\begin{minted}[linenos,escapeinside=@@]{rust}
pub struct Simulation<
    H: Habitat,
    G: RngCore,
    R: LineageReference<H>,
    S: LineageStore<H, R>,
    X: EmigrationExit<H, G, R, S>,
    D: DispersalSampler<H, G>,
    C: CoalescenceSampler<H, R, S>,
    T: TurnoverRate<H>,
    N: SpeciationProbability<H>,
    E: EventSampler<H, G, R, S, X, D, C, T, N>,
    I: ImmigrationEntry,
    A: ActiveLineageSampler<H, G, R, S, X, D, C, T, N, E, I>,
> { ... }
\end{minted}

    \item Since all components express their dependencies in their type signature, and are combined inside the \texttt{\textcolor{blue}{\textbf{Simulation}}} type, the \texttt{\textcolor{blue}{\textbf{Simulation}}} is \textit{correct-by-construction}. Only component combinations which satisfy all the inter-linking trait requirements can be statically composed into a simulation object using the Builder pattern.

    \item Since the entire \texttt{\textcolor{blue}{\textbf{Simulation}}} type and all component traits are generic, the Rust compiler specialises and monomorphises the generic code at compile time. This allows \texttt{rustc} to perform cross-component inlining optimisations, producing code on par with manually constructed special implementations.
\end{enumerate}

\noindent This statically checked component system builds on earlier work on statically checking component interactions \cite{Puntigam2003, Paola2000, Waignier2008, Eide2002} and creating software composition systems which are \textit{correct-by-construction} \cite{Genssler2001}. While these approaches have extended type systems and compilers to verify any inter-component assumption, the simulation system in this project is an implementation based only on the type and trait system of the Rust language. Therefore, it can only check compatibility conditions that the component developers have defined. More general functional correctness checks are instead delegated to Hoare triple specifications, which are specified using the \texttt{contracts} crate \cite{RustContracts} and checked in debug builds dynamically at runtime.

Lastly, it is worth noting that while the statically checked component system offers flexibility and compiler guarantees for the component implementations, any update to the core system is a breaking change that causes compilation errors and requires refactoring of all affected components. Therefore, this design should only be used once the system core has stabilised after prototyping.

\section{The Reporter Analysis System} \label{design-reporter}

The second design goal of the simulation package is to make it easily and efficiently analysable. In particular, the analysis should be separate from the core simulation and its components. Furthermore, analysis procedures should plug in live during the simulation to allow debugging and avoid creating unnecessary intermediate data structures.

\label{design-reporter-filter} The reporter analysis system consists of events and reporters. All events are generated to describe a change in the state of the model that is being simulated. The core idea of the reporter analysis system is to use Rust's type system to help the Rust compiler produce analysis-specific optimised code. In particular, not every analysis requires all events. For instance, we only need to know about speciation events to measure the species richness. By putting this information into the type of an analysis reporter, the simulation can select different code paths at compile time. We can then guarantee that only the required events are ever produced, buffered and communicated. The Rust-provided ability to compile specialised (vs generic) code paths is particularly useful to minimise data storage and movement on resource-constrained platforms such as the GPU. Languages that only support runtime dispatch or require manual inlining, e.g. Java, can only switch to the optimal code path at runtime, which consumes resources unnecessarily. \\

\noindent The remainder of this section goes over the details of this design. The following three types of events can occur in the biodiversity simulation:
\begin{itemize}
    \item \textbf{Dispersal Event:} An individual has dispersed to a new location. Dispersal can include coalescence with the parent of this individual.

\label{design-events} \begin{minted}[linenos,escapeinside=@@]{rust}
pub struct DispersalEvent {
    pub global_reference: GlobalLineageReference, // who? - individual ID
    pub origin: IndexedLocation,    // where? - previous (old) location
    pub target: IndexedLocation,    //        - current (new) location
    pub event_time: NegativeF64,    // when?  - this event
    pub prior_time: NonPositiveF64, //        - this individual's last event
    pub interaction: LineageInteraction, // with? - maybe coalescence parent
}

pub enum LineageInteraction {
    @None@,                                // definitely without coalescence
    Maybe,                               // coalescence undetermined
    Coalescence(@\textcolor{blue}{\textbf{GlobalLineageReference}}@), // definitely with coalescence
}
\end{minted}

    \item \textbf{Speciation Event:} An individual has speciated, thereby originating a new species that all of its descendants are a part of (at least in a reverse-time coalescence simulation). \texttt{\textcolor{blue}{\textbf{Speciation-}}} \texttt{\textcolor{blue}{\textbf{Event}}}s are similar to \texttt{\textcolor{blue}{\textbf{DispersalEvent}}}s but lack the \texttt{target} and \texttt{interaction} fields.

    \item \textbf{Progress:} This event lists the amount of remaining work, i.e. how many individuals have not yet finished simulating.
\end{itemize}

\noindent All of the events are exposed through an event stream which is passed to external user-defined analysis reporters. These reporters must all implement the following (simplified) \texttt{\textcolor{yellow}{\textbf{Reporter}}} trait:

\begin{minted}[linenos,escapeinside=@@]{rust}
pub trait @\textcolor{yellow}{\textbf{Reporter}}@ {
    type ReportSpeciation: Boolean;
    type ReportDispersal: Boolean;
    type ReportProgress: Boolean;

    fn report_speciation(
        &mut self,
        speciation: &MaybeUsed<SpeciationEvent, Self::ReportSpeciation>,
    );

    fn report_dispersal(
        &mut self,
        dispersal: &MaybeUsed<DispersalEvent, Self::ReportDispersal>,
    );

    fn report_progress(
        &mut self,
        remaining: &MaybeUsed<u64, Self::ReportProgress>,
    );
}
\end{minted}

\noindent The \texttt{Reporter} trait also has two optional \texttt{initialise()} and \texttt{finalise()} methods, which are used to set up and clean up any external data structures. Let us now focus on three strongly typed \texttt{report\_*()} methods, which all have similar signatures that encode the following:

\begin{quote}
    The method takes in an \mintinline[escapeinside=@@]{rust}{@\textcolor{blue}{\textbf{MaybeUsed}}@<*Event, Self::Report*>} event which can either be a \mintinline[escapeinside=@@]{rust}{@\textcolor{blue}{\textbf{Used}}@<*Event>} or is an \mintinline[escapeinside=@@]{rust}{@\textcolor{blue}{\textbf{Ignored}}@<*Event>}. The \texttt{Self::Report*} type parameter, which resolves to either a \texttt{\textcolor{blue}{\textbf{False}}} (ignored) or \texttt{\textcolor{blue}{\textbf{True}}} (used) type, specifies which of these two possible parameter types the method takes.
\end{quote}

\noindent It is worth noting that the chosen event parameter type enforces its semantics. While a \mintinline[escapeinside=@@]{rust}{@\textcolor{blue}{\textbf{Used}}@<*Event>} is transparent and can be dereferenced to reveal its internal \texttt{*Event}, the \mintinline[escapeinside=@@]{rust}{@\textcolor{blue}{\textbf{Ignored}}@<*Event>} type is opaque and does not support any operations. \\

\noindent For example, a reporter which measures biodiversity, i.e. counts the number of speciation events whilst ignoring dispersal and progress events, can be implemented as follows:

\begin{minted}[linenos,escapeinside=@@]{rust}
pub struct BiodiversityReporter { biodiversity: u64 }

impl Reporter for BiodiversityReporter {
    @\textbf{\textcolor{red}{\texttt{impl\_report!}}}@(@\textcolor{blue}{\texttt{speciation}}@(&mut self, speciation: Used) {
        self.biodiversity += 1;
    });

    @\textbf{\textcolor{red}{\texttt{impl\_report!}}}@(@\textcolor{blue}{\texttt{dispersal}}@(&mut self, dispersal: Ignored) {});

    @\textbf{\textcolor{red}{\texttt{impl\_report!}}}@(@\textcolor{blue}{\texttt{progress}}@(&mut self, remaining: Ignored) {});
}
\end{minted}

\noindent \label{design-reporter-sorted}The reporters receive all events in order, sorted lexicographically by their \texttt{event\_time} and \texttt{prior\_} \texttt{time}. As the reporters are not part of the simulation component system, they are not part of the type signature of a simulation. However, the \texttt{simulation.simulate(reporter)} method, which is used to run a simulation, is generic over the type of the reporter. Here, the \texttt{reporter} parameter is a mutable reference to any kind of reporter, including statically combined (\texttt{\textcolor{blue}{\textbf{ReporterCombinator}}}), filtered (\texttt{\textcolor{blue}{\textbf{FilteredReporter}}}) or user-defined data types implementing the \texttt{\textcolor{yellow}{\textbf{Reporter}}} trait.

The analysis reporter system is further extended by a plugin system (\texttt{necsim/plugins}). With this system, reporters can be implemented and compiled into separate dynamic libraries, which are then dynamically loaded at runtime. Throughout the simulation, a proxy reporter (\texttt{\textcolor{blue}{\textbf{Reporter-}}} \texttt{\textcolor{blue}{\textbf{PluginVec}}}) is used to forward all events to the user-loaded analysis routines. This feature, in particular, together with the static event production optimisation, has been invaluable in quickly developing and launching the analyses in \cref{analysis-evaluation}.

\section{The Parallelisation Architecture} \label{design-parallelism}

The third design goal of the simulation package is to make it easily parallelisable. There are two types of parallelisation the core simulation system could support: internal and semi-internal parallelism. For internal parallelism, every component in the simulation system would have to be parallelised individually. While this approach can support specific platforms or use cases, it also reduces the reusability of the components and increases code duplication. Instead, the simulation system is designed to be semi-internally parallelised:

\begin{itemize}
    \item The core system is sequential and requires no internal knowledge of parallelisation.
    \item Two components, the \texttt{\textcolor{yellow}{\textbf{EmigrationExit}}} and \texttt{\textcolor{yellow}{\textbf{ImmigrationEntry}}}, are designated as adapters between the local sub-simulation and the outside world.
    \item The simulation can be parallelised by partitioning the domain of the model. For instance, all individuals currently inhabiting a designated part of the landscape are assigned to one specific partition.
    \item The partitioning system provides the architecture to build parallelisation algorithms that link multiple simulation systems together through inter-partition migration.
\end{itemize}

\noindent The partitioning system is represented in code by the \texttt{\textcolor{yellow}{\textbf{Partitioning}}} and the \texttt{\textcolor{yellow}{\textbf{LocalPartition}}} traits. The former provides the entry point to instantiate the partitioning backend. It can then be queried for global information such as the number of partitions and the rank of the local partition. Most importantly, the \texttt{\textcolor{yellow}{\textbf{Partitioning}}} also provides a method to instantiate its corresponding \texttt{\textcolor{yellow}{\textbf{LocalPartition}}} type. The \texttt{\textcolor{yellow}{\textbf{LocalPartition}}} trait provides the functionality that is used to link different partitions together and implement various parallelisation algorithms:

\begin{itemize}
    \item The \texttt{get\_reporter()} method provides access to the partitioning-specific reporter. While an implementation might simply return the reporter built from the \texttt{\textcolor{yellow}{\textbf{ReporterContext}}} here, it can also communicate and combine events from collaborating partitions.
    \item The \texttt{migrate\_individuals()} method allows different partitions to exchange individuals that have moved to a different subdomain of the simulation.
    \item The \texttt{reduce\_vote\_continue()} method is an example of one of multiple helper methods which allow the partitions to collaborate by voting on global information.
    \item The \texttt{wait\_for\_termination()} method ensures that each local partition can only complete its part of the simulation when it is safe to do so. In particular, it must ensure that no migrating individuals are lost because of data races, and that there is no state in which the simulation can get stuck because of a rogue partition.
\end{itemize}

\noindent In combination, the \texttt{\textcolor{yellow}{\textbf{LocalPartition}}} trait and the \texttt{\textcolor{yellow}{\textbf{EmigrationExit}}} and \texttt{\textcolor{yellow}{\textbf{ImmigrationEntry}}} components provide the architecture for semi-internal parallelisation. \Cref{implementation} explains in detail how this system is used to implement several message passing parallelisation algorithms.

\chapter{Implementation and Parallelisation} \label{implementation}

\Cref{independence} and \cref{architecture} have introduced the novel Independent algorithm and the fundamental architectural design of the \texttt{necsim-rust} simulation package. This chapter introduces the implementation and parallelisation strategies that we evaluate in \cref{analysis-evaluation}. In particular, \cref{implementation-independent} explains how the Independent algorithm can be used as a replacement for the traditionally monolithic \texttt{necsim} and Gillespie algorithms, which require global knowledge. Next, \cref{implementation-mpi} describes how MPI can be used as a partitioning backend, while \cref{implementation-parallelisation} introduces the parallelisation strategies we have implemented. \Cref{implementation-cuda} shows how the embarrassingly parallel nature of the Independent algorithm can be exploited on the GPU. Finally, the \texttt{rustcoalescence} command-line tool, which allows the user to launch the most common simulations, is presented in \cref{implementation-rustcoalescence}.

\section{Implementing the Independent Algorithm} \label{implementation-independent}

The Independent algorithm is embarrassingly parallel and trades dependencies between individuals for redundant simulation work. While the traditional monolithic algorithms simulate all individuals simultaneously using globally consistent knowledge, the Independent algorithm simulates every individual on their own. This fundamental difference creates the following challenges:
\begin{enumerate}
    \item \textbf{Redundancy:} Since the Independent algorithm does not detect coalescence internally, it simulates both the child and parent individual after their collision. We must minimise this duplication of work to allow the Independent algorithm to compete at all.
    \item \textbf{Event Ordering:} Since every individual is simulated on its own, events from different individuals are \textbf{not} produced in sorted order. However, in \cref{design-reporter-sorted} we have guaranteed that analysis reporters receive events in sorted order. Therefore, we must sort all events first.
    \item \textbf{Coalescence:} The monolithic algorithms report both move and coalescence events as dispersals. Since the Independent algorithm does not detect coalescence internally, it cannot provide this distinction. Still, we must design the algorithm to allow any analysis reporters to reconstruct coalescence events from the sorted event stream.
\end{enumerate}

\subsection{Deduplicating Redundant Individuals} \label{implementation-deduplication}

To simulate multiple individuals, the Independent algorithm is applied to each of them. Instead of simulating each individual in one go, they can also each be simulated a few steps at a time:

\begin{minted}[linenos,escapeinside=@@]{python}
def simulate_independent(remaining_individuals, landscape, rng, reporter):
    while @\textcolor{green}{Some}@(individual) = remaining_individuals.pop():
        if not independent_algorithm(
            individual, rng, landscape, reporter, early_stop=(steps >= 10)
        ).has_speciated():
            remaining_individuals.append(individual)
\end{minted}

\noindent In this outer loop, we do know about the other individuals that we are simulating locally. Therefore, we can deduplicate individuals that have already coalesced in this loop. Since redundant work does not affect the correctness of the algorithm, we can use a deduplication mechanism that does not detect all duplicates immediately. To be precise, we use a fixed-size direct-mapped cache to check if any other individual has recently followed the same trajectory.

After $N$ steps, different individuals have sampled their $N^{th}$ event at very different time points. Therefore, checking for duplication of the individuals themselves would be very inefficient. Instead, we use an individual's minimum speciation sample as a cheaper measure of duplication. This minimum speciation sample is the triple $(s_{min}, t_{min}, (x, y, i)_{min})$ of the minimum random number $s_{min} \sim \textrm{U}(0, 1)$ that the individual has drawn to check for speciation, as well as the time and location at which $s_{min}$ was drawn. Since an individual's location and time entirely determine the random numbers generated in the Independent algorithm, a parent and child individual that have coalesced eventually draw the same minimum speciation sample. The minimum is reset periodically to reduce the time until the next minimum is drawn and duplication is detected.

\subsection{Sorting Events using the Water-Level Algorithm} \label{implementation-water-log}

\begin{figure}[h]
    \centering
    \includegraphics[width=1.0\textwidth]{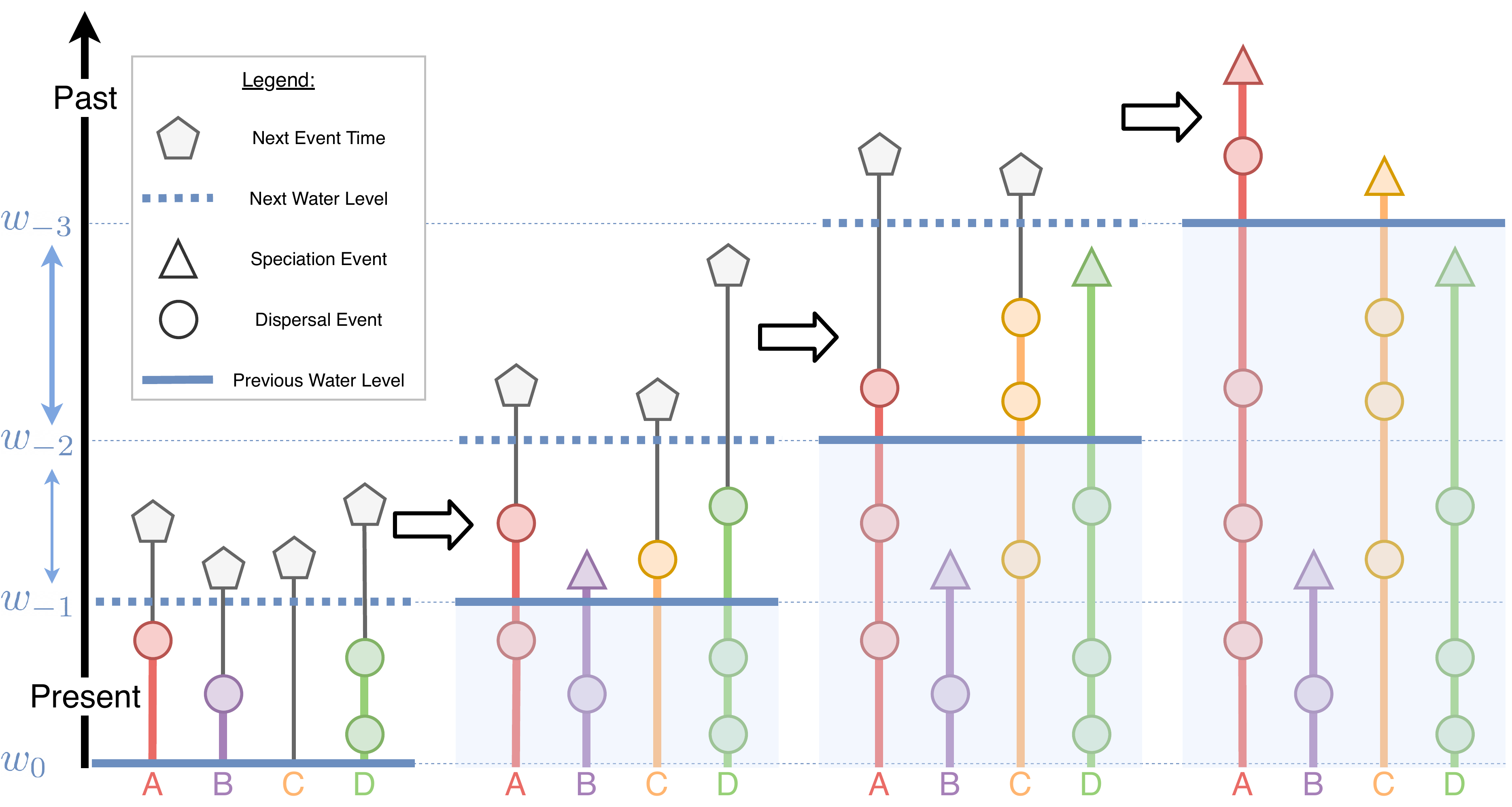}
    \caption{The Water-Level algorithm ($W = 4$) incrementally produces events in sorted order.}
    \label{fig:independent-water-level}
\end{figure}

In the Independent algorithm, every individual produces their events in sorted order. Across individuals, we need to reimpose this ordering manually. We might na\"{i}vely produce all events, store them in a buffer, sort them and finally report them. However, this approach is only practical for a small number of events, i.e. a small number of individuals and a high speciation probability $\nu$. Instead, we use the incremental Water-Level algorithm shown in \Cref{fig:independent-water-level} and \Cref{water-level-algorithm}. This algorithm uses the total event rate of all remaining individuals to predict how far the simulation should advance to generate roughly $W$ events during the next iteration. Then, \textbf{all} events that occur between the previous and this new water level time, $w_i \geq t_{event} > w_{i-1}$, (\textbf{and only those}) are produced, buffered, sorted, and reported. As the invariant `all events below the water level $w_i$ have already been reported' is maintained throughout, all events are reported in sorted order.

\noindent A partitioned simulation provides a similar challenge. While the different partitions could communicate their events and sort and report them on a root node, this would incur a significant communication overhead. Furthermore, partitions would then no longer be simulated in complete isolation without communication, a new feature that the novel Independent algorithm adds. Therefore, partitioned simulations instead write their events to an event log on disk, consisting of multiple internally sorted files. Note that the static analysis event filter described in \cref{design-reporter-filter} is used to reduce the number of events that are saved to disk. The combined global event stream can then be analysed in a subsequent event replay, which merge-sorts the different event logs.

\subsection{Reporting Coalescence Events Independently} \label{implementation-events}

The Independent algorithm does not detect coalescence events. However, analysis reporters that were written for the traditional monolithic algorithms should also work with the Independent algorithm with minimal changes. Therefore, we have designed the Independent algorithm to provide the following guarantees, allowing coalescence to be detected in the event stream:

\begin{enumerate}
    \item As events are reported in sorted order, duplicate events from duplicate individuals are directly adjacent in the event stream. Only individuals who have coalesced report duplicate events.
    \item Therefore, event reporters can deduplicate the event stream in constant space and time by remembering the previous event in the stream.
    \item The deduplication mechanism described in \cref{implementation-deduplication} can only detect duplicate individuals once both have sampled the next event. We guarantee that this event, which directly follows coalescence, is always reported for both the parent and child individual.
    \item If events were sorted by the time of the next event, coalescence could be detected immediately. A dispersal event with coalescence would then be adjacent to the parent's prior event as both events would share the time of the next event, the parent's birth. However, this sorting order would be unintuitive, require buffering one event per individual, and would unfairly disadvantage the monolithic algorithms, which produce events in natural sorted order.
    \item Events are sorted lexicographically by event time first and prior event time second. Amongst the duplicates of the event that follows coalescence, the parent's duplicate comes before the child's. Therefore, event reporters can determine the parent and time of the coalescence in constant space and time by remembering the first event in the current stream of duplicates.
    \item Duplicate events, which come more than one event after coalescence, all have the same prior event time, which clearly identifies them as events without coalescence.
\end{enumerate}

\section{The MPI partitioning backend} \label{implementation-mpi}

The simulation partitioning system in \texttt{necsim-rust} is primarly designed for message-passing parallelism. Specifically, the \texttt{\textcolor{yellow}{\textbf{LocalPartition}}} trait, which must be implemented by a partitioning backend, connects the \texttt{\textcolor{yellow}{\textbf{EmigrationExit}}} and \texttt{\textcolor{yellow}{\textbf{ImmigrationEntry}}} components of different parallel partitions. We use the \textbf{M}essage \textbf{P}assing \textbf{I}nterface from \cref{background-mpi} as an example backend.

The MPI partitioning backend directly maps the process ranks inside the MPI universe into simulation partition ranks, which collaborate on the simulation. For reporting, the process with rank $0$ is designated as the root process. Every process uses a local event log to write its events to disk. Suppose now that we know at compile time that the analysis reporters require progress events. In this case, non-root processes use immediate (non-blocking) point-to-point sends to periodically forward their local progress to the root. The root process uses immediate probing receives to capture and combine the progress updates, which it then forwards live to the reporter. Crucially, no live reporting occurs on the non-root processes, and progress events are only reported live on the root node. It is also worth noting that the simulation internally keeps a migration balance to ensure that the global progress is not affected by migrations between processes.

Every process has one immigrating individual inbox and one outgoing emigrating individual postbox for every other partition. The processes use immediate synchronous point-to-point messages to send emigrating individuals to their target process. We use non-blocking sends with synchronous receipts to ensure that the migration target process must have received the immigrating individuals before the origin process can send the next ones, whilst also not blocking on these message sends. The synchronous receipts are essential to avoid losing migrating individuals in race conditions. Migrations are also rate-limited and are sent in batches, both to improve performance. However, any parallelisation algorithm can also force the immediate delivery of migrating individuals, e.g. when there is no local work left to do.

The synchronisation between the different simulation partitions is a crucial feature in most parallelisation algorithms. The MPI partitioning backend translates the global voting methods provided by the \texttt{\textcolor{yellow}{\textbf{LocalPartition}}} trait into synchronous universe-wide reduction operations. A similar approach is used to implement the \mintinline{rust}{wait_for_termination()} method:
\begin{itemize}
    \item A partition must wait if there are any unprocessed outgoing or incoming migrating individuals. The \texttt{\textcolor{yellow}{\textbf{LocalPartition}}} does not synchronise in this case.
    \item A partition asynchronously votes to wait iff it has sent off emigrating individuals to another partition since the last vote to terminate. This nay vote avoids the following race condition: \begin{enumerate}
        \item Partition $P_{target}$ finishes its local work and votes to terminate.
        \item $P_{origin}$ sends an emigrating individual off to $P_{target}$.
        \item $P_{target}$ receives the immigrating individual from $P_{origin}$.
        \item $P_{origin}$ finishes its local work and votes to terminate.
        \item The vote to terminate succeeds, and $P_{origin}$ quits the simulation.
        \item $P_{target}$ still runs, then tries to send off an emigrating individual back to $P_{origin}$.
        \item Since $P_{origin}$ has already terminated, the send fails, and the simulation is deadlocked.
    \end{enumerate}
    \item The method only allows a partition to terminate if all partitions have voted to terminate in this round, no individuals are still in flight, and there were no migrations during this round.
\end{itemize}

\section{The Simulation Parallelisation Strategies} \label{implementation-parallelisation}

\subsection{The Monolithic Algorithms} \label{implementation-monolithic-parallelisation}

This project implements and compares three different monolithic algorithms: the \textbf{Classical} coalescence algorithm (\ref{background-neutral-simulation}) that \texttt{necsim} (\ref{background-necsim}) implements, and the ``Next-Reaction'' \textbf{Gillespie} and \textbf{SkippingGillespie} algorithms described in \cref{independence-coalescence-gillespie} and \Cref{gillespie-algorithm}. The \textbf{SkippingGillespie} algorithm skips no-coalescence dispersal events within the same deme. All three of these algorithms use a consistent global state to check for coalescence, which must be synchronised across parallel partitions. The \textbf{Classical} coalescence algorithm is shown below for reference:

\begin{minted}[linenos]{python}
def initialise_classical(simulation):
    return simulation.landscape.generate_current_population()

def simulate_classical(individuals, time, landscape, rng, reporter):
    while len(individuals) > 0:
        # Sample the time of the next event (assuming a constant event rate)
        time -= rng.exp(landscape.lambda * len(individuals))

        # Sample the next individual uniformly (event rate is homogeneous)
        individual = individuals.remove(rng.randint(len(individuals)))

        # Sample and report the individual's next event
        if rng.sample_random_event(landscape.nu(individual.location)):
            reporter.report(time, individual.speciate())
        else:
            individual.disperse(landscape, rng)

            parent = landscape.individual_at(individual.location, individuals)

            # Check for coalescence with another individual
            if parent is not None:
                reporter.report(time, individual.coalescence(parent))
            else:
                reporter.report(time, individual.move())

                individuals.append(individual)
\end{minted}

\subsubsection{The Lockstep Strategy} \label{implementation-monolithic-lockstep}
The simplest monolithic parallelisation strategy is pessimistic and runs all partitions in sequential lockstep (e.g. to split RAM usage). Specifically, every partition peeks ahead in time by one local event. Then, all partitions synchronously vote on the global next event. Only the partition with this next event is allowed to perform the next step. Afterwards, all partitions synchronously participate in the possible migration following this step. The entire process repeated until none of the partitions has any individuals left to simulate. Please refer to \Cref{monolithic-lockstep-algorithm} for a detailed implementation.

\subsubsection{The Optimistic Strategy} \label{implementation-monolithic-optimistic}

The Lockstep strategy is pessimistic as it assumes that there is a migration between partitions at every step. However, the opposite can be assumed as well. Our Optimistic strategy is inspired by Bauer's examination \cite{Bauer2015} of how individual Gillespie simulation steps can be executed optimistically in parallel on multiple cores (\ref{background-gillespie-hpc}). However, this implementation expands the idea of exploiting the potential independence between partitions to larger time frames. For instance, every partition might be an island far away from all other habitat. In that case, there would actually be very few or no migrations between partitions. Therefore, it should be much more efficient to run the partitions independently for longer periods.

The Optimistic strategy, as shown in \Cref{monolithic-optimistic-algorithm}, takes this gamble and bets that no migrations will occur in the time interval $s_i \geq t_{event} > s_{i-1}$. Every partition simulates independently from $s_{i}$ until $s_{i-1}$, when migrations are finally communicated. If there were no migrations, the bet has paid off, and the process is continued for $(s_{i-2}; s_{i-1}]$. However, if migrations did occur, all partitions must roll their simulations back to $s_i$. Then, every partition bets that \textbf{only} the already known migrations will occur in the time interval. The process is then repeated until the bet finally succeeds. It is worth noting that this strategy has to roll back once for every migration that occurs.

\subsubsection{The Optimistic Lockstep Strategy} \label{implementation-monolithic-optimistic-lockstep}

This strategy provides a compromise between pessimistic and optimistic parallelisation. Every partition independently simulates until its first emigration event. Then, all partitions vote on the first global emigration. Every partition whose emigration would come later in the event stream first rolls back. Then, it advances its rollback state to just before the next global emigration event. While this strategy also requires one rollback for every migration, it checks its bets and advances its safe rollback state more frequently than the purely optimistic strategy.

\subsubsection{The Averaging Strategy} \label{implementation-gillespie-averaging}

The Gillespie algorithm was originally designed for simulating well-mixed chemical reactants. Ridwan, Krishnan and Dhar parallelise Gillespie's ``Direct'' Method \cite{Ridwan2004} by (1) partitioning the simulation domain and (2) periodically averaging over all partitions to restore the well-mixedness (\ref{background-gillespie-hpc}). In a biodiversity simulation, however, individuals are not necessarily uniformly distributed across space and may show location-dependent behaviour. Therefore, we cannot simply globally average out individuals. However, we can average out inter-partition migrations if the partitions are assumed to be closed systems in between the averaging time points:

\begin{enumerate}
    \item[(a)] Dispersal probabilities are renormalised such that dispersal only occurs within each partition, making them closed systems. Some randomly picked individuals then randomly migrate to different partitions at every averaging point to account for this change to the dispersal distribution. However, this approach is challenging to implement. Which individuals migrate when from where to which other partition can be location-dependent and must support any arbitrary dispersal kernel. Furthermore, this method underestimates biodiversity, as coalescence events occur more frequently in the confined closed partitions.
    \item[(b)] In the second approach, emigrations occur as usual. However, their corresponding immigration events are delayed and pushed together in time such that they only occur at the averaging time points. In practice, individuals enter a metaphysical state during emigration, which they only exit at the next averaging point. As coalescence events cannot occur to these emigrating individuals, this method overestimates biodiversity. However, this approach does use the correct location-dependent dispersal probabilities and is easier to implement.
\end{enumerate}

\noindent For simplicity, we have only implemented and analysed the second option in this project.

\subsection{The Independent Algorithm} \label{implementation-independent-parallelisation}

In the Independent algorithm, every individual is simulated independently. In particular, the trajectory of every individual is independent of all other individuals. It can be perfectly reproduced irrespective of when or how often it is computed. Therefore, partitions are not required to synchronise with each other or communicate at all. However, if we introduce some communication between partitions, we can use the deduplication cache described in \cref{implementation-deduplication} to deduplicate individuals across partitions. This trade-off between redundancy and communication is explored in several parallelisation strategies for the Independent algorithm, which all share the same outer structure:

\begin{minted}[linenos]{python}
def simulate_independent_parallel(landscape, rng, parallelism, event_log):
    # Generate only the individuals assigned to this partition
    individuals = landscape.generate_current_population(parallelism.rank())

    # Synchronise and loop while the simulation has not finished globally
    while not parallelism.all_done(len(individuals)):
        while len(slow_individuals) > 0:
            simulated_individual = independent_algorithm(
                individuals.pop(), rng, landscape,
                reporter=event_log,
            )

            # Perform migration between the simulation's partitions
            if simulated_individual.has_emigrated():
                parallelism.emigrate(simulated_individual)

            for immigrant in parallelism.immigrants():
                individuals.append(immigrant)
\end{minted}

\noindent This project implements the following semi-internal parallelisation strategies:

\begin{enumerate}
    \item \textbf{Individuals:} The set of individuals is partitioned with no regard to their starting locations. Since every partition is assigned a fixed subset of individuals, no communication between partitions occurs (apart from bookkeeping like progress reporting to the root process).
    \item \textbf{Landscape:} The landscape is partitioned such that every partition is responsible for modelling the individuals living in a spatially coherent \cite{Harlacher2012} subset of the landscape. When individuals disperse between locations assigned to different partitions, they also have to migrate between sub-simulations. In contrast to the monolithic parallelisation algorithms, the partitions are not required to perform the migrations of the independent individuals in any particular order, which reduces the amount of synchronisation that is needed.
    \item \textbf{Probabilistic:} The landscape is partitioned like in the \textbf{Landscape} variant, and every location is assigned to a particular partition. However, each emigrating individual only emigrates to a different sub-simulation with probability $p$. This parameter $p$ allows us to explore the trade-off between communication and redundancy and analyse hybrid methods. If $p = 1$, this approach is equivalent to the \textbf{Landscape} variant, while it is similar to the \textbf{Individuals} variant if $p = 0$. Note that the choice of whether to physically migrate to a different sub-simulation is deterministic. Therefore, duplicate individuals all physically migrate simultaneously.
\end{enumerate}

\noindent Additionally, this project also offers two external parallelisation strategies: \textbf{IsolatedIndividuals} and \textbf{IsolatedLandscape}. These strategies initially partition the set of all individuals just like their namesakes. However, after initialisation, every individual remains in the sub-simulation it was originally assigned to, and no physical migration occurs. Both variants are parameterised by the number of partitions and rank of the sub-simulation that is to be simulated. Therefore, they can be employed when communication between different partitions is impossible, e.g. in a batch system.

\section{The Independent Algorithm on the GPU} \label{implementation-cuda}

We have designed the Independent algorithm to be embarrassingly parallel. Therefore, it directly maps onto the data parallel SIMD architecture of GPUs: every individual is simulated independently on a separate CUDA thread. This method can be seen as both internal and external parallelism: Every thread simulates an entirely independent single-individual simulation, all of which are part of the same model. The CPU is only used to launch batches of individuals, forward events from the GPU to reporters on the CPU, and perform the individual deduplication. \\

\noindent Rust support for CUDA is still in its infancy. \Cref{background-cuda-kernel} has introduced the existing solutions to implement and launch CUDA kernels, all of which require a significant amount of \mintinline{rust}{unsafe} Rust code. However, we want to ensure that (most) component implementations written for the CPU can also be safely used on the GPU. Therefore, we have developed a new Rust crate called \texttt{rust-cuda} as part of this project, which allows us to safely build a simulation on the CPU, safely transfer it to the GPU, and safely get the results back for reporting on the CPU. Specifically, \texttt{rust-cuda} provides the automatically derivable \texttt{\textcolor{yellow}{\textbf{RustToCuda}}} and \texttt{\textcolor{yellow}{\textbf{CudaAsRust}}} traits which enable the safe \textbf{transfer} of a Rust data structure with heap allocations, e.g. the simulation, from the CPU to the GPU. This trait internally takes care of all memory allocation and movement. The \texttt{\textcolor{yellow}{\textbf{LendToCuda}}} and \texttt{\textcolor{yellow}{\textbf{BorrowFromRust}}} traits can then be used to safely \textbf{share} a data structure between the CPU and GPU. One crucial aspect of Rust's memory safety is its mutability and borrowing system, which is explained in \Cref{appendix:rust-safety}. \texttt{rust-cuda} extends the borrowing system to the GPU:
\begin{enumerate}
    \item Data structures that the CPU immutably lends to CUDA are also immutable on the GPU.
    \item Data structures that the CPU mutably lends to CUDA are also mutable on the GPU:
    \begin{enumerate}
        \item Every thread gets its own mutable shallow copy of the data structure to avoid aliasing conflicts. Shallow updates \textbf{are not} communicated back to the CPU.
        \item Any heap-allocated parts of the data structure are shared mutably between all threads on the GPU. Therefore, the kernel code is only safe iff there are no race conditions with any aliased accesses. Heap updates \textbf{are} communicated back to the CPU.
    \end{enumerate}
    \item \texttt{rust-cuda} also provides a special helper \texttt{\textcolor{blue}{\textbf{CudaExchangeBuffer}}}\mintinline{rust}{<T>} struct which is optimised for frequent data movement between the CPU and GPU. Data structures containing this type are wrapped inside a \texttt{\textcolor{blue}{\textbf{ExchangeWithCudaWrapper}}}\mintinline{rust}{<T>} struct, which uses a finite state machine of types to guarantee that the data is only owned by either the CPU or GPU.
\end{enumerate}

\noindent \textbf{Remark:} Generic types and functions are a crucial feature of Rust (\ref{background-rust-generic}). At the time of writing, the Rust compiler only specialises generic functions across CPU libraries but does not yet support generic GPU kernels. However, this project's simulation architecture has been designed around a generic component system to easily support any combination of simulation scenarios, algorithms and component implementations. As a workaround, this project uses a custom linker wrapper that detects the specialised kernel types that the CPU code requires and coordinates the CUDA kernel compilation to produce the requested specialised versions. This system could be integrated with \texttt{rust-cuda} in future work and provide support for safe generic CUDA kernels in Rust. \\

\noindent CUDA threads in the same thread block have to share resources amongst each other (\Cref{appendix:cuda-recources}). In this project, the high register usage of the simulation kernel limits the maximum launch capacity of the GPU. However, many of these registers just store simulation parameters which are known at launch time and constant throughout the kernel's execution. These constants could be placed into constant memory by manually playing compiler and adapting the kernel code. Instead, we have implemented an optional lightweight \textbf{J}ust-\textbf{I}n-\textbf{T}ime compilation step which inserts the launch-time constants into the existing PTX code. When a CUDA kernel is launched, the CUDA driver first compiles the kernel from PTX assembly to the GPU-specific binary. It is then able to perform constant propagation, reduce the register usage, and increase the launch capacity:

\begin{table}[h]
\begin{tabular}{l|l|l|l|}
\cline{2-4}
scenario / reporting                    & \multicolumn{1}{|c|}{\textbf{progress events only}}           & \multicolumn{1}{|c|}{\textbf{+speciation events}}             & \multicolumn{1}{|c|}{\textbf{all events types}}              \\ \hline
\multicolumn{1}{|c|}{\textbf{NonSpatial}}        & \multicolumn{1}{|c|}{$70(896) \rightarrow 54(1024)$} & \multicolumn{1}{|c|}{$70(896) \rightarrow 54(1024)$} & \multicolumn{1}{|c|}{$75(768) \rightarrow 54(1024)$} \\ \hline
\multicolumn{1}{|c|}{\textbf{SpatiallyImplicit}} & \multicolumn{1}{|c|}{$77(768) \rightarrow 53(1024)$} & \multicolumn{1}{|c|}{$77(768) \rightarrow 53(1024)$} & \multicolumn{1}{|c|}{$87(640) \rightarrow 56(1024)$} \\ \hline
\multicolumn{1}{|c|}{\textbf{AlmostInfinite}}    & \multicolumn{1}{|c|}{$68(896) \rightarrow 57(1024)$} & \multicolumn{1}{|c|}{$68(896) \rightarrow 57(1024)$} & \multicolumn{1}{|c|}{$74(768) \rightarrow 56(1024)$} \\ \hline
\multicolumn{1}{|c|}{\textbf{SpatiallyExplicit}} & \multicolumn{1}{|c|}{$88(640) \rightarrow 54(1024)$} & \multicolumn{1}{|c|}{$88(640) \rightarrow 54(1024)$} & \multicolumn{1}{|c|}{$90(640) \rightarrow 60(1024)$} \\ \hline
\end{tabular}
\caption{Register usage (and maximum number of threads per block) without and with PTX JIT. Results are device-dependent and were generated on a GeForce GTX 1080 with CUDA 11.2.}
\label{table:implementation-cuda-jit}
\end{table}

\section{Command-Line Interface} \label{implementation-rustcoalescence}

In addition to the the \texttt{necsim-rust} simulation library, we have also implemented a user-friendly command-line interface, which is called \texttt{rustcoalescence}. This front-end initialises the most common simulation scenarios and algorithms using the component system provided by \texttt{necsim-rust}.

\texttt{pycoalescence} and the \texttt{necsim} library use the GDAL geospatial library \cite{GDAL} to load spatially-explicit GeoTif map files. However, GDAL requires a non-trivial installation process that does not fit into Rust's user-friendly dependency management system \texttt{cargo}. Therefore, \texttt{rustcoalescence} avoids this heavy dependency and instead reimplements the required functionality to load habitat and dispersal map files. It also provides different map loading modes to patch up minor inconsistencies and improve compatibility with \texttt{necsim}'s weaker validation checks.

\texttt{rustcoalescence}'s most significant feature is the configuration options it exposes to the end user. These options configure the provided algorithms, load analysis reporters as dynamic plugins, launch partitioned simulations and replay existing event logs. A traditional one-dimensional command line interface proved to be insufficient to represent all options. Instead, \texttt{rustcoalescence} accepts a configuration string written in the strongly typed \textbf{R}usty \textbf{O}bject \textbf{N}otation \cite{RON} format:

\begin{minted}[escapeinside=@@]{rust}
@:\textcolor{blue}{~}\textdollar@ @\textbf{rustcoalescence}@ simulate @'@(
    speciation: 0.01,
    sample: 1.0,
    seed: 42,

    algorithm: Independent(step_slice: 10),

    scenario: SpatiallyExplicit(
        habitat: "maps/madingley/fg0size12/habitat.tif",
        dispersal: "maps/madingley/fg0size12/dispersal.tif",
    ),

    reporters: [ @\textcolor{blue}{\textbf{Plugin}}@(
        library: "libnecsim_plugins_common.so",
        reporters: [ @\textcolor{blue}{\textbf{Progress}}@(), @\textcolor{blue}{\textbf{Counter}}@(), @\textcolor{blue}{\textbf{Biodiversity}}@() ],
    ) ],
)@'@
\end{minted}

\noindent The configuration format is directly defined by the scenarios and algorithms and their strongly typed parameters. For instance, the speciation probability $\nu$ is of the reified type \texttt{\textcolor{blue}{\textbf{ClosedUnitF64}}}, which can only store values in $[0.0; 1.0]$. When \texttt{rustcoalescence} parses the configuration, all parameters are immediately fully validated by the type initialisation code. This design separates the responsibility of parameter definition and validation from the CLI and reduces code duplication. Please refer to \Cref{appendix:rustcoalescence-examples} for more examples of how to use \texttt{rustcoalescence}.

\chapter{Evaluation} \label{analysis-evaluation}

This chapter analyses and evaluates the different algorithms that we have implemented in \texttt{necsim-} \texttt{rust}. First, \cref{analysis-implementation-correctness} and \cref{analysis-functional-correctness} cover the correctness checks that \texttt{necsim-rust} has undergone. In particular, \cref{analysis-rng} evaluates the quality of the stream of random numbers that the primeable PRNG (\ref{independence-hashing-rng}) produces in the Independent algorithm. Second, \cref{analysis-event-performance} analyses the performance of the event generation, while \cref{analysis-algorithm-sweetspot} finds sweet spot parameters for all algorithms. Fourth, the scalability of all five algorithms and their parallelisation strategies is evaluated in \cref{analysis-algorithm-scalability}. Finally, \cref{analysis-limit} explores how much further this project has pushed the limit of neutral biodiversity simulations. We have provided all analysis scripts, results and (many more) graphs in the \texttt{analysis/} folder of this project.

\section{Implementation Correctness} \label{analysis-implementation-correctness}

The Rust programming language was purposefully chosen for this project as it provides additional safety guarantees and simplifies verification. We have employed the compiler to check assumptions in our simulation component system (\ref{architecture-component-system}) and reified assumptions types such as \texttt{\textcolor{blue}{\textbf{PositiveF64}}}.

At the time of writing (commit \href{https://github.com/MomoLangenstein/necsim-rust/tree/2e352ff}{\#2e352ff}), the core architecture of \texttt{necsim-rust}, called \texttt{necsim-} \texttt{core}, contains 26 preconditions and 59 postconditions. Additionally, the component implementations contain 37 preconditions and 34 postconditions. These Hoare triples are checked in debug builds at runtime using the \texttt{contracts} crate \cite{RustContracts}. One of the most checked assumptions is that a given location is contained within the habitat.

Overall, the \texttt{necsim-rust} library has been tested extensively using both dynamically-checked Hoare triples and statistical system analysis, which is described in \cref{analysis-functional-correctness}. However, unit testing has so far only been applied sparsely.

\section{Statistical Correctness} \label{analysis-functional-correctness}

\subsection{Randomness in the Independent Algorithm} \label{analysis-rng}

The Independent algorithm uses a \texttt{\textcolor{yellow}{\textbf{PrimeableRng}}}. As the \texttt{prime()} method frequently resets the PRNG's inner state, it is important to test whether it still produces high-quality random numbers.

\subsubsection{PRNG construction and Avalanching} \label{analysis-prime-rng}

\begin{figure}[h]
    \centering
    \minipage{0.5\textwidth}
    \includegraphics[width=\linewidth]{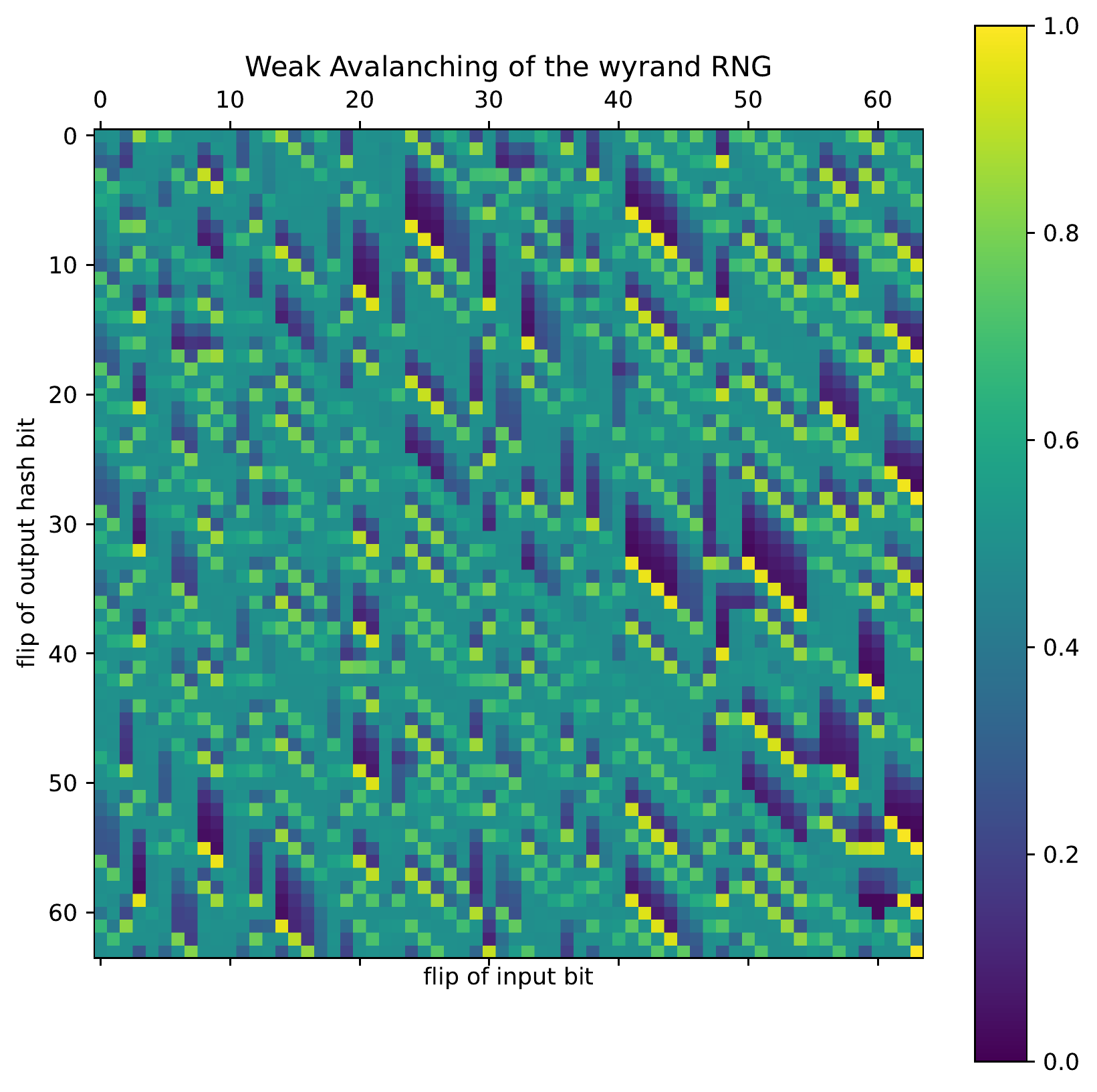}
    \endminipage\hfill
    \minipage{0.5\textwidth}
    \includegraphics[width=\linewidth]{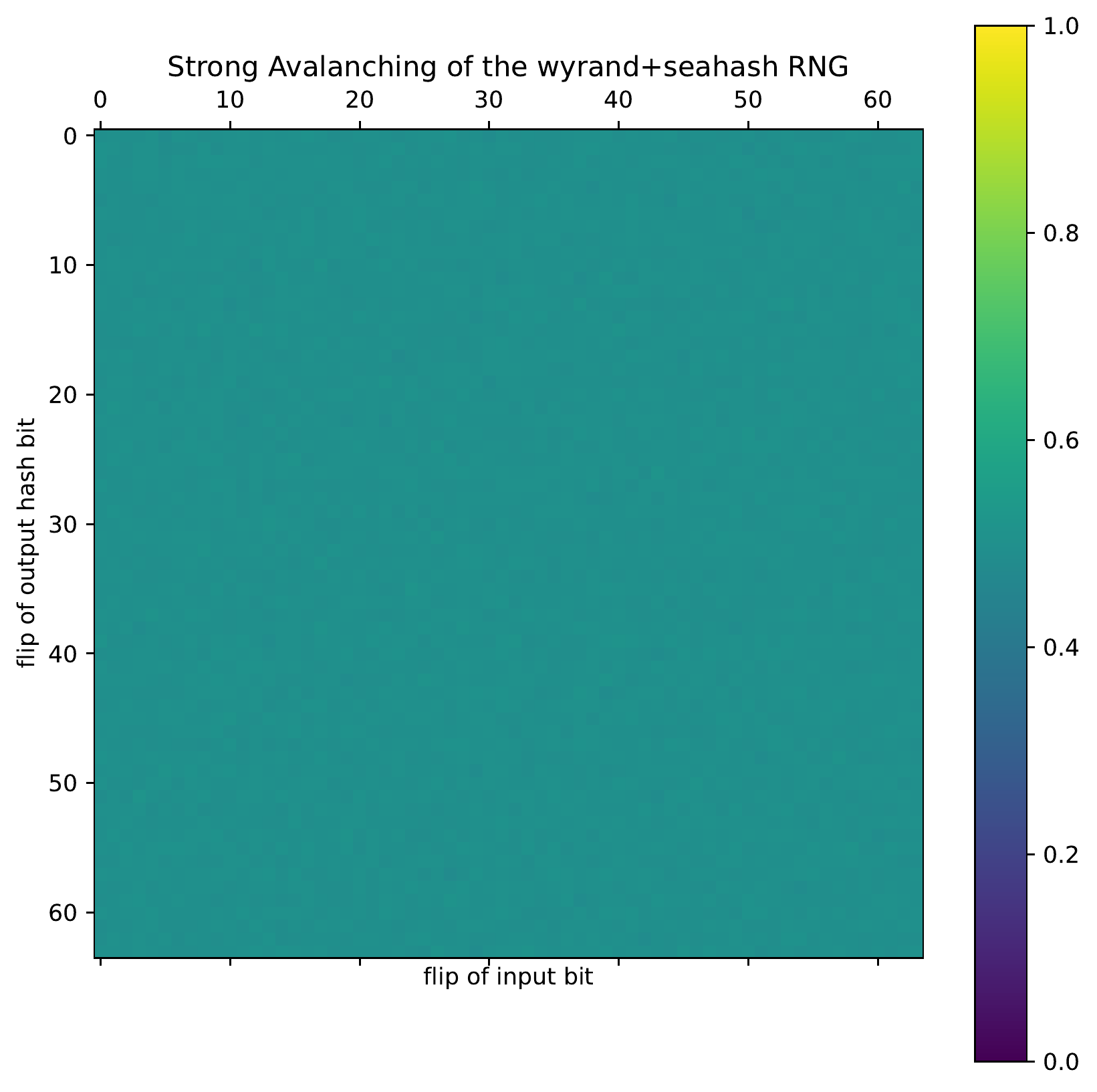}
    \endminipage\hfill
    \caption{Avalanche diagrams for the \texttt{next()} function of the primable PRNG. On the left, the weak avalaching using the raw WyRand state update and output function is shown. The right plot displays the strong avalanching of using the seahash diffusion as an additional output function.}
    \label{fig:avalanche-update}
\end{figure}

The default PRNG that the Independent algorithm uses is based on the WyHash function, WyRand RNG \cite{Wang2021} and their Rust implementation \cite{Romero2020}. In particular, the \texttt{prime()} method uses \texttt{wyhash(} \texttt{seed, indexed\_location.map\_to\_u64(), time.map\_to\_u64())} to reset the RNG's internal 64 bit state. The \texttt{next()} method then uses this internal state as a counter that is hashed again to produce the next random sample. While WyRand passes both the TestU01 \cite{LEcuyer2007} and Practrand \cite{PractRand} testsuites \cite{Wang2021}, its state update function only uses a weak hash function. The left avalanching plot (\ref{background-avalanche}) in \Cref{fig:avalanche-update} clearly shows some non-uniform patterns. To improve the avalanching, we have added seahash's diffusion function \cite{Ticki2020} as an additional output function to \texttt{necsim-rust}'s PRNG implementation. Our \texttt{next()} function has clearly improved avalanching behaviour, as the right avalanching plot in \Cref{fig:avalanche-update} shows. The \texttt{prime()} function immediately fulfils the strict avalanche criterion (\ref{background-avalanche}) without requiring any improvements.

\subsubsection{Tests of Randomness}

The \texttt{rustcoalescence} command-line interface uses two pseudo-random number generators. We use the 64bit \texttt{SetseqDXsM12864} RNG from the PCG family for the monolithic algorithms, since it supports independent per-partition streams. For the Independent algorithm, the custom primeable WyRand + Seahash RNG combination (\ref{analysis-prime-rng}) is used. Both PCG \cite{Lemire2017} and WyRand \cite{Wang2021} have been shown to pass the TestU01 \cite{LEcuyer2007} and PractRand \cite{PractRand} test suites. For this project, we test four different PRNG variants using the TestU01 \cite{LEcuyer2007}, PractRand \cite{PractRand}, Dieharder \cite{dieharder} and Ent \cite{ENT} randomness test suites:
\vspace{-1em} \begin{table}[ht]
\begin{tabular}{l|c|l|c|c|l|l|}
\cline{2-7}
                                       & \textbf{ENT}           & \multicolumn{1}{c|}{\textbf{die-}} & \multicolumn{1}{c|}{\textbf{Pract}}     & \multicolumn{3}{c|}{\textbf{TestU01}}                                                            \\ \cline{5-7}
                                       & \multicolumn{1}{c|}{\textbf{10MB x 1000}}   & \multicolumn{1}{c|}{\textbf{harder}}                                        & \multicolumn{1}{c|}{\textbf{Rand}}           & \multicolumn{1}{c|}{\textbf{Small}}         & \multicolumn{1}{c|}{\textbf{Crush}} & \multicolumn{1}{c|}{\textbf{Big}} \\ \hline
\multicolumn{1}{|c|}{\textbf{(1) PCG next}}  & p=52.3\% (pass)         & 143/144             & 512GB         & 10/10             & 94/96             & 159/160            \\ \hline
\multicolumn{1}{|c|}{\textbf{(2a) W+S next}} & p=60.6\% (pass)         & 142/144             & 512GB         & 10/10             & 96/96             & 159/160            \\ \hline
\multicolumn{1}{|c|}{\textbf{(2b) W+S prime}} & p=57.4\% (pass)         & 143/144       & 512GB         & 10/10             & 94/96             & 160/160            \\ \hline
\multicolumn{1}{|c|}{\textbf{(2c) W+S prime, fixed}} & p=91.2\% (weak)         & 139/144             & 256GB         & 10/10             & 95/96             & 159/160            \\ \hline
\end{tabular}
    \caption{Test results from four randomness test suites. For ENT, only the p-values are analysed. Since TestU01 is designed for 32bit PRNGs only, the lower and upper 32bits are tested separately. Failures from both are added up, regardless if both halves displayed the same failure or not.}
    \label{analysis-randomness-tests}
\end{table}

\begin{enumerate}
    \item \textbf{PCG:} Brooks' \texttt{SetseqDXsM12864} PCG RNG implementation \cite{Brooks2021} is tested as-is. As previously observed by Lemire \cite{Lemire2017}, it passes the tests, though with a minimal number of failures.
    \item \textbf{WyRand+SeaHash:} The improved PRNG that is implemented in \texttt{necsim-rust} is tested:
    \begin{enumerate}
        \item as-is. As previously observed by Wang \cite{Wang2021}, it passes the tests, though with a minimal number of failures. Note that a small number of tests may sometimes randomly fail.
        \item as a priming PRNG. Here, we analyse the stream of random numbers that a single individual dispersing inside an infinite landscape generates. An individual might have to reprime several times with the same location-time-tuple to find all events inside a repriming interval. The analysis ensures that these deliberate sample repetitions are not analysed. The analysis shows that the random stream produced by an individual is of equivalent quality to one produced by PCG or WyRand.
        \item as a priming PRNG without dispersal. This analysis stress tests the quality of the PRNG when part of its effective state, i.e. its reprime location, is fixed. We observe that both the ENT and PractRand test suites are able to discover some weaknesses. Nevertheless, the RNG still performs very well even in this stress test.
    \end{enumerate}
\end{enumerate}

\subsubsection{Correlation}

The primeable PRNG receives highly correlated location-time inputs. As a sanity check, we check that the PRNG still produces uncorrelated random numbers by comparing the 100-draw random streams from four individuals across 1000 seeds. The individuals start at locations $(0, 0) - (1, 1)$ of an infinite landscape with either no or $\textrm{N}^{2}(0, 100^{2})$ dispersal. In both cases, the Spearman cor-relation coefficients do \textbf{not} show a statistically significant correlation for a $10\%$ significance level.

\subsection{Event Statistics}

The quality of the generated randomness and correctness of the algorithms can be checked by analysing the properties of the simulation events. In the following analyses, we check the null hypothesis $H_0$ that `an observed statistic matches our expectation' by comparing both distributions using a Kolmogorov-Smirnov \cite{KolmogorovSmirnov} (for continuous CDFs) or Chi-square Goodness of Fit \cite{ChiSquared} (for discrete CDFs) test. Each test produces a p-value which describes the probability that the observed values were obtained under $H_0$, i.e. if the simulation works as expected. If $H_0$ holds, the p-values from independent repetitions of this test should be uniformly distributed over $[0; 1]$. We perform a meta-analysis of these tests and use Fisher's method \cite[p.~103]{Fisher1934} to calculate a combined p-value. If this combined p-value is not in $[5\%; 95\%]$, we reject $H_0$ at the $10\%$ significance level.

\subsubsection{Turnover and Speciation Times}

In a scenario with a constant turnover rate $\lambda_{x, y} = \lambda_{const}$, the inter-event times for any individual (\ref{design-events}) must be distributed exponentially with rate $\lambda_{const}$. The \textbf{Classical}, \textbf{Gillespie} and \textbf{Independent} algorithms all pass with 1000-repetition p-values of $19.8\%$, $11.2\%$ and $22.7\%$, respectively. As an additional test for consistency, we also tested the \textbf{SkippingGillespie} algorithm, which we expect to fail as it skips events and consequently has longer inter-event times. Indeed, fails with $p \approx 3.3\%$.

If the speciation probability is constant as well, the times from the start of the simulation to the speciation events must be distributed exponentially with rate $\lambda_{const} \cdot \nu_{const}$. However, we cannot simply measure the speciation times of any simulation to test this property. When two individuals coalesce, only the parent is simulated afterwards. Therefore, only the parent has the opportunity to speciate and report its speciation time. To avoid this bias towards short speciation times, we disable dispersal in this analysis. Now, all algorithms pass this test with 1000-repetition p-values of $24.0\%$ (\textbf{Classical}), $46.3\%$ (\textbf{Gillespie}), $65.0\%$ (\textbf{SkippingGillespie}) and $75.2\%$ (\textbf{Independent}).

\subsubsection{Uniform Dispersal}

In the \texttt{necsim-rust} simulation, the \texttt{\textcolor{yellow}{\textbf{DispersalSampler}}} component samples jumps between locations. Inside the deme at each location, both movement and coalescence are always sampled uniformly. In the non-spatial scenario (\ref{background-neutral-scenarios}), dispersal is even uniform over all indexed locations. Therefore, we can check that in a long-running simulation with few coalescences, every location and every index within each deme is visited with equal probability. All algorithms pass this test with 1000-repetition p-values of $24.1\%$ (\textbf{Classical}), $34.2\%$ (\textbf{Gillespie}), $18.5\%$ (\textbf{SkippingGillespie}) and $14.7\%$ (\textbf{Independent}).

\subsubsection{Normal Dispersal}

In the spatially explicit scenario (\ref{background-neutral-scenarios}), individuals live on an infinite landscape and disperse with $\textrm{N}^{2}(0, \sigma^{2})$. In this correctness test, the distribution of all dispersal moves is compared against a discretised normal distribution. For simplicity, we analyse each axis of the 2d dispersal separately. All algorithms pass this test with 1000-repetition p-values of $(30.1\%, 79.5\%)$ (\textbf{Classical}), $(44.1\%, 42.6\%)$ (\textbf{Gillespie}), $(42.1\%, 53.5\%)$ (\textbf{SkippingGillespie}) and $(88.6\%, 38.6\%)$ (\textbf{Independent}).

\subsection{Neutral Scenarios}

\Cref{background-neutral-scenarios} has introduced the non-spatial, spatially implicit and spatially explicit scenarios. We now use their analytical solutions (\Cref{appendix:neutral-scenarios}) to verify that the simulation produces the expected species richness. As we use species richness to measure biodiversity (\Cref{appendix:neutral-scenarios}), we only refer to `biodiversity' from now on. We have designed \texttt{necsim-rust} such that the algorithms are implemented separately from the scenarios. Due to this modularity, we avoid checking all combinations and instead test them separately: If every algorithm correctly simulates one scenario, and one algorithm simulates every scenario correctly, then all algorithms simulate all scenarios correctly.

\subsubsection{Non-Spatial}

The non-spatial scenario is the only model whose analytical solution, \Cref{eq:non-spatial}, can be reproduced correctly for all parameter values. Therefore, we use it to verify the correctness of all algorithms and their parameters. Specifically, the biodiversity measurements from independently seeded repetitions of the same model should approach a Normal distribution whose mean converges to the expected analytical solution. All algorithms and parameters pass this test.

\subsubsection{Spatially Implicit} \label{analysis-scenario-spatially-implicit}

The spatially implicit scenario models migration from an external static non-spatial metacommunity to a local non-spatial community. Crucially, the analytical solution for this scenario, \Cref{eq:spatially-implicit}, assumes that the metacommunity is of infinite size. As this assumption cannot be fulfilled, a fixed size metacommunity has to be used. However, a finite dynamic metacommunity might have the same effect as an infinite static one. While the static approach assumes that the metacommunity does not change over time, a dynamic metacommunity is simulated simultaneously with the local community. We have implemented both approaches to test this hypothesis.

\vspace{-0.5em} \begin{figure}[h]
    \centering
    \includegraphics[width=1.0\textwidth]{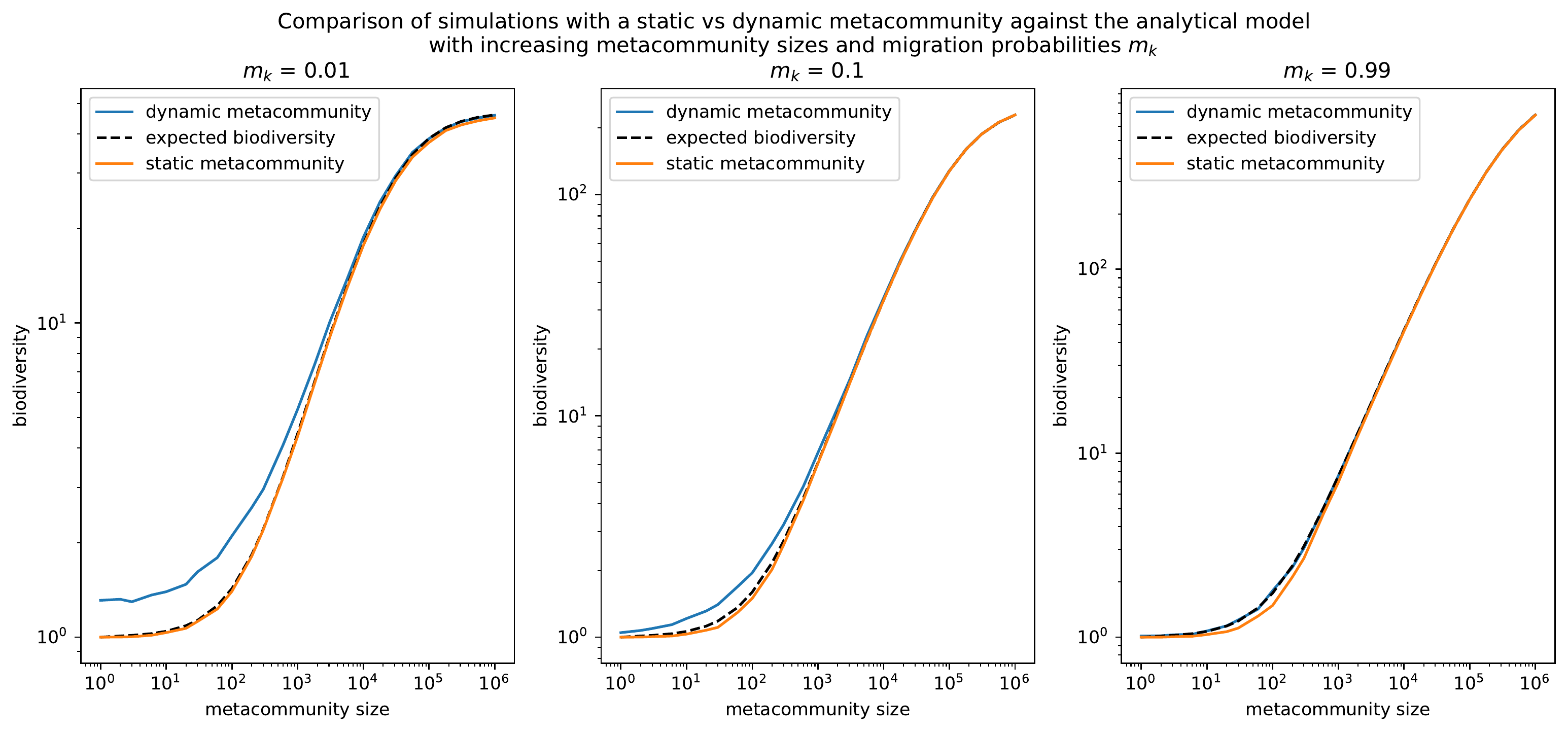}
    \caption{Comparison of the static and dynamic spatially implicit simulations for a local community of size $J = 10^{3}$ and speciation in the metacommity with probability $\nu_m = 0.1$. The dynamic metacommunity only gives the closer-to-analytical solution for small $J_m$ and $m_k \rightarrow 1$.}
    \label{fig:spatially-implicit-scenario}
\end{figure}

\noindent \Cref{fig:spatially-implicit-scenario} compares both metacommunity simulations with the analytical solution. When the metacommunity is large or the migration probability $m(A) = m_k$ is low, the metacommunity appears to be effectively infinite to the reverse-migrating individuals. Both approaches correctly simulate the model in this case. For small metacommunities and low $m_k$, the static method converges to the expected biodiversity while the dynamic method overestimates it. However, for small metacommunities and $m_k \rightarrow 1$, the dynamic method converges to the expected biodiversity while the static method underestimates it. This observation invites future work to extend the analytical solution such that it can also be applied to finite metacommunities.

\subsubsection{Spatially Explicit}

In the spatially explicit model, individuals perform $\textrm{N}^{2}(0, \sigma^{2})$ dispersal on an infinite landscape. \Cref{eq:spatially-explicit} describes this scenario's approximate solution. For large $\sigma$, the simulation produces the expected approximate results. For low $\sigma$, however, the simulated biodiversity spikes above the expected value \cite{Rosindell2007}. This behaviour is a fully expected artefact of the discretised normal dispersal in the simulation. While the model assumes a continuous landscape with sub-integer dispersal, individuals can only disperse between integer locations in the simulation. For $\sigma \rightarrow 0$, each location then becomes a closed, island-like deme and gives rise to unique endemic species\footnote{Private communication from Samuel Thompson suggests to approximate this effect by correcting the speciation probability to account for dispersal events that do not leave the current location. Since these same-deme dispersals occur roughly with probability $P(\textrm{Rayleigh}(\sigma) < 0.5)$, we get an increased speciation probability of $\nu'(\sigma) = 1 - {(1 - \nu)}^{e^{-\frac{1}{2 \cdot 2^2 \cdot \sigma^{2}}}}$}.

\subsection{Convergence}

\begin{figure}[h]
    \centering
    \minipage{0.5\textwidth}
    \includegraphics[width=\linewidth]{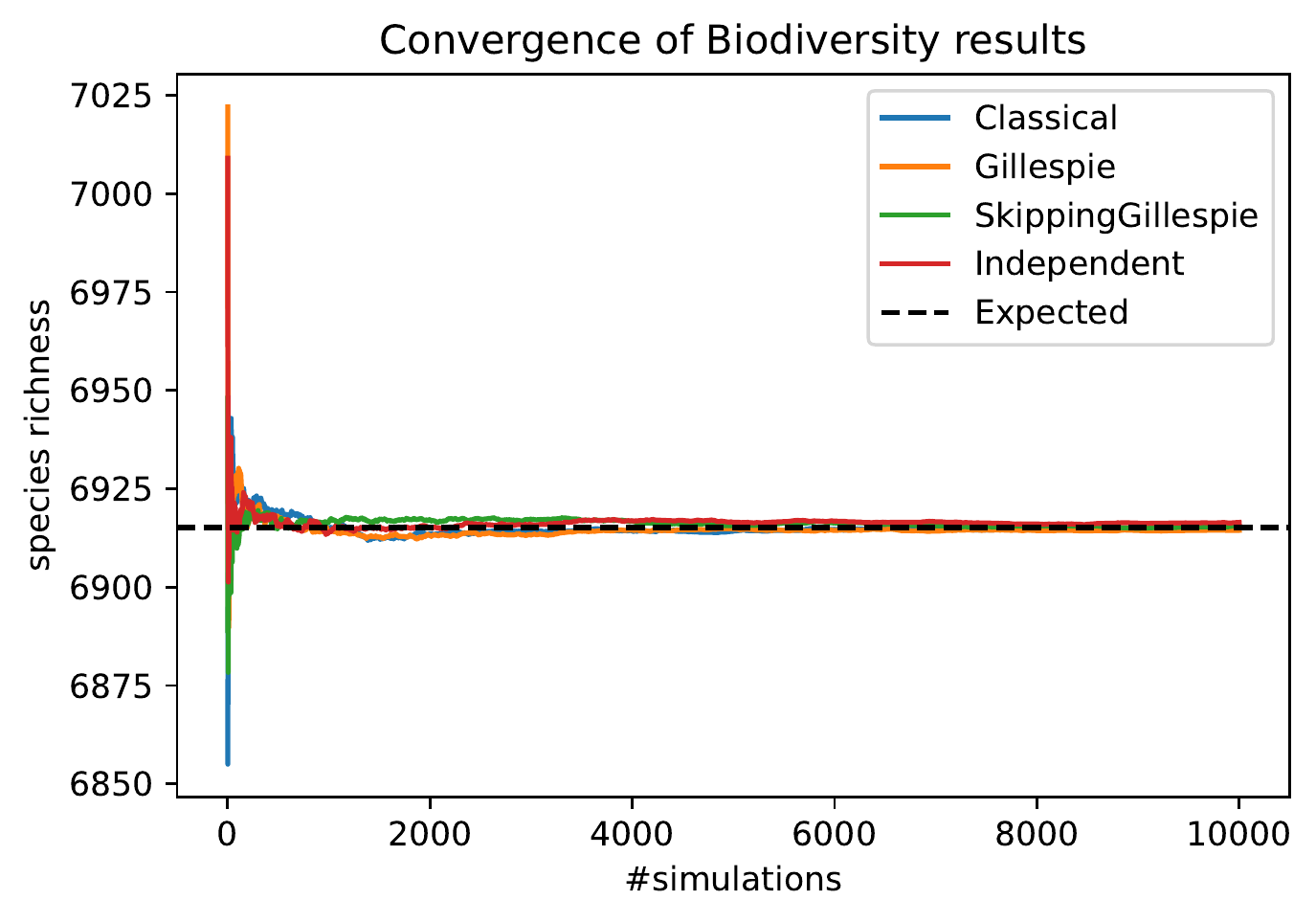}
    \endminipage\hfill
    \minipage{0.5\textwidth}
    \includegraphics[width=\linewidth]{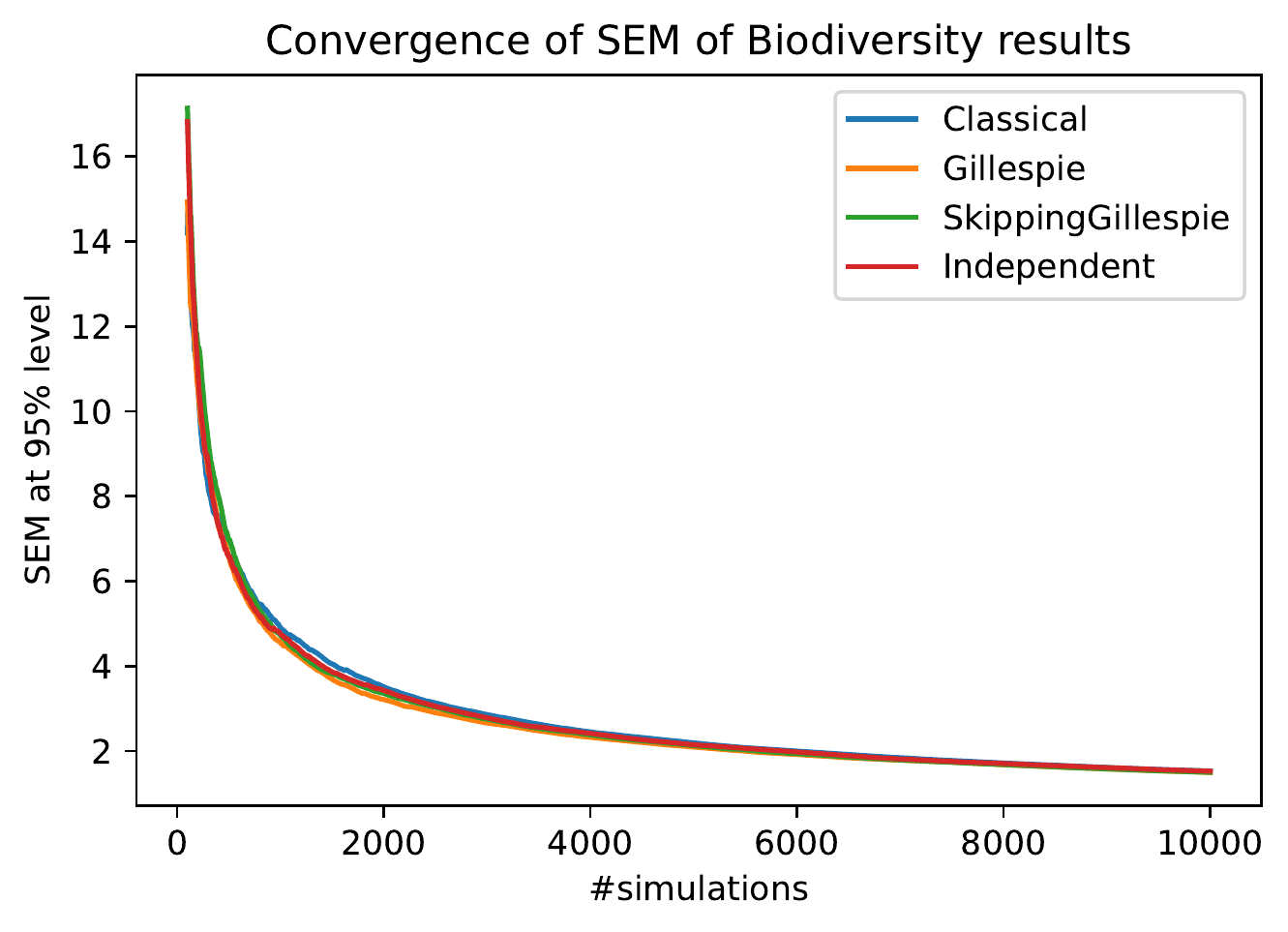}
    \endminipage\hfill
    \caption{Convergence behaviour of the four algorithms for the $\textrm{NonSpatial}(J=10^{6}, \nu=10^{-3})$ scenario. While the left graph plots the convergence of the mean species richness as more results come in, the right graph displays the decreasing standard error of the mean.}
    \label{fig:convergence}
\end{figure}

While all algorithms converge to the correct results, they might have different convergence speeds. As the biodiversity simulation is a Monte Carlo simulation that requires multiple runs to produce a reliable result, different convergence speeds would affect the effective performance of the different algorithms. For this analysis, each algorithm simulates $10,000$ independent instances of the $\textrm{NonSpatial}(J=10^{6}, \nu=10^{-3})$ scenario. As \Cref{fig:convergence} shows, all algorithms display very similar if not equivalent convergence behaviour. In particular, the standard errors of their means converge at almost identical rates. Therefore, all algorithms provide indistinguishable convergence speeds.

\vspace{-1em} \section{Event Generation Performance} \label{analysis-event-performance}

In this section, we provide a baseline performance evaluation of event generation in the simulation. All tests were performed on a machine with 16GB of RAM and an Intel(R) Xeon(R) CPU E5-1620 v2 @ 3.70GHz processor, which has four cores with two threads each. It also has an NVIDIA GeForce GTX 1080 with 8GB of RAM, 2560 CUDA cores, CUDA 11.2 and compute capability 6.1.

\subsection{Independent Exponential Inter-Event Time Sampling} \label{analysis-exponential-sampling}

\begin{figure}[h]
    \centering
    \minipage{0.5\textwidth}
    \includegraphics[width=\linewidth]{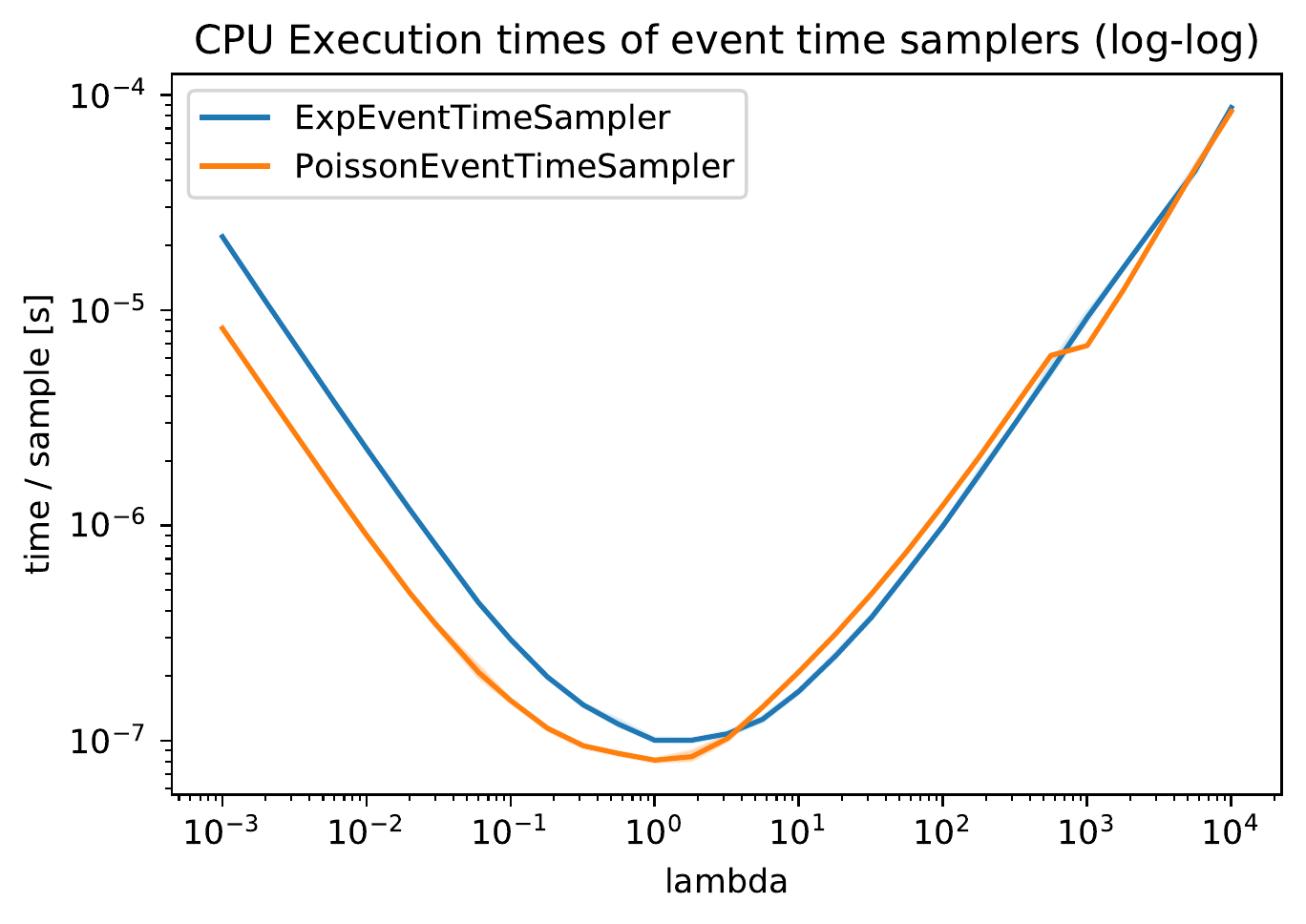}
    \endminipage\hfill
    \minipage{0.5\textwidth}
    \includegraphics[width=\linewidth]{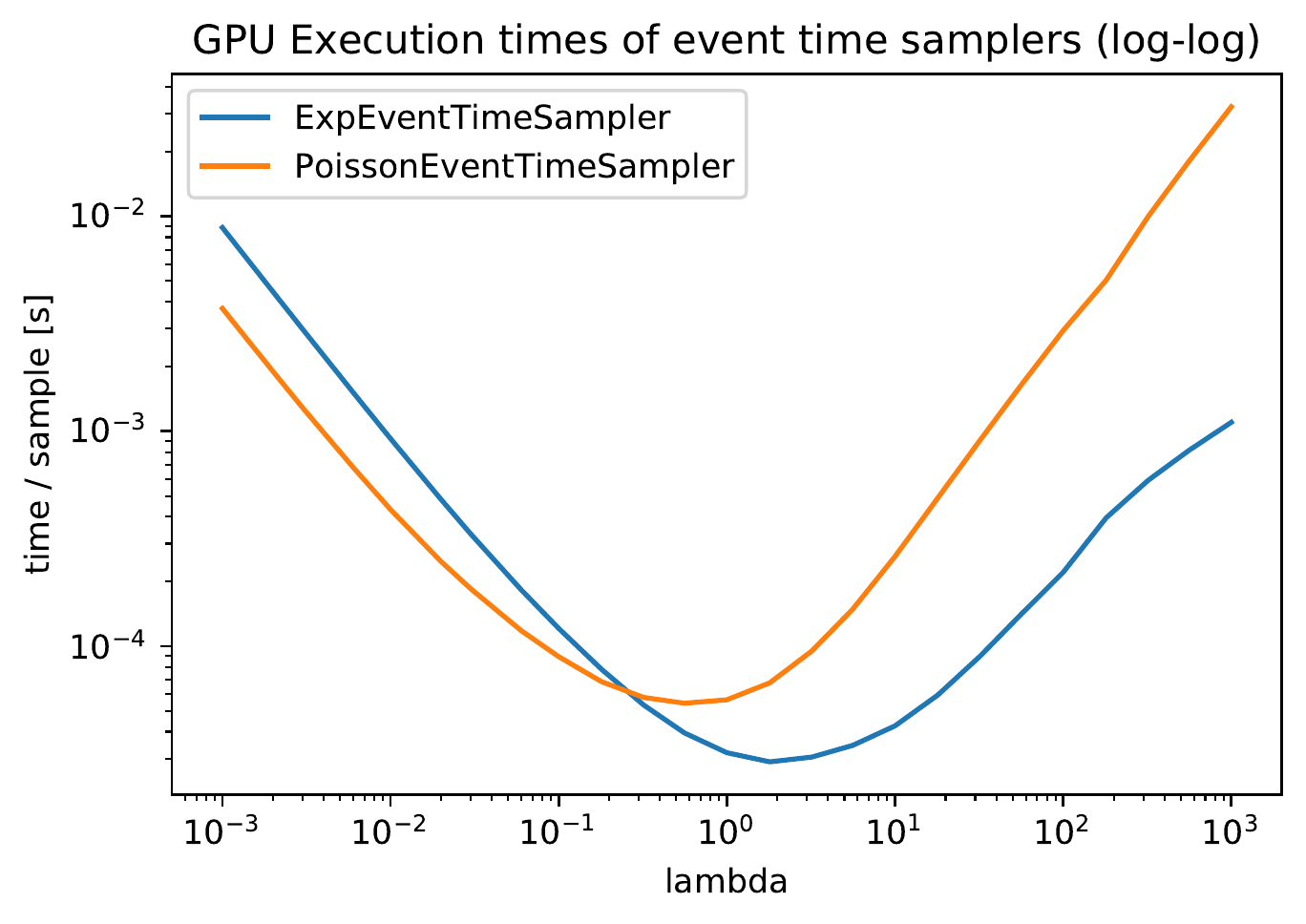}
    \endminipage\hfill
    \caption{Comparison of the performance of the Exponential- and Poisson- based inter-event samplers on the CPU and GPU. On the CPU, the \texttt{\textcolor{blue}{\textbf{PoissonEventTimeSampler}}} provides the overall fastest performance. For turnover rates $\lambda \geq 4$, the \texttt{\textcolor{blue}{\textbf{ExpEventTimeSampler}}} is slightly faster. On the GPU, the \texttt{\textcolor{blue}{\textbf{ExpEventTimeSampler}}} is significantly faster for turnover rates $\lambda > 0.3$.}
    \label{fig:exponential}
\end{figure}

\Cref{independence-exponential-generations} introduces two methods to independently sample exponential inter-event times using a \texttt{\textcolor{yellow}{\textbf{PrimeableRng}}}. They are implemented in the \texttt{\textcolor{blue}{\textbf{ExpEventTimeSampler}}} and \texttt{\textcolor{blue}{\textbf{PoissonEventTime-}}} \texttt{\textcolor{blue}{\textbf{Sampler}}}. Both of these algorithms are mathematically equivalent but computationally different. While the Exponential method has to sample one exponential random number for each event time it checks, the Poisson method only evaluates the $e^{x}$ function once. \Cref{fig:exponential} shows that the Poisson method provides the overall fastest inter-event sampling on the CPU for $\lambda \approx 1.0$.

It is also worth noting that the Poisson method switches from Devroye's linear Poisson sampler \cite[p.~505]{Devroye1986} to the $\lim_{\lambda\to\infty} \textrm{Poi}(\lambda) \approx \textrm{N}(\lambda, \lambda)$ approximation around $\lambda \approx 745$ to avoid an \mintinline{rust}{f64} underflow. This change in sampling is actually visibile in the CPU's performance plot in \Cref{fig:exponential}.

\noindent On the GPU, however, the Exponential method is faster for turnover rates $\lambda > 0.3$ and provides the best execution time at $\lambda \approx 2.0$. Why is the optimal method different for both platforms? Both methods have the same complexity, which is linear with the expected number of reprimes and the number of events between reprimes. However, while the Exponential method can be implemented as a single loop, the Poisson method requires nested loops. As GPUs favour uniform control flow and punish non-uniform branching, the \texttt{\textcolor{blue}{\textbf{ExpEventTimeSampler}}} outperforms the \texttt{\textcolor{blue}{\textbf{PoissonEventTimeSampler}}} on the GPU. For low turnover rates, i.e. on average zero or one event per reprime interval, the inner loop to sample the Poisson distribution (\Cref{appendix:geometric-poisson-sampling}) has almost uniform branching, and the Exponential method is faster again.

\Cref{fig:exponential} also shows that the inter-event time sampling is almost three orders of magnitude slower on the GPU than on the CPU. Thus, a GPU-specific primeable PRNG, which does not rely on emulated 64bit arithmetic \cite[pp.~117-120]{CUDACpp}, might be able to reduce this performance gap.

\subsection{Event Reporting Performance Baseline} \label{analysis-event-reporting}

Our \texttt{necsim-rust} simulation library reports events (\ref{design-reporter}), which we can analyse in several ways:
\begin{enumerate}
    \item The \texttt{simulation.simulate(reporter)} method can be specialised for a specific analysis reporter type. This fixes the analysis at compile time and allows \texttt{rustc} to inline the analysis.
    \item Instead, we can use our dynamic plugin system, which loads analysis reporters at runtime as dynamic libraries. In this approach, the compiler can at best produce an optimised code path based on the selection of event types that the simulation has to generate and report.
    \item The events can also be written to an on-disk event log, postponing analysis to a future replay.
\end{enumerate}
To analyse events produced on the GPU using the above methods, our implementation first moves them to the CPU. Therefore, we also analyse the CPU-GPU data movement. In this test, we simulate $J = 10^5$ individuals with $\nu = 10^{-3}$ without dispersal. Consequently, we should produce $10^8$ events on average. We also record and analyse one GPU execution using the NVIDIA Nsight Systems profiler:
\vspace{-1em} \begin{table}[h]
\begin{tabular}{l|l|l|l|}
\cline{2-4}
                                                     & \multicolumn{1}{|c|}{\textbf{progress events only}} & \multicolumn{1}{|c|}{\textbf{+speciation events}} & \multicolumn{1}{|c|}{\textbf{all event types}} \\ \hline
\multicolumn{1}{|c|}{\textbf{CPU Compiled Analysis}} & \multicolumn{1}{|c|}{51.45s $\pm$ 0.44s}                & \multicolumn{1}{|c|}{51.55s $\pm$ 0.40s}              & \multicolumn{1}{|c|}{53.24s $\pm$ 0.58s}            \\ \hline
\multicolumn{1}{|c|}{\textbf{CPU Analysis Plugins}}  & \multicolumn{1}{|c|}{51.62s $\pm$ 0.52s}                & \multicolumn{1}{|c|}{52.0s $\pm$ 0.49s}               & \multicolumn{1}{|c|}{55.18s $\pm$ 0.57s}            \\ \hline
\multicolumn{1}{|c|}{\textbf{CPU Event Log}}     & \multicolumn{1}{|c|}{51.64s $\pm$ 0.51s}                & \multicolumn{1}{|c|}{51.74s $\pm$ 0.36s}              & \multicolumn{1}{|c|}{75.05s $\pm$ 0.48s}            \\ \cline{2-4}
\multicolumn{1}{|c|}{Event Log size}                 & \multicolumn{1}{|c|}{no event log}                                & \multicolumn{1}{|c|}{3.95MB $\pm$ 313.59kB}           & \multicolumn{1}{|c|}{5.57GB $\pm$ 17.33MB}          \\ \hline
\multicolumn{1}{|c|}{\textbf{GPU Analysis Plugins}}  & \multicolumn{1}{|c|}{0.64s $\pm$ 0.01s}                 & \multicolumn{1}{|c|}{0.64s $\pm$ 0.01s}               & \multicolumn{1}{|c|}{25.82s $\pm$ 0.24s}            \\ \cline{2-4}
\multicolumn{1}{|c|}{memcpy/launch (\%GPU)}                 & \multicolumn{1}{|c|}{655kB (2.7\%)}                                & \multicolumn{1}{|c|}{983kB (3.9\%)}           & \multicolumn{1}{|c|}{92.7MB (81.7\%)}          \\ \cline{2-4}
\multicolumn{1}{|c|}{kernel|memcpy (\%CPU)}                 & \multicolumn{1}{|c|}{67.95\% | too small}                                & \multicolumn{1}{|c|}{65.48\% | 0.60\%}           & \multicolumn{1}{|c|}{2.78\% | 11.59\%}          \\ \hline
\end{tabular}
\caption{Performance of different reporting methods and increasing event requirements analysed over 160 runs. Reporting dispersal events greatly slows down the GPU and the event log.}
\label{analysis-reporting-performance}
\end{table}

\noindent First, \Cref{analysis-reporting-performance} shows that compiling the analysis into the simulation is the best reporting strategy on the CPU. However, it is only $1-4\%$ faster than using the more user-friendly dynamic analysis plugins that we can easily swap out within seconds. Second, writing just speciation events to an on-disk event log has a similar performance penalty as using plugins. However, logging dispersal events incurs a $45\%$ runtime increase and requires an enormous amount of disk space.

The GPU massively outperforms all CPU analysis methods if we only analyse progress and speciation events. However, reporting dispersal events increases the execution time forty-fold. Where does this performance hit come from? On the GPU, reporting dispersal increases the memory transfers by two orders of magnitude, which consume 81.7\% of the GPU's execution time. However, these transfers only take up 11.6\% of the overall execution time on the CPU. The actual bottleneck is sorting all events in the Water-Level algorithm, which takes up 74.6\% of the execution time. To put it simply, the CPU just cannot keep up with the GPU's event production.

\subsection{Event Throughput Baseline} \label{analysis-event-throughput}

We now analyse the event production throughput of all single-machine algorithms across three scenarios. In order to get a fair neutral baseline, we simulate and \texttt{perf} only $J = 10^6$ individuals:

\begin{table}[h]
\begin{tabular}{l|l|l|l|}
\cline{2-4}
                                                 & \multicolumn{1}{|c|}{\textbf{Non-Spatial}}        & \multicolumn{1}{|c|}{\textbf{Spatially Explicit}}    & \multicolumn{1}{|c|}{\textbf{Spatially Explicit Map}} \\
                                                 & \multicolumn{1}{|c|}{with $deme = 100$}        & \multicolumn{1}{|c|}{with $\textrm{N}^2(0, 10^2)$}    & \multicolumn{1}{|c|}{with \texttt{fg0size12} (\ref{analysis-algorithm-sweetspot})} \\ \hline
\multicolumn{1}{|c|}{\textbf{Classical}}         & \multicolumn{1}{|c|}{$3271 \pm 136$}                  & \multicolumn{1}{|c|}{$2711 \pm 25$}                  & \multicolumn{1}{|c|}{$3680 \pm 119$}                  \\ \cline{2-4}
\multicolumn{1}{|c|}{Bottleneck (\%CPU)}             & \multicolumn{1}{|c|}{25\% individual pop}                  & \multicolumn{1}{|c|}{23\% dispersal}                  & \multicolumn{1}{|c|}{15\% individual pop} \\ \hline
\multicolumn{1}{|c|}{\textbf{Gillespie}}         & \multicolumn{1}{|c|}{$1084 \pm 9$}                   & \multicolumn{1}{|c|}{$883 \pm 17$}                   &  \multicolumn{1}{|c|}{$1224 \pm 17$}                  \\ \cline{2-4}
\multicolumn{1}{|c|}{Bottleneck (\%CPU)}             & \multicolumn{1}{|c|}{24\% heapify}                  & \multicolumn{1}{|c|}{32\% heapify}                  & \multicolumn{1}{|c|}{22\% heapify} \\ \hline
\multicolumn{1}{|c|}{\textbf{SkippingGillespie}} & \multicolumn{1}{|c|}{$1053 \pm 9$}                   & \multicolumn{1}{|c|}{$865 \pm 6$}                   & \multicolumn{1}{|c|}{$1153 \pm 12$}                   \\ \cline{2-4}
\multicolumn{1}{|c|}{Bottleneck (\%CPU)}             & \multicolumn{1}{|c|}{24\% heapify}                  & \multicolumn{1}{|c|}{31\% heapify}                  & \multicolumn{1}{|c|}{50\% simulation setup} \\ \hline
\multicolumn{1}{|c|}{\textbf{Independent (CPU)}}       & \multicolumn{1}{|c|}{$1488 \pm 21$}                   & \multicolumn{1}{|c|}{$1424 \pm 11$}                   & \multicolumn{1}{|c|}{$2838 \pm 16$}                   \\ \cline{2-4}
\multicolumn{1}{|c|}{Raw Throughput}             & \multicolumn{1}{|c|}{$2634 \pm 37$}                  & \multicolumn{1}{|c|}{$2167 \pm 15$}                  & \multicolumn{1}{|c|}{$2839 \pm 16$}                  \\ \cline{2-4}
\multicolumn{1}{|c|}{Bottleneck (\%CPU)}             & \multicolumn{1}{|c|}{41\% event sorting}                  & \multicolumn{1}{|c|}{33\% event sorting}                  & \multicolumn{1}{|c|}{41\% event sorting} \\ \hline
\multicolumn{1}{|c|}{\textbf{Independent (GPU)}}              & \multicolumn{1}{|c|}{$427 \pm 5$}                   & \multicolumn{1}{|c|}{$593 \pm 7$}                   & \multicolumn{1}{|c|}{$2707 \pm 17$}                   \\ \cline{2-4}
\multicolumn{1}{|c|}{Raw Throughput}             & \multicolumn{1}{|c|}{$895 \pm 11$}                   & \multicolumn{1}{|c|}{$1026 \pm 11$}                   & \multicolumn{1}{|c|}{$2708 \pm 17$}                   \\ \cline{2-4}
\multicolumn{1}{|c|}{Bottleneck (\%CPU)}             & \multicolumn{1}{|c|}{33\% event memcpy}                  & \multicolumn{1}{|c|}{29\% event memcpy}                  & \multicolumn{1}{|c|}{37\% event sorting} \\ \hline
\end{tabular}
\caption{Event throughput in $10^{3} \times events / s$ for different algorithms and scenarios with $J=10^6$ individuals and $\nu = 10^{-3}$ over 160 runs. For the \textbf{Independent} algorithm, the raw CPU and GPU throughput (before event deduplication) is also listed. The \textbf{Classical} algorithm is fastest.}
\label{analysis-deduplication-throughput}
\end{table}

\noindent \Cref{analysis-deduplication-throughput} shows that the \textbf{Classical} algorithm has the highest baseline throughput of all algorithms. The \textbf{Independent} algorithm on the CPU comes in second. We can infer from its raw throughput that more than a third of all reported events in densely populated models were duplicates, while it almost reached full throughput on the sparse spatially explicit map. In third and fourth place, we have the \textbf{Gillespie} and \textbf{SkippingGillespie} algorithms, respectively. Since both require additional work to calculate event rates and queue event times, they have the lowest throughput on the CPU. Finally, there is the \textbf{Independent} algorithm on the GPU, which comes in last. This low performance is not unexpected since we analyse dispersal events to compute the throughput, which first have to be moved to and sorted on the CPU (\Cref{analysis-reporting-performance}). Like on the CPU, the GPU's raw throughput shows that around half of all reported events were duplicates in dense scenarios.

We can further observe that all CPU algorithms struggle with the spatially explicit model. In this scenario, each location only houses one individual that performs $\textrm{N}^2$ dispersal. In contrast, all implementations perform best on the sparsely populated spatially explicit map. Finally, the \textbf{Independent} algorithm's performance is limited by event reporting on the CPU.

\section{Configuration Sweetspot Analysis} \label{analysis-algorithm-sweetspot}

Most algorithms and parallelisation strategies implemented in \texttt{rustcoalescence} are configurable. In this section, we explore the effects of all parameters on the simulation performance and try to find the sweet spots. We also provide valuable insight into the tradeoff between data dependencies and redundancy. While most options can be considered independently, many indirectly interact and perform differently depending on the model we are simulating. We have only had the time to explore each option separately instead of their exponentially many combinations. Therefore, these sweet spots should be considered good starting points for further tuning.

We have performed all sweet spot tests on machines in the Imperial College London Research Computing Service's HPC cluster, which have 1TB of RAM and provide two sockets with AMD EPYC 7742 64-core processors @ 2.95GHz and two threads per core. As all parallel MPI simulations were run on single large nodes, we report the lower bound effect of communication costs. CUDA simulations use NVIDIA Quadro P1000 GPUs, which have CUDA 11.0, 4GB of RAM, 640 CUDA cores and compute capability 6.1. To compromise between the breadth of our analyses and the statistical quality of each singular one, we have repeated every experiment at least ten times. All graphs plot both the median result curve and the standard deviation interval around the mean. Finally, we are only analysing speciation events. Note that this means that the GPU implementation does not incur the significant performance drop from memory movement and event sorting (\ref{analysis-event-reporting}).

\begin{figure}[h]
    \centering
    \includegraphics[width=1.0\textwidth]{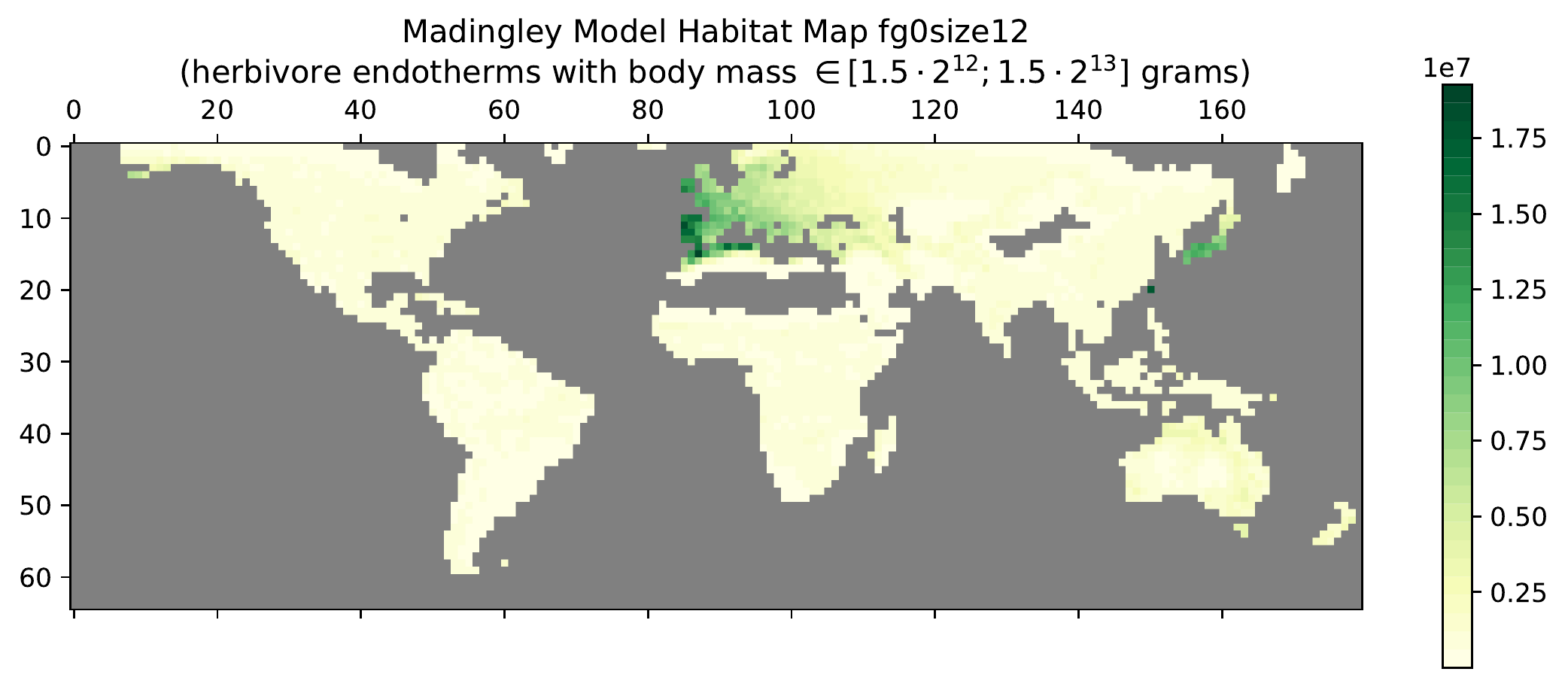}
    \caption{Example habitat map which shows the global distribution of herbivore (plant-eating) endotherm (`warm-blooded') animals that weigh between $6.1$kg and $12.3$kg, e.g. koalas.}
    \label{fig:madingley}
\end{figure}

\noindent To evaluate the real-life performance of the algorithms, we test them on spatially explicit habitat and dispersal maps from an existing dataset. Specifically, we use the \texttt{fg0size12} map (\Cref{fig:madingley}) created by Hintzen \cite{HintzenPhd} using the Madingley Model \cite{Harfoot2014}, a general ecosystems model which is able to produce realistic species distribution patterns on a global scale. This map has $4.2 \cdot 10^{9}$ individuals spread over just $180 \times 64$ locations. We quasi-randomly subsample this population for experiments with smaller domains. Note that since this map has very large demes, which are only sparsely populated in most analyses, we expect the \textbf{SkippingGillespie} algorithm to perform well.

\subsection{The Parallelised Monolithic Algorithms}

While the sequential monolithic algorithms have no configuration options, the \textbf{OptimisticLockstep} and \textbf{Averaging} strategies are both parameterised by their synchronisation frequencies.

The \textbf{OptimisticLockstep} strategy (\ref{implementation-monolithic-optimistic-lockstep}) scales very badly on this highly connected map, so we only tested it for $5,000$ up to $20,000$ individuals. Overall, it was fastest for $delta\_sync \approx 20.0$. With an increasing number of partitions, this sweet spot moves higher, while more individuals lower it. It is also worth noting that increasing the number of partitions further slows down this strategy.

\begin{figure}[h]
    \centering
    \includegraphics[width=1.0\textwidth]{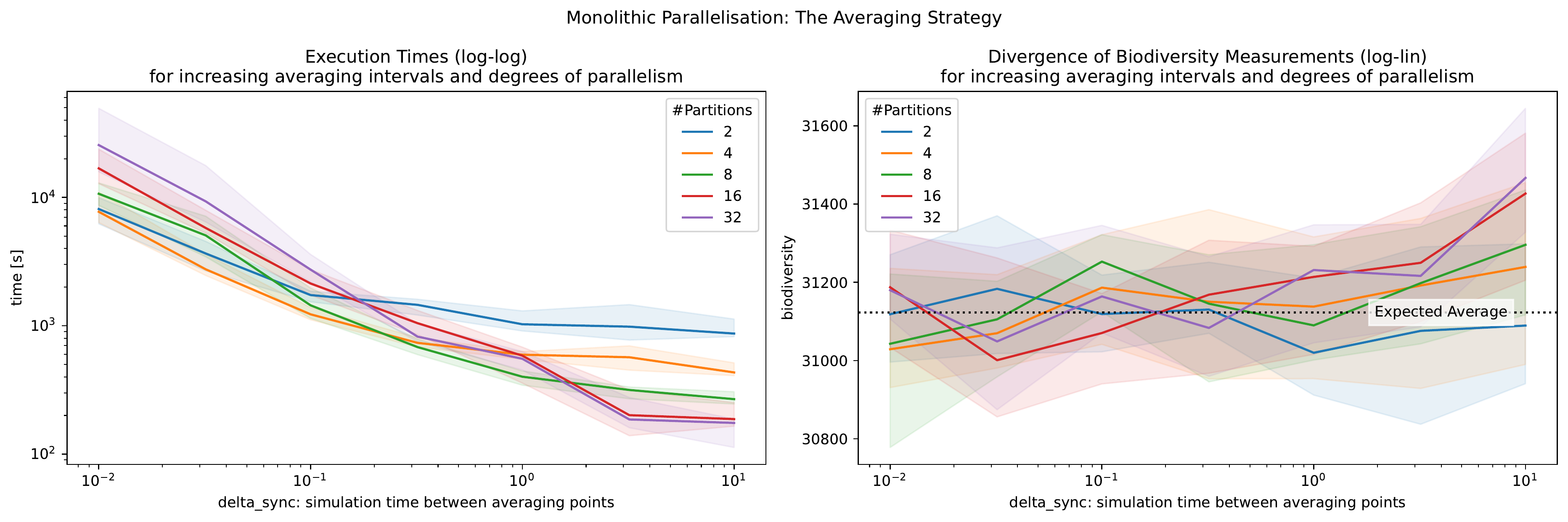}
    \caption{Execution time (left) and biodiversity (right) for different averaging intervals in the parallelised \textbf{SkippingGillespie} algorithm for $J = 10^{8}$ individuals, $\nu = 10^{-6}$ and $\lambda = 0.5$. As the averaging interval and the number of partitions increase, execution time decreases but the approximate biodiversity measurements also increasingly diverge from the expected value.}
    \label{fig:monolithic-averaging}
\end{figure}

The \textbf{Averaging} strategy (\ref{implementation-gillespie-averaging}) trades off accuracy for performance. \Cref{fig:monolithic-averaging} shows the execution times and biodiversity for increasing averaging intervals and degrees of parallelism. As expected, more partitions and longer independent intervals increase performance but result in an increasing overestimate of biodiversity. An averaging interval of $delta\_sync \approx 1.0$ provides a good compromise between performance and very little approximation error -- at least for $\lambda_{x, y} = 0.5$.

\subsection{The Independent Algorithm on the CPU} \label{analysis-independent-sweetspot}

We test the Independent algorithm's (\ref{implementation-independent}) parallelism-independent parameters on a single CPU:

\noindent \begin{minipage}{0.5\textwidth} \begin{itemize}
    \item \textbf{delta\_t:} The global sweet spot for the RNG repriming interval length lies around $2.0$. For $\lambda = 0.5$, the event rate per repriming interval becomes $1.0$, matching the sweet spot that was observed in \cref{analysis-exponential-sampling} for the \texttt{\textcolor{blue}{\textbf{PoissonEventTimeSampler}}} on the CPU.
    \item \textbf{dedup\_capacity:} We choose an individual deduplication cache that has one space for every individual. While lower capacities reduce performance, larger caches only provide diminishing returns as RAM usage and random-access costs increase. This highlights the importance of some local deduplication, i.e. that simulating individuals entirely communication-free is not practical when processing many individuals.
\end{itemize} \end{minipage} \hfill
\begin{minipage}{0.5\textwidth}
    \centering
    \includegraphics[width=\textwidth]{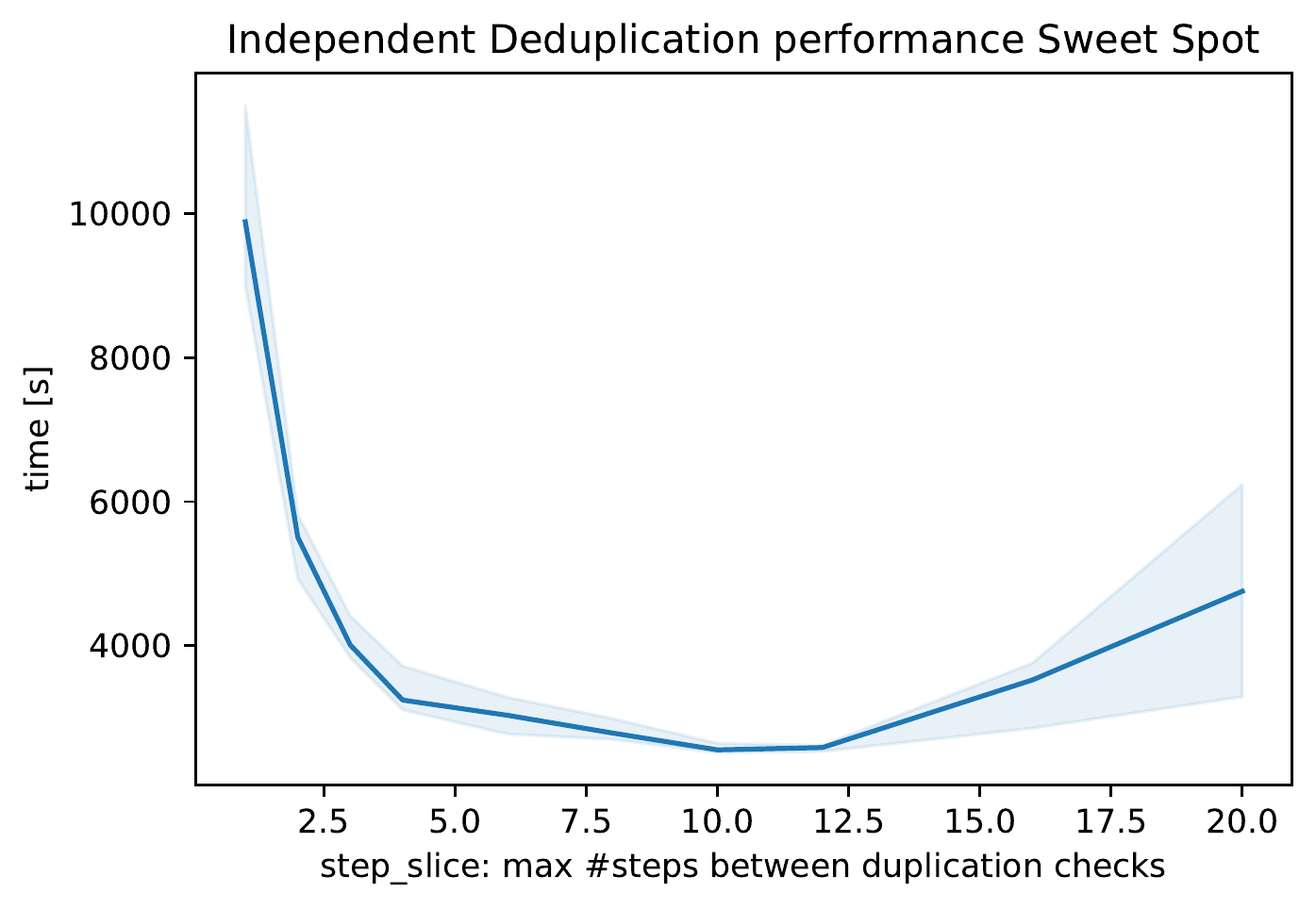}
    \captionof{figure}{Execution times for increasing simulation step intervals between checking for individual duplication. For $J = 10^{6}$ individuals and $\nu = 10^{-6}$, the sweetspot is at 10/11 steps.}
    \label{fig:independent-deduplication}
\end{minipage}
\vspace{-12pt} \begin{itemize}
    \item \textbf{step\_slice:} The step slice sets the (maximum) number of steps each individual simulates independently before checking the deduplication cache and resetting its minimum speciation sample (\ref{implementation-deduplication}). \Cref{fig:independent-deduplication} shows that performing ten steps independently is optimal, even though this also means coalesced individuals often duplicate the next ten events at least.
    \item \textbf{event\_slice:} This parameter specifies the average number of events that will be produced, buffered and sorted on every iteration of the Water-Level algorithm (\ref{implementation-water-log}). The noisy analysis data only suggests that the sweet spot lies between $J$ and $10J$. Smaller buffers have significantly worse performance, and more sorting increases execution times for larger ones. Note that if we only analyse speciation events, a buffer that is smaller by a factor of $\nu$ suffices.
\end{itemize}

\subsection{The Probabilistically Communicating Independent Algorithm}

\begin{figure}[h]
    \centering
    \includegraphics[width=1.0\textwidth]{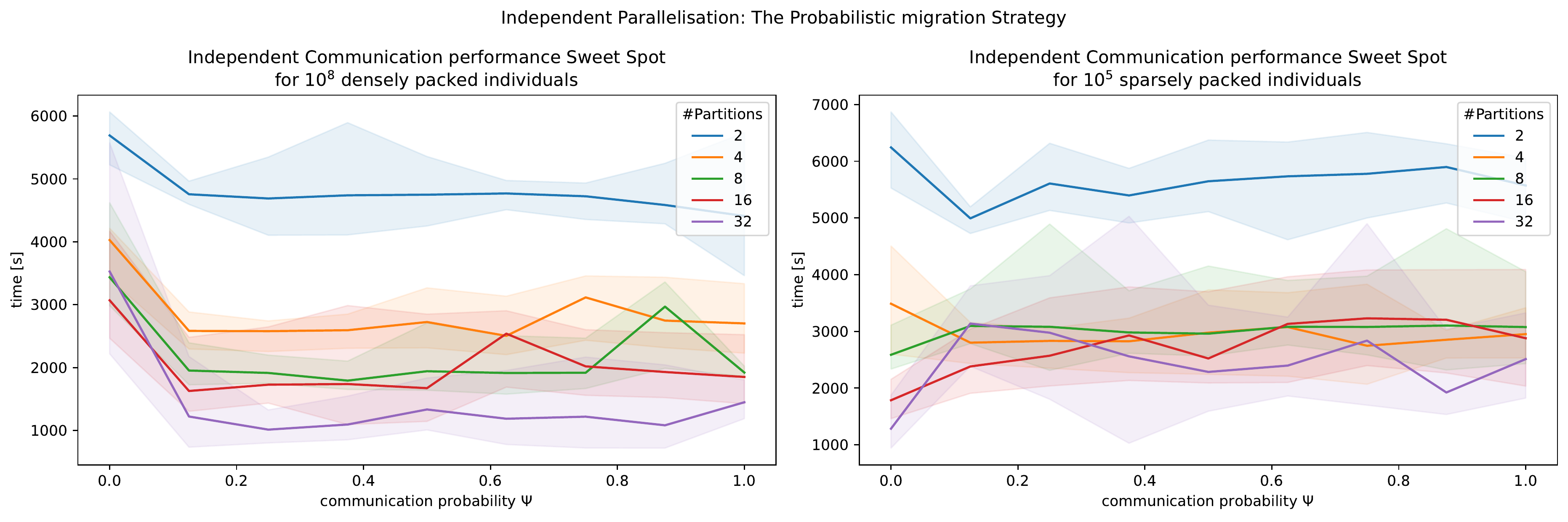}
    \caption{Execution times for different communication probabilities $\Psi$ in the parallelised \textbf{Independent} algorithm for $\nu = 10^{-6}$. The left graph shows that for a large number of densely-packed individuals, adding some communication is beneficial. For fewer and sparser individuals and larger degrees of parallelism (right), less communication is beneficial.}
    \label{fig:independent-probabilistic}
\end{figure}

The \textbf{Independent} algorithm's \textbf{Probabilistic} parallelisation strategy (\ref{implementation-independent-parallelisation}) is parameterised by the communication probability $\Psi$. \Cref{fig:independent-probabilistic} shows the simulation's performance for increasing $\Psi$ and parallelism. From the left graph, we can see that adding in some communication is beneficial when there are many densely packed individuals that frequently coalesce. Using $\Psi > 25\%$ does not appear to provide any further benefits, however. In the right graph, we approximate the later stage of the simulation using fewer, sparsely distributed individuals. Here, we can see that less commu-nication improves performance when there are $\geq 8$ partitions. Intuitively, most individuals do not coalesce at this stage of the simulation, and it is better to simulate some redundant individuals instead of communicating their migrations. Overall, we choose $\Psi = 25\%$ as the sweet spot.

\vspace{-6pt} \subsection{The Independent Algorithm on a CUDA GPU} \label{analysis-cuda-sweetspot}

The GPU version of the Independent algorithm (\ref{implementation-cuda}) is configured similarly to the CPU version:

\noindent \begin{minipage}{0.45\textwidth}
    \begin{itemize}
    \item \textbf{delta\_t:} As predicted by \cref{analysis-exponential-sampling}, the GPU's sweet spot is larger than the CPU's and lies around $3.0$.
    \item \textbf{dedup\_capacity:} The size $0.1 J$ is best.
    \item \textbf{step\_slice:} The GPU favours longer independent intervals of $150$ steps.
    \item \textbf{event\_slice:} On the GPU, a larger buffer of size $20 J$ is optimal as it too enables longer independent intervals.
    \item \textbf{block\_size $\times$ grid\_size:} \Cref{fig:cuda-block-grid} shows that launching $64 \times 64$ threads is optimal, which suggests that each thread is using the integer compute resources of multiple threads.
\end{itemize}
\end{minipage} \hfill
\begin{minipage}{0.55\textwidth}
    \centering
    \includegraphics[width=1.0\textwidth]{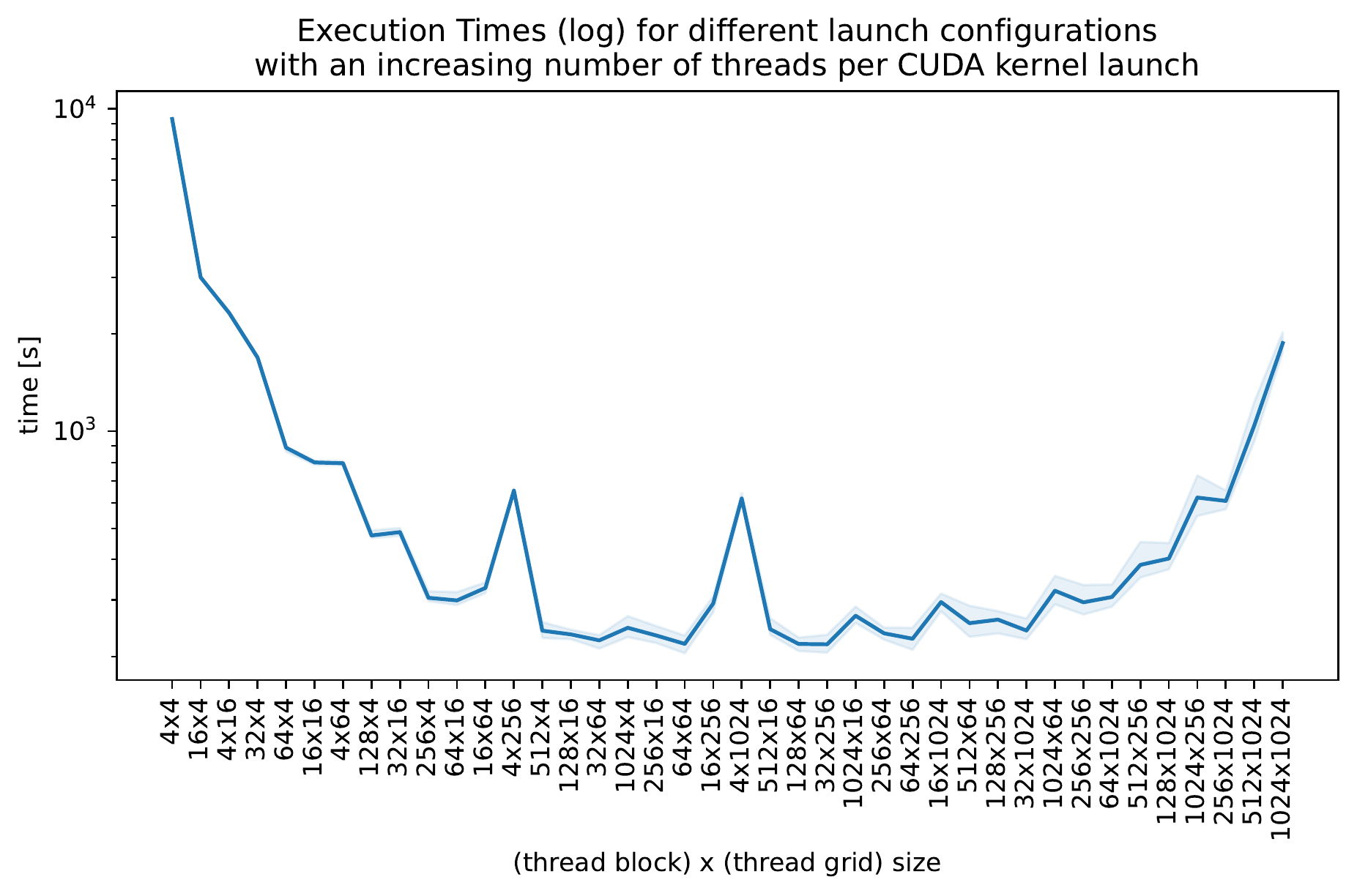}
    \captionof{figure}{Execution times for different CUDA launch configurations for $J = 10^{6}$ and $\nu = 10^{-6}$. Launching only $32 - 128$ threads/block is optimal.}
    \label{fig:cuda-block-grid}
\end{minipage}
\vspace{-12pt} \begin{itemize}
  \item \textbf{ptx\_jit:} As \Cref{table:implementation-cuda-jit} shows, JIT compiling simulation parameter reduces register pressure.
\end{itemize}

\section{Scalability Analysis} \label{analysis-algorithm-scalability}

In this section, we analyse the scaling of all algorithms for three different parameters:
\begin{enumerate}
  \item The \textbf{speciation probability} $\nu$, which is inversely proportional to the time until speciation.
  \item The \textbf{number of individuals} $J$, which describes the size of the simulation domain. Since we are subsampling the \texttt{fg0size12} map, it also affects the density of individuals.
  \item The \textbf{number of partitions} that the simulation is split into and which run in parallel.
\end{enumerate}
We use the same HPC analysis setup and sweet spot configurations from \cref{analysis-algorithm-sweetspot}.

\subsection{Simulation Domain and Speciation Probability} \label{analysis-domain-speciation-scalability}

\begin{figure}[h]
    \centering
    \minipage{0.5\textwidth}
    \includegraphics[width=\linewidth]{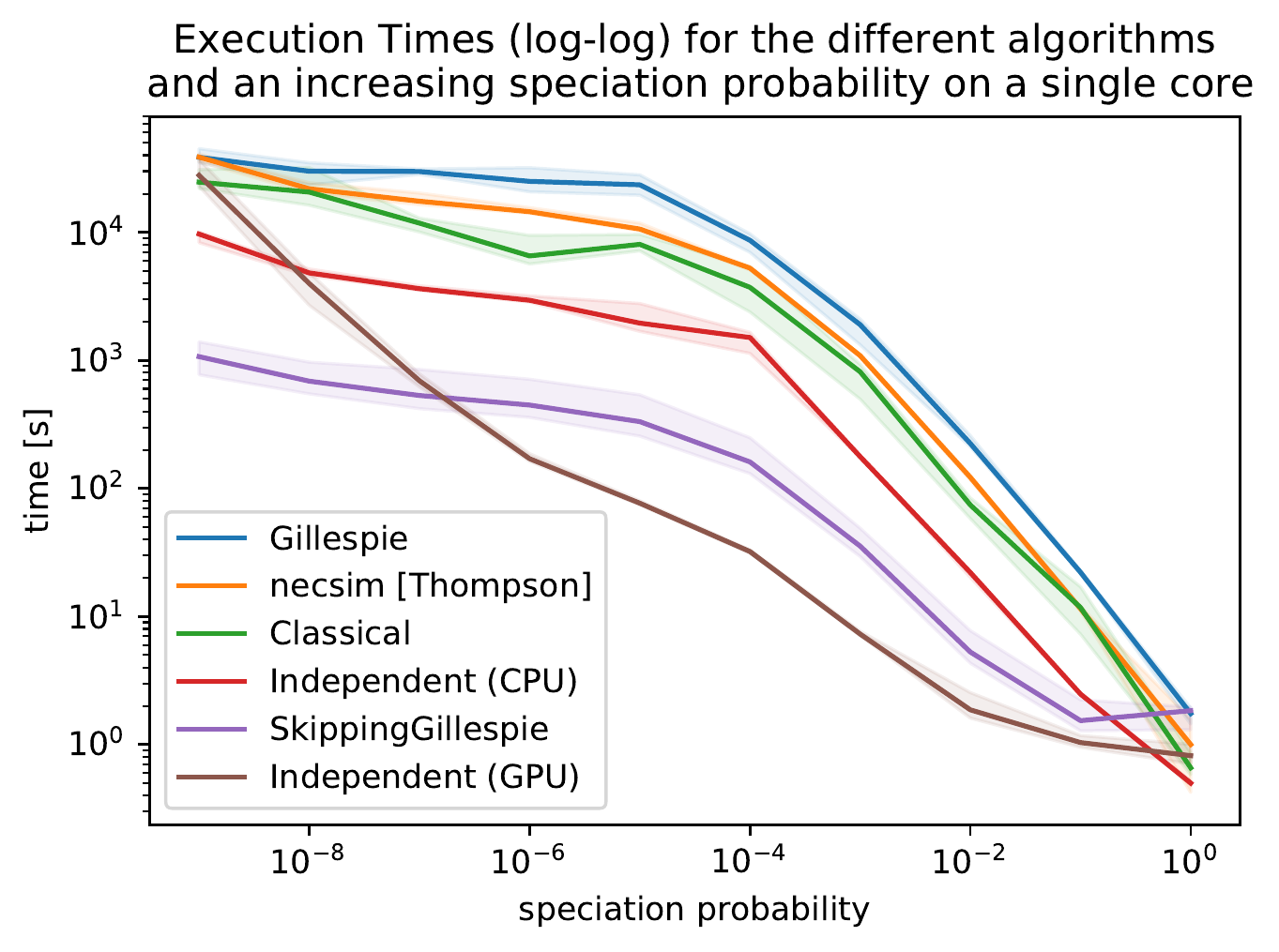}
    \endminipage\hfill
    \minipage{0.5\textwidth}
    \includegraphics[width=\linewidth]{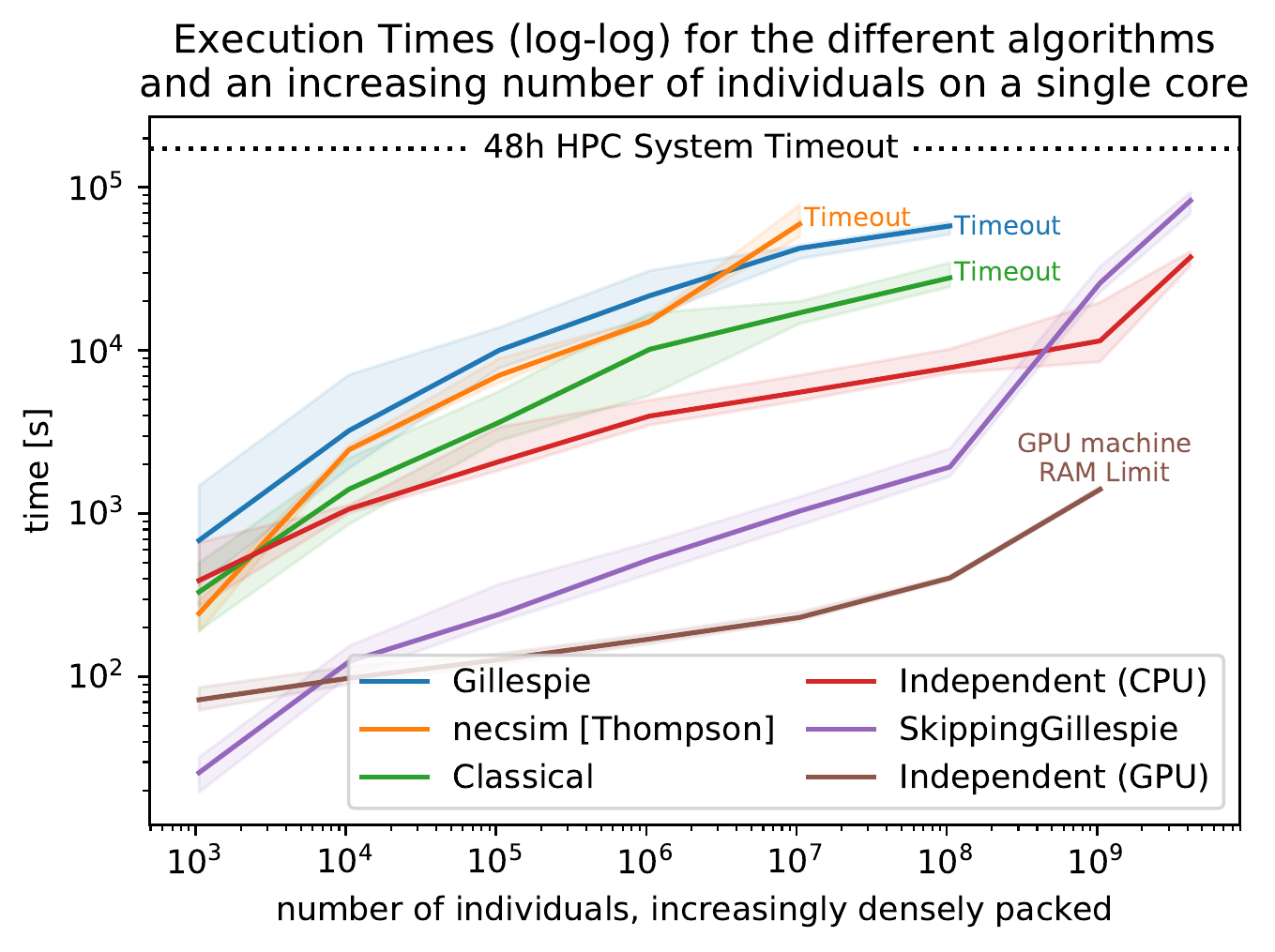}
    \endminipage\hfill
    \caption{Execution times for $J = 10^{6}$ individuals and decreasing speciation probabilities (left), and increasing simulation domains with $\nu = 10^{-6}$ (right). The monolithic \textbf{Skipping-} \textbf{Gillespie} algorithm performs best for lower $\nu$ when it can skip many events, while the \textbf{Independent} algorithm on the CPU and GPU is fastest for large numbers of individuals.}
    \label{fig:speciation-domain-scalability}
\end{figure}

First, we analyse the single-machine scaling of all algorithms for increasing numbers of individuals, and for decreasing speciation probabilities. This analysis includes the sequential \textbf{Independent}, \textbf{Classical}, \textbf{Gillespie} and \textbf{SkippingGillespie} algorithms, the original \texttt{\textbf{necsim}} simulation, and the single-GPU implementation of the \textbf{Independent} algorithm. \Cref{fig:speciation-domain-scalability} plots the results of both analyses. Note that the domain analysis is missing some results due to timeouts or RAM limits.

\noindent In the left graph, we observe that all CPU algorithms, including the \texttt{\textbf{necsim}} simulation, display very similar scaling for decreasing speciation probabilities. In the first of two phases, where $\nu \geq 10^{-4}$, the execution times scale approximately linearly, as getting through $\nu^{-1}$ steps for all individuals dominates performance. In the second phase for $\nu < 10^{-4}$, the execution times grow much more slowly as the speciation probability decreases. Now that the individuals have more time to coalesce, the pruning of duplicate individuals (\ref{background-coalescence-pruning}, \ref{implementation-deduplication}) offsets the longer waiting times until speciation.

\Cref{fig:speciation-domain-scalability} also shows that the \textbf{SkippingGillespie} algorithm performs best for low speciation probabilities. Since we are only simulating $0.025\%$ of all individuals from the \texttt{fg0size12} Madingley Map, the algorithm can skip most events. On the CPU, the \textbf{Independent} algorithm is the second fastest, while the \textbf{Classical} algorithm, the original \texttt{\textbf{necsim}} simulation, and the \textbf{Gillespie} algorithm come in third to fifth, respectively. Despite its lower baseline throughput \cref{analysis-event-throughput}, the \textbf{Independent} algorithm performs better than the \textbf{Classical} variant, which may be caused by the increased spatial locality of simulating each individual independently for several steps.

The GPU implementation of the \textbf{Independent} algorithm outperforms all CPU simulations for $10^{-6} \leq \nu \leq 0.1$ (for $\nu \approx 1.0$, the CUDA kernel launch overhead dominates the execution time). However, the GPU implementation does not prune duplicate individuals as efficiently\footnote{To amortise the high kernel launch costs, The GPU simulates thousands of individuals simultaneously for long step slices (\ref{analysis-cuda-sweetspot}). Therefore, it also has less frequent deduplication checks on the CPU, and deduplication lags further behind.}. In fact, since its curve never flattens, it is on track to be the slowest simulation for $\nu < 10^{-9}$. \\

\noindent The right graph in \Cref{fig:speciation-domain-scalability} plots how the single-machine algorithms scale for a growing number of increasingly densely packed individuals. First, we observe that the performance ranking of the algorithms is very similar. Second, the CPU algorithms again have similar scaling behaviour. In the first phase for $J < 10^{8}$, most individuals added by a higher sampling percentage coalesce quickly. As the number of discovered species approaches the landscape's full species richness, the execution times also appear to converge to some upper bound. However, for $J > 10^{8}$, the curves' slopes become much steeper again. In this second phase, the algorithms' memory usage grows significantly. While the \textbf{SkippingGillespie} algorithm uses 9GB to simulate $J = 10^{8}$ individuals, it requires 119GB and 474GB for $J = 10^{9}$ and $J = 4.2 \cdot 10^{9}$, respectively. By comparison, the \textbf{Independent} algorithm uses 7GB, 66GB, and 256GB, respectively.

We can further observe that for larger domains, the \textbf{Independent} algorithm on the CPU gets increasingly faster compared to the \textbf{Classical} algorithm. This finding supports our hypothesis that the \textbf{Independent} algorithm benefits from its higher spatial locality. \Cref{fig:speciation-domain-scalability} also shows that the \textbf{Independent} algorithm outperforms the \textbf{SkippingGillespie} algorithm for $J > 10^{9}$. In addition to its higher RAM usage, the \textbf{SkippingGillespie} struggles to skip events. When (almost) all $4.2 \cdot 10^{9}$ individuals are simulated, the model is densely populated and very few events can be skipped (\ref{independence-event-skipping}, \ref{implementation-monolithic-parallelisation}). Instead, the \textbf{Independent} algorithm's higher throughput pays off. Finally, the \textbf{Independent} algorithm on the GPU again outperforms all CPU algorithms. However, for increasingly densely-packed individuals, the gap is shortened due to the GPU's reduced deduplication. \\

\noindent There are three key findings from both of these scalability analyses:
\begin{enumerate}
  \item The \textbf{SkippingGillespie} algorithm provides the fastest CPU performance in scenarios where it can skip most events. However, it does not scale well to large simulations.
  \item For large, densely populated systems, the \textbf{Independent} algorithm outperforms the \textbf{SkippingGillespie} algorithm on the CPU. In future work, one could switch from the former to the latter have both early-simulation domain scalability and late-simulation event skipping.
  \item The \textbf{Independent} algorithm on the GPU provides the fastest algorithm for medium speciation probabilities. However, its performance suffers when pruning duplicate individuals is crucial for performance. Therefore, future work might investigate deduplication on the GPU itself.
\end{enumerate}

\subsection{Comparison of Parallelisation Strategies} \label{analysis-parallelism}

Finally, this section compares the execution times of all monolithic and independent parallelisation strategies. The first observation is that the \textbf{Lockstep}, \textbf{OptimisticLockstep} and \textbf{Optimistic} strategies (\ref{implementation-monolithic-parallelisation}) for the monolithic algorithms all failed to complete $10^8$ individuals within 24h.

\begin{figure}[ht]
    \centering
    \includegraphics[width=1.0\textwidth]{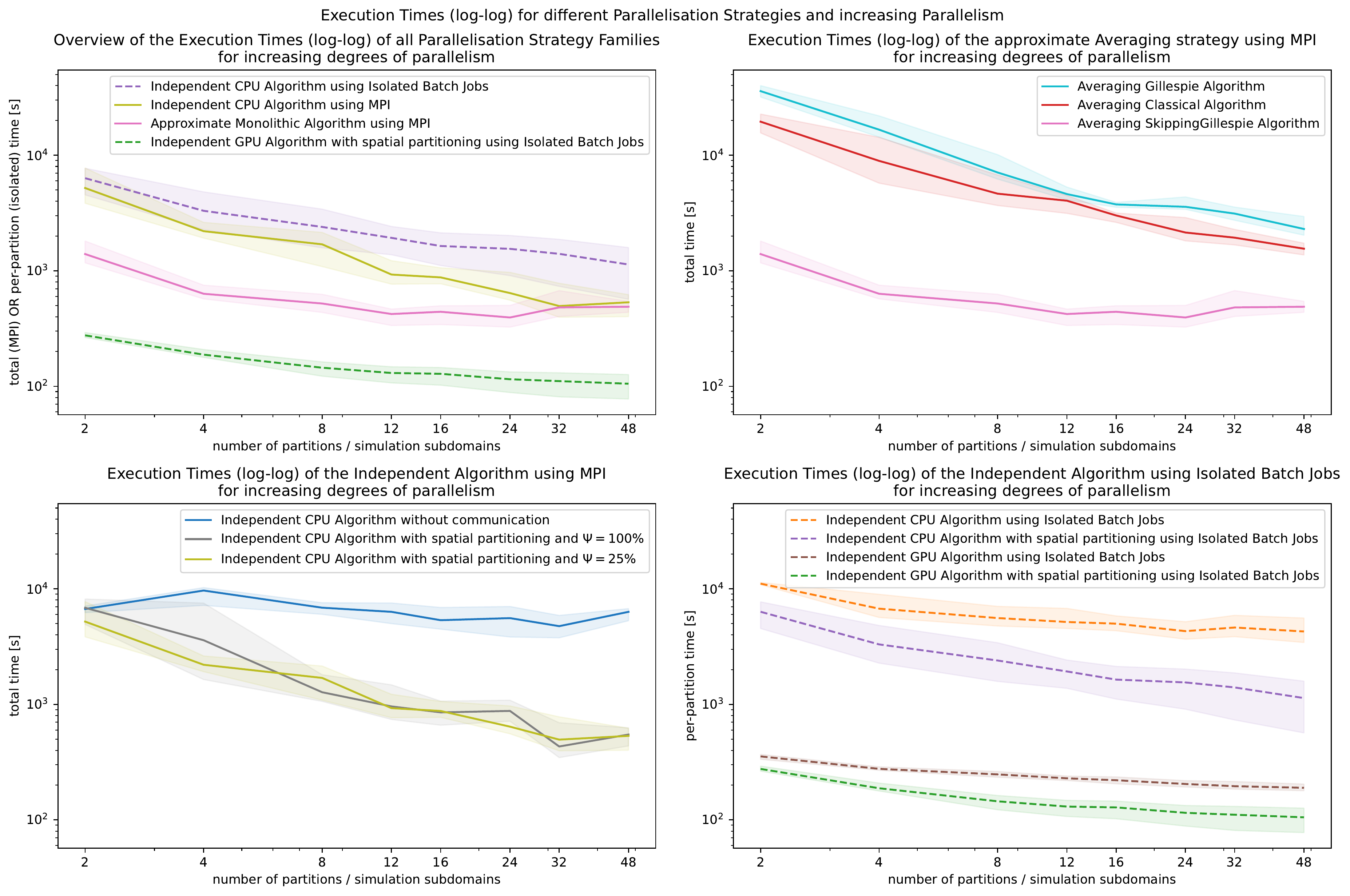}
    \caption{Execution times of different parallelisation strategies and increasing degrees of parallelism for $J = 10^{8}$ individuals and $\nu = 10^{-6}$. On the CPU, the approximate averaging \textbf{SkippingGillespie} algorithm performs best, while the \textbf{Independent} algorithm comes in second.}
    \label{fig:parallelism-scalability}
\end{figure}

The approximate \textbf{Averaging} strategy (\ref{implementation-gillespie-averaging}), on the other hand, performs very well. The top right plot in \Cref{fig:parallelism-scalability} shows that, as expected, the \textbf{SkippingGillespie} algorithm is still the fastest monolithic algorithm. Next, we analyse the parallelised \textbf{Independent} algorithm on the CPU using MPI. The bottom left graph shows that adding communication and dividing the landscape such that nearby locations are in the same partition both boost performance significantly. Third, we compare the isolated independent batch jobs in the bottom right graph. As expected, using spatial partitioning and running the jobs on a GPU both significantly improve performance.

The top left plot in \Cref{fig:parallelism-scalability} compares the performance of the different parallelisation strategy families, resulting in the following insights. If we only have access to a few CPU cores, then the approximate \textbf{Averaging} \textbf{SkippingGillespie} algorithm (\ref{implementation-gillespie-averaging}) performs best. For higher degrees of parallelism, however, the \textbf{Probabilistic} \textbf{Independent} algorithm (\ref{implementation-independent-parallelisation}) performs at least as well. If we do not want to allocate multiple machines or require crash tolerance, the isolated batch jobs provide a great compromise between local deduplication and inter-partition independence. They are also only two times slower than their corresponding MPI implementations. Finally, it is also worth noting that none of the methods scales very well with the increasing parallelism and domain decomposition. Therefore, future work should reduce the overheads of (1) individual deduplication, (2) event reporting, and (3) communication to achieve better performance.

\section{Biodiversity Simulation Limit Analysis} \label{analysis-limit}

Finally, we analyse how far our project has pushed the capabilities of biodiversity simulation:

\noindent \begin{minipage}{0.5\textwidth}
    \setlength{\parskip}{1em}
    \setlength{\parindent}{1pc} \Cref{fig:speciation-limit} plots the performance of the \textbf{SkippingGillespie} algorithm for very low speciation probabilities. It shows that there is a third stage to the scaling curve from \cref{analysis-domain-speciation-scalability}. For $\nu < 10^{-9}$, with the few remaining individuals failing to coalesce towards the end, simulation times are dominated by the speciation timescale, $\nu^{-1}$.

    \setlength{\parindent}{1pc} {We also test how far we can now push the simulation domain. We use the \texttt{fg0size8} Madingley map, which has $7.1 \cdot 10^{10}$ individuals, and run isolated \textbf{Independent} partitions with $J = 10^{8}$ each on the HPC batch system. During the combined replay analysis, we then count the total and raw number of speciation events to measure the degree of duplication between partitions. For $J = 10^{9}, 10^{10} \text{ and } 7.1 \cdot 10^{10}$, this factor\parfillskip=0pt\par}
\end{minipage} \hfill
\begin{minipage}{0.5\textwidth}
    \centering
    \includegraphics[width=1.0\textwidth]{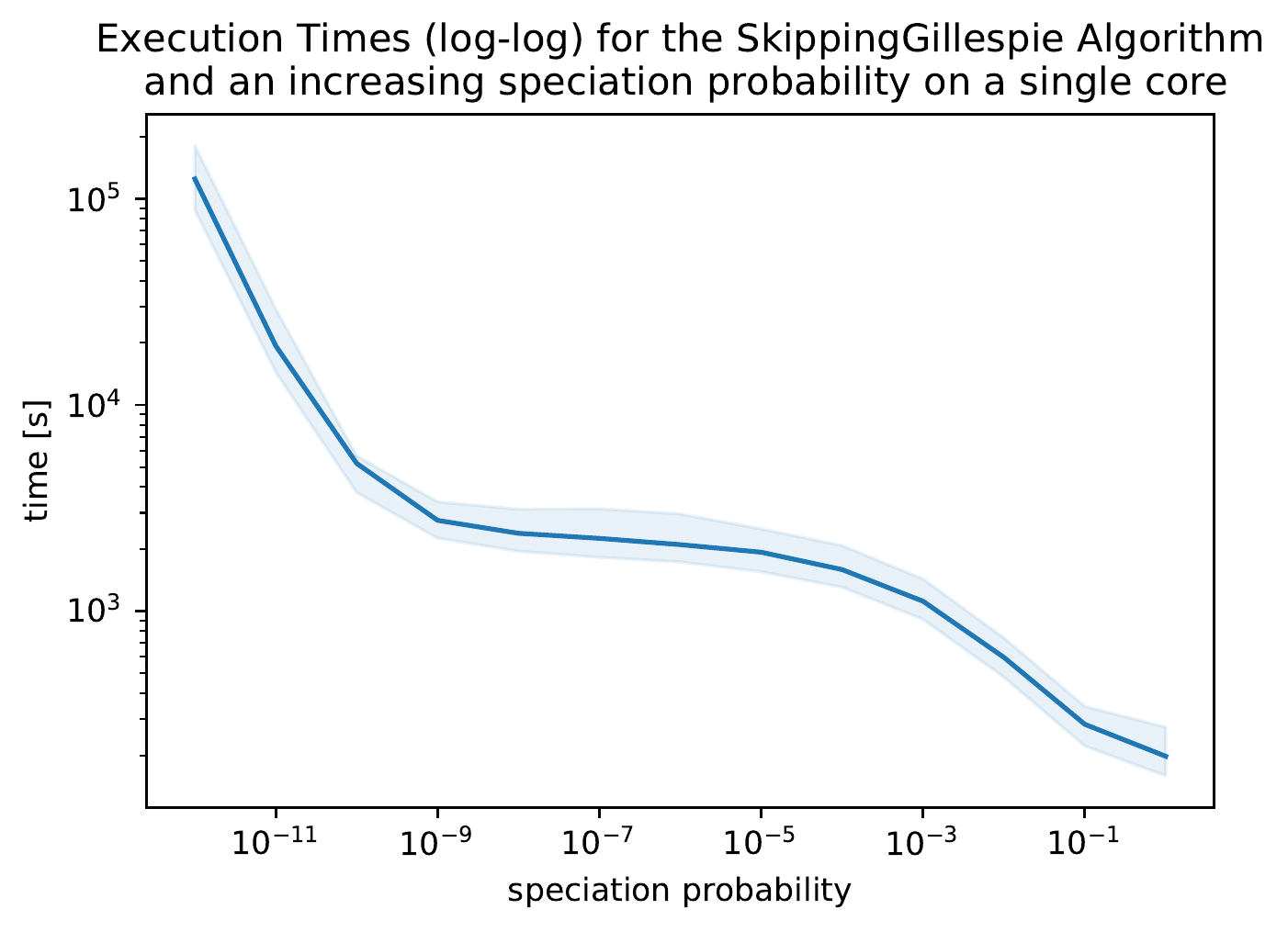}
    \captionof{figure}{Performance limit of the \textbf{SkippingGillespie} algorithm for $J = 10^{8}$ and very low speciation probabilities.}
    \label{fig:speciation-limit}
\end{minipage}
\noindent rises from $164.19\%$ to $453.93\%$ and $1,602.39\%$ respectively. Despite this high redundancy, each of the 710 partitions only takes one hour to simulate. In an HPC batch system, simulating such a large model only takes a few hours. Since batch jobs fail independently, this approach is resilient.

\vspace{6px}\noindent So how far have we pushed the limits of the biodiversity simulation? While \texttt{necsim} takes more than 48 hours to simulate $10^{8}$ individuals for $\nu = 10^{-6}$, the \textbf{SkippingGillespie} algorithm takes only one hour. We can now even simulate $\nu = 10^{-12}$ within just 27 hours. More importantly, the \textbf{Independent} algorithm has freed us from allocating a single large cluster. Instead, we can now use isolated batch jobs and, therefore, robustly and reproducibly scale up to any domain size.

\chapter{Conclusion and Future Work} \label{conclusion}

\section{Summary and Conclusions}

This project set out to expand the applicability of the neutral biodiversity simulation \texttt{necsim} by parallelising it. As a first step, we have translated the classical coalescence algorithm into a Gillespie simulation (\ref{independence-coalescence-gillespie}). We have also applied existing parallelisation strategies such as optimistic parallelism and averaging over independent subdomains (\ref{implementation-monolithic-parallelisation}). In particular, we show that the Gillespie algorithm can skip some events, allowing it to excel at very low speciation probabilities (\ref{analysis-domain-speciation-scalability}). However, parallelising the simulation with traditional methods has one fundamental problem. As there are data dependencies between different parallel simulation partitions, they must be simulated concurrently and synchronise to keep a consistent global state. Therefore, we have to allocate multiple CPUs and large amounts of RAM simultaneously.

We have reformulated the biodiversity simulation algorithm so we can simulate each individual independently and in isolation. This concept is presented in the novel Independent algorithm (\ref{independence-independent-algorithm}), which we implement on the CPU (\ref{implementation-independent}) and GPU (\ref{implementation-cuda}). Specifically, we expand the concepts of parallel, reproducible random number generators to a primeable RNG (\ref{independence-hashing-rng}). This PRNG draws random numbers deterministically based only on environmental factors such as an individual's location and time. Since the simulation requires exponentially distributed inter-event times, we also present two methods to sample the next event time (\ref{independence-exponential-generations}), which also ensure that individuals can coalesce independently (\ref{independence-align-rng}). We statistically validate both the PRNG and the sampling methods in our analysis (\ref{analysis-functional-correctness}).

Our novel Independent algorithm trades off communication for redundancy. We explore this tradeoff both for local individual deduplication (\ref{analysis-independent-sweetspot}) and global inter-partition migration (\ref{analysis-parallelism}). While every partition can be simulated in isolation, we can also split the landscape and migrate individuals between sub-simulations (\ref{implementation-independent-parallelisation}). During our analysis, we have found that a hybrid model that only communicates a fraction of all migrations performs best on the CPU. Furthermore, the GPU implementation shows the potential to outperform all other methods consistently.

Most significantly, however, the Independent algorithm can also perform partial simulations. With this reproducibility, a simulation can be run as isolated independent batch jobs, which use fewer resources and are scheduled more favourably on HPC systems. We have been able to simulate $10$ simulations with $7.1 \cdot 10^{10}$ individuals each within just a few hours. If we are only interested in part of an ecosystem, we can now simulate just the individuals of interest whilst also maintaining consistency with the not-yet simulated individuals. Overall, the Independent algorithm provides a competitive performance and significant quality of life improvements to scientists, who can now always add on just one more individual to the simulation without wasting prior computation. \\

\noindent This thesis set out to answer whether the saved communication costs outweigh the additional costs of redundant computation. We have found that local individual deduplication is crucial to get a competitive sequential performance (\ref{analysis-independent-sweetspot}). Across partitions, however, the freedom of not having to communicate has allowed us to scale up the simulation and improve its parallel performance. Thus, overall, communication-free parallelism is worth the redundancy.

\section{A Rustacean's Advice}

In this project, we have extensively used the Rust programming language to encode assumptions in the type system. There have been several instances where the compiler helped us to catch subtle assumption-breaking bugs. Our most prominent use of strongly typed guarantees is the simulation component system (\ref{fig:simulation-components}). From the experience of designing and implementing this system, we provide the following advice for future work:
\begin{enumerate}
    \item The strict guarantees of a strongly typed system offer enormous flexibility to prototype component implementations with fewer worries about forgetting to fulfil Hoare triples.
    \item This design is not at all suited for prototyping, however. Instead, type guarantees should only be introduced in fundamental parts of a system that do not change very often, as any modification of the component system breaks the compilation of most source files.
\end{enumerate}

\section{Future Work}

\noindent The \texttt{necsim} library (\ref{background-necsim}), which this project is based on, includes several quality-of-life features for ecological analysis that are not implemented in this project. Additionally, \texttt{necsim} also supports temporally varying landscapes, which enables simulating gradual habitat loss. Particularly the Independent algorithm could easily be adapted to extend the functionality of \texttt{necsim-rust}. \\

\noindent This thesis has introduced the novel Independent algorithm and implemented and parallelised it on the CPU and GPU. There are several opportunities to improve its performance further:
\begin{enumerate}
    \item \textbf{Redundancy and Deduplication:} The algorithm fundamentally trades off data dependencies for redundancy. Both the Water-Level algorithm (\ref{implementation-water-log}) and deduplication cache (\ref{implementation-deduplication}) are used to locally detect redundant work. Even still, around $50\%$ of local events are duplicates, and redundancy across partitions is even higher (\ref{analysis-limit}). For instance, reinforcement learning might be used to dynamically trade off the costs and benefits of deduplication.
    \item \textbf{Event Reporting:} Events in a parallel simulation are currently logged to disk so we can later replay them. However, the storage footprint of dispersal events is enormous (\ref{analysis-reporting-performance}). The reporting system could be extended to include reporters that can perform local analyses and later combine the partial results. Pushing such commutative and associative reporters onto the GPU could reduce data transfers (\ref{analysis-event-reporting}). In general, these reporters could circumvent most event sorting costs that limit the Independent algorithm's performance (\ref{analysis-event-throughput}).
    \item \textbf{Random Number Generation:} The CPU and GPU currently use the same primable PRNG (\ref{analysis-prime-rng}), which is based on 64bit and 128bit integer multiplications. A GPU-specific RNG that does not use these large emulated integer types (\ref{analysis-exponential-sampling}) could increase the low sweet spot thread block occupancy (\ref{analysis-cuda-sweetspot}). Furthermore, it would be worth exploring if scrambled (\ref{background-rng-scrambling}) quasi-random number generators (\ref{background-pseudo-random}) might make the simulation converge faster.
    \item \textbf{Parallelisation:} The component system developed for \texttt{necsim-rust} is built to easily support different parallelisation strategies and backends, e.g. a multi-GPU variant.
\end{enumerate}

\noindent Beyond ecology, the Independent algorithm can also be applied to several other fields such as particle transport simulations and population genetics. In particular, Jerome Kelleher, who leads the development of the \texttt{msprime} genetic simulator (\ref{background-msprime}), and his colleague Yan Wong, have expressed an interest in collaborating to bring our algorithmic approach into \texttt{msprime}. This collaboration would extend the application of this thesis into population genetics and biomedical research. \\

\noindent Most importantly, however, we hope that this project will prove helpful to conservation and environmental protection research. This Computer Science thesis has always been motivated by the dire need to protect our environment and the species that make it habitable for us.

\nocite{*}
\sloppy
\newrefcontext[sorting=none]
\printbibliography[category=cited, heading=bibintoc, title={Reference List}, segment=\therefsegment]
\newrefcontext[sorting=nty]
\printbibliography[notcategory=cited, heading=bibintoc, title={Further Reading}, segment=\therefsegment]

@article {Thompson2019,
    author = {Thompson, Samuel E. D. and Chisholm, Ryan A. and Rosindell, James},
    title = {Characterising extinction debt following habitat fragmentation using neutral theory},
    journal = {Ecology Letters},
    volume = {22},
    number = {12},
    pages = {2087-2096},
    doi = {10.1111/ele.13398},
    url = {https://onlinelibrary.wiley.com/doi/abs/10.1111/ele.13398},
    eprint = {https://onlinelibrary.wiley.com/doi/pdf/10.1111/ele.13398},
    year = {2019}
}

@article {Ceballose2015,
	author = {Ceballos, Gerardo and Ehrlich, Paul R. and Barnosky, Anthony D. and Garc{\'i}a, Andr{\'e}s and Pringle, Robert M. and Palmer, Todd M.},
	title = {Accelerated modern human{\textendash}induced species losses: Entering the sixth mass extinction},
	volume = {1},
	number = {5},
	elocation-id = {e1400253},
	year = {2015},
	doi = {10.1126/sciadv.1400253},
	publisher = {American Association for the Advancement of Science},
	URL = {https://advances.sciencemag.org/content/1/5/e1400253},
	eprint = {https://advances.sciencemag.org/content/1/5/e1400253.full.pdf},
	journal = {Science Advances}
}

@article {Woodley2019,
    author = {Woodley, Stephen and Locke, Harvey and Laffoley, Dan and MacKinnon, Kathy and Sandwith, Trevor and Smart, Jane},
    title = {A Review of Evidence for Area-based Conservation Targets for the Post-2020 Global Biodiversity Framework},
    journal = {Parks Journal},
    volume = {25},
    number = {2},
    pages = {31-46},
    doi = {10.2305/IUCN.CH.2019.PARKS-25-2SW2.en},
    year = {2019}
}

@online {aichitargets,
    author = {{Convention on Biological Diversity}},
    title = {\textit{Aichi Biodiversity Targets}},
    url = {https://www.cbd.int/sp/targets/},
    urldate = {2021-01-06}
}

@phdthesis {ThompsonPhd,
    author = {Thompson, Samuel Enrico Dale},
    title = {\textit{Theory and models of biodiversity in fragmented landscapes}},
    school = {Imperial College London},
    year = {2019},
    doi = {10.25560/76299},
}

@article {Thompson2020,
    author = {Thompson, Samuel E. D. and Chisholm, Ryan A. and Rosindell, James},
    title = {pycoalescence and rcoalescence: Packages for simulating spatially explicit neutral models of biodiversity},
    journal = {Methods in Ecology and Evolution},
    volume = {11},
    number = {10},
    pages = {1237-1246},
    doi = {10.1111/2041-210X.13451},
    url = {https://besjournals.onlinelibrary.wiley.com/doi/abs/10.1111/2041-210X.13451},
    eprint = {https://besjournals.onlinelibrary.wiley.com/doi/pdf/10.1111/2041-210X.13451},
    year = {2020}
}

@article {Mrema2020,
    author = {Mrema (ed.), Elizabeth},
    title = {Analysis of the contribution of targets established by parties and progress towards the Aichi Biodiversity Targets},
    journal = {Convention on Biological Diversity},
    year = {2020},
    url = {https://www.cbd.int/doc/c/f1e4/ab2c/ff85fe53e210872a0ceffd26/sbi-03-02-add2-en.pdf},
    urldate = {2021-01-06}
}

@article {RSPB2020,
    title = {A lost decade for nature},
    journal = {Royal Society for the Protection of Birds},
    year = {2020},
    url = {https://ww2.rspb.org.uk/Images/A LOST DECADE FOR NATURE_tcm9-481563.pdf},
    urldate = {2020-12-13}
}

@unpublished {RustContracts,
    author = {karoffel},
    title = {\textit{contracts}},
    note = {A crate implementing ``Design by Contract'' via procedural macros.},
    version = {0.6.0},
    license = {Mozilla Public License 2.0},
    url = {https://docs.rs/contracts/0.6.0/contracts/},
    urldate = {2021-01-06},
}

@article {Mehran2017,
    author = {Karimzadeh, Mehran and Hoffman, Michael M},
    title = {Top considerations for creating bioinformatics software documentation},
    journal = {Briefings in Bioinformatics},
    volume = {19},
    number = {4},
    pages = {693-699},
    year = {2017},
    month = {01},
    issn = {1477-4054},
    doi = {10.1093/bib/bbw134},
    url = {https://doi.org/10.1093/bib/bbw134},
    eprint = {https://academic.oup.com/bib/article-pdf/19/4/693/25193101/bbw134.pdf},
}

@article {Ferre2001,
    author = {Ferré, Xavier and Juristo, Natalia and Windl, Helmut and Constantine, Larry},
    journal = {IEEE Software},
    title = {Usability basics for software developers},
    year = {2001},
    volume = {18},
    number = {1},
    pages = {22-29},
    doi = {10.1109/52.903160},
    ISSN = {1937-4194},
    month = {01}
}

@book {Klabnik2019,
    author    = {Klabnik, Steve and Nichols, Carol},
    title     = {\textit{The Rust Programming Language (Covers Rust 2018)}},
    publisher = {no starch press},
    year      = {2019},
    edition   = {2},
    month     = {08},
    isbn      = {9781718500440},
    url       = {https://doc.rust-lang.org/stable/book/},
    urldate   = {2021-01-07},
}

@book {Rustonomicon,
    author    = {{The Rust Project Developers}},
    title     = {\textit{The Rustonomicon}},
    url       = {https://doc.rust-lang.org/nomicon/},
    urldate   = {2021-01-07},
}

@unpublished {RustPtxBuilder,
    author = {Zariaiev, Denys},
    title = {\textit{rust-ptx-builder}},
    note = {NVPTX build helper},
    version = {0.5.3},
    license = {MIT License},
    url = {https://docs.rs/ptx-builder/0.5.3/ptx_builder/},
    urldate = {2021-01-06},
}

@unpublished {RustPtxLinker,
    author = {Zariaiev, Denys},
    title = {\textit{rust-ptx-linker}},
    note = {LLVM NVPTX bitcode linker for Rust without external system dependencies},
    version = {0.9.1},
    license = {MIT License},
    url = {https://crates.io/crates/ptx-linker},
    urldate = {2021-01-06},
}

@unpublished {RustaCUDA,
    author = {Heisler, Brook},
    title = {\textit{RustaCUDA}},
    note = {High-level Interface to NVIDIA® CUDA™ Driver API in Rust},
    version = {0.1.2},
    license = {Apache-2.0 License OR MIT License},
    url = {https://docs.rs/rustacuda/0.1.2/rustacuda/},
    urldate = {2021-01-06},
}

@unpublished {Accel,
    author = {Teramura, Toshiki},
    title = {\textit{Accel}},
    note = {GPGPU Framework for Rust},
    license = {Apache-2.0 License OR MIT License},
    url = {https://docs.rs/accel/0.3.1/accel/},
    urldate = {2021-01-06},
}

@online {RustNVPTX,
    author = {Aparicio, Jorge},
    title = {\textit{NVPTX backend metabug}},
    url = {https://github.com/rust-lang/rust/issues/38789},
    urldate = {2021-01-06}
}

@online {RustPtxAbi,
    author = {Aparicio, Jorge},
    title = {\textit{Tracking issue for the "ptx-kernel" ABI}},
    url = {https://github.com/rust-lang/rust/issues/38788},
    urldate = {2021-01-06}
}

@misc {Grover2009,
    author       = {Grover, Vinod and Kerr, Andrew and Lee, Sean},
    title        = {PLANG: Translating NVIDIA PTX language to LLVM IR Machine},
    howpublished = {In: LLVM. \textit{2009 LLVM Developers' Meeting}},
    month        = {10},
    year         = {2009},
    url = {https://llvm.org/devmtg/2009-10/Grover_PLANG.pdf},
    urldate = {2021-01-07}
}

@misc {Holewinski2011,
    author       = {Holewinski, Justin},
    title        = {PTX Back-End: GPU Programming with LLVM},
    howpublished = {In: LLVM. \textit{2011 LLVM Developers' Meeting}},
    month        = {11},
    year         = {2011},
    url = {https://llvm.org/devmtg/2011-11/Holewinski_PTXBackend.pdf},
    urldate = {2021-01-07}
}

@misc {Harris2012,
    author       = {Harris, Mark},
    title        = {Compiling Parallel Languageswith the NVIDIA Compiler SDK},
    howpublished = {In: NVIDIA. \textit{GPU Technology Conference 2012}},
    month        = {05},
    year         = {2012},
    url = {https://on-demand.gputechconf.com/supercomputing/2012/presentation/SB019-Harris-Compiling-Parallel-Languages-Compiler-SDK.pdf},
    urldate = {2021-01-07}
}

@manual {CUDARuntimeAPI,
    title        = {\textit{CUDA Runtime API}},
    organization = {NVIDIA},
    edition      = {v11.2.0},
    month        = {07},
    year         = {2019},
    url          = {https://docs.nvidia.com/cuda/pdf/CUDA_Runtime_API.pdf},
    urldate      = {2021-01-07}
}

@manual {CUDAVolta,
    title        = {\textit{Tuning CUDA Applications for Volta}},
    organization = {NVIDIA},
    edition      = {v11.2},
    month        = {12},
    year         = {2020},
    url          = {https://docs.nvidia.com/cuda/pdf/Volta_Tuning_Guide.pdf},
    urldate      = {2021-01-07}
}

@manual {CUDACpp,
    title        = {\textit{CUDA C++ Programming Guide}},
    organization = {NVIDIA},
    edition      = {v11.2},
    month        = {05},
    year         = {2021},
    url          = {https://docs.nvidia.com/cuda/pdf/CUDA_C_Programming_Guide.pdf},
    urldate      = {2021-01-07}
}

@manual {nvcc,
    title        = {\textit{CUDA Compiler Driver NVCC}},
    organization = {NVIDIA},
    edition      = {v11.2},
    month        = {12},
    year         = {2020},
    url          = {https://docs.nvidia.com/cuda/pdf/CUDA_Compiler_Driver_NVCC.pdf},
    urldate      = {2021-01-07}
}

@manual {ptxisa,
    title        = {\textit{Parallel Thread Execution ISA}},
    organization = {NVIDIA},
    edition      = {v7.2},
    month        = {12},
    year         = {2020},
    url          = {https://docs.nvidia.com/cuda/pdf/ptx_isa_7.2.pdf},
    urldate      = {2021-01-07}
}

@online {EU2020,
    author = {{European Commission}},
    title = {\textit{Explanatory note on ``exclusive focus on civil applications''}},
    url = {http://ec.europa.eu/research/participants/portal/doc/call/h2020/h2020-bes-2015/1645164-explanatory_note_on_exclusive_focus_on_civil_applications_en.pdf},
    urldate = {2021-01-08}
}

@online {RustLicense,
    author = {{Rust Team}},
    title = {\textit{Licenses}},
    url = {https://www.rust-lang.org/policies/licenses},
    urldate = {2021-01-08}
}

@book {FosterIan1995,
    publisher = {Addison-Wesley},
    isbn = {0201575949},
    year = {1995},
    title = {\textit{Designing and building parallel programs: concepts and tools for parallel software engineering}},
    language = {eng},
    address = {Reading, MA; Wokingham},
    author = {Foster, Ian},
    lccn = {94003661},
    url = {https://edoras.sdsu.edu/~mthomas/docs/foster/Foster_Designing_and_Building_Parallel_Programs.pdf},
    urldate = {2021-06-01}
}

@conference {Cleve1986,
    author       = {Moler, Cleve},
    title        = {Matrix Computation on Distributed Memory Multiprocessors},
    booktitle    = {Hypercube Multiprocessors, Proceedings of the First Conference on Hypercube Multiprocessors},
    year         = {1986},
    editor       = {Heath, Michael T.},
    pages        = {181-195},
    organization = {Oak Ridge National Laboratory},
    isbn         = {9780898712094}
}

@misc {Cleve2007,
    author       = {{ANL Training}},
    title        = {\textit{Evolution of MATLAB | Cleve Moler, MathWorks}},
    howpublished = {[Video]},
    year         = {2007},
    url = {https://youtu.be/YGpUKWKo410},
    urldate = {2021-01-08}
}

@online {Cleve2013,
    author = {Moler, Cleve},
    title = {\textit{The Intel Hypercube, part 2, reposted}},
    url = {https://blogs.mathworks.com/cleve/2013/11/12/the-intel-hypercube-part-2-reposted/},
    year = {2013},
    month = {11},
    day = {12},
    urldate = {2021-06-02}
}

@inproceedings {Kunz2012,
    author = {Kunz, Georg and Schemmel, Daniel and Gross, James and Wehrle, Klaus},
    booktitle = {2012 ACM/IEEE/SCS 26th Workshop on Principles of Advanced and Distributed Simulation},
    title = {Multi-level Parallelism for Time- and Cost-Efficient Parallel Discrete Event Simulation on GPUs},
    year = {2012},
    pages = {23-32},
    doi = {10.1109/PADS.2012.27},
    ISSN= {1087-4097},
    month={07}
}

@online {Christophe2014,
    author = {{Christophe}},
    title = {\textit{Internal and external parallelism}},
    url = {https://stackoverflow.com/a/27346561},
    date = {2014-12-07},
    urldate = {2021-01-08}
}

@inproceedings {Porobic2012,
    title = {OLTP on Hardware Islands},
    booktitle = {Proceedings of the VLDB Endowment (PVLDB)},
    author = {Porobic, Danica and Pandis, Ippokratis and Branco, Miguel and Tözün, Pınar and Ailamaki, Anastasia},
    year = {2012},
    eprint={1208.0227},
    archivePrefix={arXiv},
    primaryClass={cs.DB}
}

@online {Sanglard2020,
    author = {Sanglard, Fabien},
    title = {\textit{A history of NVidia Stream Multiprocessor}},
    url = {https://fabiensanglard.net/cuda/index.html},
    date = {2020-05-02},
    urldate = {2021-01-08}
}

@misc {Tamasi2008,
    author       = {Tamasi, Tony},
    title        = {The Evolution of Computer Graphics},
    howpublished = {In: NVIDIA. \textit{Nvision 08}},
    month        = {08},
    year         = {2008},
    url = {https://www.nvidia.com/content/nvision2008/tech_presentations/Technology_Keynotes/NVISION08-Tech_Keynote-GPU.pdf},
    urldate = {2021-01-08}
}

@inproceedings{MPI1993,
    author = {{The MPI Forum, CORPORATE}},
    title = {MPI: A Message Passing Interface},
    year = {1993},
    isbn = {0818643404},
    publisher = {Association for Computing Machinery},
    address = {New York, NY, USA},
    url = {https://doi.org/10.1145/169627.169855},
    doi = {10.1145/169627.169855},
    booktitle = {Proceedings of the 1993 ACM/IEEE Conference on Supercomputing},
    pages = {878–883},
    numpages = {6},
    location = {Portland, Oregon, USA},
    series = {Supercomputing '93}
}

@article {MPIv4,
    author  = {{Message Passing Interface Forum}},
    title   = {\textit{MPI: A Message-Passing Interface Standard}},
    year    = {2020},
    note    = {v4.0 Draft},
    url     = {https://www.mpi-forum.org/docs/drafts/mpi-2020-draft-report.pdf},
    urldate = {2021-01-08}
}

@article {MPIv3.1,
    author  = {{Message Passing Interface Forum}},
    title   = {\textit{MPI: A Message-Passing Interface Standard}},
    year    = {2015},
    note    = {v3.1},
    url     = {https://www.mpi-forum.org/docs/mpi-3.1/mpi31-report.pdf},
    urldate = {2021-01-09}
}

@inproceedings {Salmon2011,
    author = {Salmon, John K. and Moraes, Mark A. and Dror, Ron O. and Shaw, David E.},
    title = {Parallel Random Numbers: As Easy as 1, 2, 3},
    year = {2011},
    isbn = {9781450307710},
    publisher = {Association for Computing Machinery},
    address = {New York, NY, USA},
    url = {https://doi.org/10.1145/2063384.2063405},
    doi = {10.1145/2063384.2063405},
    booktitle = {Proceedings of 2011 International Conference for High Performance Computing, Networking, Storage and Analysis},
    articleno = {16},
    numpages = {12},
    location = {Seattle, Washington},
    series = {SC '11}
}

@article {Rosindell2008,
    author = {Rosindell, James and Wong, Yan and Etienne, Rampal S},
    journal = {Ecological Informatics},
    pages = {259-271},
    title = {A coalescence approach to spatial neutral ecology},
    volume = {3},
    year = {2008},
    issn = {1574-9541},
    doi = {10.1016/j.ecoinf.2008.05.001},
    url = {http://www.sciencedirect.com/science/article/pii/S1574954108000265}
}

@book {Hubbell2001,
    isbn = {9780691021287},
    url = {http://www.jstor.org/stable/j.ctt7rj8w},
    author = {Hubbell, Stephen P},
    publisher = {Princeton University Press},
    title = {\textit{The Unified Neutral Theory of Biodiversity and Biogeography (MPB-32)}},
    year = {2001},
    urldate = {2021-01-13}
}

@article {Perkel2020,
    title   = {Why scientists are turning to Rust},
    author  = {Perkel, Jeffrey M.},
    journal = {Nature},
    year    = {2020},
    number  = {588},
    pages   = {185-186},
    month   = {12},
    doi     = {10.1038/d41586-020-03382-2}
}

@misc {Hoare2010,
    author       = {Hoare, Graydon},
    title        = {Project Servo - Technology from the past come to save the future from itself},
    howpublished = {In: \textit{Mozilla Annual Summit 2010}},
    month        = {07},
    year         = {2010},
    url = {http://venge.net/graydon/talks/intro-talk-2.pdf},
    urldate = {2021-01-14}
}

@online {Rust2015,
    author = {{The Rust Core Team}},
    title = {\textit{Announcing Rust 1.0}},
    url = {https://blog.rust-lang.org/2015/05/15/Rust-1.0.html},
    year = {2015},
    month = {05},
    urldate = {2021-01-14}
}

@online {Rust2018,
    author = {{The Rust Core Team}},
    title = {\textit{Announcing Rust 1.31 and Rust 2018}},
    url = {https://blog.rust-lang.org/2018/12/06/Rust-1.31-and-rust-2018.html},
    year = {2018},
    month = {12},
    urldate = {2021-01-14}
}

@online {Rust1.52,
    author = {{The Rust Release Team}},
    title = {\textit{Announcing Rust 1.52.0}},
    url = {https://blog.rust-lang.org/2021/05/06/Rust-1.52.0.html},
    year = {2021},
    month = {05},
    urldate = {2021-05-16}
}

@online {Rust2021,
    author = {Bos, Mara and Matsakis, Niko and Levick, Ryan},
    title = {\textit{The Plan for the Rust 2021 Edition}},
    url = {https://blog.rust-lang.org/2021/05/11/edition-2021.html},
    year = {2021},
    month = {05},
    urldate = {2021-05-16}
}

@unpublished {StackOverflow2020,
    author = {{Stack Overflow}},
    title  = {\textit{2020 Developer Survey}},
    year   = {2020},
    url    = {https://insights.stackoverflow.com/survey/2020},
    urldate = {2021-01-14}
}

@inproceedings {Schaerli2003,
    author = {Sch{\"a}rli, Nathanael and Ducasse, St{\'e}phane and Nierstrasz, Oscar and Black, Andrew P.},
    editor = "Cardelli, Luca",
    title = {Traits: Composable Units of Behaviour},
    booktitle = {ECOOP 2003 -- Object-Oriented Programming},
    year = {2003},
    publisher = {Springer Berlin Heidelberg},
    address = {Berlin, Heidelberg},
    pages = {248-274},
    isbn = {978-3-540-45070-2},
    doi = {10.1007/978-3-540-45070-2_12}
}

@article {Jung2017,
    author = {Jung, Ralf and Jourdan, Jacques-Henri and Krebbers, Robbert and Dreyer, Derek},
    title = {RustBelt: Securing the Foundations of the Rust Programming Language},
    year = {2017},
    issue_date = {January 2018},
    publisher = {Association for Computing Machinery},
    address = {New York, NY, USA},
    volume = {2},
    number = {POPL},
    url = {https://doi.org/10.1145/3158154},
    doi = {10.1145/3158154},
    journal = {Proc. ACM Program. Lang.},
    month = {12},
    articleno = {66},
    numpages = {34}
}

@phdthesis {JungPhD,
    author = {Jung, Ralf},
    title = {\textit{Understanding and evolving the Rust programming language}},
    school = {Universit{\"a}t des Saarlandes},
    year = {2020},
    doi = {10.22028/D291-31946},
}

@book {Bhattacharjee2020,
%     author    = {Bhattacharjee, Joydeep},
%     title     = {\textit{Practical Machine Learning with Rust}},
%     publisher = {Apress},
%     year      = {2020},
%     address   = {Berkeley, CA},
%     isbn      = {978-1-4842-5120-1},
%     doi       = {10.1007/978-1-4842-5121-8},
% }

@article {Jung2018,
    author = {Jung, Ralf and Jourdan, Jacques-Henri and Krebbers, Robbert and Dreyer, Derek},
    title = {RustBelt: Securing the Foundations of the Rust Programming Language},
    year = {2017},
    issue_date = {January 2018},
    publisher = {Association for Computing Machinery},
    address = {New York, NY, USA},
    volume = {2},
    number = {POPL},
    url = {https://doi.org/10.1145/3158154},
    doi = {10.1145/3158154},
    journal = {Proc. ACM Program. Lang.},
    month = {12},
    articleno = {66},
    numpages = {34}
}

@article{Astrauskas2019,
    author = {Astrauskas, Vytautas and M\"{u}ller, Peter and Poli, Federico and Summers, Alexander J.},
    title = {Leveraging Rust Types for Modular Specification and Verification},
    year = {2019},
    issue_date = {October 2019},
    publisher = {Association for Computing Machinery},
    address = {New York, NY, USA},
    volume = {3},
    number = {OOPSLA},
    url = {https://doi.org/10.1145/3360573},
    doi = {10.1145/3360573},
    journal = {Proc. ACM Program. Lang.},
    month = {10},
    articleno = {147},
    numpages = {30}
}

@inproceedings {Evans2020,
    author = {Evans, Ana Nora and Campbell, Bradford and Soffa, Mary Lou},
    title = {Is Rust Used Safely by Software Developers?},
    year = {2020},
    isbn = {9781450371216},
    publisher = {Association for Computing Machinery},
    address = {New York, NY, USA},
    url = {https://doi.org/10.1145/3377811.3380413},
    doi = {10.1145/3377811.3380413},
    booktitle = {Proceedings of the ACM/IEEE 42nd International Conference on Software Engineering},
    pages = {246–257},
    numpages = {12},
    location = {Seoul, South Korea},
    series = {ICSE '20}
}

@unpublished {MIRAI,
    author = {{Facebook Inc.}},
    title = {\textit{MIRAI}},
    note = {An abstract interpreter for the Rust compiler's mid-level intermediate representation.},
    license = {MIT License},
    url = {https://github.com/facebookexperimental/MIRAI},
    urldate = {2021-01-14},
}

@inbook {Rosindell2021,
    place = {Cambridge},
    series = {Ecology, Biodiversity and Conservation},
    title = {The Species–Area Relationships of Ecological Neutral Theory},
    doi = {10.1017/9781108569422.016},
    booktitle = {The Species–Area Relationship: Theory and Application},
    publisher = {Cambridge University Press},
    author = {Rosindell, James and Chisholm, Ryan A.},
    editor = {Matthews, Thomas J. and Triantis, Kostas A. and Whittaker, Robert J.Editors},
    year = {2021},
    pages = {259–288},
    collection = {Ecology, Biodiversity and Conservation}
}

@article {Etienne2005,
    author = {Etienne, Rampal S. and Alonso, David},
    title = {A dispersal-limited sampling theory for species and alleles},
    journal = {Ecology Letters},
    volume = {8},
    number = {11},
    pages = {1147-1156},
    doi = {https://doi.org/10.1111/j.1461-0248.2005.00817.x},
    url = {https://onlinelibrary.wiley.com/doi/abs/10.1111/j.1461-0248.2005.00817.x},
    eprint = {https://onlinelibrary.wiley.com/doi/pdf/10.1111/j.1461-0248.2005.00817.x},
    year = {2005}
}

@article {Chisholm2016,
    author = {Chisholm, Ryan A. and Fung, Tak and Chimalakonda, Deepthi and O'Dwyer, James P.},
    title = {Maintenance of biodiversity on islands},
    journal = {Proceedings of the Royal Society B: Biological Sciences},
    volume = {283},
    number = {1829},
    pages = {20160102},
    year = {2016},
    doi = {10.1098/rspb.2016.0102},
    url = {https://royalsocietypublishing.org/doi/abs/10.1098/rspb.2016.0102},
    eprint = {https://royalsocietypublishing.org/doi/pdf/10.1098/rspb.2016.0102}
}

@article {Vallade2003,
    title = {Analytical solution of a neutral model of biodiversity},
    author = {Vallade, M. and Houchmandzadeh, B.},
    journal = {Phys. Rev. E},
    volume = {68},
    issue = {6},
    pages = {061902},
    numpages = {5},
    year = {2003},
    month = {12},
    publisher = {American Physical Society},
    doi = {10.1103/PhysRevE.68.061902},
    url = {https://link.aps.org/doi/10.1103/PhysRevE.68.061902},
    eprint = {https://hal.archives-ouvertes.fr/hal-00976654/document}
}

@article {Rosindell2010,
    author = {Rosindell, James and Cornell, Stephen J. and Hubbell, Stephen P. and Etienne, Rampal S.},
    title = {Protracted speciation revitalizes the neutral theory of biodiversity},
    journal = {Ecology Letters},
    volume = {13},
    number = {6},
    pages = {716-727},
    doi = {10.1111/j.1461-0248.2010.01463.x},
    url = {https://onlinelibrary.wiley.com/doi/abs/10.1111/j.1461-0248.2010.01463.x},
    eprint = {https://onlinelibrary.wiley.com/doi/pdf/10.1111/j.1461-0248.2010.01463.x},
    year = {2010}
}

@article {ODwyer2018,
    author = {O'Dwyer, James P. and Cornell, Stephen J.},
    title = {Cross-scale neutral ecology and the maintenance of biodiversity},
    journal = {Scientific Reports},
    year = {2018},
    month = {07},
    day = {05},
    volume = {8},
    number = {1},
    pages = {10200},
    issn = {2045-2322},
    doi = {10.1038/s41598-018-27712-7},
    url = {https://doi.org/10.1038/s41598-018-27712-7}
}

@online {Shaaban2014,
%    author = {Shaaban, Muhammad},
%    title = {\textit{Considerations in Parallel Program Creation Steps for Performance}},
%    url = {http://meseec.ce.rit.edu/cmpe655-fall2014/655-10-16-2014.pdf},
%    date = {2014-10-16},
%    urldate = {2021-01-15}
%}

@online {Gehringer2002,
%    author = {Gehringer, Edward F.},
%    title = {\textit{Lecture 6: Architecture of Parallel Computers: Steps in parallelization}},
%    url = {https://people.engr.ncsu.edu/efg/506/s02/lectures/notes/lec6.pdf},
%    date = {2002-01-29},
%    urldate = {2021-01-15}
%}

@online {Jain2014,
%    author = {Jain, Amit},
%    title = {\textit{Embarrassingly Parallel Computations}},
%    url = {http://cs.boisestate.edu/~amit/teaching/530/handouts/ep.pdf},
%    year = {2014},
%    urldate = {2021-01-15}
%}

@online {Weiss2017,
%    author = {Jain, Amit},
%    title = {\textit{CSci 493.65 Parallel Computing: Chapter 3 Parallel Algorithm Design}},
%    url = {http://compsci.hunter.cuny.edu/~sweiss/course_materials/csci493.65/lecture_notes/chapter03.pdf},
%    year = {2017},
%    urldate = {2021-01-15}
%}

@article {Gillespie1976,
    title = {A general method for numerically simulating the stochastic time evolution of coupled chemical reactions},
    journal = {Journal of Computational Physics},
    volume = {22},
    number = {4},
    pages = {403-434},
    year = {1976},
    issn = {0021-9991},
    doi = {10.1016/0021-9991(76)90041-3},
    url = {http://www.sciencedirect.com/science/article/pii/0021999176900413},
    author = {Gillespie, Daniel T}
}

@article {Gillespie2007,
    author = {Gillespie, Daniel T.},
    title = {Stochastic Simulation of Chemical Kinetics},
    journal = {Annual Review of Physical Chemistry},
    volume = {58},
    number = {1},
    pages = {35-55},
    year = {2007},
    doi = {10.1146/annurev.physchem.58.032806.104637},
    note ={PMID: 17037977},
    URL = {https://doi.org/10.1146/annurev.physchem.58.032806.104637},
    eprint = {https://doi.org/10.1146/annurev.physchem.58.032806.104637}
}

@article {Yang2004,
    author = {Cao, Yang and Li, Hong and Petzold, Linda},
    title = {Efficient formulation of the stochastic simulation algorithm for chemically reacting systems},
    journal = {The Journal of Chemical Physics},
    volume = {121},
    number = {9},
    pages = {4059-4067},
    year = {2004},
    doi = {10.1063/1.1778376},
    url = {https://doi.org/10.1063/1.1778376},
    eprint = {https://doi.org/10.1063/1.1778376}
}

@article {McCollum2006,
    title = {The sorting direct method for stochastic simulation of biochemical systems with varying reaction execution behavior},
    journal = {Computational Biology and Chemistry},
    volume = {30},
    number = {1},
    pages = {39-49},
    year = {2006},
    issn = {1476-9271},
    doi = {10.1016/j.compbiolchem.2005.10.007},
    url = {http://www.sciencedirect.com/science/article/pii/S1476927105001088},
    author = {McCollum, James M. and Peterson, Gregory D. and Cox, Chris D. and Simpson, Michael L. and Samatova, Nagiza F.}
}

@article{Gibson2000,
    author = {Gibson, Michael A. and Bruck, Jehoshua},
    title = {Efficient Exact Stochastic Simulation of Chemical Systems with Many Species and Many Channels},
    journal = {The Journal of Physical Chemistry A},
    volume = {104},
    number = {9},
    pages = {1876-1889},
    year = {2000},
    doi = {10.1021/jp993732q},
    url = {https://doi.org/10.1021/jp993732q},
    eprint = {https://doi.org/10.1021/jp993732q}
}

@article {Gillespie2001,
    author = {Gillespie, Daniel T.},
    title = {Approximate accelerated stochastic simulation of chemically reacting systems},
    journal = {The Journal of Chemical Physics},
    volume = {115},
    number = {4},
    pages = {1716-1733},
    year = {2001},
    doi = {10.1063/1.1378322},
    url = {https://doi.org/10.1063/1.1378322},
    eprint = {https://doi.org/10.1063/1.1378322}
}

@article {Gillespie1977,
    author = {Gillespie, Daniel T.},
    title = {Exact stochastic simulation of coupled chemical reactions},
    journal = {The Journal of Physical Chemistry},
    volume = {81},
    number = {25},
    pages = {2340-2361},
    year = {1977},
    doi = {10.1021/j100540a008},
    url = {https://doi.org/10.1021/j100540a008},
    eprint = {https://doi.org/10.1021/j100540a008}
}

@article {Masuda2018,
%   title = {A Gillespie Algorithm for Non-Markovian Stochastic Processes},
%   volume = {60},
%   issn = {1095-7200},
%   url = {https://doi.org/10.1137/16M1055876},
%   doi = {10.1137/16m1055876},
%   number = {1},
%   journal = {SIAM Review},
%   publisher = {Society for Industrial \& Applied Mathematics (SIAM)},
%   author = {Masuda, Naoki and Rocha, Luis E. C.},
%   year = {2018},
%   month = {01},
%   pages = {95–115}
% }

@techreport {ONeill2014,
    title = {\textit{PCG: A Family of Simple Fast Space-Efficient Statistically Good Algorithms for Random Number Generation}},
    author = {O'Neill, Melissa E.},
    institution = {Harvey Mudd College},
    address = {Claremont, CA},
    number = {HMC-CS-2014-0905},
    year = {2014},
    month = {09},
    url = {https://www.cs.hmc.edu/tr/hmc-cs-2014-0905.pdf},
    urldate = {2021-01-17},
}

@misc {ONeill2017,
    title = {\textit{Too Big to Fail}},
    author = {O'Neill, Melissa E.},
    year = {2017},
    month = {08},
    day = {20},
    url = {https://www.cs.hmc.edu/tr/hmc-cs-2014-0905.pdf},
    urldate = {2021-05-16},
}

@misc {Dammertz2010,
    author       = {Dammertz, Holger},
    title        = {Massively Parallel Random Number Generators},
    howpublished = {In: \textit{NVIDIA GTC 2010}},
    month        = {10},
    year         = {2010},
    url = {https://www.nvidia.com/content/GTC-2010/pdfs/2136_GTC2010.pdf},
    urldate = {2021-01-17}
}

@inproceedings {Claessen2013,
    author = {Claessen, Koen and Pa\l{}ka, Micha\l{} H.},
    title = {Splittable Pseudorandom Number Generators Using Cryptographic Hashing},
    year = {2013},
    isbn = {9781450323833},
    publisher = {Association for Computing Machinery},
    address = {New York, NY, USA},
    url = {https://doi.org/10.1145/2503778.2503784},
    doi = {10.1145/2503778.2503784},
    booktitle = {Proceedings of the 2013 ACM SIGPLAN Symposium on Haskell},
    pages = {47–58},
    numpages = {12},
    location = {Boston, Massachusetts, USA},
    series = {Haskell '13}
}

@techreport {Schoo2005,
    title = {\textit{A Survey and Empirical Comparison of Modern Pseudo-Random Number Generators for Distributed Stochastic Simulations}},
    author = {Schoo, Marcus and Pawlikowski, Krzysztof and McNickle, Donald C.},
    institution = {University of Canterbury},
    address = {Christchurch, New Zealand},
    number = {TR-CSSE 03/05},
    year = {2005},
    url = {https://www.cosc.canterbury.ac.nz/research/reports/TechReps/2005/tr_0503.pdf},
    urldate = {2021-01-17},
}

@article {LEcuyer2001,
    title = {On the performance of birthday spacings tests with certain families of random number generators},
    journal = {Mathematics and Computers in Simulation},
    volume = {55},
    number = {1},
    pages = {131-137},
    year = {2001},
    note = {The Second IMACS Seminar on Monte Carlo Methods},
    issn = {0378-4754},
    doi = {10.1016/S0378-4754(00)00253-6},
    url = {http://www.sciencedirect.com/science/article/pii/S0378475400002536},
    author = {L’Ecuyer, Pierre and Simard, Richard}
}

@misc {Widynski2020,
    title = {\textit{Squares: A Fast Counter-Based RNG}},
    author = {Widynski, Bernard},
    year = {2020},
    eprint={2004.06278},
    archivePrefix={arXiv},
    primaryClass={cs.DS}
}

@misc{Widynski2017,
    title = {\textit{Middle Square Weyl Sequence RNG}},
    author = {Widynski, Bernard},
    year = {2020},
    eprint={1704.00358},
    archivePrefix={arXiv},
    primaryClass={cs.CR}
}

@article {LEcuyer2007,
    author = {L'Ecuyer, Pierre and Simard, Richard},
    title = {TestU01: A C Library for Empirical Testing of Random Number Generators},
    year = {2007},
    issue_date = {August 2007},
    publisher = {Association for Computing Machinery},
    address = {New York, NY, USA},
    volume = {33},
    number = {4},
    issn = {0098-3500},
    url = {https://doi.org/10.1145/1268776.1268777},
    doi = {10.1145/1268776.1268777},
    journal = {ACM Trans. Math. Softw.},
    month = {08},
    articleno = {22},
    numpages = {40}
}

@article {LEcuyer2017,
    title = {Random numbers for parallel computers: Requirements and methods, with emphasis on GPUs},
    journal = {Mathematics and Computers in Simulation},
    volume = {135},
    pages = {3-17},
    year = {2017},
    note = {Special Issue: 9th IMACS Seminar on Monte Carlo Methods},
    issn = {0378-4754},
    doi = {10.1016/j.matcom.2016.05.005},
    url = {http://www.sciencedirect.com/science/article/pii/S0378475416300829},
    author = {L’Ecuyer, Pierre and Munger, David and Oreshkin, Boris and Simard, Richard}
}

@inproceedings {Neves2012,
    author = {Neves, Samuel and Araujo, Filipe},
    editor = {Wyrzykowski, Roman and Dongarra, Jack and Karczewski, Konrad and Wa{\'{s}}niewski, Jerzy},
    title = {Fast and Small Nonlinear Pseudorandom Number Generators for Computer Simulation},
    booktitle = {Parallel Processing and Applied Mathematics},
    year = {2012},
    publisher = {Springer Berlin Heidelberg},
    address = {Berlin, Heidelberg},
    pages = {92-101},
    isbn = {978-3-642-31464-3},
    doi = {10.1007/978-3-642-31464-3},
    eprint = {https://link.springer.com/chapter/10.1007%2F978-3-642-31464-3_10},
}

@article {Coddington1996,
    author = {Coddington, Paul D.},
    title = {Random number generators for parallel computers},
    journal = {The NHSE Review},
    year = {1996},
    volume = {2},
    url = {http://citeseerx.ist.psu.edu/viewdoc/summary?doi=10.1.1.84.545},
    urldate = {2021-01-17},
}

@article {Tan2002,
    title = {The PLFG parallel pseudo-random number generator},
    journal = {Future Generation Computer Systems},
    volume = {18},
    number = {5},
    pages = {693-698},
    year = {2002},
    note = {ICCS2001},
    issn = {0167-739X},
    doi = {10.1016/S0167-739X(02)00034-1},
    url = {http://www.sciencedirect.com/science/article/pii/S0167739X02000341},
    author = {Tan, Chih Jeng Kenneth}
}

@misc {Jun2018,
    author       = {Jun, Soon Yung and Apostolakis, John},
    title        = {Parallel Random Number Generators:  VecRNG},
    howpublished = {In: \textit{2018 Geant4 Collaboration Meeting}},
    month        = {08},
    year         = {2018},
    address      = {Lund, Sweden},
    url = {https://indico.cern.ch/event/727112/contributions/3097294/attachments/1708255/2753368/VRNG-Geant42018.pdf},
    urldate = {2021-01-17}
}

@incollection {Bradley2011,
    title = {Chapter 16 - Parallelization Techniques for Random Number Generators},
    editor = {Hwu, Wen-mei W.},
    booktitle = {GPU Computing Gems Emerald Edition},
    publisher = {Morgan Kaufmann},
    address = {Boston},
    pages = {231-246},
    year = {2011},
    series = {Applications of GPU Computing Series},
    isbn = {978-0-12-384988-5},
    doi = {10.1016/B978-0-12-384988-5.00016-4},
    url = {http://www.sciencedirect.com/science/article/pii/B9780123849885000164},
    author = {Bradley, Thomas and {du Toit}, Jacques and Tong, Robert and Giles, Mike and Woodhams, Paul}
}

@article {Eddy1990,
    title = {Random number generators for parallel processors},
    journal = {Journal of Computational and Applied Mathematics},
    volume = {31},
    number = {1},
    pages = {63-71},
    year = {1990},
    issn = {0377-0427},
    doi = {10.1016/0377-0427(90)90336-X},
    url = {http://www.sciencedirect.com/science/article/pii/037704279090336X},
    author = {Eddy, William F.}
}

@article {Burley2020,
    author =       {Burley, Brent},
    title =        {Practical Hash-based Owen Scrambling},
    year =         {2020},
    month =        {12},
    day =          {29},
    journal =      {Journal of Computer Graphics Techniques (JCGT)},
    volume =       {10},
    number =       {4},
    pages =        {1-20},
    url =          {http://jcgt.org/published/0009/04/01/},
    issn =         {2331-7418},
    urldate =      {2021-01-17}
}

@article {Laine2011,
    author = {Laine, Samuli and Karras, Tero},
    title = {Stratified Sampling for Stochastic Transparency},
    journal = {Computer Graphics Forum},
    volume = {30},
    number = {4},
    pages = {1197-1204},
    doi = {10.1111/j.1467-8659.2011.01978.x},
    url = {https://onlinelibrary.wiley.com/doi/abs/10.1111/j.1467-8659.2011.01978.x},
    eprint = {https://onlinelibrary.wiley.com/doi/pdf/10.1111/j.1467-8659.2011.01978.x},
    year = {2011}
}

@article{Owen2003,
    author = {Owen, Art B.},
    title = {Variance with Alternative Scramblings of Digital Nets},
    year = {2003},
    issue_date = {October 2003},
    publisher = {Association for Computing Machinery},
    address = {New York, NY, USA},
    volume = {13},
    number = {4},
    issn = {1049-3301},
    url = {https://doi.org/10.1145/945511.945518},
    doi = {10.1145/945511.945518},
    journal = {ACM Trans. Model. Comput. Simul.},
    month = {10},
    pages = {363–378},
    numpages = {16}
}

@inproceedings {Owen1995,
    author = {Owen, Art B.},
    editor = {Niederreiter, Harald and Shiue, Peter Jau-Shyong},
    title = {Randomly Permuted (t,m,s)-Nets and (t, s)-Sequences},
    booktitle = {Monte Carlo and Quasi-Monte Carlo Methods in Scientific Computing},
    year = {1995},
    publisher = {Springer New York},
    address = {New York, NY},
    pages = {299-317},
    isbn = {978-1-4612-2552-2},
    doi = {10.1007/978-1-4612-2552-2_19}
}

@article {Matousek1998,
    title = {On the L2-Discrepancy for Anchored Boxes},
    journal = {Journal of Complexity},
    volume = {14},
    number = {4},
    pages = {527-556},
    year = {1998},
    issn = {0885-064X},
    doi = {10.1006/jcom.1998.0489},
    url = {http://www.sciencedirect.com/science/article/pii/S0885064X98904897},
    author = {Matoušek, Jiřı́}
}

@article {Kollig2002,
    author = {Kollig, Thomas and Keller, Alexander},
    title = {Efficient Multidimensional Sampling},
    journal = {Computer Graphics Forum},
    volume = {21},
    number = {3},
    pages = {557-563},
    doi = {10.1111/1467-8659.00706},
    url = {https://onlinelibrary.wiley.com/doi/abs/10.1111/1467-8659.00706},
    eprint = {https://onlinelibrary.wiley.com/doi/pdf/10.1111/1467-8659.00706},
    year = {2002}
}

@misc {Weinzierl2000,
    title = {\textit{Introduction to Monte Carlo methods}},
    author = {Weinzierl, Stefan},
    year = {2000},
    eprint={hep-ph/0006269},
    archivePrefix={arXiv}
}

@article {Morokoff1994,
    author = {Morokoff, William J. and Caflisch, Russel E.},
    title = {Quasi-Random Sequences and Their Discrepancies},
    journal = {SIAM Journal on Scientific Computing},
    volume = {15},
    number = {6},
    pages = {1251-1279},
    year = {1994},
    doi = {10.1137/0915077},
    url = {https://doi.org/10.1137/0915077},
    eprint = {https://doi.org/10.1137/0915077}
}

@article {Sobol1967,
    title = {On the distribution of points in a cube and the approximate evaluation of integrals},
    journal = {USSR Computational Mathematics and Mathematical Physics},
    volume = {7},
    number = {4},
    pages = {86-112},
    year = {1967},
    issn = {0041-5553},
    doi = {10.1016/0041-5553(67)90144-9},
    url = {http://www.sciencedirect.com/science/article/pii/0041555367901449},
    author = {Sobol', Ilya Meyerovich}
}

@inproceedings {Chi2007,
    author = {Chi, Hongmei and Mascagni, Michael},
    editor= {Shi, Yong and {van Albada}, Geert Dick and Dongarra, Jack and Sloot, Peter M. A.},
    title = {Efficient Generation of Parallel Quasirandom Faure Sequences Via Scrambling},
    booktitle = {Computational Science -- ICCS 2007},
    year = {2007},
    publisher = {Springer Berlin Heidelberg},
    address = {Berlin, Heidelberg},
    pages = {723-730},
    isbn = {978-3-540-72584-8},
    doi = {10.1007/978-3-540-72584-8_96}
}

@article {Schmid2001,
    title = {Techniques for parallel quasi-Monte Carlo integration with digital sequences and associated problems},
    journal = {Mathematics and Computers in Simulation},
    volume = {55},
    number = {1},
    pages = {249-257},
    year = {2001},
    note = {The Second IMACS Seminar on Monte Carlo Methods},
    issn = {0378-4754},
    doi = {10.1016/S0378-4754(00)00268-8},
    url = {http://www.sciencedirect.com/science/article/pii/S0378475400002688},
    author = {Schmid, Wolfgang Ch. and Uhl, Andreas}
}

@unpublished {Roberts2018,
    author       = {Roberts, Martin},
    title        = {\textit{The Unreasonable Effectiveness of Quasirandom Sequences}},
    month        = {04},
    year         = {2018},
    url          = {http://extremelearning.com.au/unreasonable-effectiveness-of-quasirandom-sequences/},
    urldate = {2021-01-18}
}

@misc {Smith2013,
    author       = {Smith, Jenny},
    title        = {Protected areas: origins, criticisms and contemporary issues for outdoor recreation},
    howpublished = {\textit{Centre for Environment and Society Research}},
    year         = {2013},
    isbn         = {978-1-904839-65-1},
    number       = {15},
    url          = {https://bcuassets.blob.core.windows.net/docs/CESR_Working_Paper_15_2013_Smith.pdf},
    urldate      = {2020-12-13}
}

@article {Rosindell2012,
    title = {The case for ecological neutral theory},
    journal = {Trends in Ecology \& Evolution},
    volume = {27},
    number = {4},
    pages = {203-208},
    year = {2012},
    issn = {0169-5347},
    doi = {10.1016/j.tree.2012.01.004},
    eprint = {http://www.sciencedirect.com/science/article/pii/S0169534712000237},
    url = {http://www.sciencedirect.com/science/article/pii/S0169534712000237},
    author = {Rosindell, James and Hubbell, Stephen P. and He, Fangliang and Harmon, Luke J. and Etienne, Rampal S.}
}

@article {Leigh2007,
    author = {Leigh, Egbert Giles Jr},
    title = {Neutral theory: a historical perspective},
    journal = {Journal of Evolutionary Biology},
    volume = {20},
    number = {6},
    pages = {2075-2091},
    doi = {10.1111/j.1420-9101.2007.01410.x},
    url = {https://onlinelibrary.wiley.com/doi/abs/10.1111/j.1420-9101.2007.01410.x},
    eprint = {https://onlinelibrary.wiley.com/doi/pdf/10.1111/j.1420-9101.2007.01410.x},
    year = {2007}
}

@article {Yahav2017,
    author = {Yahav, Shem-Tov and Danino, Matan and Shnerb, Nadav M.},
    title = {Solution of the spatial neutral model yields new bounds on the Amazonian species richness},
    journal = {Scientific Reports},
    year = {2017},
    month = {02},
    day = {17},
    volume = {7},
    number = {1},
    pages = {42415},
    issn = {2045-2322},
    doi = {10.1038/srep42415},
    url = {https://doi.org/10.1038/srep42415}
}

@article {Fisher1943,
    issn = {00218790, 13652656},
    url = {http://www.jstor.org/stable/1411},
    author = {Fisher, R. A. and Corbet, A. Steven and Williams, C. B.},
    journal = {Journal of Animal Ecology},
    number = {1},
    pages = {42-58},
    publisher = {Wiley, British Ecological Society},
    title = {The Relation Between the Number of Species and the Number of Individuals in a Random Sample of an Animal Population},
    volume = {12},
    year = {1943},
    doi = {10.2307/1411}
}

@article {Beck2010,
%     author = {Beck, Jan and Schwanghart, Wolfgang},
%     title = {Comparing measures of species diversity from incomplete inventories: an update},
%     journal = {Methods in Ecology and Evolution},
%     volume = {1},
%     number = {1},
%     pages = {38-44},
%     keywords = {alpha diversity, effective number of species, Shannon’s entropy, simulation, species richness, undersampling bias},
%     doi = {10.1111/j.2041-210X.2009.00003.x},
%     url = {https://besjournals.onlinelibrary.wiley.com/doi/abs/10.1111/j.2041-210X.2009.00003.x},
%     eprint = {https://besjournals.onlinelibrary.wiley.com/doi/pdf/10.1111/j.2041-210X.2009.00003.x},
%     abstract = {Summary 1. Measuring biodiversity quantitatively is a key component to its investigation, but many methods are known to be biased by undersampling (i.e. incomplete inventories), a common situation in ecological field studies. 2. Following a long tradition of comparing measures of alpha diversity to judge their usefulness, we used simulated data to assess bias of nine diversity measures – some of them proposed fairly recently, such as estimating true species richness depending on the completeness of inventories (Brose, U. \& Martinez, N.D. Oikos (2004) 105, 292), bias-corrected Shannon diversity (Chao, A. \& Shen, T.-J. Environmental and Ecological Statistics (2003) 10, 429), while others are commonly applied (e.g. Shannon’s entropy, Fisher’s α) or long known but rarely used (estimating Shannon’s entropy from Fisher’s α). 3. We conclude that the ‘effective number of species’ based on bias-corrected Shannon’s entropy is an unbiased estimator of diversity at sample completeness c. >0·5, while below that it is still less biased than, e.g., estimated species richness (Brose, U. \& Martinez, N.D. Oikos (2004) 105, 292). 4. Fisher’s α cannot be tested with the same rigour because it cannot measure the diversity of completely inventoried communities, and we present simulations illustrating this effect when sample completeness approaches high values. However, we can show that Fisher’s α produces relatively stable values at low sample completeness (an effect previously shown only in empirical data), and we tentatively conclude that it may still be considered a good (possibly superior) measure of diversity if completeness is very low.},
%     year = {2010}
% }

@article{Moran1958,
    title = {Random processes in genetics},
    volume = {54},
    doi = {10.1017/S0305004100033193},
    number = {1},
    journal = {Mathematical Proceedings of the Cambridge Philosophical Society},
    publisher = {Cambridge University Press},
    author = {Moran, P. A. P.},
    year = {1958},
    pages = {60–71}
}

@article{Wright1942,
    author = {Wright, Sewall},
    journal = {Bulletin of the American Mathematical Society},
    month = {04},
    number = {4},
    pages = {223-246},
    publisher = {American Mathematical Society},
    title = {Statistical genetics and evolution},
    url = {https://projecteuclid.org:443/euclid.bams/1183504254},
    volume = {48},
    year = {1942},
    urldate = {2021-01-19}
}

@online {Taylor2015,
%    author = {Taylor, Jay},
%    title = {\textit{Moran Model}},
%    url = {https://math.la.asu.edu/~jtaylor/teaching/Spring2015/APM504/lectures/Moran.pdf},
%    date = {2015-02-17},
%    urldate = {2021-01-19}
%}

@mastersthesis {Ravela2010,
  author       = {Ravela, Srikar Chowdary},
  title        = {\textit{Comparison of Shared memory based parallel programming models}},
  school       = {Blekinge Institute of Technology},
  year         = {2010},
  address      = {Ronneby, Sweden},
  url          = {https://www.diva-portal.org/smash/get/diva2:830690/FULLTEXT01.pdf},
  urldate      = {2021-01-19}
}

@article {Kelleher2016,
%    author = {Kelleher, Jerome and Etheridge, Alison M and McVean, Gilean},
%    journal = {PLOS Computational Biology},
%    publisher = {Public Library of Science},
%    title = {Efficient Coalescent Simulation and Genealogical Analysis for Large Sample Sizes},
%    year = {2016},
%    month = {05},
%    volume = {12},
%    url = {https://doi.org/10.1371/journal.pcbi.1004842},
%    pages = {1-22},
%    abstract = {Author Summary Our understanding of the distribution of genetic variation in natural populations has been driven by mathematical models of the underlying biological and demographic processes. A key strength of such coalescent models is that they enable efficient simulation of data we might see under a variety of evolutionary scenarios. However, current methods are not well suited to simulating genome-scale data sets on hundreds of thousands of samples, which is essential if we are to understand the data generated by population-scale sequencing projects. Similarly, processing the results of large simulations also presents researchers with a major challenge, as it can take many days just to read the data files. In this paper we solve these problems by introducing a new way to represent information about the ancestral process. This new representation leads to huge gains in simulation speed and storage efficiency so that large simulations complete in minutes and the output files can be processed in seconds.},
%    number = {5},
%    doi = {10.1371/journal.pcbi.1004842}
%}

@article {Arenas2007,
%    author = {Arenas, Miguel and Posada, David},
%    title = {Recodon: Coalescent simulation of coding DNA sequences with recombination, migration and demography},
%    journal = {BMC Bioinformatics},
%    year = {2007},
%    month = {11},
%    day = {20},
%    volume = {8},
%    number = {1},
%    pages = {458},
%    abstract = {Coalescent simulations have proven very useful in many population genetics studies. In order to arrive to meaningful conclusions, it is important that these simulations resemble the process of molecular evolution as much as possible. To date, no single coalescent program is able to simulate codon sequences sampled from populations with recombination, migration and growth.},
%    issn = {1471-2105},
%    doi = {10.1186/1471-2105-8-458},
%    url = {https://doi.org/10.1186/1471-2105-8-458}
%}

@online {Pavlidis2013,
%    author = {Pavlidis, Pavlos},
%    title = {\textit{Introduction to Coalescent}},
%    url = {https://cme.h-its.org/exelixis/web/teaching/lectures/lecture14.pdf},
%    year = {2013},
%    month = {02},
%    urldate = {2021-01-19}
%}

@book {Last2017,
    place = {Cambridge},
    series = {Institute of Mathematical Statistics Textbooks},
    title = {\textit{Lectures on the Poisson Process}},
    doi = {10.1017/9781316104477},
    publisher = {Cambridge University Press},
    author = {Last, Günter and Penrose, Mathew},
    year = {2017},
    collection = {Institute of Mathematical Statistics Textbooks},
    eprint = {https://www.math.kit.edu/stoch/~last/seite/lectures_on_the_poisson_process/media/lastpenrose2017.pdf}
}

@book {Chiu2013,
    title = {\textit{Stochastic geometry and its applications}},
    author = {Chiu, Sung Nok and Stoyan, Dietrich and Kendall, Wilfrid S and Mecke, Joseph},
    year = {2013},
    publisher = {John Wiley \& Sons},
    edition = {3},
    doi = {10.1002/9781118658222},
    isbn = {9780470664810}
}

@misc {PiR8,
    title = {Proving that the discrete exponential distribution is geometric distribution},
    author = {{$\pi r 8$}},
    howpublished = {\textit{Mathematics Stack Exchange}},
    note = {(version: 2017-01-08)},
    url = {https://math.stackexchange.com/q/2087674},
    urldate = {2021-01-20}
}

@online {Brereton2015,
    author = {Brereton, Tim},
    title = {\textit{Methods of Monte Carlo Simulation}},
    url = {https://www.uni-ulm.de/fileadmin/website_uni_ulm/mawi.inst.110/lehre/ws15/MonteCarloMethods/Lecture_Notes.pdf},
    year = {2015},
    urldate = {2021-01-20}
}

@book {Thomopoulos2017,
  author    = {Thomopoulos, Nick T.},
  title     = {\textit{Statistical Distributions}},
  publisher = {Springer, Cham},
  year      = {2017},
  doi       = {10.1007/978-3-319-65112-5}
}

@book {Gupta2010,
  author    = {Gupta, Arjun K. and Zeng, Wei-Bin and Wu, Yanhong},
  title     = {\textit{Probability and Statistical Models}},
  publisher = {Birkhäuser Boston},
  year      = {2010},
  isbn      = {978-0-8176-4986-9},
  doi       = {10.1007/978-0-8176-4987-6}
}

@online {Takahara2017,
    author = {Takahara, Glen},
    title = {\textit{The Exponential Distribution}},
    url = {https://mast.queensu.ca/~stat455/lecturenotes/set4.pdf},
    year = {2017},
    urldate = {2021-01-20}
}

@inproceedings {Webster1986,
    author = {Webster, A. F. and Tavares, S. E.},
    editor = {Williams, Hugh C.},
    title = {On the Design of S-Boxes},
    booktitle = {Advances in Cryptology --- CRYPTO '85 Proceedings},
    year = {1986},
    publisher = {Springer Berlin Heidelberg},
    address = {Berlin, Heidelberg},
    pages = {523-534},
    isbn = {978-3-540-39799-1},
    doi = {10.1007/3-540-39799-X_41}
}

@online {Sateesan2020,
    author = {Sateesan, Arish},
    title = {\textit{Analyze your hash functions: The Avalanche Metrics Calculation}},
    url = {https://arishs.medium.com/analyze-your-hash-functions-the-avalanche-metrics-calculation-767b7445ee6f},
    year = {2020},
    month = {07},
    day = {06},
    urldate = {2021-01-20}
}

@online {Ticki2016,
    author = {Ticki},
    title = {\textit{Designing a good non-cryptographic hash function}},
    url = {http://ticki.github.io/blog/designing-a-good-non-cryptographic-hash-function/},
    year = {2016},
    month = {11},
    day = {04},
    urldate = {2021-01-20}
}

@article {Estébanez2014,
    author = {Estébanez, César and Saez, Yago and Recio, Gustavo and Isasi, Pedro},
    title = {AUTOMATIC DESIGN OF NONCRYPTOGRAPHIC HASH FUNCTIONS USING GENETIC PROGRAMMING},
    journal = {Computational Intelligence},
    volume = {30},
    number = {4},
    pages = {798-831},
    doi = {10.1111/coin.12033},
    url = {https://onlinelibrary.wiley.com/doi/abs/10.1111/coin.12033},
    eprint = {https://onlinelibrary.wiley.com/doi/pdf/10.1111/coin.12033},
    year = {2014}
}

@inbook {vanderLeest2012,
    author = {{van der Leest}, Vincent and {van der Sluis}, Erik and Schrijen, Geert-Jan and Tuyls, Pim and Handschuh, Helena},
    editor = {Naccache, David},
    title = {Efficient Implementation of True Random Number Generator Based on SRAM PUFs},
    booktitle = {Cryptography and Security: From Theory to Applications: Essays Dedicated to Jean-Jacques Quisquater on the Occasion of His 65th Birthday},
    year = {2012},
    publisher = {Springer Berlin Heidelberg},
    address = {Berlin, Heidelberg},
    pages = {300-318},
    isbn = {978-3-642-28368-0},
    doi = {10.1007/978-3-642-28368-0_20},
    url = {https://doi.org/10.1007/978-3-642-28368-0_20}
}

@article {IEEE2019,
    author = {IEEE},
    journal = {IEEE Std 754-2019 (Revision of IEEE 754-2008)},
    title = {IEEE Standard for Floating-Point Arithmetic},
    year = {2019},
    doi = {10.1109/IEEESTD.2019.8766229}
}

@online {Vigna2019,
    author = {Vigna, Sebastiano and Blackman, David},
    title = {\textit{xoshiro / xoroshiro generators and the PRNG shootout}},
    url = {http://prng.di.unimi.it/},
    year = {2019},
    urldate = {2021-01-20}
}

@online {Campbell2014,
    author = {Campbell, Taylor R},
    title = {\textit{Uniform random floats:  How to generate a double-precision floating-point number in [0, 1] uniformly at random given a uniform random source of bits.}},
    url = {http://mumble.net/~campbell/tmp/random_real.c},
    year = {2014},
    urldate = {2021-01-20}
}

@article {Box1958,
    author = {Box, George Edward Pelham and Muller, Mervin Edgar},
    doi = {10.1214/aoms/1177706645},
    journal = {Annals of Mathematical Statistics},
    month = {06},
    number = {2},
    pages = {610-611},
    publisher = {The Institute of Mathematical Statistics},
    title = {A Note on the Generation of Random Normal Deviates},
    url = {https://doi.org/10.1214/aoms/1177706645},
    volume = {29},
    year = {1958}
}

@book {Devroye1986,
    title = {\textit{Non-Uniform Random Variate Generation}},
    author = {Devroye, Luc},
    publisher = {Springer, New York, NY},
    year = {1986},
    doi = {10.1007/978-1-4613-8643-8},
    url = {http://www.nrbook.com/devroye/Devroye_files/chapter_ten.pdf},
    address = {New York, NY},
    isbn = {978-1-4613-8645-2},
    urldate = {2021-04-29}
}

@article {Weyl1916,
    author = {Weyl, Hermann},
    title = {{\"U}ber die Gleichverteilung von Zahlen mod. Eins},
    journal = {Mathematische Annalen},
    year = {1916},
    month = {09},
    day = {01},
    volume = {77},
    number = {3},
    pages = {313-352},
    issn = {1432-1807},
    doi = {10.1007/BF01475864},
    url = {https://doi.org/10.1007/BF01475864}
}

@article{Niederreiter1978,
    title = {Quasi-Monte Carlo methods and pseudo-random numbers},
    author = {Niederreiter, Harald},
    journal = {Bulletin of the American Mathematical Society},
    volume = {84},
    number = {6},
    pages = {957-1041},
    year = {1978},
    doi = {10.1090/S0002-9904-1978-14532-7 }
}

@article{vonNeumann1951,
    title = {Various Techniques Used in Connection with Random Digits},
    author = {{von Neumann}, John},
    journal = {National Bureau of Standards Applied Math Series},
    volume = {12},
    pages = {36-38},
    year = {1951},
    url = {https://dornsifecms.usc.edu/assets/sites/520/docs/VonNeumann-ams12p36-38.pdf},
    urldate = {2021-01-20}
}

@article {Komarov2012,
    author = {Komarov, Ivan and D'Souza, Roshan M.},
    journal = {PLOS ONE},
    publisher = {Public Library of Science},
    title = {Accelerating the Gillespie Exact Stochastic Simulation Algorithm Using Hybrid Parallel Execution on Graphics Processing Units},
    year = {2012},
    month = {11},
    volume = {7},
    url = {https://doi.org/10.1371/journal.pone.0046693},
    pages = {1-9},
    number = {11},
    doi = {10.1371/journal.pone.0046693}
}

@article {Nobile2014,
    author = {Nobile, Marco S. and Cazzaniga, Paolo and Besozzi, Daniela and Pescini, Dario and Mauri, Giancarlo},
    journal = {PLOS ONE},
    publisher = {Public Library of Science},
    title = {cuTauLeaping: A GPU-Powered Tau-Leaping Stochastic Simulator for Massive Parallel Analyses of Biological Systems},
    year = {2014},
    month = {03},
    volume = {9},
    url = {https://doi.org/10.1371/journal.pone.0091963},
    pages = {1-20},
    number = {3},
    doi = {10.1371/journal.pone.0091963}
}

@inproceedings {Macchiarulo2008,
    author = {Macchiarulo, Luca},
    booktitle = {2008 30th Annual International Conference of the IEEE Engineering in Medicine and Biology Society},
    title = {A massively parallel implementation of Gillespie algorithm on FPGAs},
    year = {2008},
    pages = {1343-1346},
    doi = {10.1109/IEMBS.2008.4649413},
    issn = {1558-4615},
    month = {08}
}

@misc{Bauer2015,
   author = {Bauer, Pavol},
   institution = {Uppsala University, Division of Scientific Computing},
   title = {Parallelism and efficiency in discrete-event simulation},
   series = {IT licentiate theses / Uppsala University, Department of Information Technology},
   issn = {1404-5117},
   number = {2015-004},
   year = {2015},
   url = {http://urn.kb.se/resolve?urn=urn\%3Anbn\%3Ase\%3Auu\%3Adiva-264756},
   urldate = {2021-01-20}
}

@article {Liang2007,
%     author = {Liang, Liming and Zöllner, Sebastian and Abecasis, Gonçalo R.},
%     title = {GENOME: a rapid coalescent-based whole genome simulator},
%     journal = {Bioinformatics},
%     volume = {23},
%     number = {12},
%     pages = {1565-1567},
%     year = {2007},
%     month = {04},
%     abstract = {Summary: GENOME proposes a rapid coalescent-based approach to simulate whole genome data. In addition to features of standard coalescent simulators, the program allows for recombination rates to vary along the genome and for flexible population histories. Within small regions, we have evaluated samples simulated by GENOME to verify that GENOME provides the expected LD patterns and frequency spectra. The program can be used to study the sampling properties of any statistic for a whole genome study.Availability: The program and C++ source code are available online at http://www.sph.umich.edu/csg/liang/genome/Contact:lianglim@umich.eduSupplementary information: Supplementary data are available at Bioinformatics online.},
%     issn = {1367-4803},
%     doi = {10.1093/bioinformatics/btm138},
%     url = {https://doi.org/10.1093/bioinformatics/btm138},
%     eprint = {https://academic.oup.com/bioinformatics/article-pdf/23/12/1565/606251/btm138.pdf},
% }

@article {Etienne2004,
    author = {Etienne, Rampal S. and Olff, Han},
    title = {A novel genealogical approach to neutral biodiversity theory},
    journal = {Ecology Letters},
    volume = {7},
    number = {3},
    pages = {170-175},
    doi = {10.1111/j.1461-0248.2004.00572.x},
    url = {https://onlinelibrary.wiley.com/doi/abs/10.1111/j.1461-0248.2004.00572.x},
    eprint = {https://onlinelibrary.wiley.com/doi/pdf/10.1111/j.1461-0248.2004.00572.x},
    year = {2004}
}

@article{Freudenstein2016,
    author = {Freudenstein, John V. and Broe, Michael B. and Folk, Ryan A. and Sinn, Brandon T.},
    title = {Biodiversity and the Species Concept—Lineages are not Enough},
    journal = {Systematic Biology},
    volume = {66},
    number = {4},
    pages = {644-656},
    year = {2016},
    month = {10},
    issn = {1063-5157},
    doi = {10.1093/sysbio/syw098},
    url = {https://doi.org/10.1093/sysbio/syw098},
    eprint = {https://academic.oup.com/sysbio/article-pdf/66/4/644/17844478/syw098.pdf},
}

@inproceedings{Dittamo2009,
    author = {Dittamo, Cristian and Cangelosi, Davide},
    booktitle = {2009 International Conference on Computer Modeling and Simulation},
    title = {Optimized Parallel Implementation of Gillespie's First Reaction Method on Graphics Processing Units},
    year = {2009},
    pages = {156-161},
    doi={10.1109/ICCMS.2009.42}
}

@inproceedings{Ridwan2004,
    author = {Ridwan, Azmi Mohamed and Krishnan, Arun and Dhar, Pawan},
    editor = {Bubak, Marian and {van Albada}, Geert Dick and Sloot, Peter M. A. and Dongarra, Jack},
    title = {A Parallel Implementation of Gillespie's Direct Method},
    booktitle = {Computational Science - ICCS 2004},
    year = {2004},
    publisher = {Springer Berlin Heidelberg},
    address = {Berlin, Heidelberg},
    pages = {284-291},
    isbn = {978-3-540-24687-9},
    doi = {10.1007/978-3-540-24687-9_36}
}

@article {Sen2008,
    title = {Golden ratio in science, as random sequence source, its computation and beyond},
    journal = {Computers \& Mathematics with Applications},
    volume = {56},
    number = {2},
    pages = {469-498},
    year = {2008},
    issn = {0898-1221},
    doi = {10.1016/j.camwa.2007.06.030},
    url = {http://www.sciencedirect.com/science/article/pii/S089812210800031X},
    author = {Sen, S. K. and Agarwal, Ravi P.}
}

@inproceedings {Holk2013,
    author = {Holk, Eric and Pathirage, Milinda and Chauhan, Arun and Lumsdaine, Andrew and Matsakis, Nicholas D.},
    booktitle = {2013 IEEE International Symposium on Parallel Distributed Processing, Workshops and Phd Forum},
    title = {GPU Programming in Rust: Implementing High-Level Abstractions in a Systems-Level Language},
    year = {2013},
    pages = {315-324},
    doi = {10.1109/IPDPSW.2013.173}
}

@online {Stormark2005,
    author = {Stormark, Kristian},
    title = {\textit{An introduction to Quasi Monte Carlo methods}},
    url = {https://nambafa.com/depot/fag/prosjekt.pdf},
    year = {2005},
    urldate = {2021-01-20}
}

@inproceedings {Klenk2007,
    author = {Klenk, Benjamin and Fröening, Holger and Eberle, Hans and Dennison, Larry},  booktitle = {2017 IEEE International Parallel and Distributed Processing Symposium (IPDPS)},
    title = {Relaxations for High-Performance Message Passing on Massively Parallel SIMT Processors},
    year = {2017},
    pages = {855-865},
    doi = {10.1109/IPDPS.2017.94}
}

@article {Arjunan2020,
    author = {Arjunan, Satya N.V. and Miyauchi, Atsushi and Iwamoto, Kazunari and Takahashi, Koichi},
    title = {pSpatiocyte: a high-performance simulator for intracellular reaction-diffusion systems},
    journal = {BMC Bioinformatics},
    year = {2020},
    month = {01},
    day = {29},
    volume = {21},
    number = {1},
    pages = {33},
    issn = {1471-2105},
    doi = {10.1186/s12859-019-3338-8},
    url = {https://doi.org/10.1186/s12859-019-3338-8}
}

@online {Patel2019,
%     author = {Patel, Amit},
%     title = {\textit{Wraparound square tile maps on a sphere}},
%     url = {https://www.redblobgames.com/x/1938-square-tiling-of-sphere/},
%     year = {2005},
%     month = {09},
%     day = {16},
%     urldate = {2021-01-20}
% }

@article {Mancy2013,
    title = {Discrete and continuous time simulations of spatial ecological processes predict different final population sizes and interspecific competition outcomes},
    journal = {Ecological Modelling},
    volume = {259},
    pages = {50-61},
    year = {2013},
    issn = {0304-3800},
    doi = {10.1016/j.ecolmodel.2013.03.013},
    url = {http://www.sciencedirect.com/science/article/pii/S0304380013001701},
    author = {Mancy, Rebecca and Prosser, Patrick and Rogers, Simon}
}

@online {Lemire2019,
    author = {Lemire, Daniel},
    title = {\textit{The fastest conventional random number generator that can pass Big Crush?}},
    url = {https://lemire.me/blog/2019/03/19/the-fastest-conventional-random-number-generator-that-can-pass-big-crush/},
    year = {2019},
    month = {03},
    day = {19},
    urldate = {2021-01-20}
}

@misc {Brandt2019,
%     title = {\textit{A Technique for Finding Optimal Program Launch Parameters Targeting Manycore Accelerators}},
%     author = {Brandt, Alexander and Mohajerani, Davood and Maza, Marc Moreno and Paudel, Jeeva and Wang, Lin-Xiao},
%     year = {2019},
%     eprint={1906.00142},
%     archivePrefix={arXiv},
%     primaryClass={cs.DC}
% }

@article {Wu2013,
    author = {Wu, Wei and Qi, Fengbin and He, Wangquan and Wang, Shanshan},
    journal = {Tsinghua Science and Technology},
    title = {CUDA's mapped memory to support I/O functions on GPU},
    year = {2013},
    volume = {18},
    number = {6},
    pages = {588-598},
    doi = {10.1109/TST.2013.6678904}
}

@online {Defazio2016,
    author = {Defazio, Aaron},
    title = {\textit{Weighted random sampling with replacement with dynamic weights}},
    url = {https://www.aarondefazio.com/adefazio-weighted-sampling.pdf},
    year = {2016},
    month = {02},
    day = {14},
    urldate = {2021-01-20}
}

@phdthesis {ChiPhD,
    author = {Chi, Hongmei},
    title = {\textit{Scrambled Quasirandom Sequences and Their Applications}},
    school = {Florida State University},
    year = {2004},
    url = {http://purl.flvc.org/fsu/fd/FSU_migr_etd-3823},
    urldate = {2021-01-20}
}

@online {Occil,
    author = {Occil, Peter},
    title = {\textit{Random Number Generator Recommendations for Applications}},
    url = {https://peteroupc.github.io/random.html},
    urldate = {2021-01-20}
}

@article {Slepoy2008,
    author = {Slepoy, Alexander and Thompson, Aidan P. and Plimpton, Steven J.},
    title = {A constant-time kinetic Monte Carlo algorithm for simulation of large biochemical reaction networks},
    journal = {The Journal of Chemical Physics},
    volume = {128},
    number = {20},
    pages = {205101},
    year = {2008},
    doi = {10.1063/1.2919546}
}

@inproceedings {Puntigam2003,
    author = {Puntigam, Frank},
    booktitle = {Eighth International Workshop on Component-Oriented Programming},
    title = {State Information in Statically Checked Interfaces},
    year = {2003},
    url = {https://www.researchgate.net/profile/Franz-Puntigam/publication/247639898_State_information_in_statically_checked_interfaces/links/54bd0a130cf218d4a1690416/State-information-in-statically-checked-interfaces.pdf},
    urldate = {2021-05-03},
}

@inproceedings {Waignier2008,
    author = {Waignier, Guillaume and Sriplakich, Prawee and Le Meur, Anne-Fran{\c{c}}oise and Duchien, Laurence},
    editor = {Czarnecki, Krzysztof and Ober, Ileana and Bruel, Jean-Michel and Uhl, Axel and V{\"o}lter, Markus},
    title = {A Model-Based Framework for Statically and Dynamically Checking Component Interactions},
    booktitle = {Model Driven Engineering Languages and Systems},
    year = {2008},
    publisher = {Springer Berlin Heidelberg},
    address = {Berlin, Heidelberg},
    pages = {371-385},
    isbn = {978-3-540-87875-9},
    doi = {10.1007/978-3-540-87875-9_27},
}

@inproceedings {Genssler2001,
    author = {Genßler, Thomas and Zeidler, Christian and {Forschungszentrum Informatik}},
    title = {Rule-Driven Component Composition for Embedded Systems},
    booktitle = {Intl. Conf. on Software Engineering (ICSE): Workshop on ComponentBased Software Engineering},
    year = {2001},
    url = {http://scg.unibe.ch/archive/pecos/public_documents/Gens01a.pdf},
    urldate = {2021-05-03},
}

@inproceedings {Eide2002,
    author = {Eide, Eric and Reid, Alastair and Regehr, John and Lepreau, Jay},
    title = {Static and Dynamic Structure in Design Patterns},
    year = {2002},
    isbn = {158113472X},
    publisher = {Association for Computing Machinery},
    address = {New York, NY, USA},
    url = {https://doi.org/10.1145/581339.581367},
    doi = {10.1145/581339.581367},
    booktitle = {Proceedings of the 24th International Conference on Software Engineering},
    pages = {208–218},
    numpages = {11},
    location = {Orlando, Florida},
    series = {ICSE '02},
}

@article {Paola2000,
    author = {Inverardi, Paola and Wolf, Alexander L. and Yankelevich, Daniel},
    title = {Static Checking of System Behaviors Using Derived Component Assumptions},
    year = {2000},
    issue_date = {July 2000},
    publisher = {Association for Computing Machinery},
    address = {New York, NY, USA},
    volume = {9},
    number = {3},
    issn = {1049-331X},
    url = {https://doi.org/10.1145/352591.352593},
    doi = {10.1145/352591.352593},
    journal = {ACM Trans. Softw. Eng. Methodol.},
    month = {06},
    pages = {239–272},
    numpages = {34},
}

@unpublished {RON,
    author = {{RON developers}},
    title = {\textit{Rusty Object Notation}},
    note = {RON is a simple readable data serialization format that looks similar to Rust syntax.},
    license = {MIT License, Apache License},
    url = {https://github.com/ron-rs/ron},
    urldate = {2021-05-14},
}

@manual {GDAL,
    title = {{GDAL/OGR} Geospatial Data Abstraction software Library},
    author = {{GDAL/OGR contributors}},
    organization = {Open Source Geospatial Foundation},
    year = {2021},
    url = {https://gdal.org},
    urldate = {2021-05-14},
}

@online {PractRand,
    author = {Doty-Humphrey, Chris},
    title = {\textit{Practically Random}},
    year = {2019},
    month = {10},
    url = {https://sourceforge.net/projects/pracrand/},
    urldate = {2021-05-16}
}

@online {dieharder,
    author = {Brown, Robert G.},
    title = {\textit{Dieharder: A Random Number Test Suite}},
    year = {2020},
    month = {06},
    url = {https://webhome.phy.duke.edu/~rgb/General/dieharder.php},
    urldate = {2021-05-16}
}

@online {ENT,
    author = {Brown, Robert G.},
    title = {\textit{ENT: A Pseudorandom Number Sequence Test Program}},
    year = {2008},
    month = {01},
    url = {https://www.fourmilab.ch/random/},
    urldate = {2021-05-16}
}

@inproceedings {Lehmer1951,
    booktitle = {Proceedings of a Second Symposium on Large-Scale Digital Calculating Machinery},
    title = {Mathematical Methods in Large-scale Computing Units},
    author = {Lehmer, Derrick H},
    publisher = {The Annals of the Computation Laboratory of Harvard University},
    volume = {26},
    pages = {141-146},
    year = {1951},
    url = {http://www.bitsavers.org/pdf/harvard/Proceedings_of_a_Second_Symposium_on_Large-Scale_Digital_Calculating_Machinery_Sep49.pdf},
    urldate = {2021-05-16}
}

@article {Dominic2020,
    doi = {10.1371/journal.pgen.1008619},
    author = {Nelson, Dominic and Kelleher, Jerome and Ragsdale, Aaron P. and Moreau, Claudia and McVean, Gil and Gravel, Simon},
    journal = {PLOS Genetics},
    publisher = {Public Library of Science},
    title = {Accounting for long-range correlations in genome-wide simulations of large cohorts},
    year = {2020},
    month = {05},
    volume = {16},
    url = {https://doi.org/10.1371/journal.pgen.1008619},
    pages = {1-12},
    number = {5},
}

@inbook {Kelleher2020,
    author = {Kelleher, Jerome and Lohse, Konrad},
    editor = {Dutheil, Julien Y.},
    title = {Coalescent Simulation with msprime},
    booktitle = {Statistical Population Genomics},
    year = {2020},
    publisher = {Springer US},
    address = {New York, NY},
    pages = {191-230},
    isbn = {978-1-0716-0199-0},
    doi = {10.1007/978-1-0716-0199-0_9},
    url = {https://doi.org/10.1007/978-1-0716-0199-0_9}
}

@article {Kelleher2018,
    doi = {10.1371/journal.pcbi.1006581},
    author = {Kelleher, Jerome AND Thornton, Kevin R. AND Ashander, Jaime AND Ralph, Peter L.},
    journal = {PLOS Computational Biology},
    publisher = {Public Library of Science},
    title = {Efficient pedigree recording for fast population genetics simulation},
    year = {2018},
    month = {11},
    volume = {14},
    url = {https://doi.org/10.1371/journal.pcbi.1006581},
    pages = {1-21},
    number = {11},
}

@article {Kelleher2013,
    author = {Kelleher, Jerome and Barton, Nicholas H. and Etheridge, Alison M.},
    title = {Coalescent simulation in continuous space},
    journal = {Bioinformatics},
    volume = {29},
    number = {7},
    pages = {955-956},
    year = {2013},
    month = {02},
    issn = {1367-4803},
    doi = {10.1093/bioinformatics/btt067},
    url = {https://doi.org/10.1093/bioinformatics/btt067},
    eprint = {https://academic.oup.com/bioinformatics/article-pdf/29/7/955/17344595/btt067.pdf},
}

@article {Mudalige2013,
    title = {Design and initial performance of a high-level unstructured mesh framework on heterogeneous parallel systems},
    journal = {Parallel Computing},
    volume = {39},
    number = {11},
    pages = {669-692},
    year = {2013},
    issn = {0167-8191},
    doi = {10.1016/j.parco.2013.09.004},
    url = {https://www.sciencedirect.com/science/article/pii/S0167819113001166},
    author = {G.R. Mudalige and M.B. Giles and J. Thiyagalingam and I.Z. Reguly and C. Bertolli and P.H.J. Kelly and A.E. Trefethen}
}

@inproceedings {Harlacher2012,
    author = {Harlacher, Daniel F. and Klimach, Harald and Roller, Sabine and Siebert, Christian and Wolf, Felix},
    booktitle = {2012 IEEE 26th International Parallel and Distributed Processing Symposium Workshops   PhD Forum},
    title = {Dynamic Load Balancing for Unstructured Meshes on Space-Filling Curves},
    year = {2012},
    pages = {1661-1669},
    doi = {10.1109/IPDPSW.2012.207}
}

@inproceedings {Bertolli2003,
    author = {Bertolli, C. and Betts, A. and Loriant, N. and Mudalige, G. R. and Radford, D. and Ham, D. A. and Giles, M. B. and Kelly, P. H. J.},
    editor = {Kasahara, Hironori and Kimura, Keiji},
    title = {Compiler Optimizations for Industrial Unstructured Mesh CFD Applications on GPUs},
    booktitle = {Languages and Compilers for Parallel Computing},
    year = {2013},
    publisher = {Springer Berlin Heidelberg},
    address = {Berlin, Heidelberg},
    pages = {112-126},
    doi = {10.1007/978-3-642-37658-0_8},
    isbn = {978-3-642-37658-0}
}

@article {Hepburn2012,
%     author = {Hepburn, Iain and Chen, Weiliang and Wils, Stefan and De Schutter, Erik},
%     title = {STEPS: efficient simulation of stochastic reaction--diffusion models in realistic morphologies},
%     journal = {BMC Systems Biology},
%     year = {2012},
%     month = {05},
%     day = {10},
%     volume = {6},
%     number = {1},
%     pages = {36},
%     abstract = {Models of cellular molecular systems are built from components such as biochemical reactions (including interactions between ligands and membrane-bound proteins), conformational changes and active and passive transport. A discrete, stochastic description of the kinetics is often essential to capture the behavior of the system accurately. Where spatial effects play a prominent role the complex morphology of cells may have to be represented, along with aspects such as chemical localization and diffusion. This high level of detail makes efficiency a particularly important consideration for software that is designed to simulate such systems.},
%     issn = {1752-0509},
%     doi = {10.1186/1752-0509-6-36},
%     url = {https://doi.org/10.1186/1752-0509-6-36}
% }

@inproceedings {Field2002,
    author = {Field, A. J. and Kelly, P. H. J. and Hansen, T. L.},
    editor = {Monien, Burkhard and Feldmann, Rainer},
    title = {Optimising Shared Reduction Variables in MPI Programs},
    booktitle = {Euro-Par 2002 Parallel Processing},
    year = {2002},
    publisher = {Springer Berlin Heidelberg},
    address = {Berlin, Heidelberg},
    pages = {630-639},
    doi = {10.1007/3-540-45706-2_87},
    isbn = {978-3-540-45706-0}
}

@article {Stawinoga2018,
    author = {Stawinoga, Nicolai and Field, Tony},
    title = {Predictable Thread Coarsening},
    year = {2018},
    issue_date = {June 2018},
    publisher = {Association for Computing Machinery},
    address = {New York, NY, USA},
    volume = {15},
    number = {2},
    issn = {1544-3566},
    url = {https://doi.org/10.1145/3194242},
    doi = {10.1145/3194242},
    journal = {ACM Trans. Archit. Code Optim.},
    month = {06},
    articleno = {23},
    numpages = {26}
}

@unpublished {Wang2021,
    author = {Wang, Yi},
    title = {\textit{wyhash}},
    year = {2021},
    note = {wyhash and wyrand are the ideal 64-bit hash function and PRNG respectively.},
    license = {Unlicense},
    url = {https://github.com/wangyi-fudan/wyhash},
    urldate = {2021-05-19}
}

@unpublished {Romero2020,
    author = {Romero, Diego Barrios},
    title = {\textit{wyhash-rs}},
    year = {2020},
    note = {Rust implementation of the wyhash algorithm by Wang Yi.},
    license = {MIT or Apache2.0},
    url = {https://github.com/eldruin/wyhash-rs},
    urldate = {2021-05-19}
}

@online {Ticki2020,
    author = {Ticki},
    title = {\textit{SeaHash: A bizarrely fast hash function}},
    url = {https://github.com/redox-os/tfs/tree/master/seahash},
    year = {2020},
    urldate = {2021-01-20}
}

@misc {Bhattacharjee2018,
    title = {\textit{A Search for Good Pseudo-random Number Generators: Survey and Empirical Studies}},
    author = {Bhattacharjee, Kamalika and Maity, Krishnendu and Das, Sukanta},
    year = {2018},
    eprint = {1811.04035},
    archivePrefix = {arXiv},
    primaryClass = {cs.CR},
}

@unpublished {Cameron2021,
    author = {Cameron, Nick},
    title = {\textit{Rust For Systems Programmers}},
    year = {2021},
    note = {A Rust tutorial for experienced C and C++ programmers.},
    license = {MIT or Apache2.0},
    url = {https://github.com/nrc/r4cppp},
    urldate = {2021-05-20}
}

@article {Bloom1970,
    author = {Bloom, Burton H.},
    title = {Space/Time Trade-Offs in Hash Coding with Allowable Errors},
    year = {1970},
    issue_date = {July 1970},
    publisher = {Association for Computing Machinery},
    address = {New York, NY, USA},
    volume = {13},
    number = {7},
    issn = {0001-0782},
    url = {https://doi.org/10.1145/362686.362692},
    doi = {10.1145/362686.362692},
    journal = {Commun. ACM},
    month = {06},
    pages = {422–426},
    numpages = {5}
}

@online {Skarupke2018,
    author = {Skarupke, Malte},
    title = {\textit{Fibonacci Hashing: The Optimization that the World Forgot (or: a Better Alternative to Integer Modulo)}},
    url = {https://probablydance.com/2018/06/16/fibonacci-hashing-the-optimization-that-the-world-forgot-or-a-better-alternative-to-integer-modulo/},
    year = {2018},
    month = {06},
    day = {16},
    urldate = {2021-01-06}
}

@online {Wang2007,
    author = {Wang, Thomas},
    title = {\textit{Integer Hash Function}},
    url = {https://gist.github.com/badboy/6267743},
    year = {2007},
    month = {03},
    urldate = {2021-05-22}
}

@online {floppsy,
    author = {Cris},
    title = {\textit{floppsy}},
    url = {https://github.com/dosyago/floppsy},
    year = {2020},
    month = {04},
    urldate = {2021-05-22}
}

@online {Mulvey2015,
    author = {Mulvey, Bret},
    title = {\textit{Hash Functions}},
    url = {https://papa.bretmulvey.com/post/124027987928/hash-functions},
    year = {2015},
    month = {07},
    day = {13},
    urldate = {2021-05-22}
}

@online {Wellons2018,
    author = {Wellons, Chris},
    title = {\textit{Prospecting for Hash Functions}},
    url = {https://nullprogram.com/blog/2018/07/31/},
    year = {2018},
    month = {07},
    day = {31},
    urldate = {2021-01-06}
}

@article {Denk2020,
	author = {Denk, Jonas and Martis, Stephen and Hallatschek, Oskar},
	title = {Chaos may lurk under a cloak of neutrality},
	volume = {117},
	number = {28},
	pages = {16104-16106},
	year = {2020},
	doi = {10.1073/pnas.2010120117},
	publisher = {National Academy of Sciences},
	issn = {0027-8424},
	URL = {https://www.pnas.org/content/117/28/16104},
	eprint = {https://www.pnas.org/content/117/28/16104.full.pdf},
	journal = {Proceedings of the National Academy of Sciences}
}

@phdthesis {HintzenPhd,
    author = {Hintzen, Rogier Eduard},
    title = {\textit{Mechanistic models of global biodiversity}},
    school = {Imperial College London},
    year = {2019},
    doi = {10.25560/76499},
}

@article {Harfoot2014,
    doi = {10.1371/journal.pbio.1001841},
    author = {Harfoot, Michael B. J. and Newbold, Tim and Tittensor, Derek P. and Emmott, Stephen and Hutton, Jon and Lyutsarev, Vassily and Smith, Matthew J. and Scharlemann, Jörn P. W. and Purves, Drew W.},
    journal = {PLOS Biology},
    publisher = {Public Library of Science},
    title = {Emergent Global Patterns of Ecosystem Structure and Function from a Mechanistic General Ecosystem Model},
    year = {2014},
    month = {04},
    volume = {12},
    url = {https://doi.org/10.1371/journal.pbio.1001841},
    pages = {1-24},
    number = {4},
}

@online {Lemire2017,
    author = {Lemire, Daniel},
    title = {\textit{Testing non-cryptographic random number generators: my results}},
    url = {https://lemire.me/blog/2017/08/22/testing-non-cryptographic-random-number-generators-my-results/},
    year = {2017},
    month = {08},
    day = {22},
    urldate = {2021-05-25}
}

@online {Brooks2021,
    author = {Brooks, Jeb},
    title = {\textit{pcg\_rand}},
    url = {https://docs.rs/pcg_rand/0.13.0/pcg_rand/type.Pcg64.html},
    year = {2021},
    month = {02},
    urldate = {2021-05-25}
}

@inbook {KolmogorovSmirnov,
    title = {Kolmogorov-Smirnov Test},
    booktitle = {The Concise Encyclopedia of Statistics},
    year = {2008},
    publisher = {Springer New York},
    address = {New York, NY},
    pages = {283-287},
    isbn = {978-0-387-32833-1},
    doi = {10.1007/978-0-387-32833-1_214},
    url = {https://doi.org/10.1007/978-0-387-32833-1_214}
}

@inbook {ChiSquared,
    title = {Chi-square Goodness of Fit Test},
    booktitle = {The Concise Encyclopedia of Statistics},
    year = {2008},
    publisher = {Springer New York},
    address = {New York, NY},
    pages = {72-76},
    isbn = {978-0-387-32833-1},
    doi = {10.1007/978-0-387-32833-1_55},
    url = {https://doi.org/10.1007/978-0-387-32833-1_55}
}

@book {Fisher1934,
    title = {\textit{Statistical methods for research workers}},
    author = {Fisher, Ronald Aylmer},
    year = {1934},
    edition = {5},
    publisher = {Oliver and Boyd, Edinburgh and London},
    url = {http://www.haghish.com/resources/materials/Statistical_Methods_for_Research_Workers.pdf},
    urldate = {2021-06-02}
}

@article {Callaghane2021,
% 	author = {Callaghan, Corey T. and Nakagawa, Shinichi and Cornwell, William K.},
% 	title = {Global abundance estimates for 9,700 bird species},
% 	volume = {118},
% 	number = {21},
% 	elocation-id = {e2023170118},
% 	year = {2021},
% 	doi = {10.1073/pnas.2023170118},
% 	publisher = {National Academy of Sciences},
% 	abstract = {For the fields of ecology, evolutionary biology, and conservation, abundance estimates of organisms are essential. Quantifying abundance, however, is difficult and time consuming. Using a data integration approach integrating expert-derived abundance estimates and global citizen science data, we estimate the global population of 9,700 bird species (\~{}92\% of all extant bird species). We conclude that there are many rare species, highlighting the need to continue to refine global population estimates for all taxa and the role that global citizen science data can play in this effort.Quantifying the abundance of species is essential to ecology, evolution, and conservation. The distribution of species abundances is fundamental to numerous longstanding questions in ecology, yet the empirical pattern at the global scale remains unresolved, with a few species{\textquoteright} abundance well known but most poorly characterized. In large part because of heterogeneous data, few methods exist that can scale up to all species across the globe. Here, we integrate data from a suite of well-studied species with a global dataset of bird occurrences throughout the world{\textemdash}for 9,700 species (\~{}92\% of all extant species){\textemdash}and use missing data theory to estimate species-specific abundances with associated uncertainty. We find strong evidence that the distribution of species abundances is log left skewed: there are many rare species and comparatively few common species. By aggregating the species-level estimates, we find that there are \~{}50 billion individual birds in the world at present. The global-scale abundance estimates that we provide will allow for a line of inquiry into the structure of abundance across biogeographic realms and feeding guilds as well as the consequences of life history (e.g., body size, range size) on population dynamics. Importantly, our method is repeatable and scalable: as data quantity and quality increase, our accuracy in tracking temporal changes in global biodiversity will increase. Moreover, we provide the methodological blueprint for quantifying species-specific abundance, along with uncertainty, for any organism in the world.eBird data are freely available to download from https://ebird.org/data/download. Population estimates were extracted from refs. 62 and 63 and http://datazone.birdlife.org/home. Range maps used to adjust population areas are available from BirdLife International. All trait data are freely available through the sources mentioned above. Code and necessary data to reproduce our analyses are available from Zenodo, https://doi.org/10.5281/zenodo.4723365 (99).},
% 	issn = {0027-8424},
% 	URL = {https://www.pnas.org/content/118/21/e2023170118},
% 	eprint = {https://www.pnas.org/content/118/21/e2023170118.full.pdf},
% 	journal = {Proceedings of the National Academy of Sciences}
% }

@inproceedings {Coulier2018,
    author = {Coulier, Adrien and Hellander, Andreas},
    booktitle = {2018 IEEE 14th International Conference on e-Science (e-Science)},
    title = {Orchestral: A Lightweight Framework for Parallel Simulations of Cell-Cell Communication},
    year = {2018},
    pages = {168-176},
    doi = {10.1109/eScience.2018.00032}
}

@article {Jeschke2008,
    author = {Jeschke, Matthias and Ewald, Roland and Park, Alfred and Fujimoto, Richard and Uhrmacher, Adelinde M.},
    title = {A Parallel and Distributed Discrete Event Approach for Spatial Cell-Biological Simulations},
    year = {2008},
    issue_date = {March 2008},
    publisher = {Association for Computing Machinery},
    address = {New York, NY, USA},
    volume = {35},
    number = {4},
    issn = {0163-5999},
    url = {https://doi.org/10.1145/1364644.1364652},
    doi = {10.1145/1364644.1364652},
    journal = {SIGMETRICS Perform. Eval. Rev.},
    month = {03},
    pages = {22–31},
    numpages = {10}
}

@article {Li2010,
%     author = {Li, Hong and Petzold, Linda},
%     title = {Efficient Parallelization of the Stochastic Simulation Algorithm for Chemically Reacting Systems On the Graphics Processing Unit},
%     journal = {The International Journal of High Performance Computing Applications},
%     volume = {24},
%     number = {2},
%     pages = {107-116},
%     year = {2010},
%     doi = {10.1177/1094342009106066},
%     url = {https://doi.org/10.1177/1094342009106066},
%     eprint = {https://doi.org/10.1177/1094342009106066},
%     abstract = {The small number of some reactant molecules in biological systems formed by living cells can result in dynamical behavior which cannot be captured by traditional deterministic models. In such a problem, a more accurate simulation can be obtained with discrete stochastic simulation (Gillespie’s stochastic simulation algorithm — SSA). Many stochastic realizations are required to capture accurate statistical information of the solution. This carries a very high computational cost. The current generation of graphics processing units (GPU) is well-suited to this task. In this paper we describe our implementation and present some computational experiments illustrating the power of this technology for this important and challenging class of problems.}
% }

@inproceedings {LEcuyer2017History,
    author = {L'Ecuyer, Pierre},
    booktitle = {2017 Winter Simulation Conference (WSC)},
    title = {History of uniform random number generation},
    year = {2017},
    pages = {202-230},
    doi = {10.1109/WSC.2017.8247790}
}

@article {Phillips2011,
    title = {Pseudo-random number generation for Brownian Dynamics and Dissipative Particle Dynamics simulations on GPU devices},
    journal = {Journal of Computational Physics},
    volume = {230},
    number = {19},
    pages = {7191-7201},
    year = {2011},
    issn = {0021-9991},
    doi = {10.1016/j.jcp.2011.05.021},
    url = {https://www.sciencedirect.com/science/article/pii/S0021999111003329},
    author = {Phillips, Carolyn L. and Anderson, Joshua A. and Glotzer, Sharon C.}
}

@inproceedings {Martineau2017,
    author = {Martineau, Matt and McIntosh-Smith, Simon},
    booktitle = {2017 IEEE International Conference on Cluster Computing (CLUSTER)},
    title = {Exploring On-Node Parallelism with Neutral, a Monte Carlo Neutral Particle Transport Mini-App},
    year = {2017},
    pages = {498-508},
    doi = {10.1109/CLUSTER.2017.83}
}

@article {Hill2015,
    author = {Hill, David R.C.},
    journal = {Computing in Science \& Engineering},
    title = {Parallel Random Numbers, Simulation, and Reproducible Research},
    year = {2015},
    volume = {17},
    number = {4},
    pages = {66-71},
    doi = {10.1109/MCSE.2015.79}
}

@article {Jun2020,
	doi = {10.1088/1742-6596/1525/1/012054},
	url = {https://doi.org/10.1088/1742-6596/1525/1/012054},
	year = {2020},
	month = {04},
	publisher = {{IOP} Publishing},
	volume = {1525},
	pages = {012054},
	author = {Jun, S Y and Canal, P and Apostolakis, J and Gheata, A and Moneta, L},
	title = {Vectorization of random number generation and reproducibility of concurrent particle transport simulation},
	journal = {Journal of Physics: Conference Series}
}

@article {Lawrie2017,
    author = {Lawrie, David S},
    title = {Accelerating Wright–Fisher Forward Simulations on the Graphics Processing Unit},
    journal = {G3 Genes|Genomes|Genetics},
    volume = {7},
    number = {9},
    pages = {3229-3236},
    year = {2017},
    month = {09},
    issn = {2160-1836},
    doi = {10.1534/g3.117.300103},
    url = {https://doi.org/10.1534/g3.117.300103},
    eprint = {https://academic.oup.com/g3journal/article-pdf/7/9/3229/37151490/g3journal3229.pdf},
}

@article {Lang2017,
    title = {An embarrassingly parallel algorithm for random walk simulations on random fractal structures},
    journal = {Journal of Computational Science},
    volume = {19},
    pages = {1-10},
    year = {2017},
    issn = {1877-7503},
    doi = {10.1016/j.jocs.2016.11.014},
    url = {https://www.sciencedirect.com/science/article/pii/S1877750316304057},
    author = {Lang, Jens and Prehl, Janett}
}

@article {Halton1989,
    title = {Pseudo-random trees: Multiple independent sequence generators for parallel and branching computations},
    journal = {Journal of Computational Physics},
    volume = {84},
    number = {1},
    pages = {1-56},
    year = {1989},
    issn = {0021-9991},
    doi = {10.1016/0021-9991(89)90180-0},
    url = {https://www.sciencedirect.com/science/article/pii/0021999189901800},
    author = {Halton, John H.}
}

@article {Rosindell2007,
    author = {Rosindell, James and Cornell, Stephen J.},
    title = {Species–area relationships from a spatially explicit neutral model in an infinite landscape},
    journal = {Ecology Letters},
    volume = {10},
    number = {7},
    pages = {586-595},
    doi = {10.1111/j.1461-0248.2007.01050.x},
    url = {https://onlinelibrary.wiley.com/doi/abs/10.1111/j.1461-0248.2007.01050.x},
    eprint = {https://onlinelibrary.wiley.com/doi/pdf/10.1111/j.1461-0248.2007.01050.x},
    year = {2007}
}

@misc {EthicsChecklist,
    author = {Lancaster, Thomas E.},
    title = {\textit{Ethics Process -- Ethics Checklist UG}},
    year = {2021},
    month = {01},
    day = {13},
    url = {https://wiki.imperial.ac.uk/download/attachments/250285119/ethics-checklist-ug.xlsx?version=1},
    urldate = {2021-06-08}
}

@article{White2014,
    author = {White, J. Wilson and Rassweiler, Andrew and Samhouri, Jameal F. and Stier, Adrian C. and White, Crow},
    title = {Ecologists should not use statistical significance tests to interpret simulation model results},
    journal = {Oikos},
    volume = {123},
    number = {4},
    pages = {385-388},
    doi = {https://doi.org/10.1111/j.1600-0706.2013.01073.x},
    url = {https://onlinelibrary.wiley.com/doi/abs/10.1111/j.1600-0706.2013.01073.x},
    eprint = {https://onlinelibrary.wiley.com/doi/pdf/10.1111/j.1600-0706.2013.01073.x},
    year = {2014}
}

@misc{OttoPortner2021,
    author       = {Otto-Portner, Hans and
                    Scholes, Bob and
                    Agard, John and
                    Archer, Emma and
                    Arneth, Almut and
                    Bai, Xuemei and
                    Barnes, David and
                    Burrows, Michael and
                    Chan, Lena and
                    Cheung, Wai Lung (William) and
                    Diamond, Sarah and
                    Donatti, Camila and
                    Duarte, Carlos and
                    Eisenhauer, Nico and
                    Foden, Wendy and
                    Gasalla, Maria A. and
                    Handa, Collins and
                    Hickler, Thomas and
                    Hoegh-Guldberg, Ove and
                    Ichii, Kazuhito and
                    Jacob, Ute and
                    Insarov, Gregory and
                    Kiessling, Wolfgang and
                    Leadley, Paul and
                    Leemans, Rik and
                    Levin, Lisa and
                    Lim, Michelle and
                    Maharaj, Shobha and
                    Managi , Shunsuke and
                    Marquet, Pablo A. and
                    McElwee, Pamela and
                    Midgley, Guy and
                    Oberdorff, Thierry and
                    Obura, David and
                    Osman Elasha, Balgis and
                    Pandit, Ram and
                    Pascual, Unai and
                    Pires, Aliny P. F. and
                    Popp, Alexander and
                    Reyes-García, Victoria and
                    Sankaran, Mahesh and
                    Settele, Josef and
                    Shin, Yunne-Jai and
                    Sintayehu, Dejene W. and
                    Smith, Peter and
                    Steiner, Nadja and
                    Strassburg, Bernardo and
                    Sukumar, Raman and
                    Trisos, Christopher and
                    Val, Adalberto Luis and
                    Wu, Jianguo and
                    Aldrian, Edvin and
                    Parmesan, Camille and
                    Pichs-Madruga, Ramon and
                    Roberts, Debra C. and
                    Rogers, Alex D. and
                    Díaz, Sandra and
                    Fischer, Markus and
                    Hashimoto, Shizuka and
                    Lavorel, Sandra and
                    Wu, Ning and
                    Ngo, Hien},
    title        = {\textit{Scientific outcome of the IPBES-IPCC co-sponsored
                     workshop on biodiversity and climate change}},
    month        = {06},
    year         = {2021},
    publisher    = {Zenodo},
    address      = {Bonn, Germany},
    organization = {IPBES secretariat},
    version      = {1},
    doi          = {10.5281/zenodo.4923212},
    url          = {https://doi.org/10.5281/zenodo.4923212}
}
\fussy

\addtocontents{toc}{\protect\contentsline{part}{\protect Appendices}{}{}}
\appendix
\addtocontents{toc}{\protect\setcounter{tocdepth}{0}}
\chapter{Ethical Discussion} \label{ethical-discussion}

\section*{Preface and Project Objective} \label{ethical-objective}

The following evaluation of Ethical Issues is based on the departmental Ethics Checklist \cite{EthicsChecklist}. In particular, it only goes into detail for questions that we have not answered with `No'. \\

\noindent This project is focused on individual-based probabilistic biodiversity simulations. It uses reproducible random number generation to parallelise such a simulation efficiently. The choice of biodiversity simulations as a case study is not accidental. On the contrary, this project's secondary goal is to help ecologists and conservationists make higher-resolution predictions of biodiversity loss more efficiently. These models are helpful both in theoretical ecology and practical conservation work, e.g. to provide guidance such that protected areas, which are crucial to protect endangered species, are placed most effectively.

\setcounter{section}{5} \section{Dual Use}
\subsection*{Does your project have an exclusive civilian application focus?}

According to an explanatory note published by the European Commission Explanatory on the ``exclusive focus on civil applications'' \cite{EU2020}, the project's objective determines if the project has an exclusive civilian application focus. As described in \Cref{ethical-objective}, this project aims to aid the civilian protection of biodiversity.

\setcounter{section}{7} \section{Legal Issues}
\subsection*{Will your project use or produce software for which there are copyright licensing implications?}

We have developed the \texttt{necsim-rust} simulation library and its command-line frontend, called \texttt{rustcoalescence}, for this project. Both are written in the Rust Programming Language, which is dual-licensed under the Apache License, Version 2.0, and the MIT license \cite{RustLicense}. Since the tool is being designed to be used in active research, \texttt{necsim-rust} is also dual-licensed under the Apache License, Version 2.0, and the MIT license. \\

\noindent At the time of writing (commit \href{https://github.com/MomoLangenstein/necsim-rust/tree/2e352ff}{\#2e352ff}), the project depends on the following Rust crates, which are sorted by their licenses:

\begin{enumerate}
    \item \textbf{0BSD} OR \textbf{Apache-2.0} OR \textbf{MIT} (1): \texttt{adler}
    \item \textbf{Apache-2.0} (2): \texttt{clang-sys}, \texttt{pcg\_rand}
    \item \textbf{Apache-2.0} OR \textbf{Apache-2.0 WITH LLVM-exception} OR \textbf{MIT} (1): \texttt{wasi}
    \item \textbf{Apache-2.0} OR \textbf{MIT} (102): \texttt{addr2line}, \texttt{aes}, \texttt{ahash}, \texttt{anyhow}, \texttt{autocfg}, \texttt{backtrace}, \texttt{base64}, \texttt{bitflags}, \texttt{build-probe-mpi}, \texttt{cc}, \texttt{cexpr}, \texttt{cfg-if}, \texttt{cipher}, \texttt{cpufeatures}, \texttt{cuda-config}, \texttt{cuda-} \texttt{driver-sys}, \texttt{custom\_derive}, \texttt{env-logger}, \texttt{erased-serde}, \texttt{failure}, \texttt{failure\_derive}, \\ \texttt{fallible-iterator}, \texttt{fallible-streaming-iterator}, \texttt{getrandom}, \texttt{gimli}, \texttt{glob}, \texttt{hashbrown}, \texttt{hashlink}, \texttt{heck}, \texttt{hermit-abi}, \texttt{humantime},  \texttt{indexmap}, \texttt{jpeg-decoder}, \texttt{lazy\_static}, \texttt{lazycell}, \texttt{libc}, \texttt{libffi}, \texttt{libffi-sys}, \texttt{libm}, \texttt{log}, \texttt{mpi}, \texttt{mpi-sys}, \texttt{num-traits}, \texttt{object}, \texttt{once-cell}, \texttt{opaque-debug}, \texttt{peeking\_take\_while}, \texttt{pest}, \texttt{pkg-config}, \texttt{ppv-lite86}, \texttt{proc-macro-error}, \texttt{proc-macro-error-attr}, \texttt{proc-macro2}, \texttt{quick-error}, \texttt{quote}, \texttt{rand}, \texttt{rand\_chacha}, \texttt{rand\_} \texttt{core}, \texttt{rand\_hc}, \texttt{regex}, \texttt{regex-syntax}, \texttt{remove\_dir\_all}, \texttt{ron},  \texttt{rustacuda}, \texttt{rustacuda\_core}, \texttt{rustacuda\_derive}, \texttt{rustc-demangle}, \texttt{rustc-hash}, \texttt{rustc\_version}, \texttt{semver}, \texttt{semver-parser}, \texttt{serde}, \texttt{serde\_derive}, \texttt{serde\_derive\_state}, \texttt{serde\_path\_to\_error}, \texttt{serde\_state}, \texttt{shlex}, \texttt{smallvec}, \texttt{structopt}, \texttt{structopt-derive}, \texttt{syn}, \texttt{tempfile}, \texttt{thiserror}, \texttt{thiserror-impl}, \texttt{toml}, \texttt{thread\_local}, \texttt{toml}, \texttt{tynm}, \texttt{typed-builder}, \texttt{typenum}, \texttt{ucd-trie}, \texttt{unicode-} \\ \texttt{segmentation}, \texttt{unicode-width}, \texttt{unicode-xid}, \texttt{vcpkg}, \texttt{vec\_map}, \texttt{version\_check}, \texttt{weezl}, \\ \texttt{winapi}, \texttt{winapi-i686-pc-windows-gnu}, \texttt{winapi-x86\_64-pc-windows-gnu}, \texttt{wyhash}
    \item \textbf{Apache-2.0} OR \textbf{MIT} OR \textbf{Zlib} (1): \texttt{miniz\_oxide}
    \item \textbf{BSD-3-Clause} (1): \texttt{bindgen}
    \item \textbf{BSL-1.0} (1): \texttt{xxhash-rust}
    \item \textbf{CC0-1.0} (1): \texttt{abort\_on\_panic}
    \item \textbf{ISC} (1): \texttt{libloading}
    \item \textbf{LGPL-3.0} (1): \texttt{priority-queue}
    \item \textbf{MIT} (24): \texttt{ansi\_term}, \texttt{array2d}, \texttt{atty}, \texttt{bincode}, \texttt{byte-unit}, \texttt{clap}, \texttt{conv}, \texttt{float\_next\_after}, \texttt{generic-array}, \texttt{libsqlite3-sys}, \texttt{make-cmd}, \texttt{memoffset}, \texttt{nom}, \texttt{ptx-builder}, \texttt{redox\_syscall}, \texttt{rusqlite}, \texttt{seahash}, \texttt{slab}, \texttt{strsim}, \texttt{synstructure}, \texttt{textwrap}, \texttt{tiff}, \texttt{utf8-width}, \texttt{which}
    \item \textbf{MIT} OR \textbf{Unlicense} (5): \texttt{aho-corasick}, \texttt{byteorder}, \texttt{memchr}, \texttt{termcolor}, \texttt{winapi-util}
    \item \textbf{MPL-2.0} (2): \texttt{colored}, \texttt{contracts}
    \item \textbf{N/A} (1): \texttt{mpi-derive} \textit{(part of the \texttt{rsmpi} package that is licensed using \textbf{Apache-2.0} OR \textbf{MIT})}
    \item \textbf{Zlib} (1): \texttt{nanorand}
\end{enumerate}
Furthermore, if the CLI \texttt{rustcoalescence} is compiled with support for CUDA (by enabling the \texttt{rustcoalescence-algorithms-cuda} feature flag), the tool is also \underline{dynamically linked} with the local CUDA driver. If the CLI is compiled with MPI support (by enabling the \texttt{necsim-partitioning} \texttt{-mpi} feature flag), the tool is also \underline{dynamically linked} with the local MPI installation. \\

\noindent To the best of our knowledge, there is no licensing conflict.

\chapter{Analytical solutions for three Neutral scenarios} \label{appendix:neutral-scenarios}

The Neutral Theory of Biodiversity \cite{Hubbell2001} can be used to describe several model scenarios. Ecologists have developed species-area relationships (SARs) to describe how the species richness of a landscape depends on its area. In this section, we summarise three scenarios and their (sometimes approximate) analytical solutions \cite{Rosindell2021}. We use the following list of symbols:
\begin{itemize}
    \item $S$ is the steady-state species richness of the scenario.
    \item $A$ is the area of the landscape that the individuals live on.
    \item $\rho$ is the density of individuals living in one unit area. It is assumed to be constant over the landscape for the scenarios that we describe.
    \item $\nu$ is the per-capita per-generation speciation probability.
    \item $J = \rho A$ is the size of the community of individuals.
    \item $\digamma(z)$ is the digamma function.
\end{itemize}

\section{Non-Spatial}

The non-spatial scenario describes a closed and well-mixed community of individuals, e.g. an island. Dispersal inside this community is homogeneous, i.e. the dispersal kernel is the uniform distribution over all possible locations. The following exact analytical solution exists to calculate the species richness in this scenario \cite[Eq.~7]{Vallade2003}\cite[Eq.~1]{Rosindell2021}:
\begin{equation} \label{eq:non-spatial}
    S(A) = \theta(A) \cdot \left( \digamma(\theta(A) + J) - \digamma(\theta(A)) \right)
\end{equation}
where $\theta(A) = \frac{(J - 1) \cdot \nu}{1 - \nu}$\footnote{Note that $(J - 1)$ implies that a child can never replace its parent directly. If replacement is allowed, $(J)$ must be used instead.}, as defined by Vallade and Houchmandzadeh \cite{Vallade2003}. $\theta(A)$ is confusingly named the same as Hubbell's fundamental biodiversity number, which is defined as $\theta = 2 J \nu$ \cite[p.~126]{Hubbell2001}. In the limit of large $J$ and small $\nu$, $2 \cdot \theta(A)$ approximates Hubbell's $\theta$. For consistency, we do not use Hubbell's $\theta$ \footnote{Interestingly, Hubbel showed that in the limit of large $J$, $\theta$ is asymptotically identical to Fisher's $\alpha$ \cite{Fisher1943}, a crucial biodiversity index \cite[p.~126]{Hubbell2001}.}.

Rosindell and Chisholm have shown that the non-spatial species richness can also be approximated with its limit under large $\theta(A)$ \cite[Eq.~2]{Rosindell2021}:
\begin{equation} \label{eq:non-spatial-limit}
    S(A) \sim \theta(A) \cdot \log{\left( \frac{1}{\nu} \right)}
\end{equation}

\subsection{Protracted Speciation} \label{appendix:protracted-speciation}

In most neutral models, including the non-spatial scenario, speciation is modelled as an instantaneous point process, meaning every speciation event creates a new and unique species. Consequently, the present state might contain some very young species, which only consist of a small number of individuals.

Rosindell et al. addressed this limitation of point mutation speciation by introducing protracted speciation \cite{Rosindell2010}. With protracted speciation, the speciation process is instead assumed to take $\tau$ generations to complete. Specifically, a lineage must survive for at least $\tau$ generations after a speciation event for the speciation to be counted. Therefore, speciation events that would occur between present-time $t_{0}$ and $t_{-\tau}$ can be ignored as they would not have had time to complete.

To calculate the species richness in this scenario, we can reuse the non-spatial formula \Cref{eq:non-spatial} and rescale $\nu$ to $\nu' = \frac{\mu}{1 + \tau}$. Here, $\mu$ is the original speciation probability, now called speciation-initiation probability.

\section{Spatially Implicit}

The spatially implicit scenario expands the non-spatial model by adding migration. There now is a small local community of size $J$ and a larger metacommunity of size $J_m$. The primary source of biodiversity in the local community is migration from the metacommunity. The migration probability function $m(A)$ describes the per capita probability that migration from the metacommunity to the local community occurs. As this migration is assumed to dominate speciation, speciation is ignored in the local community, i.e. $\nu_{l} = 0$ \cite{Rosindell2021}. The species richness of the spatially implicit scenario can be calculated as \cite{Chisholm2016, Etienne2005}:
\begin{equation} \label{eq:spatially-implicit}
    S(A) = \theta \cdot \left(- \digamma(\theta) + \frac{1}{(\gamma(A))_{J}} \sum_{j = 0}^{J} \mid s(J, j) \mid (\gamma(A))^{j} \digamma (J + \theta) \right)
\end{equation}
where $\gamma(A) = \frac{(J - 1) \cdot m(A)}{1 - m(A)}$, $\theta = \frac{(J_m - 1) \cdot \nu_m}{1 - \nu_m}$, $s(n, m)$ is a Stirling number of the first kind and $(a)_{n}$ is a Pochhammer symbol.

This formula can be approximated for computational efficiency under the assumption that the number of lineages, i.e. the number of original ancestors, on the local community is constant at the equilibrium \cite[Eq.~2.5]{Chisholm2016}:
\begin{equation}
    S(A) = \theta \cdot \left( \digamma \left( \theta + \gamma(A) \cdot \left( \digamma(\gamma(A) + J) - \digamma(\gamma(A)) \right) \right) - \digamma(\theta) \right)
\end{equation}
Rosindell and Chisholm have shown that this formula can again be approximated by its limit in the case of high diversity \cite[Eq.~8]{Rosindell2021}:
\begin{equation}
    S(A) \sim \theta \cdot \log{\left( 1 - \frac{\gamma(A)}{\theta} \cdot \log(m(A)) \right)}
\end{equation}
In the case where $m(A) \rightarrow 1$, the spatially implicit model can be thought of as a random sample of $J$ migrations from the metacommunity. In that case, the local biodiversity is expected to be \cite[Eq.~9]{Rosindell2021}:
\begin{equation}
    S(A) \sim \theta \cdot \log{\left( 1 + \frac{J - 1}{\theta} \right)}
\end{equation}

\section{Spatially Explicit}

The neutral model can also describe spatially explicit scenarios in which the individual's behaviour \textbf{is} affected by its location on the landscape. This scenario requires an explicit description of habitat distribution and the dispersal kernel across the landscape. The analytical solutions of the previous two scenarios have calculated the species richness on an island. In the spatially explicit scenario with an infinite landscape, we now have to introduce a small survey area $A$ -- only the present-time species identities of individuals in the sample area count towards the sampled biodiversity.

If the size of the entire landscape $A_L \rightarrow \infty$, $\rho = 1$ \footnote{$\rho = 1$ is without loss of generality: as long as $\rho = \text{const.}$, the coordinate system of the landscape can simply be rescaled.}, and dispersal occurs according to a normal distribution $\textrm{N}^{2}(0, {\sigma}^{2})$, then there exists the following SAR \cite{ODwyer2018}:
\begin{equation}
    S(A) = S_{contig}(A, \nu, \sigma^{2}) \sim \sigma^2 \cdot \Psi \left( \frac{A}{\sigma^2}, \nu \right)
\end{equation}
where $\Psi(A^*, \nu)$ is the Preston function. The Preston function is defined as the ``SAR for contiguous circular sample areas taken from an infinite world whose dynamics follow a non-zero-sum version of the spatial neutral model [...], with a bivariate normal dispersal kernel having $\sigma$ = 1'' \cite{Rosindell2021}, which can be evaluated using a coalescence-based simulation of the spatially explicit neutral model. There also exists the following approximation to the Preston function \cite{ODwyer2018}:
\begin{equation}
    \Psi(A^*, \nu) \approx \nu_{eff} A^* + \frac{2 \pi \sqrt{\frac{A^*}{\pi}} (1 - \nu_{eff}) I_1 \left( \sqrt{\frac{A^*}{\pi}} \right)}{\frac{1}{\sqrt{v_{eff}}} I_1 \left( \sqrt{\frac{A^*}{\pi}} \right) \frac{K_0 \left( \sqrt{\frac{v_{eff} A^*}{\pi}} \right)}{K_1 \left( \sqrt{\frac{v_{eff} A^*}{\pi}} \right)} + I_0 \left( \sqrt{\frac{A^*}{\pi}} \right)}
\end{equation}
where $\nu_{eff} = \frac{\nu \cdot \log \left( \frac{1}{\nu} \right)}{1 - \nu}$, and $I_i(z)$ and $K_i(z)$ are the modified Bessel functions. \\

The spatially explicit scenario can further be expanded by considering habitat loss and fragmentation. First, there is the case where habitat has been removed randomly, such that only $h = \frac{A_{e}}{A_{max}}$ of the focal area $A_{max}$ remains habitable, which is $A_{e}$. If $h = 100\%$, i.e. no habitat was lost, the average distance between adjacent habitable cells remains $1$. Otherwise, it shrinks to $\sqrt{\frac{A_{e}}{A_{max}}}$. We can think about the remaining habitat as dots on a balloon. If the balloon is compressed, the dots come closer together, and dispersal on the balloon scales proportionally. Thompson, Chisholm and Rosindell have shown how to apply the spatially explicit formula using this metaphor \cite[p.~2090]{Thompson2019}:
\begin{equation} \label{eq:spatially-explicit}
    S_{random}(A_{max}, A_{e}, \nu, {\sigma}^{2}) = S_{contig}(A_{e}, \nu, \frac{A_{e}}{A_{max}} \sigma^{2}) \sim \frac{A_{e}}{A_{max}} \sigma^2 \cdot \Psi \left( \frac{A_{max}}{\sigma^2}, \nu \right)
\end{equation}

For the other case, where habitat loss has not been uniform, Thompson, Chisholm and Rosindell introduced the effective connectivity metric $c_{e}^{2} = h \cdot \sigma_{e}^{2}$ which replaces $\sigma^{2}$ \cite[p.~2090]{Thompson2019}:
\begin{equation}
    S_{contig}(A_{e}, \nu, c_{e}^{2}) \sim c_{e}^{2} \cdot \Psi \left( \frac{A_{e}}{c_{e}^{2}}, \nu \right)
\end{equation}
where $\sigma_{e}^{2}$ is the mean effective dispersal over the landscape. $\sigma_{e}^{2}$ can be estimated as $\sigma_{e}^{2} \approx \mu_{n}^{2} \cdot \frac{2}{n \cdot \pi}$ where $\mu_{n}^{2}$ is the mean dispersal distance after $n$ normal dispersal jumps on the landscape. In practice, $\mu_{n}^{2}$ can be calculated by averaging over many random dispersal walks from different locations on the landscape with $n = \nu^{-1}$. On an (almost) continuous landscape with $h \rightarrow 1$, $\sigma_{e}^{2} \rightarrow \sigma^{2}$ of the Gaussian dispersal kernel. On an island, where the continuous formula does not apply, $\sigma_{e}^{2} \rightarrow 0$ as the island gets smaller.

Thompson, Chisholm and Rosindell have also shown that the random uniform habitat loss represents the best-case scenario for biodiversity loss with the same $h$ \cite[p.~2093]{Thompson2019}.

\chapter{Sampling Probability Distributions} \label{appendix:distribution-sampling}

\Cref{background-rng} has summarised how random integers can be generated. This part discusses how we can transform a random bitstring into a standard uniform $\textrm{U}(0, 1)$, and how we can sample the $\textrm{U}(a, b)$, $\textrm{Exp}(\lambda)$, $\textrm{Geo}(p)$, $\textrm{Poi}(\lambda)$ and $\textrm{N}(\mu, \sigma^{2})$ distributions\footnote{We do not cover rejection sampling \cite{vonNeumann1951} in this appendix.}.

\section{Standard Uniform in IEEE 754 Floating point}

Random number generators, especially pseudo-random generators, usually produce random fixed-length bitstrings. The IEEE 754 standard \cite{IEEE2019} defines the floating-point format, which is used by most computers today. For simplicity, we assume that our RNG produces 64 random bits and that the $\textrm{U}(0, 1)$ sample is returned in 64-bit double-precision. However, this procedure can also be generalised and applied to other formats.

IEEE 754 double-precision floating point-numbers consist of one sign bit, eleven exponent bits, and 53 mantissa (digits after the binary point) bits, one of which is implicit \cite{IEEE2019}. The xoshiro / xoroshiro RNG generates uniform samples by only filling the mantissa with 53 random bits \cite{Vigna2019}:
\begin{minted}{c}
static inline double to_uniform(uint64_t x) {
    return (x >> 11) * 0x1.0p-53
}
\end{minted}
This conversion produces values in $\{0, \frac{1}{2^{53}}, ..., \frac{2^{53} - 1}{2^{53}}\}$, i.e. $[0, 1)$ approximately. While this algorithm runs in $O(1)$, it is important to note that it does not sample all available floating point values in $[0, 1]$, which would require $O(2^{|exponent|})$ \cite{Campbell2014}.

\section{Inverse Transform: Uniform and Exponential} \label{appendix:inverse-transform}

Given a probability distribution $X$ with cdf $F_{X}(x)$, the inverse transform method can be used to sample $X$ using only one standard uniform sample $U \sim \textrm{U}(0, 1)$ \cite[pp.~28-19]{Devroye1986}:
\begin{equation}
    X = F_{X}^{-1}(U)
\end{equation}
In practice, the inverse transform method is only used when the inverse of the cdf, $F_{x}^{-1}$, can be calculated easily. This is the case for the uniform $\textrm{U}(a, b)$ and exponential $\textrm{Exp}(\lambda)$ distributions:
\begin{equation}
    \begin{split}
        F_{\textrm{U}(a,b)}(x) &= \begin{cases}
            0 & \text{for } x \le a \\
            \frac{x - a}{b - a} & \text{for } a \leq x \leq b \\
            1 & \text{for } x \ge b
        \end{cases} \\
        F_{\textrm{U}(a,b)}^{-1}(u) &= a + (b - a) \cdot u \\
    \end{split}
\end{equation}
\begin{equation}
    \begin{split}
        F_{\textrm{Exp}(\lambda)}(x) &= 1 - e^{- \lambda \cdot x} \\
        F_{\textrm{Exp}(\lambda)}^{-1}(u) &= - \frac{\log(u)}{\lambda}
    \end{split}
\end{equation}
We can also use the fact that $\textrm{Geo}(p) = \lfloor \textrm{Exp}(-\log(1 - p)) \rfloor$ (\ref{background-exp-poi-properties}) to sample the geometric distribution using the exponential distribution.

\section{Iterative Sampling: Geometric and Poisson} \label{appendix:geometric-poisson-sampling}

The Geometric distribution $\textrm{Geo}(p)$ describes the number of times a Bernoulli trial $B(p)$ fails before its first success. A trivial method to sample $\textrm{Geo}(p)$ involves simply counting the number of standard uniforms $U_{i}$ that have to be drawn until $U_{i} \leq p$ \cite[p.~498]{Devroye1986}. The constant-time approach discussed above in \Cref{appendix:inverse-transform} can also be used instead.

Similarly, we can also sample the Poisson distribution $\textrm{Poi}(\lambda)$ in $O(\textrm{Poi}(\lambda))$ time using iteration. Since the inter-event times of a homogeneous Poisson process are exponentially distributed (\ref{background-exp-poi-properties}), we can count the number of $X_{i} \sim \textrm{Exp}(\lambda)$ that have to be summed up until $\sum_{i = 1}{X_{i}} \geq 1$ \cite[p.~503]{Devroye1986}. If $\textrm{Exp}(\lambda)$ is sampled using the inverse transform, the repeated evaluation of the logarithm can be avoided and replaced with one exponential function \cite[p.~504]{Devroye1986}:
\begin{equation}
    \begin{split}
        \sum_{i = 1}{\textrm{Exp}(\lambda)} = \sum_{i = 1}{-\frac{\log(U_{i})}{\lambda}} = -\frac{\sum_{i = 1}{\log(U_{i})}}{\lambda} = -\frac{\log(\prod_{i = 1}{U_{i}})}{\lambda} &\geq 1 \\
        \text{or equivalently } \prod_{i = 1}{U_{i}} &\leq e^{-\lambda}
    \end{split}
\end{equation}
Devroye further optimised the sampling by linearly searching over the Poisson distribution's cdf. This inverse-transfrom variant only requires one standard uniform sample \cite[p.~505]{Devroye1986}:
\begin{minted}[linenos,escapeinside=@@]{python}
def sample_poisson(@lambda@):
    X = 0

    U = U(0, 1); acc = exp(-@lambda@); prod = acc

    while U > acc:
        X += 1

        prod *= @lambda@ / X; acc += prod

    return X
\end{minted}

\section{Box-Muller Transform: Normal Distribution}

In 1958, Box and Muller presented a simple transformation to calculate two independent standard normal variables, $X_1$ and $X_2$, using two standard uniforms, $U_1$ and $U_2$ \cite{Box1958}:
\begin{equation}
    \begin{split}
        X_1 &= \sqrt{-2 \cdot \log(U_1)} \cdot \cos(2 \pi U_2) \\
        X_2 &= \sqrt{-2 \cdot \log(U_1)} \cdot \sin(2 \pi U_2)
    \end{split}
\end{equation}
While $X_1, X_2 \sim \textrm{N}(0, 1)$, both standard normal random variables can easily be transformed to sample any normal distribution \cite[p.~379]{Devroye1986}:
\begin{equation}
    \textrm{N}(\mu, \sigma^{2}) = \sigma \cdot \textrm{N}(0, 1) + \mu
\end{equation}

\chapter{Rust's Basic Type System Syntax} \label{appendix:rust-syntax}

Rust is a statically typed language that statically enforces its type system at compile time. The type system is split up into primitive and custom types. Scalar types are primitive single-value types which store:
\vspace{-1em}
\begin{itemize}
    \item integers: \mintinline{rust}{u8}, \mintinline{rust}{i8}, \mintinline{rust}{u16}, \mintinline{rust}{i16}, \mintinline{rust}{u32}, \mintinline{rust}{i32}, \mintinline{rust}{u64}, \mintinline{rust}{i64}, \mintinline{rust}{usize}, \mintinline{rust}{isize}
    \item floating point numbers: \mintinline{rust}{f32}, \mintinline{rust}{f64}
    \item booleans: \mintinline{rust}{bool}
    \item Unicode characters: \texttt{\textcolor{purple}{\textbf{char}}}
    \item unit type: \mintinline{rust}{()} (similar to \mintinline{c}{void})
\end{itemize}
\vspace{-1em}
In addition to scalar types, there are also four primitive compound types. Arrays \mintinline{rust}{[T; N]} are a contiguous list of $N$ elements of the same type \texttt{T}. Subsequences of arrays can be referenced using the slice type \mintinline{rust}{&[T]}. The string slice \mintinline{rust}{&str} is a special case of this slice type. Last but not least, there are also tuple types such as \mintinline{rust}{(T, U, V)}.

Rust also provides three forms of custom types. Structures are structured compound types that can either be unit structs, (named) tuple structs, or C-like structs:
\begin{minted}[linenos,escapeinside=@@]{rust}
// A unit struct which does not contain any data
struct UnitStruct;

// A (named) tuple struct. In contrast to an unnamed tuple, we can implement
//  methods on this type.
struct TupleStruct(u32, f32, bool);

// A classical C-like struct with fields `primitive`, `unit` and `compound`.
struct CLikeStruct {
    primitive: @\textcolor{purple}{\textbf{char}}@,
    unit: UnitStruct,
    compound: TupleStruct,
}
\end{minted}
Furthermore, Rust has both safe union types, called enumerations, and C-like unions. Enumerations store their discriminant and only allow access to the variant which is stored. In contrast, unions can be interpreted as any of the possible variants \footnote{This reinterpretation is only possible in unsafe Rust code, which is an extension to the Rust language that allows low-level operations that the compiler cannot statically verify. Unsafe Rust code must be wrapped in \mintinline{rust}{unsafe { ... }} code blocks. Unsafe code is usually hidden behind user-facing safe interfaces, which establish the safety of an operation.}.
\begin{minted}[linenos,escapeinside=@@]{rust}
// An enumeration of value-less variants
enum Colours {
    Yellow,
    White,
    Purple,
    Black
}

// An enumeration of structured variants
enum Event {
    Speciation,
    Dispersal {
        target: Location,
        coalescence: Option<Lineage>,
    }
}

// A C-like union that stores any of three variants
union UnsafeCLikeUnion {
    variant_a: u32,
    variant_b: f32,
    variant_c: @\textcolor{purple}{\textbf{char}}@,
}
\end{minted}

\chapter{Memory Safety in Rust} \label{appendix:rust-safety}

One of the primary design goals of the Rust Programming Language is memory safety. Rust achieves this goal by introducing a strict value ownership and borrowing system that is enforced by the compiler. Jung et al. have developed a formal and machine-checked safety proof of this system for a subset of the Rust language \cite{JungPhD, Jung2017}.

\section{Ownership}

In Rust, a value can only be owned by one part of the code at once. Therefore, Rust distinguishes between copying and moving a value. A value of type \mintinline{rust}{T} can only be copied implicitly if \mintinline{rust}{T} satisfies the \texttt{\textcolor{yellow}{\textbf{Copy}}} trait, i.e. \mintinline{rust}{T: Copy}. Otherwise, the value must be moved to give ownership of the value to a different part in the code. After moving a value, it is no longer accessible in the original location:

\begin{minted}[linenos]{rust}
// A simple function which requires ownership of `value`
fn use_colour(value: Colour) {
    // This function now has ownership of `value`
    ...
    // `value` goes out of scope and is dropped
}

fn main() {
    // Create an owned colour value
    let value = Colour::Yellow;

    // Move `value` into the `use_colour` function
    use_colour(value);
    // `value` is no longer accessible at this point
}
\end{minted}

\noindent The Rust compiler can prove which part of the code owns a value at any point. Therefore, it can insert the necessary deallocation code when the value goes out of scope and is dropped. Consequently, Rust does not require a garbage collector to clean up values when they are no longer used.

\section{Borrowing}

Rust also supports lending borrows of a value to other code locations without changing the value's ownership. A value can either be borrowed mutably (\mintinline{rust}{&mut T}) once or immutably aliased (\mintinline{rust}{&T}) many times. This mutual exclusion means that there is always at most one location in the code that is allowed to modify a value, while many locations can have read-only references. Furthermore, a value can only be modified (or mutably borrowed) iff there are no immutable borrows of this value. These borrowing rules ensure that there is either mutability or aliasing. Therefore, code using immutable references can rely on the immutability of the values behind the references. For instance, we can look at the following simplified example:

\begin{minted}[linenos]{rust}
// Create an owned mutable vector (growable array) of integers
let mut numbers: Vec<u8> = vec![3, 1, 4, 1, 5, 7];

// Borrow `numbers` mutably in this line to modify the first element
numbers[5] = 9;

// Create an immutable reference to the third element
let my_number: &u8 = numbers[2];
// Create an immutable reference to the first element
let my_other_number: &u8 = numbers[0];

// COMPILER ERROR: cannot borrow `numbers` mutably while
//  it is also borrowed immutably
numbers[2] = 27;

// Use the immutable reference `my_number`
println!("My number is {}", my_number);
\end{minted}
In addition to the strictly compile-time enforced borrow rules, Rust also provides the \mintinline{rust}{RefCell<T>} helper struct in its standard library, which can check the borrowing rules dynamically at runtime.

\section{Lifetimes}

Let us now consider the case where we have to lend out a reference to another piece of code. How can we ensure that the value behind the reference is not destroyed while the reference is still in use? How can we avoid dangling references? Lifetimes are the final building block to ensure memory safety in this case. In Rust, the compiler tags every reference with the lifetime span of the underlying value. The compiler then enforces that a reference cannot outlive its value, i.e. that a reference cannot be used or stored beyond its lifetime and that the owner cannot destroy the value during the lifetime. Lifetimes are also used to check if mutable and immutable borrows overlap. Recently, the Rust compiler has been able to elide, i.e. infer, most reference lifetimes automatically. However, the programmer is always able to specify them explicitly to help the compiler prove memory safety:

\begin{minted}[linenos]{rust}
// A generic struct that contains an immutable reference to a `Vec<T>`:
//  `vector` is an immutable reference with a lifetime named `a`
//  `BorrowedVec` is generic over the lifetime 'a, which means that it
//    must not outlive 'a so that `vector` does not outlive 'a
struct BorrowedVec<'a, T> {
    vector: &'a Vec<T>
}

fn main() {
    // Create an owned immutable vector of integers
    let vector = vec![1, 2, 3, 4, 5];

    // Create a `BorrowedVec` that must not outlive `vector`
    let borrowed_vec = BorrowedVec {
        vector: &vector
    };

    // Explicitly drop the `vector` and terminate its lifetime
    std::mem::drop(vector);

    // COMPILER ERROR: `borrowed_vec` contains a reference to `vector`
    //  and must not outlive `vector`.
    println!("The 1st number of vector is {}", borrowed_vec.vector[0]);
}
\end{minted}

\noindent Rust also contains a special \mintinline{rust}{'static} lifetime that is given to references that are valid for the entire duration of a program. For instance, string literals such as \mintinline{rust}{"Hello world!"} have the explicit type signature \mintinline{rust}{&'static str}.

\section{Fearless Concurrency}

The Rust language has chosen message passing as its primary programming model for concurrency \cite[ch.~16-02]{Klabnik2019}. The Rust compiler enforces the borrowing rules across threads using the following automatically derived traits \cite[ch.~8.2]{Rustonomicon}:
\begin{enumerate}
    \item \texttt{\textcolor{yellow}{\textbf{Send}}} is required for a value of type \mintinline{rust}{T} to be sent to another thread. \mintinline{rust}{T} can only be \mintinline{rust}{Send} iff all of its components implement \mintinline{rust}{Send}. While almost all Rust types do, the non-atomic reference counter \mintinline{rust}{Rc<T>}, for instance, implements \mintinline{rust}{!Send}.
    \item \texttt{\textcolor{yellow}{\textbf{Sync}}} is required for an immutable reference of type \mintinline{rust}{&T} to be shared between multiple threads. \mintinline{rust}{T} can only be \mintinline{rust}{Sync} iff \mintinline{rust}{&T} implements \mintinline{rust}{Send}. While most Rust types are \mintinline{rust}{Sync}, some are not. For instance, \mintinline{rust}{UnsafeCell<T>}, which is Rust's helper type to implement interior mutability, is \mintinline{rust}{Send + !Sync}, while \mintinline{rust}{Rc<T>} is \mintinline{rust}{!Send + !Sync}.
\end{enumerate}
Rust labels its model as ``Fearless Concurrency'' \cite[ch.~16]{Klabnik2019} precisely because of the safety guarantees that message passing using the \texttt{\textcolor{yellow}{\textbf{Send}}} and \texttt{\textcolor{yellow}{\textbf{Sync}}} traits provide in safe Rust code.

\chapter{CUDA Resources and Memory Hierarchy} \label{appendix:cuda-recources}

All threads inside a thread block share a fixed allocation of resources, including the register file, L1 cache and shared memory \cite[p.~106]{CUDACpp}. Thus, if a CUDA kernel requires a vast number of registers or shared memory, the entire thread block might exceed the available resources. It then has to be split up into multiple sequential executions, which decreases the GPU occupancy, i.e. the percentage of utilised compute units.

\begin{figure}[h]
    \centering
    \includegraphics[width=0.8\textwidth]{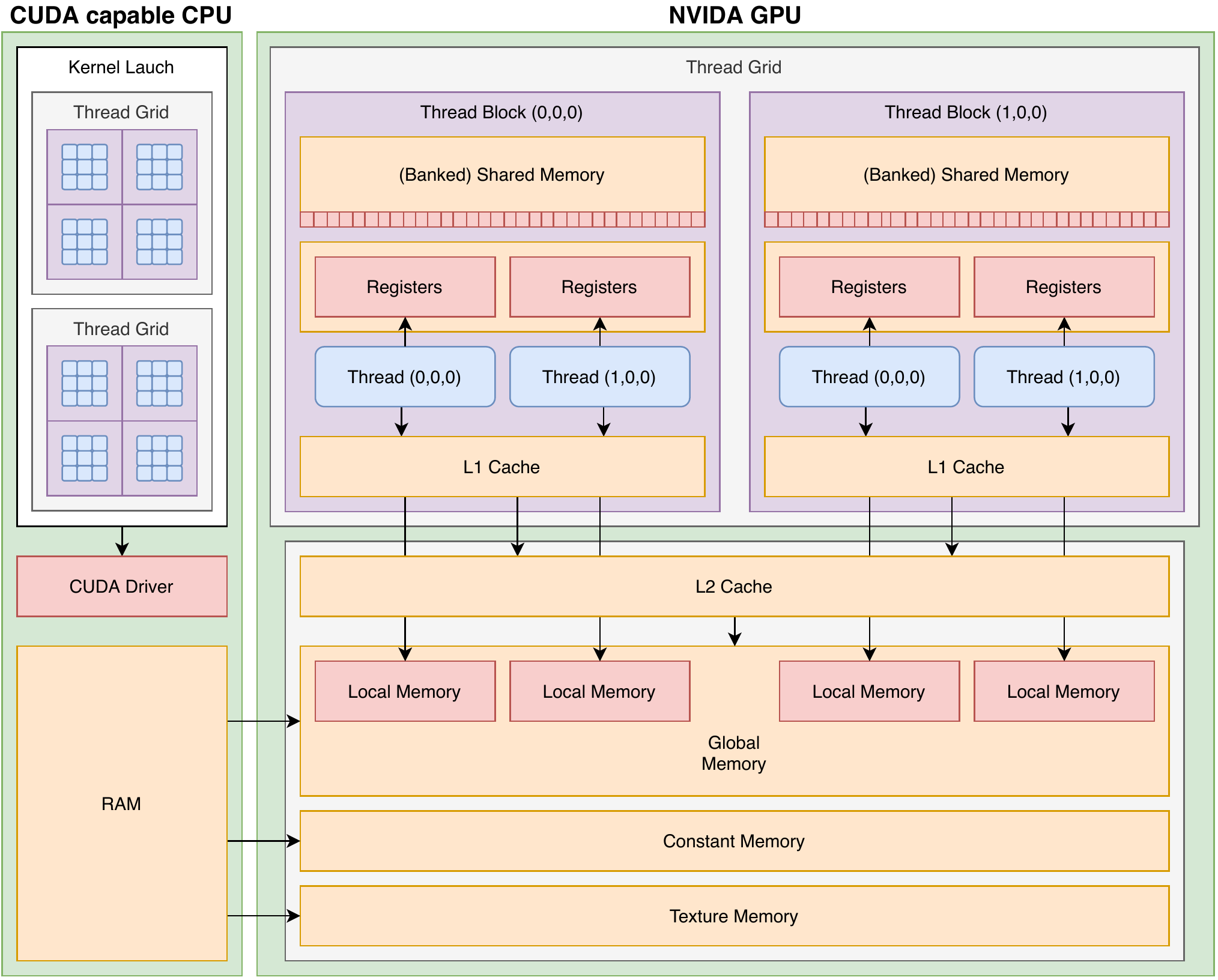}
    \caption{Overview of CUDA's logical architecture, memory hierarchy and kernel launch dimensions. Compute tasks are divided into a grid of blocks of threads. Within each block, all threads share the same register file, shared memory and L1 cache.}
    \label{fig:cuda-overview}
\end{figure}

\noindent {As \Cref{fig:cuda-overview} shows, CUDA contains a more extensive and specialised memory hierarchy than CPUs, though both have registers and on-chip L1 caches \cite[p.~2,106,311,320,324]{CUDACpp}. There is also a small amount of on-chip shared memory that all threads in a block can access. It uses a banked design that is highly optimised for the case where all 32 threads in a warp access or\parfillskip=0pt\par}

\noindent modify adjacent memory locations\footnote{The \texttt{cudaFuncSetSharedMemConfig} API function can be used to set the bank size to four or eight bytes iff the GPU does not use a fixed bank size \cite[pp.~105-106]{CUDARuntimeAPI}.}. Shared memory performance is optimal when the requested addresses do not conflict in their bank index as conflicting accesses have to be serialised. If all threads read the same value, only one thread reads and then broadcasts it \cite[pp.~115-116]{CUDACpp}.

The off-chip L2 cache, which is shared between all thread blocks, comes next in the memory hierarchy \cite[p.~2,311,315,319]{CUDACpp}. It caches both local and global memory accesses. While local memory is private to each thread, it is also located off-chip despite its misleading name. It is often used to spill registers or place stack-allocated arrays that need to be indexed dynamically \cite[p.~115]{CUDACpp}. Global memory uses the same physical backing storage but is globally accessible to all threads and the CPU. Global memory access is optimised for coalesced access on the GPU, i.e. when all threads of a warp request a continuous range of memory at once. Otherwise, the access is split up as necessary, which in the worst case means that every thread requires its own memory transaction, thereby reducing the instruction throughput \cite[pp.~113-114]{CUDACpp}.

Finally, there are constant memory, as well as texture and surface memory spaces on the GPU. Constant memory is read-only while a kernel is running. If all threads access the same address in constant memory, the access time can be amortised to perform almost as fast as register access. However, any other access patterns have to be executed sequentially \cite[p.~116]{CUDACpp}. Texture and surface memory are laid out to store 1D/2D/3D images that are accessed with spatial locality \cite[p.~116]{CUDACpp}. Furthermore, specialised texture memory access routines offer additional functionality such as floating-point coordinate addressing, wrapping, and interpolation \cite[p.~10]{CUDACpp}.

\chapter{Simulation Algorithm Pseudo-Code} \label{simulation-algorithms}

\section{The Independent Water-Level Algorithm} \label{water-level-algorithm}

\begin{minted}[linenos]{python}
def simulate_independent_waterlevel(event_buffer_size, landscape, rng, reporter):
    # Slow individuals are below the water level, fast ones are above it
    slow_individuals = landscape.generate_current_population()
    fast_individuals = []

    event_buffer = []

    water_level = 0.0

    while len(slow_individuals) > 0:
        # Increase the water level s.t. on average event_buffer_size events
        #  are produced between each water level rise
        water_level -= event_buffer_size / landscape.total_event_rate(individuals)

        while len(slow_individuals) > 0:
            # Simulate the individual until it has either finished, or
            # its next event would be above the water level
            simulated_individual = independent_algorithm(
                slow_individuals.pop(), rng, landscape,
                reporter=event_buffer.append,
                early_stop=(-next_event >= -water_level),
            )

            if not simulated_individual.has_speciated():
                fast_individuals.append(simulated_individual)

        # Sort and report all events below the water level
        for event in sorted(event_buffer.drain()):
            reporter.report(event)

        slow_individuals = fast_individuals
        fast_individuals = []
\end{minted}

\newpage

\section{The ``Next-Reaction'' Gillespie Algorithm} \label{gillespie-algorithm}

\begin{minted}[linenos]{python}
def initialise_gillespie(simulation, rng):
    event_queue = PriorityQueue()

    # Submit the initial event times at all locations
    for location, deme in (
        simulation.landscape.generate_current_population_locations()
    ):
        event_queue.push(
            -rng.exp(simulation.landscape.lambda(location) * len(deme)),
            location, deme,
        )

    return event_queue

def simulate_gillespie(event_queue, landscape, rng, reporter):
    while len(event_queue) > 0:
        event_time, location, deme = event_queue.pop()

        # Select one individual at random in the deme of the next event
        individual = deme.remove(rng.randint(len(deme)))

        # Re-enqueue the remaining deme with its next event time
        if len(deme) > 0:
            event_queue.push(
                event_time - rng.exp(
                    landscape.lambda(location) * len(deme)
                ), location, deme,
            )

        # Sample and report the individual's next event
        if rng.sample_random_event(landscape.nu(location)):
            reporter.report(event_time, individual.speciate())
        else:
            individual.disperse(landscape, rng)

            parent = landscape.individual_at(individual.location, individuals)

            # Check for coalescence with another individual
            if parent is not None:
                reporter.report(time, individual.coalescence(parent))
            else:
                reporter.report(time, individual.move())

                # Get the next event deme at the dispersal target
                _, _, deme = event_queue.remove(individual.location)
                deme.push(individual)

                # Re-enqueue the enlarged deme with its next event time
                event_queue.push(
                    event_time - rng.exp(
                        landscape.lambda(location) * len(deme)
                    ), individual.location, deme,
                )
\end{minted}

\newpage

\section{The Classical Coalescence Algorithm} \label{classical-algorithm}

\begin{minted}[linenos]{python}
def initialise_classical(simulation):
    return simulation.landscape.generate_current_population()

def simulate_classical(individuals, time, landscape, rng, reporter):
    while len(individuals) > 0:
        # Sample the time of the next event (assuming a constant event rate)
        time -= rng.exp(landscape.lambda * len(individuals))

        # Sample the next individual uniformly (event rate is homogeneous)
        individual = individuals.remove(rng.randint(len(individuals)))

        # Sample and report the individual's next event
        if rng.sample_random_event(landscape.nu(individual.location)):
            reporter.report(time, individual.speciate())
        else:
            individual.disperse(landscape, rng)

            parent = landscape.individual_at(individual.location, individuals)

            # Check for coalescence with another individual
            if parent is not None:
                reporter.report(time, individual.coalescence(parent))
            else:
                reporter.report(time, individual.move())

                individuals.append(individual)
\end{minted}

\section{The Monolithic Lockstep Algorithm} \label{monolithic-lockstep-algorithm}

\begin{minted}[linenos]{python}
def simulate_monolithic_lockstep(simulation, parallelism, event_log):
    # Initialise the monolithic algorithm task list assigned to this partition
    tasks = initialise_monolithic(
        simulation.landscape, simulation.rng, parallelism.rank()
    )

    # Synchronise and loop while the simulation has not finished globally
    while not parallelism.all_done(len(tasks)):
        next_event = simulation.peek_next_event_time(tasks)

        # Only the partition with the next event simulates the next step
        if parallelism.vote_min(next_event):
            simulate_monolithic(
                tasks, simulation.landscape, simulation.rng,
                reporter=event_log,
                early_stop=(-last_event >= -next_event),
            )

        # Migrate individuals between partitions synchronously
        parallelism.perform_emigrations(simulation.emigrations())
        simulation.immigrate(parallelism.immigrations(), tasks)
\end{minted}

\newpage

\section{The Monolithic Optimistic Algorithm} \label{monolithic-optimistic-algorithm}

\begin{minted}[linenos]{python}
def simulate_monolithic_optimistic(
    delta_sync, simulation, parallelism, event_log
):
    # Initialise the monolithic algorithm task list assigned to this partition
    tasks = initialise_monolithic_tasks(
        simulation.landscape, simulation.rng, parallelism.rank()
    )

    safe_time = 0.0

    # Synchronise and loop while the simulation has not finished globally
    while not parallelism.all_done(len(tasks)):
        # Advance the globally synchronised safe time and create a backup
        safe_time -= delta_sync
        backup = tasks, simulation

        immigration_buffer = []

        while True:
            event_buffer = []

            # Simulate this partition independently until the next safe point
            simulate_monolithic(
                tasks, simulation.landscape, simulation.rng,
                reporter=event_buffer.append,
                early_stop=(-next_event >= -safe_time),
            )

            # Migrate individuals between partitions synchronously
            parallelism.perform_emigrations(simulation.emigrations())
            immigrations = parallelism.immigrations()

            # Simulation was successful if all immigrations were expected
            if parallelism.vote_and(immigrations == immigration_buffer):
                break

            # Rollback the simulation to the last safe backup
            tasks, simulation = backup.clone()

            # Integrate the known immigrations into the simulation
            immigration_buffer = immigrations
            simulation.preregister_immigrations(immigration_buffer)

        # Report the buffered events up until the successful new safe point
        for event in event_buffer.drain():
            event_log.report(event)
\end{minted}

\chapter{\texttt{rustcoalescence} Example Usages} \label{appendix:rustcoalescence-examples}

\section{Selecting the Simulation Scenario}

\subsection{The Non-Spatial Model}

\begin{minted}[escapeinside=@@]{rust}
@:\textcolor{blue}{~}\textdollar@ @\textbf{rustcoalescence}@ simulate @'@(
    speciation: 0.001, /* speciation probability */
    sample: 1.0, /* percentage of individuals that are simulated */
    seed: 42,

    algorithm: Classical(),

    scenario: NonSpatial(
        area: (100, 100), /*   in a non-spatial model, only   */
        deme: 10,         /* `area.0 * area.1 * deme` matters */
    ),

    reporters: [ @\textcolor{blue}{\textbf{Plugin}}@(
        library: "libnecsim_plugins_common.so",
        reporters: [ @\textcolor{blue}{\textbf{Biodiversity}}@() ],
    ) ],
)@'@
\end{minted}

\subsection{The Spatially Implicit Model with a Static Metacommunity}

\begin{minted}[escapeinside=@@]{rust}
@:\textcolor{blue}{~}\textdollar@ @\textbf{rustcoalescence}@ simulate @'@(
    speciation: 0.1, /* migration probability */
    sample: 1.0,
    seed: 42,

    ...

    reporters: [ @\textcolor{blue}{\textbf{Plugin}}@(
        library: "libnecsim_plugins_metacommunity.so",
        reporters: [ @\textcolor{blue}{\textbf{Metacommunity}}@(
            metacommunity: Finite(<METACOMMUNITY>),
            seed: 42,
        ) ],
    ) ],
)@'@

> @\textcolor{blue}{\textbf{INFO}}@: @There@ were <MIGRATIONS> migrations to <ANCESTORS> ancestors on a finite
>       metacommunity of size <METACOMMUNITY> during the simulation.
\end{minted}

\noindent The \texttt{rustcoalescence} tool does not implement a static metacommunity itself. However, the \texttt{\textcolor{blue}{\textbf{Metacommunity}}} analysis reporter plugin can be used on any scenario to measure the number of migrations. Then, we can simulate a non-spatial scenario with the size of the metacommunity. To get the final overall biodiversity, we only have to sample as many individuals in the metacommunity as there are immigrant ancestors.

\begin{minted}[escapeinside=@@]{rust}
@:\textcolor{blue}{~}\textdollar@ @\textbf{rustcoalescence}@ simulate @'@(
    speciation: 0.001, /* metacommunity speciation */
    sample: <ANCESTORS / METACOMMUNITY>,
    seed: 42,

    algorithm: Classical(),

    scenario: NonSpatial(
        area: (1, 1),
        deme: <METACOMMUNITY>,
    ),

    reporters: [ @\textcolor{blue}{\textbf{Plugin}}@(
        library: "libnecsim_plugins_common.so",
        reporters: [ @\textcolor{blue}{\textbf{Biodiversity}}@() ],
    ) ],
)@'@

> @\textcolor{blue}{\textbf{INFO}}@: @The@ simulation resulted @in@ a biodiversity of <BIODIVERSITY> unique
>       species.
\end{minted}

\subsection{The Spatially Implicit Model with a Dynamic Metacommunity}

\begin{minted}[escapeinside=@@]{rust}
@:\textcolor{blue}{~}\textdollar@ @\textbf{rustcoalescence}@ simulate @'@(
    speciation: 0.001, /* metacommunity speciation */
    sample: 1.0,
    seed: 42,

    algorithm: Classical(),

    scenario: SpatiallyImplicit(
        local_area: (100, 100),
        local_deme: 10,
        meta_area: (<METACOMMUNITY>, 1),
        meta_deme: 1,
        migration: 0.1, /* migration probability */
    ),

    reporters: [ @\textcolor{blue}{\textbf{Plugin}}@(
        library: "libnecsim_plugins_common.so",
        reporters: [ @\textcolor{blue}{\textbf{Biodiversity}}@() ],
    ) ],
)@'@

> @\textcolor{blue}{\textbf{INFO}}@: @The@ simulation resulted @in@ a biodiversity of <BIODIVERSITY> unique
>       species.
\end{minted}

\subsection{The Spatially Explicit Model}

\begin{minted}[escapeinside=@@]{rust}
@:\textcolor{blue}{~}\textdollar@ @\textbf{rustcoalescence}@ simulate @'@(
    ...
    scenario: AlmostInfinite(
        radius: 564,
        sigma: 10.0,
    ),
    ...
)@'@
\end{minted}

\subsection{Spatially Explicit Scenario with Maps}

\begin{minted}[escapeinside=@@]{rust}
@:\textcolor{blue}{~}\textdollar@ @\textbf{rustcoalescence}@ simulate @'@(
    ...
    scenario: SpatiallyExplicit(
        habitat: "maps/madingley/fg0size8/habitat.tif",
        dispersal: "maps/madingley/fg0size8/dispersal.tif",
    ),
    ...
)@'@
\end{minted}

\section{Selecting the Algorithm}

\subsection{The Monolithic Algorithms}

\begin{minted}[escapeinside=@@]{rust}
@:\textcolor{blue}{~}\textdollar@ @\textbf{rustcoalescence}@ simulate @'@(
    ...
    algorithm: Classical() / @\textcolor{blue}{\textbf{Gillespie}}@() / @\textcolor{blue}{\textbf{SkippingGillespie}}@(),
    ...
)@'@
\end{minted}

\subsection{The Independent Algorithm on the CPU}

\begin{minted}[escapeinside=@@]{rust}
@:\textcolor{blue}{~}\textdollar@ @\textbf{rustcoalescence}@ simulate @'@(
    ...
    algorithm: Independent(
        delta_t: 2.0,
        step_slice: 10,
        dedup_cache: Relative(factor: 1.0),
        parallelism_mode: Monolithic(event_slice: 1000000),
    ),
    ...
)@'@
\end{minted}

\subsection{The Independent Algorithm on a CUDA GPU}

\begin{minted}[escapeinside=@@]{rust}
@:\textcolor{blue}{~}\textdollar@ @\textbf{rustcoalescence}@ simulate @'@(
    ...
    algorithm: CUDA(
        device: 0,
        ptx_jit: @\textcolor{red}{\textbf{true}}@,
        block_size: 64,
        grid_size: 64,
        delta_t: 2.0,
        step_slice: 10,
        dedup_cache: Relative(factor: 1.0),
        parallelism_mode: Monolithic(event_slice: 1000000),
    ),
    ...
)@'@
\end{minted}

\section{Selecting the Parallelisation Strategy}

\subsection{The Monolithic Algorithms, e.g. using MPI}

\subsubsection{The Lockstep / OptimisticLockstep Strategies}

\begin{minted}[escapeinside=@@]{rust}
@:\textcolor{blue}{~}\textdollar@ @\textbf{rustcoalescence}@ simulate @'@(
    ...
    algorithm: SkippingGillespie(
        parallelism_mode: Lockstep / @\textcolor{blue}{\textbf{OptimisticLockstep}}@,
    ),

    log: "event-log",
    ...
)@'@
\end{minted}

\subsubsection{The Optimistic / Averaging Strategies}

\begin{minted}[escapeinside=@@]{rust}
@:\textcolor{blue}{~}\textdollar@ @\textbf{rustcoalescence}@ simulate @'@(
    ...
    algorithm: SkippingGillespie(
        parallelism_mode: Optimistic(delta_sync: 20.0) /
                          @\textcolor{blue}{\textbf{Averaging}}@(delta_sync: 1.0),
    ),

    log: "event-log",
    ...
)@'@
\end{minted}

\subsection{The Independent Algorithm on the CPU}

\subsubsection{Parallelism with Synchronisation, e.g. using MPI}

\begin{minted}[escapeinside=@@]{rust}
@:\textcolor{blue}{~}\textdollar@ @\textbf{rustcoalescence}@ simulate @'@(
    ...
    algorithm: Independent(
        ...,
        parallelism_mode: Probabilistic(communication: 0.25),
    ),

    log: "event-log",
    ...
)@'@
\end{minted}

\subsubsection{Parallelism with Isolated Batches}

\begin{minted}[escapeinside=@@]{rust}
@:\textcolor{blue}{~}\textdollar@ @\textbf{rustcoalescence}@ simulate @'@(
    ...
    algorithm: Independent(
        ...,
        parallelism_mode: IsolatedLandscape(
            event_slice: 1000000,
            partition: Partition(rank: 2, partitions: 8),
        ),
    ),

    log: "event-log",
    ...
)@'@
\end{minted}

\subsection{The Independent Algorithm on a CUDA GPU}

\subsubsection{Parallelism with Isolated Batches}

\begin{minted}[escapeinside=@@]{rust}
@:\textcolor{blue}{~}\textdollar@ @\textbf{rustcoalescence}@ simulate @'@(
    ...
    algorithm: CUDA(
        ...,
        parallelism_mode: IsolatedLandscape(
            event_slice: 1000000,
            partition: Partition(rank: 2, partitions: 8),
        ),
    ),

    log: "event-log",
    ...
)@'@
\end{minted}

\section{Replaying a recorded Event Log}

\subsection{Replaying the Full Event Log}

\begin{minted}[escapeinside=@@]{rust}
@:\textcolor{blue}{~}\textdollar@ @\textbf{rustcoalescence}@ replay @'@(
    logs: [ "event-log/*/*" ],

    reporters: [ @\textcolor{blue}{\textbf{Plugin}}@(
        library: "libnecsim_plugins_common.so",
        reporters: [ @\textcolor{blue}{\textbf{Biodiversity}}@() ],
    ) ],
)@'@
\end{minted}

\subsection{Replaying a Partial Event Log}

\begin{minted}[escapeinside=@@]{rust}
@:\textcolor{blue}{~}\textdollar@ @\textbf{rustcoalescence}@ replay @'@(
    logs: [ "event-log/0/*", "event-log/2/6", "event-log/82/*" ],

    reporters: [ ... ],
)@'@
\end{minted}

\end{document}